\documentclass[%
 amsmath,amssymb,
 aps,
 nofootinbib,
]{revtex4}
\usepackage{comment}
\usepackage{xcolor}
\usepackage{dsfont}
\usepackage{graphicx}
\usepackage{dcolumn}
\usepackage{bm}
\usepackage{hyperref}

\def\slash#1{\setbox0=\hbox{$#1$}  
   \dimen0=\wd0     
   \setbox1=\hbox{/} \dimen1=\wd1  
   \ifdim\dimen0>\dimen1   
      \rlap{\hbox to \dimen0{\hfil/\hfil}} 
      #1     
   \else     
      \rlap{\hbox to \dimen1{\hfil$#1$\hfil}} 
      /      
   \fi}      %

\newcommand{\sumint}{\sum_{X}\hspace{-0.5cm}\int}

\usepackage{amsmath,bm}
\newcommand{\dd}{\mathrm{d}}
\newcommand{\e}{\mathrm{e}}
\newcommand{\bP}{\bm{P}}
\newcommand{\be}{\begin{equation}}
\newcommand{\ee}{\end{equation}}
\newcommand{\bea}{\begin{eqnarray}}
\newcommand{\eea}{\end{eqnarray}}
\newcommand{\nn}{\nonumber}

\begin{document}

\preprint{arXiv:XXXXX.XXXXX [hep-ph]}

\title{The Transverse Nucleon Single-Spin Asymmetry for Single-Inclusive Hadron and Jet Production at the EIC at NLO Accuracy}
\title{Transverse Nucleon Single-Spin Asymmetry for Single-Inclusive Hadron and Jet Production at NLO Accuracy}

\author{Daniel Rein}
\email{da.rein@student.uni-tuebingen.de}
\author{Marc Schlegel}
 \email{marc.schlegel@uni-tuebingen.de}
 \author{Patrick Tollk\"uhn}
 \email{patrick.tollkuehn@uni-tuebingen.de}
 \author{Werner Vogelsang}
 \email{werner.vogelsang@uni-tuebingen.de}
\affiliation{
 Institute for Theoretical Physics, University of T\"ubingen, Auf der Morgenstelle 14, D-72076 T\"ubingen, Germany
}

\date{\today}

\begin{abstract}
We investigate the single-spin asymmetry for the single-inclusive production of hadrons and jets in collisions of transversely polarized nucleons and unpolarized leptons, $\ell N^\uparrow \to (h\,\mathrm{or\,jet})X$. We compute the spin-dependent cross section within collinear twist-3 factorization in perturbative QCD at next-to-leading order (NLO) accuracy. In this approach, multiparton correlations generate a non-vanishing effect. For the present paper, we focus on correlations in the nucleon initial-state rather than in the fragmentation process. We explicitly verify that collinear twist-3 factorization is valid at the one-loop level. Our analytical results show that at NLO the relevant multiparton correlation functions in the nucleon are probed on their full support in momentum fractions. Our numerical analysis for collisions at the Electron-Ion Collider indicates that the NLO corrections can be large and are sensitive to the functional form of the twist-3 correlation functions.
\end{abstract}

\maketitle


\section{Introduction\label{sec:intro}}
The understanding of spin-related observables in highly energetic particle collisions involving hadrons remains a crucial pursuit in the research area of Quantum Chromodynamics (QCD). It is well-known in general that the study of so-called \emph{single-spin asymmetries} (SSA) offers insight into the partonic spin structure of hadrons, particularly the nucleon, as well as into the underlying QCD mechanisms that generate the SSA. Among the various emerging spin observables, the SSAs of transversely polarized nucleons for single-inclusive events in high-energy collisions are particularly interesting. A great amount of experimental data for such SSAs has been gathered in proton-proton collisions since the 1970s at Argonne National Lab, FermiLab and the Relativistic Heavy-Ion Collider (RHIC) for all sorts of single-inclusive final states, 
such as hadrons
\cite{Bunce:1976yb,Adams:1991rw,Krueger:1998hz,Allgower:2002qi,Adams:2003fx,Adler:2005in,Lee:2007zzh,Abelev:2008af,Arsene:2008mi,Adamczyk:2012xd,Adare:2013ekj,Adare:2014qzo,STAR:2020nnl}, jets \cite{Adamczyk:2012qj,Bland:2013pkt,STAR:2020nnl} and photons \cite{PHENIX:2021irw}. Some of the transverse spin effects that were measured in these experiments turn out to be 
large, reaching values of about $10\% - 20\%$ even at higher center-of-mass (c.m.) energies. This alone is reason enough to get to the ground of the QCD mechanisms behind
these large effects.

On the theoretical side, one can analyze the transverse nucleon spin asymmetries in single-inclusive high-energy processes in the framework of perturbative QCD using the so-called \emph{collinear twist-3} factorization approach. This method may be viewed as an extension of the widely used collinear factorization approach that is typically applied to cross sections of high-energy processes involving unpolarized particles. In order to describe such power-suppressed observables $\--$ often synonymously called \emph{sub-leading twist} or just \emph{twist-3} observables $\--$  in the framework of collinear twist-3 factorization one has to work with different types of hadronic QCD matrix elements that are more complex than the commonly used quark/gluon parton distribution functions (PDF) or fragmentation functions (FF) entering the usual factorization formulas for unpolarized cross sections. The matrix elements that generate the nonzero transverse spin effects in the collinear twist-3 formalism can be considered as two-parton and three-parton correlation functions. They may appear in the initial state as correlation functions of partons in the nucleon, or emerge in the final state (with detected hadrons) as parton correlations in the ``parton-to-hadron'' fragmentation process. 

The collinear twist-3 approach has been applied to the aforementioned transverse SSAs in polarized proton collisions with single-inclusive final states in Refs.~\cite{Efremov:1981sh,Efremov:1983eb,Efremov:1984ip,Qiu:1991pp,Qiu:1991wg,Qiu:1998ia,Eguchi:2006qz,Kouvaris:2006zy,Eguchi:2006mc,Koike:2009ge,Kang:2011hk,Metz:2012ct,Beppu:2013uda}. However, due to the many contributions that appear at twist-3 for these ``purely-QCD'' observables, the theoretical calculations are quite complex. As a result, most of the works \cite{Efremov:1981sh,Efremov:1983eb,Efremov:1984ip,Qiu:1991pp,Qiu:1991wg,Qiu:1998ia,Eguchi:2006qz,Kouvaris:2006zy,Eguchi:2006mc,Koike:2009ge,Kang:2011hk,Metz:2012ct,Beppu:2013uda} deal with the SSA at leading-order (LO) accuracy in perturbative QCD only.

The situation is somewhat simpler for single-inclusive final states in polarized {\it lepton}-nucleon collisions, where the interaction between the lepton and the nucleon is mediated by exchange of an electroweak gauge boson, typically by a virtual photon. One may study the same single-inclusive final states as for proton collisions, i.e., single-inclusive leptons, hadrons, jets, or photons. 

The single-inclusive production of leptons, $\ell N\to \ell X$, the \emph{deep-inelastic scattering} (DIS) process, is one of the benchmark reactions in hadronic physics. Its analysis has led to the development of the parton model of the nucleon, and it is $\--$ to the present day $\--$ substantial for our understanding of the nucleon properties in terms of partonic degrees of freedom. As far as the observable of interest for this paper, that is, the transverse nucleon single-spin asymmetry, is concerned, experimental measurements in DIS have been performed by the HERMES experiment \cite{HERMES:2009hsi} as well as at Jefferson Lab (JLAB) \cite{Katich:2013atq}. On the theoretical side, pQCD investigations of the transverse nucleon SSA within the collinear twist-3 factorization approach have been reported in Refs.~\cite{Metz:2006pe,Afanasev:2007ii,Metz:2012ui,Schlegel:2012ve}. One peculiarity of this particular observable is that a non-vanishing SSA appears in QED perturbation theory only at a higher perturbative order $\alpha_{\mathrm{em}}^3$ in the QED fine structure constant $\alpha_{\mathrm{em}}\simeq 1/137$ $\--$ compared to the unpolarized cross section at order $\alpha_{\mathrm{em}}^2$. As a consequence, the transverse nucleon SSA is expected to be suppressed by $\alpha_{\mathrm{em}}$, along with the usual subleading twist suppression factor $M/Q$ (where $M$ is the nucleon mass), and therefore is expected to be small. Indeed, the HERMES data \cite{HERMES:2009hsi} for the SSA for scattering off a polarized proton are consistent with zero, and even though the JLab data \cite{Katich:2013atq} (obtained with a $6\,\mathrm{GeV}$ electron beam) indicate a first non-zero SSA for a polarized $\mathrm{He}^3$ target, a measurement that shows a non-zero SSA on a polarized proton target is yet to be performed.

A process related to DIS is the single-inclusive production of hadrons $\--$ typically pions $\--$ in lepton-nucleon collisions $\ell N\to h X$. In contrast to DIS, the final-state lepton remains \emph{undetected} and so the virtuality of the exchanged photon cannot be reconstructed. Instead, a hadron is detected in the final state with a transverse momentum large enough ($\gtrsim 1\,\mathrm{GeV}$) that perturbative QCD can be applied. One may consider this process as the simplified version of single-inclusive hadron production in $pp$ collisions discussed above, with the unpolarized proton replaced by a lepton. Theoretically, the unpolarized cross section of the process $\ell N\to h X$ has been calculated in pQCD to NLO accuracy \cite{Hinderer:2015hra}. It was found 
that the NLO corrections can become quite sizable for various experimental setups.

Experimentally, the transverse nucleon SSA has been measured for single-inclusive pion and kaon production from a polarized proton target by the HERMES Collaboration \cite{Airapetian:2013bim}, and at Jefferson Lab \cite{Allada:2013nsw} for single-inclusive hadrons on a polarized $\mathrm{He}^3$ target. Theoretical calculations of various transverse-spin observables in hadron or jet production within the collinear twist-3 formalism have been presented in~\cite{Kang:2011jw,Gamberg:2014eia,Kanazawa:2014tda,Kanazawa:2015jxa,Kanazawa:2015ajw} at LO accuracy. Interestingly, the authors of Ref.~\cite{Gamberg:2014eia} confronted the LO result for the transverse nucleon SSA obtained in the collinear twist-3 formalism with the HERMES and JLab data \cite{Airapetian:2013bim,Allada:2013nsw} and observed a discrepancy of approximately a factor of two between the LO prediction and the experimental data. Just recently, it was argued in~\cite{Fitzgibbons:2024zsh} that adding the NLO corrections of~\cite{Hinderer:2015hra} to just the \emph{denominator} of the SSA (partially) reconciles this discrepancy. Based on this argument, the conjecture was made in Ref.~\cite{Fitzgibbons:2024zsh} that the NLO corrections to the LO prediction of the \emph{numerator} of the SSA are likely small, or affected by cancellations among the various contributions.
In this paper, we partially fill this knowledge gap and calculate the NLO corrections to the transverse nucleon spin-dependent cross section, i.e. the numerator of the SSA. One of the motivations for us to perform the NLO calculation presented in this paper is to verify or falsify the conjecture raised in Ref.~\cite{Fitzgibbons:2024zsh}.

In this paper, we focus on the pQCD calculation of NLO corrections to multiparton correlations within the nucleon only. For single-inclusive jet production, $\ell N^\uparrow\to \mathrm{jet}\,X$, the multiparton correlations within the nucleon are the only sources for generating a non-vanishing transverse nucleon SSA. Thus, the formulas for the SSA in single-inclusive jet production, presented in this paper, constitute the complete NLO result in perturbative QCD. They may be viewed as the first NLO result ever obtained in the collinear twist-3 formalism for a \emph{truly} single-inclusive observable. We also explicitly show that all divergences that appear at intermediate stages of the calculation eventually cancel. This cancellation implies that collinear factorization holds at the NLO level for the transverse nucleon SSA in single-inclusive jet production. In fact, it is the first time that collinear twist-3 factorization is verified at the one-loop level for a truly single-inclusive observable.

That said, we do need to be precise with the last statements made in the previous sentences, particularly with what we mean by the term \emph{truly single-inclusive}. Indeed, NLO calculations for the transverse SSA within the collinear twist-3 formalism have been performed in the past and have been reported in the literature for Drell-Yan lepton pair production in polarized proton collisions, see Ref.~\cite{Vogelsang:2009pj,Chen:2016dnp}, in semi-inclusive DIS, see Refs.~\cite{Kang:2012ns,Dai:2014ala,Yoshida:2016tfh,Chen:2017lvx,Benic:2019zvg}, and in single-inclusive polarized hyperon production in electron-positron annihilation, see Ref.~\cite{Gamberg:2018fwy}. All of these NLO calculations are undoubtedly important for the theoretical understanding of the collinear twist-3 factorization approach. However, there is an important difference between these processes compared to the single-inclusive hadron or jet production considered in this paper. In all of the processes above it is always possible (theoretically) to separate off the leptonic part from the hadronic part and describe the spin-dependent cross sections in terms of \emph{structure functions} \cite{Bacchetta:2006tn,Arnold:2008kf,Pitonyak:2013dsu}. Technically, this leads to a great simplification of the kinematics, in particular the complexity of the momentum flow in Feynman diagrams and hence the complexity of phase space integrations is much reduced. In this sense we do not count the processes mentioned above as \emph{truly single-inclusive}.

At this point, we emphasize that a final conclusion on whether NLO corrections can reconcile the factor of two discrepancy observed in Ref.~\cite{Gamberg:2014eia} has to be postponed even if the NLO results of this paper are included in a numerical analysis. The reason is that for single-inclusive hadron production our NLO results $\--$ in contrast to those for single-inclusive jet production $\--$ are not yet complete. Throughout this paper, we disregard subleading-twist multiparton correlations in \emph{fragmentation} even at LO and pretend that they vanish. This assumption is, of course, a gross oversimplification. However, a NLO calculation for multiparton correlations in fragmentation will likely be as complex as the one presented in this paper for multiparton correlations in the initial state. Furthermore, it is known that the so-called \emph{pole contributions} do not play a role in multiparton fragmentation functions \cite{Meissner:2008yf,Gamberg:2018fwy} and therefore it is likely that the methods in this paper cannot be copied and applied one-to-one to multiparton fragmentation contributions. We therefore decided to leave this problem as a future project.

We also mention that the collinear twist-3 factorization approach advocated in this paper is not the only theoretical approach to describe the transverse nucleon SSA in single-inclusive hadron or jet production. Several articles in the literature have applied the so-called \emph{Generalized Parton Model} \cite{Anselmino:1994tv,Anselmino:1999pw,Anselmino:1999gd,DAlesio:2004eso,Anselmino:2013rya,DAlesio:2017nrd} to the transverse SSA, where one takes into account a non-vanishing transverse parton momentum $k_T$ at all times. Although this approach has enjoyed considerable phenomenological success, it is not clear from a theoretical point of view whether such a factorization approach holds beyond LO for single-inclusive processes.

As a final comment in this introduction, we mention our numerical results for the transverse nucleon SSA at NLO accuracy for both single-inclusive pion and jet production in polarized electron-proton collisions. As shown in this paper one needs to know the quark-gluon-quark correlation functions $F_{FT}$ and $G_{FT}$ \emph{on their full support} in momentum fractions, in order to provide a complete NLO prediction for the transverse nucleon SSA. Currently, such information is not available in the literature. Ideally, one would want to extract the unknown functions $F_{FT}$ and $G_{FT}$ from experimental data. Unfortunately, experimental data do not exist for the transverse nucleon SSA for single-inclusive jet production, and we cannot reliably apply our NLO result for single-inclusive hadron production for that purpose to the existing HERMES and JLab data of Refs.~\cite{Airapetian:2013bim,Allada:2013nsw} because the fragmentation contributions are not yet included. However, experimental data for both single-inclusive jet and hadron final states can be expected to become available once the future Electron-Ion Collider (EIC)~\cite{Accardi:2012qut,AbdulKhalek:2021gbh,AbdulKhalek:2022hcn,Burkert:2022hjz,Abir:2023fpo} starts its operations. For this reason, we provide numerical plots for single-inclusive jet and hadron final states using the expected kinematical setup of an EIC. We impose certain models or scenarios
for the quark-gluon-quark correlation functions and use them as input for our numerical NLO plots. The purpose of these plots is to illustrate that NLO corrections can significantly modify the transverse nucleon SSA depending on the scenario we choose for the correlation functions. In other words, the transverse spin observables turn out to be quite sensitive to the specific form of the correlation functions, and future EIC data will be able to rule out some of the possible scenarios. However, a complete picture of the quark-gluon-quark correlation functions will likely not emerge from a single analysis of one or two transverse spin observables in one process alone but from a global QCD analysis of a combination of experimental data for transverse spin effects gathered from other processes at the EIC (like DIS, SIDIS or (semi)-inclusive photon production \cite{Albaltan:2019cyc,Rein:2024fns}) as well as from polarized proton collisions at RHIC. There is still a long way to go until such a complete global QCD analysis will become available; however, the methods presented in this paper open the door in this direction.

The paper is organized as follows: In Section \ref{sec:ct3} we give an overview of the methods we used to calculate analytical NLO corrections and the conceptual obstacles we had to overcome. In Section \ref{sec:AnalyticResults} we present the analytical NLO formulas for the transverse nucleon spin-dependent cross section for single-inclusive hadron and jet production. In Section \ref{sec:Numerics} we present our numerical scenarios for the transverse nucleon SSA in single-inclusive pion and jet production at the EIC. We conclude our paper in Section \ref{sec:Conclusions}.

\section{Collinear Twist-3 Factorization\label{sec:ct3}}

In this section, we present details of the approach we used to calculate the transverse SSA in single-inclusive hadron production in the scattering of unpolarized leptons off transversely polarized nucleons, $\ell(l)+N^\uparrow(P)\to h(P_h)+X$. Let $E_h \frac{\dd \sigma}{\dd^3 \bP_h }(\bm{S}_T)$ (with $E_h$ the energy of the detected hadron) denote the polarized differential cross section with $\bm{S}_T$ being the transverse spin vector of the polarized nucleon. Then, the transverse SSA is defined as
\be
A_{UT}\equiv \frac{E_h \frac{\dd \sigma}{\dd^3 \bP_h }(\bm{S}_T)-E_h \frac{\dd \sigma}{\dd^3 \bP_h }(-\bm{S}_T)}{E_h \frac{\dd \sigma}{\dd^3 \bP_h }(\bm{S}_T)+E_h \frac{\dd \sigma}{\dd^3 \bP_h }(-\bm{S}_T)}\,,\label{eq:SSA}
\ee
where the subscripts denote the polarization of the initial-state particles, $U$ for unpolarized and $T$ for transversely polarized. Moreover, let us define some useful kinematical variables. Throughout this paper, we label the four-momenta of the lepton, the nucleon and the produced hadron as $l^\mu$, $P^\mu$ and $P_h^\mu$. Their masses shall be denoted as $m_\ell$, $M$ and $M_h$, respectively. In this work, we assume that the scattering process is happening on an energy scale much larger than the masses of the particles. Hence, we will neglect $m_\ell$, $M$ and $M_h$ wherever possible and treat the four-momenta as light-like vectors $\ell^2\simeq 0$, $P^2\simeq 0$, $P_h^2\simeq 0$. We also introduce an additional four-vector $S^\mu$ that describes the spin of the nucleon with the usual normalization $S^2=-1$ and $P\cdot S=0$.

We further define the hadronic Mandelstam variables $s=(l+P)^2\simeq 2 l\cdot P$, $t=(P-P_h)^2\simeq -2P\cdot P_h$ and $u=(l-P_h)^2\simeq -2l\cdot P_h$. These kinematical variables will be used in the following. Note that $s+t+u>0$ for the single-inclusive reaction we are considering here.

\subsection{The Transverse SSA at Leading Order (LO)\label{sub:LO}}

Transverse-spin observables in single-inclusive hard processes $\--$ such as the SSA \eqref{eq:SSA} in $\ell N^\uparrow\to hX$ $\--$ can be analyzed in perturbative QCD within the \emph{collinear twist-3} factorization approach~\cite{Qiu:1991pp,Qiu:1991wg,Qiu:1998ia}. In this approach (parts of) the soft physics are typically encoded in a multipartonic non-perturbative matrix element. Such matrix elements have already been discussed exhaustively in the literature, for example in~\cite{Kanazawa:2015ajw,Koike:2019zxc,Koike:2011nx}, and an overview of all field theoretical objects that are relevant to our calculation is given in Appendix \ref{app:Def}. 
In the main text, we only give the definition of the quark-gluon-quark correlator $\--$ arguably the most important of these multiparton correlators $\--$ since it will be referred to regularly throughout this work:
\bea
\Phi_{F}^{q,\rho}(x,x^\prime)&=& \int_{-\infty}^\infty \tfrac{\dd\lambda}{2\pi} \int_{-\infty}^\infty \tfrac{\dd \mu}{2\pi} \,\e^{i\lambda x^\prime}\e^{i\mu (x-x^\prime)}\,\langle P,S|\,\bar{q}(0)\,\,ig\,G^{n\rho}(\mu n)\,\,q(\lambda n)\,|P,S\rangle\nn\\
&=&\frac{1}{2}M\,i\epsilon^{Pn\rho S}\,\slash{P}\,F^{q}(x,x^\prime) - \frac{1}{2}M\,S_T^\rho\,\slash{P}\gamma_5\,G^q(x,x^\prime)+...\,.\label{eq:DefPhiFmain}
\eea
This correlator contains two quark fields $\bar{q},\,q$ as well as the gluonic field strength tensor $G^{n\rho}\equiv n_\mu G^{\mu\rho}$ and is parametrized in terms of the $qgq$-corelation functions $F^q,\,G^q$. These functions depend on two longitudinal momentum fractions $x,\,x^\prime$ and are accompanied by other twist-3 functions which, however, are simply indicated by $\dots$, since they do not enter the observable 
we are interested in here. 
The vector $n$ is light-like, $n^2=0$, and normalized by the condition $P\cdot n=1$ but otherwise arbitrary and thus unphysical. Moreover, the Wilson lines required for gauge invariance of the matrix element are not shown in favor of readability and we define $\epsilon^{Pn\rho S}\equiv P_\mu n_\nu  S_\sigma \epsilon^{\mu\nu\rho\sigma}$ where $\epsilon$ is the totally anti-symmetric tensor with sign convention $\epsilon^{0123}=+1$. Note that we have also introduced 

the transverse part $a_T^\mu$ of any four-vector $a$ via a transverse projector:
\be
a_T^\mu\equiv g_T^{\mu\nu}a_\nu\,,\, \quad\mathrm{where}\quad g_T^{\mu \nu}\equiv g^{\mu\nu}-P^\mu n^\nu-P^\nu n^\mu\,.\label{eq:TransProjectormain}
\ee

The general collinear twist-3 formalism, applied to single-inclusive lepton-nucleon collisions, has been described in great detail in Refs.~\cite{Metz:2006pe,Kang:2011jw,Metz:2012ui,Schlegel:2012ve,Kanazawa:2014tda,Kanazawa:2015jxa,Gamberg:2014eia,Kanazawa:2015ajw,Albaltan:2019cyc}. We refrain from repeating this discussion on the general formalism and focus only on the main features specific to our calculation. However, we would like to remind the reader of some terminology often used in the context of collinear twist-3 factorization: In this formalism, various types of seemingly different contributions add up to a transverse spin observable like the SSA \eqref{eq:SSA}. Contributions to a SSA generated by two-parton correlators such as \eqref{eq:DefPhid}, \eqref{eq:DefPhidGlu} are called \emph{kinematical} twist-3 contributions, while contributions generated by three-parton correlators like \eqref{eq:DefPhiFmain} are called \emph{dynamical} twist-3 contributions. A third type of contribution, the \emph{intrinsic} twist-3 contribution, is irrelevant for this paper.

We point out that the SSA \eqref{eq:SSA} has been calculated in the collinear twist-3 formalism to LO accuracy in Refs.~\cite{Gamberg:2014eia,Kanazawa:2015ajw}. Of particular importance is the discussion in Ref.~\cite{Kanazawa:2015ajw} on the role of the
nonphysical light-cone vector $n^\mu$ entering the definition of the matrix element in \eqref{eq:DefPhiFmain} (and of related vectors entering other
contributing twist-3 correlation functions).
$n$ is not unambiguously defined. On the other hand, physical observables like cross sections or the SSA must not depend on the particular choice made for the vector.
To establish the independence of $n$ of the full result becomes a rather complex task in the twist-3 calculation beyond LO, but also provides a good
check on the correctness of the results. In fact, in Ref.~\cite{Kanazawa:2015ajw} the LO result for $\ell N^\uparrow \to hX$ was computed 
for an arbitrary $n$. For simplicity, here we just fix
\be
n^\mu = -\tfrac{2}{t}P_h^\mu \label{eq:choiceLCvectormain}
\ee
at LO and recover the result of~\cite{Kanazawa:2015ajw} (see Eq.~(67) therein). Extended to $d=4-2\varepsilon$ space-time dimensions,
we find:
\bea
E_h\frac{\dd \sigma_{\mathrm{LO}}}{\dd^{d-1} \mathbf{P}_h}(S) & = & \sigma_0(S)\,\int_{v_0}^{v_1}\dd v\int_{x_0}^1\tfrac{\dd w}{w}\,
\hat{\sigma}_{\mathrm{LO}}(v,w,\varepsilon)\,\sum_q e_q^2\left[\left(F^q - x\,(F^q)^\prime\right)(x,x)\,z^{2\varepsilon}D_1^q(z) \right]\Bigg|^{z=\frac{1-v_1}{1-v}}_{x=\frac{x_0}{w}}\nn\\
& & + h_1^q \otimes \Im[\hat{H}^q_{FU}].\label{eq:LO}
\eea
In this formula \eqref{eq:LO} we have introduced a Lorentz-invariant nucleon spin-dependent prefactor $\sigma_0(S)$ that will appear in several formulas below,
\be
\sigma_0(S) = \left(\frac{2 \alpha_{\mathrm{em}}^2}{s(-u)}\right)\,\left(\frac{4\pi\,M\,\epsilon^{lPP_hS}}{s\,(-u)}\right)\,.\label{eq:prefactor}
\ee
Note that the explicit appearance of the nucleon mass $M$ in the second factor on the r.h.s. of \eqref{eq:prefactor} leads to a (twist-3) power suppression (in comparison with the unpolarized cross section) that is typical for transverse spin observables. This means that it is the factor $\sigma_0(S)$ that will ``dilute'' the SSA \eqref{eq:SSA} if the transverse momentum of the detected hadron is large. 

Furthermore, the integration boundaries in \eqref{eq:LO} are given by the Mandelstam variables $s$, $t$, $u$, with
\be
x_0(v) = \frac{1-v}{v} \frac{u}{t}, \quad v_0 = \frac{u}{t+u},\quad v_1 = \frac{s+t}{s}\,.\label{eq:boundaries}
\ee
In addition, we introduce the LO hard partonic function $\hat{\sigma}_{\mathrm{LO}}$ as a function of the integration variables $v$, $w$ in arbitrary dimensions: 
\be
\hat{\sigma}_{\mathrm{LO}}(v,w,\varepsilon) = \frac{1+v^2-\varepsilon (1-v)^2}{(1-v)^4}\,\delta(1-w)\,.\label{eq:LOpartonic}
\ee
We note that Eq.~\eqref{eq:LO} is identical to Eq.~(67) of Ref.~\cite{Kanazawa:2015ajw} if a transformation of the integration variables $z=\frac{1-v_1}{1-v}$, $x=\tfrac{x_0}{w}$ is performed, along with $\varepsilon = 0$. 

\begin{figure*}
\centering
\includegraphics[width=0.6\textwidth]{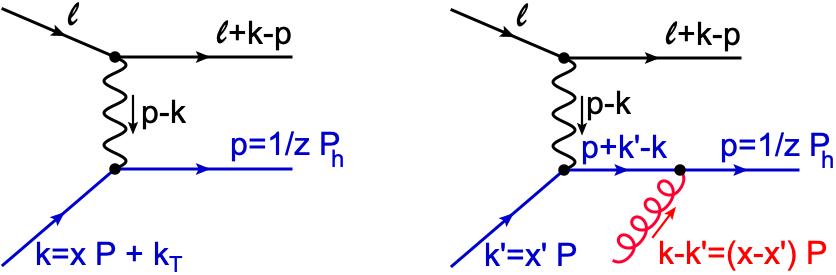}

\caption{LO Feynman diagrams that are used for the calculation of Eq.~\eqref{eq:LO}. The black line indicates the lepton line, the blue line represents a quark line. 
\textbf{Left:} \emph{Kinematical} twist-3 contribution, \textbf{Right:} \emph{Dynamical} twist-3 contribution.\label{fig:LO}}
\end{figure*}

The LO spin-dependent cross section is generated 
by two types of multiparton correlation functions:
\begin{itemize}
\item The quark-gluon-quark correlation function of a transversely polarized nucleon $F^q$, in conjunction with the ordinary twist-2 quark fragmentation function $D_1^q$ (definitions are given in ~\eqref{eq:DefPhiFmain} and ~\eqref{eq:DefFFq}). The function $F^q$ is probed in the LO formula \eqref{eq:LO} for a very specific kinematical configuration of the quark and the gluon: at $x^\prime=x$. Physically, the condition $x^\prime=x$ may be interpreted as a situation where an additional gluon attached to the hard part of the process does not carry any longitudinal momentum (see Ref.~\cite{Kanazawa:2015ajw}). This is why $F$ in this kinematical region $x^\prime=x$ 
is called a \emph{soft-gluon pole} matrix element (SGP). The SGP function $F^q(x,x)$ is also known in the literature as the \emph{Efremov-Teryaev-Qiu-Sterman} (ETQS) matrix element \cite{Efremov:1981sh,Efremov:1983eb,Efremov:1984ip,Qiu:1991pp,Qiu:1991wg}. Note that the ETQS-matrix element enters \eqref{eq:LO} not only by itself but in combination with a \emph{derivative term} $F^\prime(x,x)=\frac{\dd}{\dd x}F(x,x)$, that is, as a combination $(1-x\frac{\dd}{\dd x})F(x,x)$. 

As laid out in Ref.~\cite{Kanazawa:2015ajw}, two types of twist-3 effects generate the LO formula \eqref{eq:LO}: the \emph{kinematical} twist-3 contributions from quark-quark correlations through the Sivers function $f_{1T}^{\perp (1),q}$ (see \eqref{eq:DefPhid}), and the \emph{dynamical} twist-3 contributions from quark-gluon-quark correlations (see \eqref{eq:DefPhiFmain}). The LO diagrams for the corresponding Feynman amplitudes of the two effects are shown in Fig.~\ref{fig:LO}. 

As far as the kinematical contributions are concerned, the recipe to calculate the hard partonic factors can be briefly summarized as follows: compute the modulus squared of the Feynman amplitude for two-parton correlations (such as shown in the left LO diagram in Fig.~\ref{fig:LO}), but keep a non-zero transverse momentum $k_T$ for the incoming parton momenta. Perform all phase space and $w$ integrations such that all $\delta$-functions containing $k_T$ are integrated out. Then, perform a Taylor-expansion to first order in $k_T$, such that any dependence on $k_T$ is moved from the hard part to the \emph{first moment} $f_{1T}^{\perp (1) q}$. This approach has been used in Ref.~\cite{Kanazawa:2015ajw} at LO, but, as will be explained below, it also works at NLO.

Furthermore, \emph{dynamical} contributions to the LO formula \eqref{eq:LO} are derived from the right diagram in Fig.~\ref{fig:LO}. In general, a non-zero SSA requires an interference of the imaginary part of an amplitude with the real part of another amplitude. For the dynamical contributions an imaginary part is generated at LO by a quark propagator in Fig.~\ref{fig:LO} hitting its pole in the following way:
\be
\frac{1}{(p+k^\prime-k)^2+i\delta}=\frac{1}{(-t/z)(x^\prime - x+i\delta)}=\frac{-z}{t}\left(\frac{\mathcal{P}}{x^\prime-x}-i\pi \delta(x^\prime - x)\right)\,.\label{eq:LOSGP}
\ee
Cauchy's principal value $\mathcal{P}/(x^\prime-x)$ drops out for an SSA while the $\delta$-function precisely singles out the SGP function $F^q(x,x)$.

As explained in Ref.~\cite{Kanazawa:2015ajw}, both effects discussed above need to be combined using the identity
\be
f_{1T}^{\perp (1),q}(x) = +\pi F^q(x,x)\,.\label{eq:SiversSGP}
\ee
This relation between the first moment of the Sivers function $f_{1T}^{\perp (1),q}$ and the SGP function $F^q(x,x)$ is 
valid for lepton-nucleon scattering  and
was first derived in Ref.~\cite{Boer:2003cm}. In principle it serves as a bridge between the two types of mechanisms for transverse spin asymmetries in semi-inclusive deep-inelastic lepton-nucleon scattering (SIDIS) and single-inclusive processes such as the one we are considering here. Although the validity of the relation \eqref{eq:SiversSGP} has been questioned in Ref.~\cite{Rogers:2020tfs} we cannot stress enough the importance of this relation for the collinear twist-3 formalism. As shown in Ref.~\cite{Kanazawa:2015ajw}, it is only through \eqref{eq:SiversSGP} that the LO formula \eqref{eq:LO} is independent of the specific choice of the auxiliary light-cone vector $n^\mu$ that appears in the definitions \eqref{eq:DefPhiFmain}, \eqref{eq:DefPhid}. 
Additionally, to establish color gauge independence also in our NLO calculation, we introduce an arbitrary parameter in the metric tensor
in the gluonic polarization sums and the gluonic propagators:
\be
-g^{\mu\nu}\to -d^{\mu\nu}(k,n)\equiv -\left(g^{\mu \nu}-\kappa \,\mathcal{P}\frac{k^\mu n^\nu+k^\nu n^\mu}{k\cdot n}\right)\,,\label{eq:LCpolsummain}
\ee
where $k$ denotes the relevant four-momentum of the sum or propagator. The specific gauge is then given by the choice of the parameter $\kappa$, i.e. $\kappa=0$ for Feynman and $\kappa=1$ for the light-cone gauge. We find that NLO hard partonic factors for SGP contributions are independent of $\kappa$ only after the application of Eq.~\eqref{eq:SiversSGP},
again implying that \eqref{eq:SiversSGP} is essential for the collinear twist-3 approach.
\\
We point out that an alternative and equivalent method to calculate SGP contributions to single-spin asymmetries has been suggested in Ref.~\cite{Xing:2019ovj}, based on Ward-Takahashi identities. We also note that other approaches to a collinear twist-3 calculation of a transverse single-nucleon spin asymmetry within pQCD, e.g. Ref.~\cite{Kouvaris:2006zy}, do not strictly distinguish between \textit{kinematical} and \textit{dynamical} twist-3 effects. The reason is that the setup of the collinear twist-3 factorization in the approach of Ref.~\cite{Kouvaris:2006zy}, also \cite{Xing:2019ovj}, is directly established in Feynman gauge. In this gauge both \textit{kinematical} and \textit{dynamical} contributions are implicitly included in diagrams with a quark and a gluon in the initial state, like the right diagram in Fig.~\ref{fig:LO}, while quark-quark correlations $\--$ such as the left diagram in Fig.~\ref{fig:LO} $\--$ are not counted as extra contributions. We emphasize that eventually all approaches are equivalent.

\item In principle there is another important contribution to the LO formula \eqref{eq:LO} from twist-3 effects in the fragmentation part of the process. This is indicated in Eq.~\eqref{eq:LO} by the $h_1\otimes \Im[\hat{H}_{FU}]$ term which represents a collinear convolution of the leading twist \emph{transversity} quark distribution $h_1$ and the imaginary part of a quark-gluon-quark fragmentation function $\hat{H}_{FU}$ (see Ref.~\cite{Kanazawa:2015ajw}). Although this term may be a numerically large contribution to the SSA \eqref{eq:SSA}, we disregard this term in this paper and focus entirely on the twist-3 effects in the nucleon discussed above. The reason for this is that we can easily reduce and extract from these effects the SSA for single-inclusive jet production, $\ell N^\uparrow\to\mathrm{jet}\,X$. This is particularly simple at LO where, technically, we just need to replace the quark fragmentation function by a $\delta$ function,
\be
D_1^q(z)\to \delta(1-z)\,.\label{eq:ReplDjet}
\ee
All other fragmentation functions vanish for jet production, in particular the quark-gluon-quark fragmentation function $\Im[\hat{H}_{FU}]$. At NLO, the situation is somewhat more complicated because of soft-collinear parton radiation, and one needs to take into account the specific definition of the jet. We will discuss this procedure below in Sec.~\ref{sub:Jets}.

Effectively, in this paper, we assume that $\Im[\hat{H}_{FU}]=0$. However, we plan to investigate the NLO corrections to the $h_1\otimes \Im[\hat{H}_{FU}]$ term in Eq.~\eqref{eq:LO} in a dedicated study in the future.

\end{itemize}

\subsection{Renormalization of Multiparton Correlation Functions\label{sub:Renormalization}}

It is well known that unrenormalized \emph{bare} collinear correlation functions such as the ones given in Appendix \ref{app:Def} suffer from ultraviolet (UV) divergences that need to be subtracted. Such UV divergences emerge for matrix elements, like 
\eqref{eq:DefPhiFmain} for the quark-gluon-quark correlator, in a perturbative calculation. For example, at perturbative order $\mathcal{O}(\alpha_s)$ in the strong coupling constant, the UV divergence explicitly appears as a $1/\varepsilon$-pole in dimensional regularization. This pole may be removed by a renormalization procedure using the common $\overline{\mathrm{MS}}$-scheme.
While the method is well known for twist-2 matrix elements (the corresponding $\overline{\mathrm{MS}}$-renormalization equations for the fragmentation functions are given in Appendix \ref{app:fragmentationFunctions}), the situation is more complicated for the bare SGP function $F_{\mathrm{bare}}^q(x,x)$ in \eqref{eq:LO} as the corresponding splitting function is more involved. In Ref.~\cite{Braun:2009mi} the LO evolution equations for the functions $F$, $G$ have been studied, and one can readily read off the splitting function for the SGP function $F(x,x)$ from the evolution kernel. The kernel is given explicitly for the non-singlet case in Eq.~(43) of Ref.~\cite{Braun:2009mi}. For an individual quark flavor one also has to consider a contribution from the triple-gluon correlation functions $N(x,x^\prime),O(x,x^\prime)$ (for a definition see Eqs.~\eqref{eq:DefNF} and \eqref{eq:DefOF}), which is independent of the quark flavor and is thus canceled out in the non-singlet case. This specific contribution is given explicitly in Eq.~(107) of Ref.~\cite{Kang:2008ey}. We note that some terms of the full evolution equation are missing in Ref.~\cite{Kang:2008ey} as elaborated on in Ref.~\cite{Braun:2009mi}. With that being said, we arrive at the following formula for the renormalization of the bare SGP function of an individual quark flavor:
\bea
F_{\mathrm{bare}}^{q}(x,x,\mu) & = & F^{q,\overline{\mathrm{MS}}}(x,x,\mu) + \frac{\alpha_s(\mu)}{2\pi}\frac{S_\varepsilon}{\varepsilon}\times \nn\\
&& \int_x^1 \tfrac{\dd w}{w}\left[ P_{qq}(w)\,F^{q,\overline{\mathrm{MS}}}(\tfrac{x}{w},\tfrac{x}{w},\mu) \right.\nn\\
&&+\tfrac{N_c}{2}\left(\frac{1+w}{(1-w)_+}\,F^{q,\overline{\mathrm{MS}}}(\tfrac{x}{w},x,\mu)-\frac{1+w^2}{(1-w)_+}\,F^{q,\overline{\mathrm{MS}}}(\tfrac{x}{w},\tfrac{x}{w},\mu)\right)\nn\\
&& + \tfrac{N_c}{2}\,G^{q,\overline{\mathrm{MS}}}(\tfrac{x}{w},x,\mu)-N_c\,F^{q,\overline{\mathrm{MS}}}(x,x,\mu)\,\delta(1-w)\nn\\
&&\left. +\tfrac{1}{2N_c}\,\left((1-2w)\,F^{q,\overline{\mathrm{MS}}}(-\tfrac{1-w}{w}x,x,\mu)+G^{q,\overline{\mathrm{MS}}}(-\tfrac{1-w}{w}x,x,\mu)\right)\right.\nn\\
&&\left. + 2P_{qg}(w)\tfrac{w}{x}\left((N+O)\left(\tfrac{x}{w},\tfrac{x}{w},\mu\right)-(N-O)\left(\tfrac{x}{w},0,\mu\right)\right)\right]\,+\mathcal{O}(\alpha_s^2),\label{eq:LOsplittingfunctionTw3}
\eea
where we have introduced the renormalization/factorization scale $\mu$ and $S_\varepsilon=(4\pi)^\varepsilon/\Gamma(1-\varepsilon)$, a convenient prefactor consistent with the $\overline{\mathrm{MS}}$-scheme at NLO. $P_{qq}$ and $P_{qg}$ are the well-known LO $q\to q$ and $g\to q$ splitting functions (with $C_F=\frac{N_c^2-1}{2N_c}=\frac{4}{3}$ for $N_c=3$, and $T_R=\frac{1}{2}$),
\bea
P_{qq}(w) & = & C_F\left[\frac{1+w^2}{(1-w)_+}+\frac{3}{2}\delta(1-w)\right]\,,\nn\\
P_{qg}(w) & = & T_R\left[w^2+(1-w)^2\right]\,.
\eea
Equation~(\ref{eq:LOsplittingfunctionTw3})
indicates that the ETQS matrix element $F(x,x)$ not only mixes with itself upon renormalization, but also with $F$ in other regions of its support, e.g., $F(\tfrac{x}{w},x)$ $\--$ a so-called \emph{hard pole} contribution $\--$ as well as with the axial-vector type correlation function $G(\tfrac{x}{w},x)$ and the triple-gluon functions $(N,O)(\tfrac{x}{w},\tfrac{x}{w})$ and $(N,O)(\tfrac{x}{w},0)$. 

The leading order formula \eqref{eq:LO} does not only contain the bare ETQS matrix element $F(x,x)$. Instead, there is also a contribution from the derivative term, combined as $(1-x\tfrac{\dd}{\dd x})F(x,x)$. Hence, we need the following $\overline{\mathrm{MS}}$-subtraction:
\bea
(1-x\tfrac{\dd}{\dd x})F_{\mathrm{bare}}^{q}(x,x,\mu) & = & (1-x\tfrac{\dd}{\dd x})F^{q,\overline{\mathrm{MS}}}(x,x,\mu) + \frac{\alpha_s(\mu)}{2\pi}\frac{S_\varepsilon}{\varepsilon}\int_x^1 \tfrac{\dd w}{w}\left[ P_{qq}(w)\,\left(F^{q,\overline{\mathrm{MS}}}-\tfrac{x}{w}(F^{q,\overline{\mathrm{MS}}})^\prime\right)(\tfrac{x}{w},\tfrac{x}{w},\mu) \right.\nn\\
&&\hspace{-3cm} \left. +\tfrac{N_c}{2}\tfrac{1+w}{(1-w)_+}\,\left(F^{q,\overline{\mathrm{MS}}}-\tfrac{x}{w}\,(\partial_1F^{q,\overline{\mathrm{MS}}})-x\,(\partial_2 F^{q,\overline{\mathrm{MS}}})\right)(\tfrac{x}{w},x,\mu)-\tfrac{N_c}{2}\tfrac{1+w^2}{(1-w)_+}\,\left(F^{q,\overline{\mathrm{MS}}}-\tfrac{x}{w}(F^{q,\overline{\mathrm{MS}}})^\prime\right)(\tfrac{x}{w},\tfrac{x}{w},\mu) \right. \nn\\
&&\hspace{-3cm} \left. + \tfrac{N_c}{2}\,\left(G^{q,\overline{\mathrm{MS}}}-\tfrac{x}{w}\,(\partial_1G^{q,\overline{\mathrm{MS}}})-x\,(\partial_2 G^{q,\overline{\mathrm{MS}}})\right)(\tfrac{x}{w},x,\mu)-N_c\,\left(F^{q,\overline{\mathrm{MS}}}-x(F^{q,\overline{\mathrm{MS}}})^\prime\right)(x,x,\mu)\,\delta(1-w) \right. \nn\\
&&\hspace{-3cm} \left. +\tfrac{1}{2N_c}\left(F^{q,\overline{\mathrm{MS}}}(-\tfrac{1-w}{w}x,x,\mu) -\delta(1-w)\,F^{q,\overline{\mathrm{MS}}}(0,x,\mu)\,+\,G^{q,\overline{\mathrm{MS}}}(-\tfrac{1-w}{w}x,x,\mu) +\delta(1-w)\,G^{q,\overline{\mathrm{MS}}}(0,x,\mu)\right) \right. \nn\\
&&\hspace{-3cm} \left. -\tfrac{1}{2N_c}\left((1-2w)\,x\,[\partial_1F^{q,\overline{\mathrm{MS}}}+\partial_2F^{q,\overline{\mathrm{MS}}}](-\tfrac{1-w}{w}x,x,\mu) \,+\,x\,[\partial_1G^{q,\overline{\mathrm{MS}}}+\partial_2G^{q,\overline{\mathrm{MS}}}](-\tfrac{1-w}{w}x,x,\mu) \right)\right. \nn\\
&&\hspace{-3cm}\left.+\left(1-2w^2+\delta(1-w)\right)\,\tfrac{w}{x}\left((N^{\overline{\mathrm{MS}}}+O^{\overline{\mathrm{MS}}})\left(\tfrac{x}{w},\tfrac{x}{w};\mu\right)-(N^{\overline{\mathrm{MS}}}-O^{\overline{\mathrm{MS}}})\left(\tfrac{x}{w},0;\mu\right)\right)\right]\,+\mathcal{O}(\alpha_s^2),\label{eq:LOsplittingfunctionTw3deriv}
\eea
where we made use of a shorthand notation for the derivatives with respect to the first (second) argument of the correlation functions, $\left(\partial_1 (\partial_2)(F,G)\right)(x,x^\prime)$, which is defined in Eq.~\ref{eq:Derivatives}. Moreover, we applied integration by parts on the terms with a derivative of $N,O$.

\subsection{Setup of the NLO Calculation\label{sub:Setup}}

After all necessary definitions and discussions of the LO results in the previous sections, we turn to the actual calculation of this paper, that is, the NLO pQCD corrections to the LO formula \eqref{eq:LO} (with the assumption that $\Im[\hat{H}_{FU}]=0$). Before we discuss the results for the individual contributions and partonic channels in detail, we will describe our general strategy.

\subsubsection{Kinematical Twist-3 at NLO\label{sub:KinTw3}}

\begin{figure}[tbp]
\centering
\includegraphics[width=0.5\textwidth]{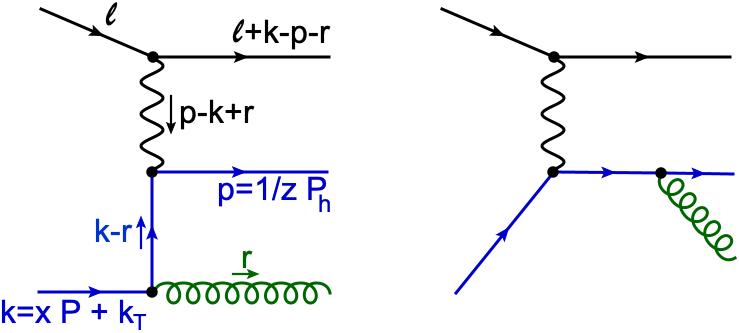}
\caption{NLO real-gluon emission diagrams in the $qg\to q$ channel relevant for the kinematical twist-3 contribution. One needs to compute the interference of the sum of these diagrams with itself.\label{fig:NLOkinqg2qreal}}
\end{figure}

Here, we would like to discuss some special features that we encounter in the calculation of kinematical twist-3 contributions at NLO. As an example of pQCD corrections, let us consider the real-gluon emission diagrams of Fig.~\ref{fig:NLOkinqg2qreal}. Squaring the sum of these diagrams leads to a partonic hard factor $\hat{\sigma}(k,p,l,l^\prime,r)$ that depends on the external momenta of the initial ($l$) and final ($l^\prime=l+k-p-r$) lepton and the quark ($k$, $p$) and the gluon (\textcolor{red}{$r$}). 

Now we perform a kinematical approximation to the initial/final quark momenta. Since the hadronization of the final quark into an observed hadron is assumed in this paper to proceed 
via the ordinary leading-twist fragmentation function $D_1^{h/q}(z)$, we may simply approximate $p^\mu \simeq 1/z\,P_h^\mu$ for the final quark momentum.

The situation is more complicated for the initial quark momentum $k^\mu$, since we are dealing with a subleading-twist kinematical effect. The usual kinematical approximation in the parton model would be $k^\mu \simeq x\,P^\mu$. Here, we also take into account a non-zero transverse momentum $k_T^\mu$ as indicated in Fig.~\ref{fig:LO}. It turns out to be favorable for an NLO calculation to enforce vanishing quark virtuality, i.e. to approximate in the kinematical twist-3 partonic hard factor,
\be
k^\mu \simeq x\,P^\mu - \tfrac{k_T^2}{2x}\,n^\mu + k_T^\mu\,.\label{eq:kinApproxktintw2}
\ee
It is not necessary to keep the term $\propto k_T^2$ since, after all loop and phase integrations have been carried out, one only expands up to linear order in $k_T$. However, keeping such a term ensures that $k^2=0$ from the very beginning, which simplifies several intermediate steps of the calculation. 

As a next step, one could perform the phase-space integration of the partonic factor $\hat{\sigma}(k,p,l,l^\prime,r)$ over the unobserved gluon momentum $r$ and the unobserved lepton momentum $l^\prime$, and then Taylor expand in the transverse quark momentum $k_T$ up to the first order (and subsequently set $k_T=0$), or vice versa. We found that the order in which phase-space integration and Taylor expansion is performed does matter. If one first expands in $k_T$ and then performs the phase space integration, the return appears to be ambiguous and often manifestly incorrect. On the other hand, we found that if we perform the phase space integration first but keep a non-zero $k_T$ at all times, then the result, again indicated by $\hat{\sigma}(x,z,k_T,P,P_h,l,\varepsilon)$ but with $l^\prime,r$ integrated out, is unambiguous.
We note however that the phase space integrations, in $d=4-2\varepsilon$ dimensions, need to be performed to all orders in $\varepsilon$, and an expansion in $\varepsilon$ has to be postponed to \emph{after} the Taylor expansion in $k_T$. Performing an \emph{exact} phase space integration is possible with the help of the methods described in Refs.~\cite{Lyubovitskij:2021ges,vanNeerven:1985xr,Beenakker:1988bq}. 

There is a unique feature that appears at NLO for twist-3 observables: having performed the phase space integration with non-zero $k_T$, we arrive at the following schematic factorization formula for the kinematical twist-3 contribution:
\be
\int \tfrac{\dd z}{z^2}\int \tfrac{\dd x}{x} \int \dd^{d-2} k_T\,\frac{\hat{\sigma}(\hat{s},\hat{t},\hat{u},k_T\cdot l_T,\mu,\varepsilon,\kappa)}{(\hat{s}+\hat{t}+\hat{u}+2\,k_T\cdot l_T)^{1+2\varepsilon}}\,\epsilon^{Pnk_TS_T}\,f_{1T}^{\perp, q}(x,k_T^2)\,D_1^q(z)\,,\label{eq:kintw3bfDkT}
\ee
where we have introduced the partonic Mandelstam variables $\hat{s}= (l+x\,P)^2=x\,s$, $\hat{t}=(x\,P-P_h/z)^2=\frac{x}{z}t$, $\hat{u}=(l-P_h/z)^2=u/z$. Since we explicitly keep the transverse quark momentum $k_T$, it appears in \eqref{eq:kintw3bfDkT} as a scalar product with the only other external transverse momentum $l_T$ of the incident lepton, thanks to our choice \eqref{eq:choiceLCvectormain} for the light-cone vector $n^\mu$. Note that the gauge parameter $\kappa$ (see \eqref{eq:LCpolsummain}) explicitly appears in $\hat{\sigma}$ so \eqref{eq:kintw3bfDkT} by itself is not gauge invariant at NLO. However, all terms proportional to $\kappa$ turn out to be proportional to $k_T$. This means that if we set $k_T=0$ right away (as we do for leading-twist observables), the gauge dependence drops out. As we will see in Sec.~\ref{sub:qg2q}, this gauge dependence also vanishes in case of our twist-3 calculation. After combining the kinematical and dynamical parts via Eq.~(\ref{eq:SiversSGP}) it is canceled by a corresponding term from the soft-gluon pole contribution. In addition, we extracted an explicit denominator $(\hat{s}+\hat{t}+\hat{u}+2\,k_T\cdot l_T)^{1+2\varepsilon}$. This denominator, for $k_T=0$, typically displays a soft singularity for $\hat{s}+\hat{t}+\hat{u}=0$ at leading twist.

However, we need to Taylor-expand the factor
\[\frac{\hat{\sigma}(\hat{s},\hat{t},\hat{u},k_T\cdot l_T,\mu,\varepsilon,\kappa)}{(\hat{s}+\hat{t}+\hat{u}+2\,k_T\cdot l_T)^{1+2\varepsilon}}\]
up to first order in $k_T$ (and subsequently set $k_T=0$), because only in this case do we obtain a non-zero collinear contribution from the first moment of the Sivers function, using
\be
\int \dd^{d-2}k_T k_T^\rho k_T^\sigma \,f_{1T}^{\perp,q}(x,k_T^2)= -M^2\,g_T^{\rho\sigma}\,f_{1T}^{\perp (1),q}(x)\,.\label{eq:FMSivers}
\ee
The Taylor expansion to first order in $k_T$ reads explicitly,
\[ \frac{-2(1+2\varepsilon)\,l_{T\sigma}\hat{\sigma}(\hat{s},\hat{t},\hat{u},\mu,\varepsilon)}{(\hat{s}+\hat{t}+\hat{u})^{2+2\varepsilon}}+ \frac{l_{T\sigma}\,\left(\frac{\partial \hat{\sigma}}{\partial(k_T\cdot l_T)}\right)(\hat{s},\hat{t},\hat{u},\mu,\varepsilon,\kappa)}{(\hat{s}+\hat{t}+\hat{u})^{1+2\varepsilon}}\,,\]
and after the change of variables $z=\frac{1-v_1}{1-v}$, $x=x_0/w$, discussed in Sec.~\ref{sub:LO}, \eqref{eq:kintw3bfDkT} assumes the following schematic form (neglecting the $v$-integration and the fragmentation functions):
\be
\int_{x_0}^1\tfrac{\dd w}{w}\,\left(\frac{\sigma_1(v,w,\varepsilon)\,f_{1T}^{\perp (1)}(\frac{x_0}{w})}{(1-w)^{2+2\varepsilon}}+\frac{\sigma_2(v,w,\varepsilon,\kappa)\,f_{1T}^{\perp (1)}(\frac{x_0}{w})}{(1-w)^{1+2\varepsilon}}\right)\,.\label{eq:kintw3Intermediate}
\ee
It is known well from leading twist NLO calculations how to 
handle the second term $\propto \,1/(1-w)^{1+2\varepsilon}$: in order to deal with the soft singularity $w\to 1$ in the integrand, one introduces the plus-distribution \eqref{eq:plusdistribution} isolating this singularity by means of the following formula:
\be
\frac{1}{(1-w)^{1+\varepsilon}}=-\frac{1}{\varepsilon}\delta(1-w)+\frac{1}{(1-w)_+}-\varepsilon\,\left(\frac{\ln(1-w)}{1-w}\right)_+ + \mathcal{O}(\varepsilon^2)\,,\label{eq:epsExpandsion}
\ee
(obviously also valid for replacements $\varepsilon\to 2\varepsilon$). The plus-distributions in this expansion are defined as follows,
\bea
\int_z^1\tfrac{\dd w}{w}\,\frac{f(w)}{(1-w)_+} & = & \int_z^1\dd w\,\frac{\frac{1}{w}f(w)-f(1)}{1-w}+f(1)\,\ln(1-z)\,.\nn\\
\int_z^1\tfrac{\dd w}{w}\,\left(\frac{\ln(1-w)}{1-w}\right)_+ \,f(w) & = & \int_z^1\dd w\,\left[\frac{\ln(1-w)}{1-w}(\tfrac{1}{w}f(w)-f(1))\right]+f(1)\,\tfrac{1}{2}\ln^2(1-z)\,.\label{eq:plusdistribution}
\eea
The $1/\varepsilon$-term in \eqref{eq:epsExpandsion} is typically canceled by virtual contributions which ensures the infra-red safety of leading twist observables.
Clearly, application of \eqref{eq:epsExpandsion} does not work for the first term in \eqref{eq:kintw3Intermediate} due to the higher power in $1/(1-w)^{2+2\varepsilon}$, which is a relic of the $k_T$-expansion for kinematical twist-3 contributions and hence a new feature compared to leading-twist NLO calculations. Our procedure of how to deal with this term is described in Appendix \ref{sub:analContinue} and is based on integration by parts. As a result of integrating by parts, a derivative term $(f_{1T}^{\perp (1)})^\prime=\pi (F^q)^\prime$ is generated as well.

\begin{figure}[tbp]
\centering
\includegraphics[width=0.2\textwidth]{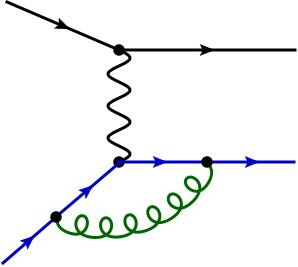}
\caption{NLO virtual vertex correction relevant for the kinematical twist-3 contribution. 
The momenta of the external lines are denoted as in Fig.~\ref{fig:LO} (left). This diagram comes in interference 
with the LO diagram in Fig.~\ref{fig:LO} (left), with a non-zero $k_T$. Also the mirror diagram needs to be taken into account.\label{fig:NLOvirkin}}
\end{figure}

The final comment of this section is devoted to the handling of the kinematical twist-3 contribution for the case with just the fragmenting quark in the final state. This situation includes the LO diagram (Fig.~\ref{fig:LO} left), as well as the NLO virtual QCD vertex correction, see Fig.~\ref{fig:NLOvirkin}. As mentioned before, the order in which integrations and $k_T$-expansions are performed matters. As for the real corrections it is important to keep $k_T$ finite (through a kinematical approximation \eqref{eq:kinApproxktintw2}) in virtual loop diagrams, integrate out all emerging $\delta$-functions that include $k_T$, and only then expand in $k_T$. Let us discuss this point using a schematic example for the NLO vertex correction, similar to \eqref{eq:kintw3bfDkT} without the fragmentation function and $z$-integration, 
\bea
&\int \tfrac{\dd x}{x} \int \dd^{d-2}k_T\,\delta\left(\hat{s}+\hat{t}+\hat{u}+2l_T\cdot k_T\right)\,\left(\sigma_0+l_T\cdot k_T\,\sigma_1\right)(\hat{s},\hat{t},\hat{u},\varepsilon)\,
\epsilon^{Pnk_TS_T}\,f_{1T}^{\perp, q}(x,k_T^2)\nn\\
\to & \int_{x_0}^\infty\tfrac{\dd w}{w}\int \dd^{d-2}k_T\,\delta\left(\tfrac{1}{w}-(1-\chi)\right)\,\left(\sigma_0+l_T\cdot k_T\,\sigma_1\right)(v,w,\varepsilon)\,\epsilon^{Pnk_TS_T}\,f_{1T}^{\perp, q}(\tfrac{x_0}{w},k_T^2)\nn\\
\to & \int \dd^{d-2}k_T\,w_0\left(\sigma_0+l_T\cdot k_T\,\sigma_1\right)(v,w_0,\varepsilon)\,\epsilon^{Pnk_TS_T}\,f_{1T}^{\perp, q}(\tfrac{x_0}{w_0},k_T^2)\,,\label{eq:kintw3bfDkTvirt}
\eea
where in the first line $\sigma_0+l_T\cdot k_T\,\sigma_1$ represents the virtual loop correction (the loop integration has been performed analytically to all orders in $\varepsilon$). This loop correction is accompanied by a $\delta$-function that already appears at LO. Both the $\delta$ function and the loop correction in principle depend on $k_T$, but in anticipation of the $k_T$ expansion we only kept terms that are at most linear in $k_T$. In the second line of \eqref{eq:kintw3bfDkTvirt} we transform the integration variable $x\to x_0/w$ as before and abbreviate $\chi = \frac{2\,t\,(l_T\cdot k_T)}{s\,u\,(1-v)}$. In the third line, we integrate the $\delta$-function with $w_0=1/(1-\chi)$. Because of the implicit dependence of the remaining integrand in the third line of \eqref{eq:kintw3bfDkTvirt} on $k_T$ via $w_0$ and $\chi$ we expand the \emph{full} integrand including the Sivers function $f_{1T}^{\perp,q}(x_0/w_0,k_T^2)$ up to the linear term in $k_T$, and use \eqref{eq:FMSivers}. Eventually, the third line of \eqref{eq:kintw3bfDkTvirt} simplifies to
\be
 \left(\sigma_1(v,w,\varepsilon)+\tfrac{2t\,\partial_w(w\,\sigma_0)(v,w,\varepsilon)}{su(1-v)}\right)\Big|_{w=1}\,f_{1T}^{\perp (1),q}(x_0)+\left(\tfrac{2t}{su(1-v)}\,\sigma_0(v,w,\varepsilon)\right)\Big|_{w=1}\,\left(-x_0\,(f_{1T}^{\perp (1),q})^\prime(x_0)\right)\,.\label{eq:kintw3IntermediateVirt}
\ee
We observe that we also obtain a derivative term in this case.

\subsubsection{Dynamical Twist-3 at NLO}
In this section we discuss QCD corrections to the LO diagram in Fig.~\ref{fig:LO} (right) generated by the SGP function $F(x,x)$. We describe special features and observations at NLO as well as our general strategy for the computation of dynamical twist-3 effects at NLO with radiation of an unobserved gluon as an example (see diagrams in Fig.~\ref{fig:NLOdynqg2qreal}). As discussed in Sec.~\ref{sub:LO}, to obtain a contribution to the SSA,
an imaginary part needs to be generated in the hard scattering. In contrast to the LO formula \eqref{eq:LO} which is generated merely by a soft-gluon pole enforced by the propagator \eqref{eq:LOSGP}, there are several other sources of imaginary parts at NLO due to the more complex phase space. We will discuss them in the following:

\begin{figure}[tbp]
\centering
\includegraphics[width=0.7\textwidth]{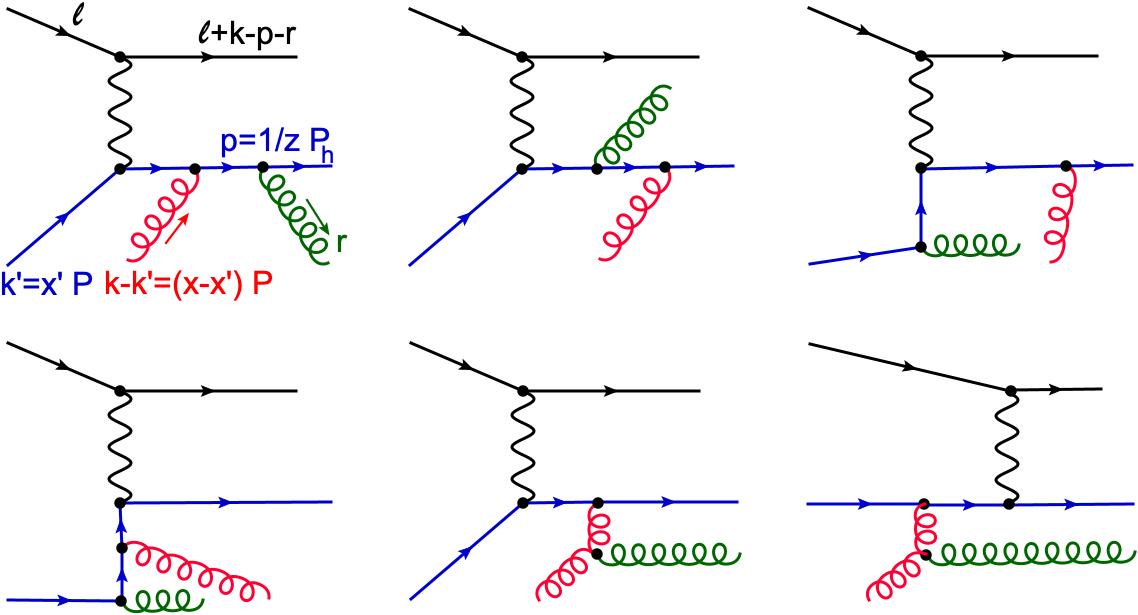}
\caption{NLO real-gluon emission diagrams in the $qg\to q$ channel relevant for the dynamical twist-3 contribution.
One needs to compute the interference of the sum of these diagrams with with the sum of the diagrams in Fig.~\ref{fig:NLOkinqg2qreal}, with $k_T=0$.\label{fig:NLOdynqg2qreal}}
\end{figure}

\paragraph{Integral contribution:}

Let us start with a general schematic collinear twist-3 factorization formula for the real corrections caused by gluon radiation in one of the possible partonic channels that we call the $qg\to q$ channel (see Fig.~\ref{fig:NLOdynqg2qreal}). In particular, the LO result Eq.~\eqref{eq:LO} is part of this channel. The notation refers to quark-gluon correlations in the initial partonic state ($qg$) and quark fragmentation in the final state ($q$). The NLO formula will receive contributions from both the quark-gluon-quark correlation functions $F$ and $G$ (in contrast to LO \eqref{eq:LO}) that can be cast into the following form,
\bea
\int\tfrac{\dd z}{z^2}\int \dd x \int_{x-1}^1 \dd x^\prime \frac{i}{x^\prime-x}\, \left[\,\hat{\sigma}^1(x,x^\prime,z)\,F^q(x,x^\prime)+\,\hat{\sigma}^5(x,x^\prime,z)\,G^q(x,x^\prime)\,\right]D_1^q(z)+\mathrm{c.c.}.&\label{eq:dynInt}
\eea
The generic partonic factors $\hat{\sigma}^{1,5}$ are interference terms of a Feynman amplitude $\mathcal{M}_{qg}$ with a collinear quark and a collinear gluon in the initial state (as in Fig.~\ref{fig:NLOdynqg2qreal}) and a Feynman amplitude $\mathcal{M}_q$ with only one collinear quark in the initial state (as in Fig.~\ref{fig:NLOkinqg2qreal}, but with $k_T=0$). We assume that the phase space integration over the undetected lepton and the radiated gluon has been performed. Hence, schematically, we have
\[\hat{\sigma}\sim \int \dd \mathrm{LIPS}_{2,\ell g}\,\mathcal{M}_{qg}\,\mathcal{M}_q^\ast\,.\]
In the integral \eqref{eq:dynInt} we have explicitly extracted an imaginary unit $i$, leaving the numerator of the partonic factors $\hat{\sigma}^{1,5}$ real. Since we always add the complex conjugate $\mathrm{c.c.}$ in \eqref{eq:dynInt}, it is clear that another imaginary unit must appear in $\hat{\sigma}^{1,5}$ in order to produce a non-zero expression. This is typically achieved by the small imaginary parts entering the denominators of propagators in Feynman diagrams. An example in LO is Eq.~\eqref{eq:LOSGP}. 

However, at NLO, complications may occur for certain (but not all) partonic propagators in the hard scattering amplitudes. As an example, let us discuss the following propagator that appears in some of the Feynman diagrams for the $qg\to q$ channel (see Fig.~\ref{fig:NLOdynqg2qreal}),

\[\frac{i}{(p-k+k^\prime+r)^2+i\delta}.\]
Here, the approximated partonic momenta $k\simeq x\,P$, $k^\prime\simeq x^\prime\,P$ and $p\simeq P_h/z$ are as in the right diagram of Fig.~\ref{fig:LO}, but $r$ refers to an additional gluon momentum that is integrated out. Naturally, one may be tempted to deal with this propagator as in Eq.~\eqref{eq:LOSGP}:
\be
\frac{1}{(p-k+k^\prime+r)^2+i\delta}=\frac{\mathcal{P}}{...}-\frac{i\,\pi}{2k\cdot r-\hat{t}}\,\delta\left(\zeta -\frac{\hat{s}+\hat{u}-2l\cdot r}{2k\cdot r-\hat{t}}\right)\,,\label{eq:PropHP}
\ee
where we introduced a useful variable $\zeta\equiv\frac{x^\prime}{x}$. Then, if one first performs the $x^\prime$ (or, equivalently, $\zeta$)-integration, the gluon momentum $r^\mu$ would enter the unknown functions $F$ and $G$ in Eq.~\eqref{eq:dynInt} implicitly as an argument for $x^\prime$. To be specific, we would obtain a generic contribution of the form
\[\int d^dr\,\delta^+(r^2)\,\hat{\sigma}(r)\,F(x,x\,\tfrac{\hat{s}+\hat{u}-2l\cdot r}{2k\cdot r-\hat{t}})\,.\]
Since the quark-gluon-quark correlation function $F$ (or $G$) is unknown, the subsequent phase space integration over the radiated gluon momentum $r^\mu$, in particular in arbitrary dimension $d$, would be very difficult to perform $\--$ although formally possible if the functions $F$ and $G$ were to be Taylor expanded in $x^\prime$. 

In order to deal with this particular complication we reverse the order of integrations and perform the phase space integration over $r^\mu$ \emph{before} we deal with the $x^\prime$-integral. To be precise, when phase space integration is performed, the light-cone momentum fractions $x$, $x^\prime$ are not constrained at all (apart from their integration boundaries in \eqref{eq:dynInt}). Hence, we call such contributions with unconstrained $x$ \emph{and} $x^\prime$ \emph{integral contributions}. Consequently, imaginary parts are not generated by a principal value and pole decomposition as in \eqref{eq:PropHP}. Instead, they emerge after the phase space integration via the appearance of complex-valued 
logarithms where we carefully take into account the imaginary part $i\delta$ of the propagator on the left side of Eq.~\eqref{eq:PropHP}. As one example, a specific logarithm that appears after phase space integration has the following form:
\bea
\ln\left(\frac{1-(B^2+C^2)/(A+i\delta)^2}{(1-B/(A+i\delta))^2}\right)&=&\ln\left|\frac{1-(B^2+C^2)/A^2}{(1-B/A)^2}\right|\nn\\
&&+i\pi \,\theta(1-\zeta)\theta(\zeta)\,\mathrm{sgn}(\zeta-\tfrac{-\hat{u}}{\hat{s}+\hat{t}})\,,\label{eq:ImLog}
\eea
 with $A=-((1-\zeta)\hat{s}+\zeta\hat{t}+\hat{u})/4$, $B=A+\frac{\zeta\hat{t}}{2}+\frac{\hat{t}\hat{u}}{2(\hat{s}+\hat{t})}$, $C=\frac{\sqrt{\hat{s}\hat{t}\hat{u}(\hat{s}+\hat{t}+\hat{u})}}{2(\hat{s}+\hat{t})}$. In \eqref{eq:ImLog} $\theta$ represents the Heaviside step function and $\mathrm{sgn}$ the sign function. We observe that the logarithm \eqref{eq:ImLog} produces an imaginary part for $0<\zeta<1$ or $0<x^\prime<x$. This is a general feature that we observe for all complex logarithms generated by the phase space integration.

 We note that we perform the phase space integration for the integral contributions in $d=4-2\varepsilon$ dimensions. Typically, the result will contain $1/\varepsilon$ poles that originate from soft and/or collinear singularities. However, since we are looking for logarithms like \eqref{eq:ImLog} producing an imaginary part, such $1/\varepsilon$ poles become irrelevant as NLO logarithms only show up at order $\mathcal{O}(\varepsilon^0)$. In this way we arrive at a result of the following form  for the integral contributions to the $qg\to q$ channel:
 \bea
&&\int_{v_0}^{v_1}\dd v\int_{x_0}^1\tfrac{\dd w}{w}\int_0^1\dd \zeta\,\left[\tilde{\sigma}^1(v,w,\zeta)\,F^q(\tfrac{x_0}{w},\zeta \tfrac{x_0}{w})+\tilde{\sigma}^5(v,w,\zeta)\,G^q(\tfrac{x_0}{w},\zeta \tfrac{x_0}{w})\right]\,D_1^q(\tfrac{1-v_1}{1-v})\,,\label{eq:dynInt2}
 \eea
where we have, as before, transformed the integration variables $x$ and $z$ to $v$ and $w$. The partonic functions $\tilde{\sigma}^{1,5}$ have a finite limit to
four dimensions and depend neither on external kinematical variables $s,\,t,\,u$, nor on the renormalization scale $\mu$.

We observe that $\tilde{\sigma}^{1,5}(v,w,\zeta)$ in general are not continuous due to the explicit appearance of sign functions in the complex logarithms as in Eq.~\eqref{eq:ImLog}. This alone would be acceptable, as the integrability of Eq.~\eqref{eq:dynInt2} would still be ensured. However, in addition we also find severe endpoint singularities of the $\zeta$-integral, as well as singularities in between the integration boundaries. To be specific, the partonic factors in the $qg\to q$ channel behave as ($f_{1,2,3}$ are some generic functions independent of $\zeta$)
\bea
\tilde{\sigma}^{1,5}(v,w,\zeta) & \xrightarrow{\zeta\to 1} & \frac{f_1(v,w)}{(1-\zeta)^2}\,,\nn\\
\tilde{\sigma}^{1,5}(v,w,\zeta) & \xrightarrow{\zeta\to 0} & \frac{f_2(v,w)}{\zeta^2}\,,\nn\\
\tilde{\sigma}^{1,5}(v,w,\zeta) & \xrightarrow{\zeta\to w} & \frac{f_3(v,w)}{(w-\zeta)^2}\,.\label{eq:dynIntPoles}
\eea
This behavior immediately indicates that the $\zeta$-integral in \eqref{eq:dynInt2} diverges. 

We deal with this feature in the following way: since we know the singular structure \eqref{eq:dynIntPoles} well, we may redefine the partonic functions as
\be
\hat{\sigma}^{1,5}_{\mathrm{Int}}(v,w,\zeta)\equiv \zeta^2(1-\zeta)^2(w-\zeta)^2\,\tilde{\sigma}^{1,5}(v,w,\zeta)\,.\label{eq:dynIntpartCS}
\ee
Indeed, the new partonic functions $\hat{\sigma}^{1,5}_{\mathrm{Int}}$ are now well-behaved and integrable. Of course, we have to compensate for 
the factor $\zeta^2(1-\zeta)^2(w-\zeta)^2$ by dividing the quark-gluon-quark correlation functions in \eqref{eq:dynInt2} accordingly as follows:
\be
(F,G)(\tfrac{x_0}{w},\zeta \tfrac{x_0}{w})\to \frac{(F,G)(\tfrac{x_0}{w},\zeta \tfrac{x_0}{w})}{\zeta^2(1-\zeta)^2(w-\zeta)^2}.
\ee
Obviously, the right-hand-side now suffers from the same divergent behavior that we tried to avoid in Eq.~\eqref{eq:dynIntPoles}.
We next try to temper the divergences by subtracting  from the numerator Taylor expansions of $(F,G)(\tfrac{x_0}{w},\zeta \tfrac{x_0}{w})$ around the three poles up to the linear term in $\zeta$. The \emph{subtracted} quark-gluon-quark correlation functions then read,
\bea
F^{qg\to q}_{\mathrm{Int}}(x_0,w,\zeta) & \equiv & \frac{1}{\zeta^2(1-\zeta)^2(w-\zeta)^2}\left[F^q(\tfrac{x_0}{w},\zeta \tfrac{x_0}{w})\right.\nn\\
&&\hspace{-3cm}+\frac{\zeta^2(1-\zeta)^2}{(1-w)^3w^3}\left((5w^2+2\zeta-3w-4w\zeta)\,F^q(\tfrac{x_0}{w},x_0)+(w-\zeta)(1-w)\,x_0(\partial_2 F^q)(\tfrac{x_0}{w},x_0)\right)\nn\\
&&\hspace{-3cm}-\frac{\zeta^2(w-\zeta)^2}{(1-w)^3}\left((5-4\zeta-3w+2w\zeta)\,F^q(\tfrac{x_0}{w},\tfrac{x_0}{w})-\frac{1}{2}(1-\zeta)(1-w)\,\tfrac{x_0}{w}(F^q)^\prime(\tfrac{x_0}{w},\tfrac{x_0}{w})\right)\nn\\
&&\hspace{-3cm}\left.-\frac{(w-\zeta)^2(1-\zeta)^2}{w^3}\left((2\zeta+w+2w \zeta)\,F^q(\tfrac{x_0}{w},0)+\zeta\,x_0(\partial_2 F^q)(\tfrac{x_0}{w},0)\right)\right]\,,\label{eq:dynIntmodF}
\eea
and
\bea
G^{qg\to q}_{\mathrm{Int}}(x_0,w,\zeta) & \equiv & \frac{1}{\zeta^2(1-\zeta)^2(w-\zeta)^2}\left[G^q(\tfrac{x_0}{w},\zeta \tfrac{x_0}{w})\right.\nn\\
&&\hspace{-3cm}+\frac{\zeta^2(1-\zeta)^2}{(1-w)^3w^3}\left((5w^2+2\zeta-3w-4w\zeta)\,G^q(\tfrac{x_0}{w},x_0)+(w-\zeta)(1-w)\,x_0(\partial_2 G^q)(\tfrac{x_0}{w},x_0)\right)\nn\\
&&\hspace{-3cm}+\frac{\zeta^2(w-\zeta)^2(1-\zeta)}{(1-w)^2}\,\tfrac{x_0}{w}(\partial_2G^q)(\tfrac{x_0}{w},\tfrac{x_0}{w})\nn\\
&&\hspace{-3cm}\left.-\frac{(w-\zeta)^2(1-\zeta)^2}{w^3}\left((2\zeta+w+2w \zeta)\,G^q(\tfrac{x_0}{w},0)+\zeta\,x_0(\partial_2 G^q)(\tfrac{x_0}{w},0)\right)\right]\,.\label{eq:dynIntmodG}
\eea
Even though not immediately obvious from Eqs.~\eqref{eq:dynIntmodF},\eqref{eq:dynIntmodG}, both modified functions $(F,G)^{qg\to q}_{\mathrm{Int}}$ turn out to be integrable over the range $0\le\zeta\le 1$ and $x_0\le w \le 1$. Hence, the triple integral,
\bea
&&\int_{v_0}^{v_1}\dd v\int_{x_0}^1\tfrac{\dd w}{w}\int_0^1\dd \zeta\,\left[\hat{\sigma}_{\mathrm{Int}}^1(v,w,\zeta)\,F_{\mathrm{Int}}^{qg\to q}(x_0,w,\zeta)+\hat{\sigma}_{\mathrm{Int}}^5(v,w,\zeta)\,G_{\mathrm{Int}}^{qg\to q}(x_0,w,\zeta)\right]\,D_1^q(\tfrac{1-v_1}{1-v})\,,\label{eq:dynInt3}
\eea
which constitutes the final formula for the \emph{integral contributions} to the $qg\to q$ channel is convergent. 

We note that there may be other ways to handle the singularities in the $\zeta$-integral. In particular, the terms proportional to the color factor $C_F$ in the partonic cross sections $\tilde{\sigma}^{1,5}(v,w,\zeta)$ are less divergent around the three $\zeta$-poles. Thus, for these terms, it would be sufficient to factor out lower powers of $1-\zeta,\,\zeta$. Interestingly, there is no divergence around $\zeta=w$ for the terms proportional to $C_F$. Consequently, the subtracted correlation functions in Eqs.~\ref{eq:dynIntmodF} and \ref{eq:dynIntmodG} as well as further parts of the calculation, i.e. the hard pole, soft-gluon pole and soft-fermion pole contributions discussed below, would change as well. 

\paragraph{Hard pole (HP), soft-gluon pole (SGP) and soft-fermion pole (SFP) contributions:\label{par:HPSGPSFP}}

Of course, the subtraction terms in Eqs.~\eqref{eq:dynIntmodF},\eqref{eq:dynIntmodG} (all terms except $(F,G)(\tfrac{x_0}{w},\zeta \tfrac{x_0}{w})$) need to be added back and handled individually. We first note that the variable $\zeta$ does not enter the subtraction terms as an argument of the functions $F$, $G$ $\--$ the latter can in principle be factored out of the $\zeta$-integral. As a result, the subtraction terms are evaluated only on a restricted part of the support of $F$, $G$. For example, we recognize in the third line of \eqref{eq:dynIntmodF} the \emph{soft-gluon pole} (SGP) function $F(\tfrac{x_0}{w},\tfrac{x_0}{w})$ and its derivative term. Here, only the diagonal support of $F$ is probed. We have already encountered such an SGP function at LO \eqref{eq:LO} and for kinematical twist-3 in Sec. \ref{sub:KinTw3} (Eqs.~\eqref{eq:kintw3Intermediate},\eqref{eq:kintw3IntermediateVirt} with Eq.~\eqref{eq:SiversSGP}). Since the SGP function $G(\tfrac{x_0}{w},\tfrac{x_0}{w})$ vanishes, only a derivative term enters the third line of Eq.~\eqref{eq:dynIntmodG}. 

Other special regions of support of the quark-gluon-quark correlation functions are probed by the subtraction terms: the so-called \emph{soft-fermion pole} (SFP) functions $F(\tfrac{x_0}{w},0)$, $G(\tfrac{x_0}{w},0)$ (and their derivatives) appear in the fourth lines of Eqs.~\eqref{eq:dynIntmodF},\eqref{eq:dynIntmodG}. The physical interpretation of an SFP matrix element $(F,G)(x,0)$ with $x^\prime=0$ is that an initial quark entering a partonic amplitude (e.g. in the right diagram of Fig.~\ref{fig:LO}) carries no longitudinal momentum, hence the name soft-fermion pole.

Lastly, the functions $F(\tfrac{x_0}{w},x_0)$, $G(\tfrac{x_0}{w},x_0)$ are probed by the subtraction terms in the second lines of Eqs.~\eqref{eq:dynIntmodF},\eqref{eq:dynIntmodG}. Such matrix elements are called \emph{hard pole} (HP) because neither the quark nor the gluon entering the partonic amplitude are soft.

Technically, for the calculation of the HP, SGP and SFP subtraction terms it is not necessary to follow the order of integration that we proposed for the integral contributions above. Since the integration variable $\zeta$, or $x^\prime$, does not appear directly in the correlation functions $F$ and $G$ in the subtraction terms 
we might as well return to the original formulation of collinear twist-3 factorization \eqref{eq:dynInt}, assuming that the 
phase space integration has not yet been performed. Then, for the subtraction terms, we may first apply the decomposition \eqref{eq:dynIntPoles}, replace $\zeta$ in the partonic cross section $\hat{\sigma}$ as well as in the $\zeta$-dependent prefactors of the subtraction terms \eqref{eq:dynIntmodF},\eqref{eq:dynIntmodG} by the kinematic fraction $\frac{\hat{s}+\hat{u}-2l\cdot r}{2k\cdot r-\hat{t}}$, and eventually perform the phase space integration. The advantage of this procedure is that we can obtain expressions for all of the subtraction terms to all orders in $\varepsilon$.

The subtraction terms \eqref{eq:dynIntmodF},\eqref{eq:dynIntmodG} are the only sources of contributions entering with HP matrix elements $F(\tfrac{x_0}{w},x_0)$, $G(\tfrac{x_0}{w},x_0)$. One may consider the expressions we obtain as the final HP results. They do carry $1/\varepsilon$ poles that point towards collinear singularities caused by gluon emission in the $qg\to q$ channel. We can explicitly show that such singularities are canceled precisely by the corresponding HP term in the renormalization formula \eqref{eq:LOsplittingfunctionTw3deriv} (third and fifth line). This will be discussed in more detail below.

The situation is more complicated for the SGP- and SFP-contributions. Not only do we encounter contributions generated indirectly through the subtraction terms 
designed for the integral contributions \eqref{eq:dynInt3} to converge, but we also obtain \emph{direct} SGP- and SFP-contributions in the partonic amplitudes, caused for example by Feynman propagators of the form
\be
\frac{1}{(r-k+k^\prime)^2+i\delta}=\frac{1}{(2k\cdot r)(\zeta-1+i\delta)}=\frac{1}{2k\cdot r}\left(\frac{\mathcal{P}}{\zeta-1}-i\pi\,\delta(\zeta-1)\right)\,,\label{eq:SGPPole}
\ee
for the direct SGP-contribution, as well as \eqref{eq:LOSGP}, and
\be
\frac{1}{(k^\prime-r)^2+i\delta}=\frac{1}{(2k\cdot r)(-\zeta+i\delta)}=-\frac{1}{2k\cdot r}\left(\frac{\mathcal{P}}{\zeta}+i\pi\,\delta(\zeta)\right)\,,\label{eq:SFPPole}
\ee
for the direct SFP-contribution. Such direct contributions need to be calculated separately. Eventually, both direct and indirect SGP- and SFP-contributions must be combined. More details will be given below.

\subsubsection{Photon-in-Lepton Contribution at NLO\label{sub:PiL}}

When performing NLO calculations for the single-inclusive hadron production process in lepton-nucleon collisions, $\ell N\to h X$, collinear singularities are encountered not only from unobserved partons in the final state, but also from the unobserved lepton if assumed massless. Upon integrating over the final-state lepton's phase space one may find a configuration where its momentum is collinear to that of the initial-state lepton. Consequently, the exchanged photon becomes quasi-real and one hits the pole of the photon propagator in this momentum configuration, which leads to a collinear singularity that appears at NLO in the form of a $1/\varepsilon$ pole.
This feature has already been observed for the unpolarized cross section in Ref.~\cite{Hinderer:2015hra}. As discussed at length in that reference, one may deal with this issue in two equivalent ways. 

In the first approach, one explicitly keeps a non-zero lepton mass $m_\ell$ throughout the NLO calculation. Consequently, the phase space integration over the unobserved parton and lepton at NLO becomes considerably more involved, especially in view of the need for an all order (in $\varepsilon$) result (see Appendix \ref{sub:analContinue}). This may be seen as a clear disadvantage of this approach. Regardless of this, one may then expand the result in the small lepton mass and keep only the $\ln(m_\ell)$-terms and those of order $\mathcal{O}(m_\ell^0)$. The non-zero lepton mass regulates the aforementioned collinear singularity, and any partonic factor $\--$ unpolarized or polarized $\--$ will take the following generic form (see \cite{Hinderer:2015hra}):
\be
\hat{\sigma}_{\mathrm{NLO}}(v,w,m_\ell,\mu)= \hat{\sigma}_{\mathrm{log}}(v,w,\mu)\,\ln\left(\tfrac{s}{m_\ell^2}\right)+\hat{\sigma}_{\mathrm{0}}(v,w,\mu)+\mathcal{O}(m_\ell^2\,\ln(m_\ell^2))\,.\label{eq:elmassexpand}
\ee
Given the small size of the lepton mass the logarithm can potentially become quite large, and its resummation to all orders may be required. However, this is beyond the scope of this work.

In the second approach one directly works out the NLO contribution for massless leptons and regulates the collinear singularity originating from quasi-real photons through dimensional regularization. However, in order to cancel the emerging $1/\varepsilon$-pole, additional contributions are needed. For such contributions one treats the quasi-real photon as a parton within the lepton and works with the photon-in-lepton distribution $f_1^{\gamma /\ell}$ \eqref{eq:Defgine}. The factorization ansatz for the transversely polarized cross section in the collinear twist-3 formalism is similar to Eq.~(23) of Ref.~\cite{Hinderer:2015hra}, for the kinematical contributions,
\bea
E_h\frac{\dd \sigma^{\mathrm{real\,\gamma}}_{\mathrm{kin}}}{\dd \mathbf{P}_h}& \propto & \int\tfrac{\dd z}{z^2}\int \dd y\int \dd x \,\delta\left(y+\tfrac{\hat{t}}{\hat{s}+\hat{u}}\right)\,f_{1,\mathrm{bare}}^{\gamma/\ell}(y)\,D_1^q(z)\,\times\nn\\
&&\hspace{2cm}\left[f_{1T}^{\perp (1),q}(x)\,\hat{\sigma}_{\mathrm{kin}}^{\mathrm{real\,\gamma}}(y,x,z)\right]\,,\label{eq:WWkintw3}
\eea
and for the dynamical contributions,
\bea
E_h\frac{\dd \sigma^{\mathrm{real\,\gamma}}_{\mathrm{dyn}}}{\dd \mathbf{P}_h}& \propto & \int\tfrac{\dd z}{z^2}\int \dd y\int \dd x\,\int_{x-1}^1\dd x^\prime \,\delta\left(y+\tfrac{\hat{t}}{\hat{s}+\hat{u}}\right)\,f_{1,\mathrm{bare}}^{\gamma/\ell}(y)\,D_1^q(z)\,\times\nn\\
&&\left[F^q(x,x^\prime)\,\hat{\sigma}_{\mathrm{dyn},1}^{\mathrm{real\,\gamma}}(y,x,x^\prime,z)+G^q(x,x^\prime)\,\hat{\sigma}_{\mathrm{dyn},5}^{\mathrm{real\,\gamma}}(y,x,x^\prime,z)\right]\,.\label{eq:WWdyntw3}
\eea
In these factorization formulas the partonic factors can be constructed by the NLO Feynman diagrams discussed above, but with the lepton lines removed and the exchanged virtual photon replaced by a real photon with collinear momentum $q^\mu=y\,l^\mu$ ($l^\mu$ being the lepton's 4-momentum). The partonic factors will be of the order $\mathcal{O}(\alpha_\mathrm{em}\alpha_s)$ and are simpler to calculate as no phase space integration is required. As discussed above, for the dynamical contribution \eqref{eq:WWdyntw3} we need to extract an imaginary part through the propagator poles. Due to the simpler structure of the partonic factors $\hat{\sigma}_{\mathrm{dyn},1,5}^{\mathrm{real\,\gamma}}$ we only encounter soft-gluon poles and soft-fermion poles that are generated for ``partonic'' photons.

After renormalization of the photon-in-lepton distribution \eqref{eq:renWW}, the contributions \eqref{eq:WWkintw3} and \eqref{eq:WWdyntw3} are of the order $\mathcal{O}(\alpha_\mathrm{em}^2\alpha_s)$ due to the fact that the renormalized photon-in-lepton distribution $f_{1}^{\gamma/\ell,\overline{\mathrm{MS}}}(x,\mu)$ in \eqref{eq:f1WW} is of order $\mathcal{O}(\alpha_{\mathrm{em}})$. Thus, the contributions \eqref{eq:WWkintw3} and \eqref{eq:WWdyntw3} can be matched to the NLO corrections from real-gluon radiation discussed above, and we find that the collinear $1/\varepsilon$-pole originating from quasi-real photons indeed cancels for every partonic channel. This serves as an important check of our calculation. Also, the artificial factorization scale $\mu$ in \eqref{eq:f1WW} drops out of the final result after the ``partonic photon'' and 
NLO corrections are merged. Eventually, expansion \eqref{eq:elmassexpand} is recovered in this way, as was already observed in \cite{Hinderer:2015hra}.

\subsection{Channel \texorpdfstring{$qg\to q$}{qg to q}: Virtual Corrections\label{sub:virt}}

Having discussed the setup of our calculation, let us now focus on the particular channels. We start with the virtual corrections to the LO diagrams. The kinematical twist-3 contribution at LO, generated by the first moment of the Sivers function $f_{1T}^{\perp (1),q}(x)$, is indicated by the left diagram of Fig.~\ref{fig:LO}. The virtual NLO correction to this diagram is shown in Fig.~\ref{fig:NLOvirkin}. This correction to the quark-photon vertex is the only virtual correction that is relevant for the kinematical twist-3 effects. The computational procedure has already been discussed in Sec. \ref{sub:KinTw3} (see Eqs.~\eqref{eq:kintw3bfDkTvirt},\eqref{eq:kintw3IntermediateVirt}). Again, we stress that the calculations can be performed to all orders in $\varepsilon$. Interestingly, the result we obtain for the NLO contribution from the virtual vertex correction in Fig.~\ref{fig:NLOvirkin} turns out to be gauge invariant in the sense that the dependence on the gauge parameter $\kappa$ (see Eq.~\eqref{eq:LCpolsummain}) vanishes. We also point out that the choice of the light-cone vector \eqref{eq:choiceLCvectormain} simplifies the loop calculation to a large extent, since for this choice $l_T$ is the only non-vanishing transverse part of the external momenta. Note that we only encounter the color factor $C_F$ in the vertex correction in Fig.~\ref{fig:NLOvirkin}.

As a general remark, since we work with the modification of Eq.~\eqref{eq:LCpolsummain} for gluonic propagators introducing denominators that do not have a quadratic dependence on the momentum, we use the following strategy to perform the loop integrals: first, we perform a Sudakov-type decomposition of the loop momentum $r^\mu$, such as
\[r^\mu = (r\cdot P)\,n^\mu+(r\cdot n)\,P^\mu+r_T^\mu\,,\]
where the transverse component is defined through \eqref{eq:TransProjectormain}.
Secondly, we divide the integration into integrals over the light-cone components and over the transverse components (in $d-2=2-2\varepsilon$ dimensions), and integrate the $r\cdot P$ component via contour integration. Third, thanks to the choice \eqref{eq:choiceLCvectormain} for the light-cone vector, the transverse integrals can be analytically computed for all $\varepsilon$. Finally, the remaining light-cone component $r\cdot n$ is integrated analytically.

\begin{figure}[tbp]
\centering
\includegraphics[width=0.8\textwidth]{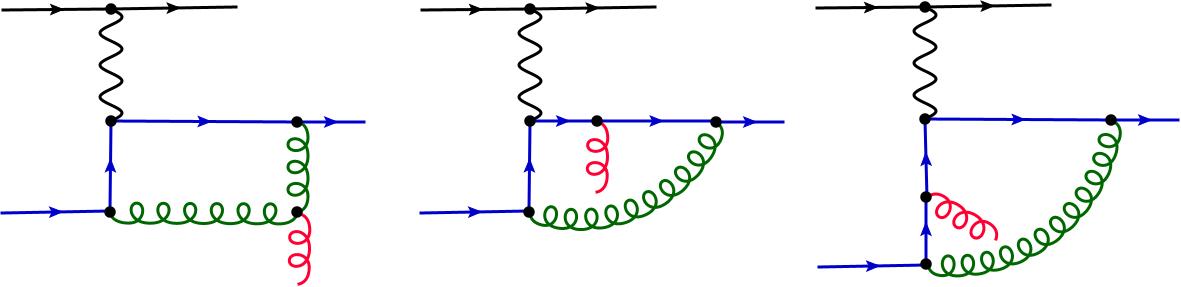}
\caption{NLO virtual box diagrams relevant for the dynamical twist-3 contribution. 
The momenta of the external lines are labeled as in Fig.~\ref{fig:LO} (right). These diagrams come in interference with the LO diagram in Fig.~\ref{fig:LO} (left), with $k_T=0$.\label{fig:NLOvirboxdyn}}
\end{figure}

\begin{figure}[tbp]
\centering
\includegraphics[width=0.8\textwidth]{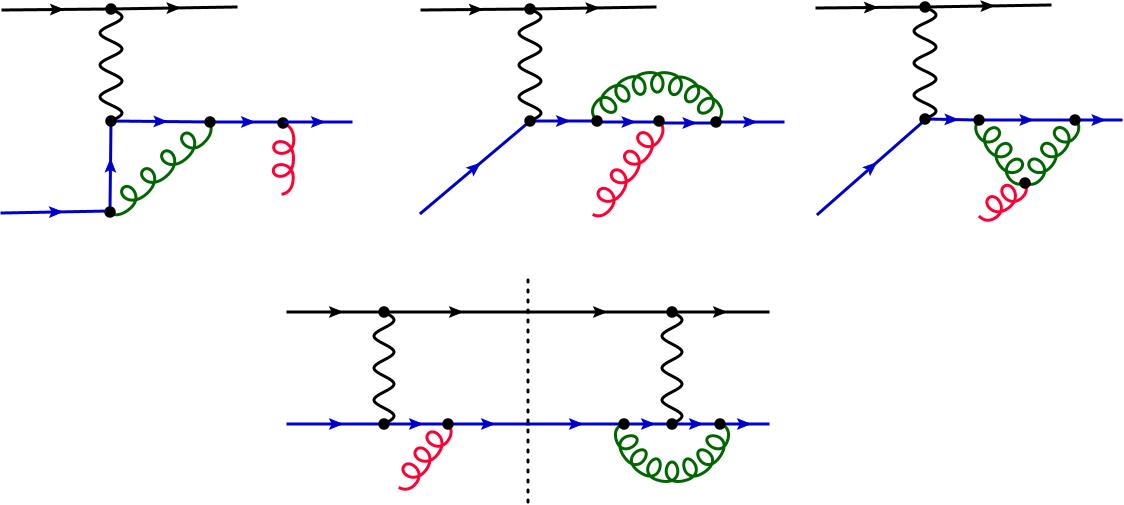}
\caption{NLO virtual vertex corrections relevant for the dynamical twist-3 contribution.  
The momenta of the external lines are labeled as in Fig.~\ref{fig:LO} (right). One needs to compute the interference of the upper diagrams 
with the LO diagram in Fig.~\ref{fig:LO} (left), with $k_T=0$. The lower diagram represents an interference of the LO dynamical twist-3 diagram of Fig.~\ref{fig:LO} (right) and the vertex correction in Fig.~\ref{fig:NLOvirkin}, with $k_T=0$. \label{fig:NLOvirvertexdyn}}
\end{figure}

\begin{figure}[tbp]
\centering
\includegraphics[width=0.3\textwidth]{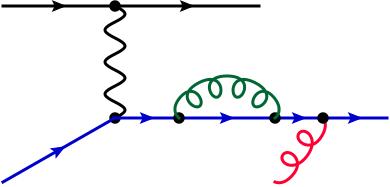}
\caption{NLO virtual self-energy diagram relevant for the dynamical twist-3 contribution. 
The momenta of the external lines are labeled as in Fig.~\ref{fig:LO} (right). This diagram comes in interference 
with the LO diagram in Fig.~\ref{fig:LO} (left), with $k_T=0$.\label{fig:NLOvirSEdyn}}
\end{figure}

Next, we focus on the virtual corrections to the LO dynamical twist-3 contribution indicated in the right diagram of Fig.~\ref{fig:LO}. The calculation of these corrections is more involved compared to that of the virtual correction for the kinematical twist-3 contribution in Fig.~\ref{fig:NLOvirkin}. In fact, there are three box diagrams shown in Fig.~\ref{fig:NLOvirboxdyn}, four vertex corrections displayed in Fig.~\ref{fig:NLOvirvertexdyn} and one self-energy correction in Fig.~\ref{fig:NLOvirSEdyn}. The different loop diagrams carry three kinds of color factors. For example, the color factor of the left box diagram in Fig.~\ref{fig:NLOvirboxdyn} is $N_c/2$ because of the three-gluon vertex, while the box diagram in the middle is proportional to $C_F-N_c/2$ and the box diagram on the right is proportional to $C_F$. 

All in all, we identify two independent color factors $C_F$ and $N_c/2$ within the loop-diagrams, and combine all virtual contributions accordingly. Eventually, we sum the virtual corrections to the kinematical and dynamical twist-3 effects by means of \eqref{eq:SiversSGP}.
Because of the more limited two-body phase space (with the final state quark momentum fixed) of the virtual corrections, the factorization formula for the dynamical twist-3 contributions (see Eq.~\eqref{eq:dynInt}) simplifies. In particular, the $x$-integration in Eq.~\eqref{eq:dynInt} can be easily performed due to the overall $\delta$-function $\delta(x-x_0)\propto\delta(1-w)$ for the LO and virtual NLO corrections.

At the same time, as discussed earlier, an imaginary part must be generated in the diagrams of Figs.~\ref{fig:NLOvirboxdyn}, \ref{fig:NLOvirvertexdyn}, \ref{fig:NLOvirSEdyn} so that the factorization formula \eqref{eq:dynInt} leads to a nonvanishing expression. Similarly to LO, such an imaginary part emerges from the decomposition \eqref{eq:LOSGP} of a quark propagator. The imaginary part of \eqref{eq:LOSGP} provides a $\delta$-function $\delta(x^\prime - x)$, and therefore, $x^\prime = x = x_0$. In this way a soft-gluon pole contribution emerges, entering as always 
with the ETQS matrix element $F(x_0,x_0)$. This is the only contribution that originates from an imaginary part of a Feynman propagator; soft-fermion poles or hard poles are not directly generated because of the restrictions on phase space.

That said, there is yet another source of an imaginary part hidden in the loop diagrams in Figs.~\ref{fig:NLOvirboxdyn}, \ref{fig:NLOvirvertexdyn}, \ref{fig:NLOvirSEdyn}. In fact, loop integrals can generate an imaginary part for certain regions within the $x^\prime$-integration in the general factorization formula \eqref{eq:dynInt}. Such an effect has been observed earlier also for other processes, for example for the transverse SSA in DIS where a two-photon exchange between an electron and a quark generates an imaginary part within a loop \cite{Metz:2006pe,Afanasev:2007ii,Schlegel:2012ve,Metz:2012ui}.

In the region $x^\prime>x=x_0$, the left and middle box diagrams in Fig.~\ref{fig:NLOvirboxdyn} as well as the upper vertex corrections in Fig.~\ref{fig:NLOvirvertexdyn} and the self-energy correction in Fig.~\ref{fig:NLOvirSEdyn} provide imaginary parts that originate from loop integration. Generally, while both light-cone fractions $x$ and $x^\prime$ are fixed by $\delta$-functions for the SGP contributions, i.e., $x=x^\prime=x_0$, the situation is different for the imaginary parts of the loop diagrams. The light-cone fraction $x=x_0$ is still fixed by the 
limited phase space, but we have to integrate over $x^\prime$, in the upper example from $x=x_0$ to $1$. We may transform $x^\prime=x_0/w$, and apply the symmetry properties of the quark-gluon-quark correlation functions \eqref{eq:Symmetry}. Interestingly, the resulting formula,
\be
\int_{v_0}^{v_1}\dd v\int_{x_0}^1\tfrac{\dd w}{w}\,D_1^q(\tfrac{1-v_1}{1-v})\,\left[\hat{\sigma}_1^{\Im,x^\prime>x}(x=x_0,x^\prime=\tfrac{x_0}{w},\varepsilon)\,F^q(\tfrac{x_0}{w},x_0)+\hat{\sigma}_5^{\Im,x^\prime>x}(x=x_0,x^\prime=\tfrac{x_0}{w},\varepsilon)\,G^q(\tfrac{x_0}{w},x_0)\right]\,,\label{eq:ImloopUp1}
\ee
has the same form as that for the 
\emph{hard-pole} contributions constituted by the subtraction terms in the second lines of both Eqs.~\eqref{eq:dynIntmodF} and \eqref{eq:dynIntmodG}. Indeed, we find that it is necessary to add the contribution \eqref{eq:ImloopUp1} to the hard-pole contributions originating from real NLO corrections to cancel all collinear $1/\varepsilon$ poles. This feature gives us much confidence in our result.

There is yet another region $x^\prime<0$ where we find non-zero imaginary parts, provided by the loop integrals of the middle and right box diagrams of Fig.~\ref{fig:NLOvirboxdyn}. We can again convert this contribution by a change of variable $x=x_0$ and $x^\prime=-\tfrac{1-w}{w}x_0$, use the symmetry features \eqref{eq:Symmetry}, and obtain the schematic formula
\be
\int_{v_0}^{v_1}\dd v\int_{x_0}^1\tfrac{\dd w}{w}\,D_1^q(\tfrac{1-v_1}{1-v})\,\left[\hat{\sigma}_1^{\Im,x^\prime<0}(x_0,-\tfrac{1-w}{w}x_0,\varepsilon)\,F^q(-\tfrac{1-w}{w}x_0,x_0)+\hat{\sigma}_5^{\Im,x^\prime<0}(x_0,-\tfrac{1-w}{w}x_0,\varepsilon)\,G^q(-\tfrac{1-w}{w}x_0,x_0)\right]\,.\label{eq:ImloopDown1}
\ee
In this expression the support of the quark-gluon-quark correlation functions is probed for negative $x=-\tfrac{1-w}{w}x_0$. It turns out that \eqref{eq:ImloopDown1} has the same form as the hard-pole contributions found in the $qq\to q$ channel. In this channel, we encounter slightly different twist-3 matrix elements (compared to \eqref{eq:DefPhiFmain} in the $qg\to q$ channel). These matrix elements may be interpreted as \emph{quark-antiquark-gluon} correlation functions. We will discuss this channel in more detail later in Sec. \ref{sub:qq2q}. At this point, we mention that, again, the contribution \eqref{eq:ImloopDown1} is crucial for cancelling the $1/\varepsilon$ poles in the hard-pole contributions to the $qq\to q$ channel.

Although the virtual corrections consist of several loop-diagrams for kinematical and dynamical twist-3 effects, we eventually arrive at a quite compact result that we present in the following formula:
\bea
E_h\frac{\dd \sigma_{\mathrm{NLO,\,virt}}}{\dd^{d-1}\mathbf{P}_h}& = & \sigma_0(S)\,\frac{\alpha_s}{2\pi}\int_{v_0}^{v_1}\dd v\int_{x_0}^1\frac{\dd w}{w}\,\sum_q\,e_q^2\,(\tfrac{1-v_1}{1-v})^{2\varepsilon}\,D_1^q(\tfrac{1-v_1}{1-v})\times\nn\\
&&\left[\hat{\sigma}_{\mathrm{virt,SGP}}(v,w,\chi_\mu,\varepsilon)\,\left((1+\varepsilon)\,F^{q}-x_0\,(F^q)^\prime\right)(x_0,x_0)\right.\nn\\
&& + \hat{\sigma}_{\mathrm{virt,\Im,x^\prime>x}}(v,w,\chi_\mu,\varepsilon)\,\left(F^{q}+G^q\right)(\tfrac{x_0}{w},x_0)\nn\\
&& \left.+ \hat{\sigma}_{\mathrm{virt,\Im,x^\prime<0}}(v,w,\chi_\mu,\varepsilon)\,\left(F^{q}+G^q\right)(-\tfrac{1-w}{w}x_0,x_0)\,\right]\,,\label{eq:NLOvirResult}
\eea
where the partonic factors can be given to all orders in $\varepsilon$, with $\chi_\mu\equiv \frac{s\,u}{t\,\mu^2}$ and $r_\Gamma\equiv \frac{\Gamma(1+\varepsilon)\Gamma^3(1-\varepsilon)}{\Gamma(1-2\varepsilon)}$,
\bea
\hat{\sigma}_{\mathrm{virt,SGP}}(v,w,\chi_\mu,\varepsilon) & = & C_F\,S_\varepsilon\,r_\Gamma\,\chi_\mu^{-\varepsilon}(1-v)^{-2\varepsilon}v^\varepsilon\,\hat{\sigma}_{\mathrm{LO}}(v,w,\varepsilon)\,\frac{-(2-\varepsilon+2\varepsilon^2)}{\varepsilon^2\,(1-2\varepsilon)},\nn\\
\hat{\sigma}_{\mathrm{virt,\Im,x^\prime>x}}(v,w,\chi_\mu,\varepsilon) & = & S_\varepsilon\,\frac{\Gamma^2(1-\varepsilon)}{\Gamma(1-2\varepsilon)}\chi_\mu^{-\varepsilon}(1-v)^{-2\varepsilon}v^\varepsilon
\frac{v(1+v)}{(1-v)^4}\,\frac{w^{1+\varepsilon}}{(1-w)^{1+\varepsilon}}\frac{N_c(1-\varepsilon-\varepsilon^2)+C_F\,\varepsilon(1+\varepsilon)}{(1-\varepsilon)(1-2\varepsilon)},\nn\\
\hat{\sigma}_{\mathrm{virt,\Im,x^\prime<0}}(v,w,\chi_\mu,\varepsilon) & = & S_\varepsilon\,\frac{\Gamma^2(1-\varepsilon)}{\Gamma(1-2\varepsilon)}\chi_\mu^{-\varepsilon}(1-v)^{-2\varepsilon}v^\varepsilon
\frac{v(1+v)}{(1-v)^4}\,\frac{w^{1+\varepsilon}}{(1-w)^{\varepsilon}}\frac{(2\,C_F-N_c)(1-\varepsilon-\varepsilon^2)}{\varepsilon\,(1-\varepsilon)(1-2\varepsilon)}.\label{eq:NLOvirtpartCS}
\eea
Interestingly, the virtual SGP correction $\hat{\sigma}_{\mathrm{virt,SGP}}$ is proportional only to the color factor $C_F$ even though several loops proportional to the other color factor $N_c$ initially also contribute (like the box diagram in Fig.~\ref{fig:NLOvirboxdyn} (left)). In other words, all terms with a color factor $N_c$ entering the virtual SGP correction eventually cancel. We also note that $\hat{\sigma}_{\mathrm{virt,SGP}}$ is proportional to $\hat{\sigma}_{\mathrm{LO}}$ of Eq.~\eqref{eq:LOpartonic} that constitutes the LO partonic factor in \eqref{eq:LO}. If we expand the remaining factor in $\hat{\sigma}_{\mathrm{virt,SGP}}$,
\[\frac{-(2-\varepsilon+2\varepsilon^2)}{\varepsilon^2\,(1-2\varepsilon)}=-\frac{2}{\varepsilon^2}-\frac{3}{\varepsilon}-8+\mathcal{O}(\varepsilon)\,,\]
we observe that the soft and collinear divergences are exactly the same compared to the ones we encounter for the unpolarized cross section (see Eq.~(8) of Ref.~\cite{Hinderer:2015hra}). This particular feature has also been observed in Ref.~\cite{Kang:2012ns} in polarized SIDIS. We note that the $-2/\varepsilon^2$-pole must cancel once the virtual and real NLO corrections are combined, and we do observe this cancellation.

The partonic cross section $\hat{\sigma}_{\mathrm{virt,\Im,x^\prime>x}}$ in \eqref{eq:NLOvirResult} yields a $1/\varepsilon$-pole indirectly through the expansion \eqref{eq:epsExpandsion} of the term $1/(1-w)^{1+\varepsilon}$. Since this pole is proportional to a $\delta(1-w)$-function, this pole may also be attributed to the SGP contribution. In any case, this $1/\varepsilon$-pole is needed to obtain a finite final result.

\subsection{Channel \texorpdfstring{$qg\to q$}{qg to q}: Real Corrections\label{sub:qg2q}}

In this section, we discuss further features that we encounter in the calculation of the real-gluon emission corrections. The relevant NLO Feynman diagrams for the kinematical twist-3 contributions generated by the first moment of the Sivers function are shown in Fig.~\ref{fig:NLOkinqg2qreal}. These kinematical contributions can be combined with the corresponding SGP contributions through \eqref{eq:SiversSGP}. As was already mentioned in Section \ref{sub:Setup}, the kinematical NLO contributions generated by real-gluon emission 
are gauge-dependent, that is, the gauge parameter $\kappa$ of \eqref{eq:LCpolsummain} explicitly appears in these expressions. This gauge dependence eventually cancels once the kinematical and SGP contributions are combined. Since the $qg\to q$ channel served as an example of the general setup of our calculation in Section \ref{sub:Setup}, most of the relevant details of the calculations have been discussed there and we refrain from repeating the discussion here.

The NLO Feynman diagrams for the real-gluon emissions within the dynamical twist-3 contributions are shown in Fig.~\ref{fig:NLOdynqg2qreal}. The procedure of how we calculate these NLO contributions has also been described in Section \ref{sub:Setup}. This procedure first involves handling the \emph{integral} contributions as in Eq.~\eqref{eq:dynInt3} as well as the subtraction terms in Eqs.~\eqref{eq:dynIntmodF},\eqref{eq:dynIntmodG}. As discussed in Sec. \ref{sub:Setup}, the subtraction terms generate HP contributions as well as SGP and SFP ones that must be combined with \emph{direct} SGP and SFP contributions.

At this point, we would like to add a remark about a technical feature within the calculation of these HP, SGP effects that does not appear in NLO calculations for leading-twist observables. In Sec.~\ref{sub:KinTw3} we already encountered an unusual $1/(1-w)^{2+2\varepsilon}$-term. Such terms are also found in the NLO calculation of the HP and SGP subtraction terms. Even more singular terms $1/(1-w)^{3+\varepsilon}$ emerge. We deal with these using analytic continuation based on integration by parts identities. The procedure is described in detail in Appendix \ref{sub:analContinue}, and it leads to first and second derivatives of the quark-gluon-quark correlation functions $F(x,x^\prime)$ and $G(x,x^\prime)$ entering our final result.

To be specific, we first extract the various $1/(1-w)^{n+\varepsilon}$ terms of a generic HP or SGP function $\sigma(v,w,\varepsilon)$ as follows:
\bea
\sigma(v,w,\varepsilon) & = & \frac{\sigma_{3}(v,\varepsilon)}{(1-w)^{3+\varepsilon}}+\frac{\sigma_{2}(v,\varepsilon)}{(1-w)^{2+\varepsilon}}+\frac{\sigma_{1}(v,\varepsilon)}{(1-w)^{1+\varepsilon}}+\sigma_{\mathrm{reg}}(v,w,\varepsilon)\,,\label{eq:1mwpoles}
\eea
where the remainder $\sigma_{\mathrm{reg}}$ is a function that is integrable over $x_0<w<1$. For the term $1/(1-w)^{1+\varepsilon}$, one can use the familiar decomposition in Eq.~\eqref{eq:epsExpandsion}. The other terms, $1/(1-w)^{2+\varepsilon}$ and $1/(1-w)^{3+\varepsilon}$, are then treated according to Appendix \ref{sub:analContinue}. The explicit formulas are also collected there. 
\subsubsection{Hard Poles \label{subsub: qgq HP}}
After applying all necessary integration by parts identities for the HP contributions,
one ends up with terms proportional to the derivatives $-x_0\,\partial_1(F,G)\left(\tfrac{x_0}{w},x_0\right)$ and $x_0^2\,\partial_1^2(F,G)\left(\tfrac{x_0}{w},x_0\right)$ of the quark-gluon-quark correlation functions. Moreover, the HP subtraction terms in \eqref{eq:dynIntmodF},\eqref{eq:dynIntmodG} also contain the derivatives $-x_0\,\partial_2(F,G)(\tfrac{x_0}{w},x_0)$ which we deal with following the prescription of Appendix \ref{sub:analContinue}. We emphasize that we already combine these HP results at this stage with the $\overline{\mathrm{MS}}$-renormalization terms (third line of Eq.~\eqref{eq:LOsplittingfunctionTw3deriv}) and the hard-pole contributions of the virtual corrections (third line of Eq.~\eqref{eq:NLOvirResult}). Furthermore, we integrate by parts the terms of our result that include derivative terms $x_0^2\,(\partial_1^2 F)(\tfrac{x_0}{w},x_0)$ and $-x_0\,(\partial_1 F)(\tfrac{x_0}{w},x_0)$. We do this for regular terms in the partonic factors, but not for distributional terms $1/(1-w)_+$, $(\tfrac{\ln(1-w)}{1-w})_+$, and $\delta(1-w)$. Useful identities regarding this kind of integration by parts are also listed in Appendix \ref{sub:analContinue}. Eventually, we explicitly observe that this procedure cancels all $1/\varepsilon$ poles except those proportional to a $\delta(1-w)$ distribution. However, we later combine such poles with the corresponding ones for soft-gluon pole contributions.
The cancellation of poles in the HP terms concerns exclusively partonic contributions with a color factor $N_c$, whereas contributions with $C_F$ do not exhibit $1/\varepsilon$-poles. The explicit result for the hard-pole contributions can be found in Eq.~\eqref{eq:Channel1HP} below.

\subsubsection{Soft-Gluon Poles \label{subsub: qgq SGP}}As a next step we focus on the \emph{direct} SGP contributions that are generated directly by propagators \eqref{eq:LOSGP},\eqref{eq:SGPPole} hitting their poles. We first discuss the contributions proportional to the color factor $N_c$ since they do not receive contributions from kinematical twist-3. The direct SGP contribution proportional to $N_c$ can be arranged in the following schematic form:
\be
\int \dd x^\prime \,i\left[\frac{\sigma_2(x)}{(x^\prime-x)(x^\prime-x+i\delta)}+\frac{\sigma_1(x,x^\prime)}{x^\prime-x+i\delta}\right]\,F(x,x^\prime)+\mathrm{c.c.}\,.\label{eq:SGPNc1}
\ee
The second term in \eqref{eq:SGPNc1} clearly sets $x^\prime = x$ in the partonic factor $\sigma_1$ and projects out the ETQS function $F(x,x)$. But what about the first term which turns out to be proportional to $\frac{\delta(x^\prime-x)}{x^\prime-x}$? We may easily replace this term with a derivative of a delta function, i.e. $-\tfrac{\dd}{\dd x^\prime}\delta(x^\prime-x)$, followed by a subsequent integration by parts. In this way a derivative term of the ETQS function $F(x,x)$ is generated, and \eqref{eq:SGPNc1} reads
\be
\frac{2\pi}{x}\,\sigma_2(x)\,(x\,(\partial_2F)(x,x))+2\pi\,\sigma_1(x,x)\,F(x,x)\,.\label{eq:SGPNc2}
\ee
The rest of the calculation of the partonic functions $\sigma_1$ and $\sigma_2$ in \eqref{eq:SGPNc2} is straightforward and follows the lines described above. Some technical details can be found again in the Appendix \ref{sub:analContinue}.

At this point, we observe an explicit cancellation of all $1/\varepsilon$-poles that contribute to the $N_c$-parts of the SGPs and their derivative terms. This cancellation is particularly involved for the $\delta(1-w)$ terms that are generated not only by the $1/(1-w)^{1+\varepsilon}$ expansion of \eqref{eq:epsExpandsion}, but also by the boundary terms of integration by parts in the calculation of the NLO hard poles.\\

\paragraph{Cancellation of collinear poles emerges for the terms proportional to $N_c$:}
For the \emph{second derivative} $F^{\prime\prime}(x,x)$ we observe an explicit nontrivial cancellation of the $1/\varepsilon$-poles for the $\delta(1-w)$ distribution once the SGP and HP contributions are added. 
The remaining collinear poles accompanied by a $1/(1-w)_+$-distribution cancel as well.

The situation is more complicated for the \emph{first derivative} $-x_0\,F^\prime(x,x)$. As far as the $\delta(1-w)$-distribution is concerned, we need to add contributions from \emph{direct} SGP and its derivative term, as well as SGP and HP \emph{subtraction terms} in \eqref{eq:dynIntmodF},\eqref{eq:dynIntmodG}. However, this is not yet enough to cancel all $1/\varepsilon$ poles. In fact, the cancellation is completed by including the renormalization term in the last line of \eqref{eq:LOsplittingfunctionTw3deriv}. For the remaining non-$\delta$-function terms it turns out that the $1/\varepsilon$-poles do not cancel even after the direct SGP terms, subtraction SGP terms and renormalization terms, and additionally the photon-in-lepton contributions (cf. Sec. \ref{sub:PiL}), have been added. However, the plus-distributions at order $1/\varepsilon$ do cancel out, and the remaining $1/\varepsilon$ pole becomes regular with respect to the integration variable $w$. Consequently, we can integrate these pole terms by parts and combine them with the ones appearing with just $F(x,x)$ (without derivative).

Eventually, several contributions to the $\delta(1-w)$-distribution accompanying the ETQS function $F(x,x)$ are accumulated from the direct SGP terms, subtraction SGP and HP terms, renormalization terms (last line of \eqref{eq:LOsplittingfunctionTw3deriv}), boundary terms of integrations by parts involving the first derivative $-x_0\,F^\prime(x,x)$, as well as the imaginary part of the loop diagrams (second line of Eq.~\eqref{eq:NLOvirtpartCS}, the $\delta(1-w)$ term). It is intriguing that the $1/\varepsilon$ poles cancel after adding so many contributions, which again gives us great confidence in our result. In the same way, the $1/\varepsilon$ poles cancel for all the non-$\delta(1-w)$ terms.\\

\paragraph{Cancellation of collinear poles emerges for the terms proportional to $C_F$:}
The complexity of computing the $C_F$ parts of the SGP NLO contributions is even larger due to the kinematical twist-3 contributions. The technical difficulties of this particular contribution have already been discussed in Sec. \ref{sub:KinTw3}. We emphasize again that the kinematical twist-3 contributions are an integral part for 
obtaining color gauge invariance of partonic cross sections. Apart from this feature, the computation of the various partonic cross sections is similar to that for the $N_c$-terms described above. In particular, we also observe a cancellation of all $1/\varepsilon$-poles. In contrast to the $N_c$ part, we need to include not only the SGP renormalization terms of the ETQS function in \eqref{eq:LOsplittingfunctionTw3deriv} but also the renormalization terms of the twist-2 fragmentation function in Eq.~\eqref{eq:renFF}. On top of that, not only do we need to include the imaginary parts of the loop diagrams (second line of Eq.~\eqref{eq:NLOvirtpartCS}, finite as $\varepsilon\to 0$), but also their SGP contributions (first line of Eq.~\eqref{eq:NLOvirtpartCS}). The inclusion of the latter terms guarantees the cancellation of the $1/\varepsilon^2$-poles as well as the cancellation of the $1/\varepsilon$-poles for the $\delta(1-w)$-distribution of the $C_F$ part of the partonic cross section accompanying the function $F(x,x)$. In addition, also the non-$\delta(1-w)$ terms can be arranged in such a way that eventually all $1/\varepsilon$ poles cancel. The full analytic NLO results for the SGP contributions, valid in $d=4$ dimensions, are presented in the next section, Eq.~\eqref{eq:Channel1SGP}.

As a last remark we mention again that we organize our analytic SGP results such that all regular, non-distributional terms in the partonic cross section for the \emph{derivative} terms are integrated by parts so that only the $\delta(1-w)$- and plus distributions remain in these cross sections. This is similar to what was done for the non-distributional HP derivative terms, and corresponding identities that we found useful for this task can again be found in Appendix \ref{sub:analContinue}.

\subsubsection{Soft-Fermion Poles \label{subsub: qgq SFP}} The partonic channel $qg\to q$ also receives contributions from soft-fermion poles. Similarly to hard and soft-gluon poles, there are SFP contributions that originate from the appropriate SFP \emph{subtraction terms} in Eqs.~\eqref{eq:dynIntmodF}, \eqref{eq:dynIntmodG}. The calculation of these partonic cross sections that accompany the SFPs $F(\tfrac{x_0}{w},0)$, $G(\tfrac{x_0}{w},0)$, and their derivative terms $-x_0\,(\partial_2 F)(\tfrac{x_0}{w},0)$, $-x_0\,(\partial_2 G)(\tfrac{x_0}{w},0)$ is straightforward. We do not meet any of the complications in the soft limit $w\to 1$. Consequently, we do not need to calculate the SFP factors to all orders in $\varepsilon$ and we can directly work with an expansion in $\varepsilon$. Additional ``direct'' contributions to the SFP partonic factors originate from propagators in the diagrams in Fig.~\ref{fig:NLOdynqg2qreal} where an imaginary SFP part is generated by a propagator such as \eqref{eq:SFPPole} hitting its pole. One also needs to take into account additional SFP contributions from the photon-in-lepton kinematics of Eq.~\eqref{eq:WWdyntw3} (see discussion in Sec. \ref{sub:PiL}). 

One peculiarity we observe is that a collinear $1/\varepsilon$-pole still remains even after the aforementioned SFP partonic factors originating from the SFP subtraction terms, ``direct'' SFP terms, and photon-in-lepton SFP terms are added, indicating that there are additional so far unaccounted contributions. Indeed, it turns out that also another partonic channel, the $qq\to q$ channel to be discussed in the next section, generates contributions proportional to the SFP matrix elements $F(\tfrac{x_0}{w},0)$, $G(\tfrac{x_0}{w},0)$. 
\begin{figure}[tbp]
\centering
\includegraphics[width=0.5\textwidth]{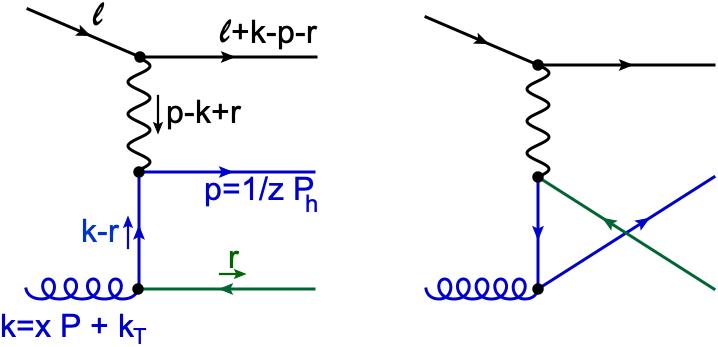}
\caption{NLO diagrams relevant for the $qq\to q$ channel
with a gluon in the initial state.
One needs to compute the interference of the sum of these diagrams with 
the sum of the diagrams in Fig.~\ref{fig:NLOdynqq2qreal}, with $k_T=0$
.\label{fig:NLOkinqq2qreal}}
\end{figure}
We combine the corresponding contributions with the SFP contributions of the $qg\to q$ channel and find that eventually all $1/\varepsilon$ poles cancel, resulting in finite and well-behaved SFP cross sections\footnote{Interestingly, a similar cancellation of SFP contributions between different partonic channels has been observed earlier in Ref.~\cite{Koike:2007dg} in SIDIS.}. Since SFP renormalization terms are present neither in~\eqref{eq:LOsplittingfunctionTw3deriv} nor in the LO cross section \eqref{eq:LO}, there is no dependence on the renormalization scale $\mu$. However, the SFP cross sections will depend on the lepton mass $m_\ell$ due to the photon-in-lepton contributions.

The partonic cross sections accompanying the SFP derivative terms $-x_0\,(\partial_2 F)(\tfrac{x_0}{w},0)$ and $-x_0\,(\partial_2 G)(\tfrac{x_0}{w},0)$ are generated only by the SFP subtraction terms \eqref{eq:dynIntmodF},\eqref{eq:dynIntmodG} and are finite.

\subsection{Channel \texorpdfstring{$qq\to q$}{qq to q}\label{sub:qq2q}}

\begin{figure}[tbp]
\centering
\includegraphics[width=0.5\textwidth]{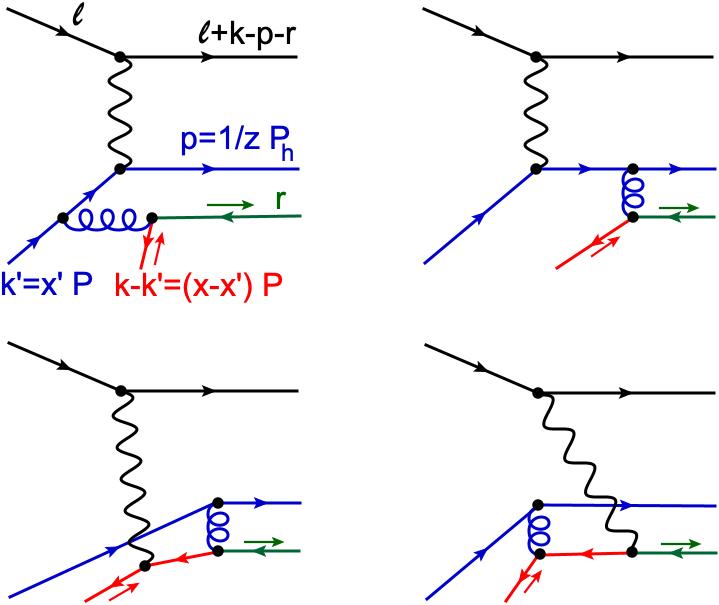}
\caption{NLO diagrams in the $qq\to q$ channel with a quark-antiquark pair in the initial state. These diagrams are relevant for the dynamical twist-3 contribution of the $qq\to q$ channel. 
One needs to compute the interference of the sum of the diagrams with the sum of the diagrams in Fig.~\ref{fig:NLOkinqq2qreal}, with $k_T=0$.\label{fig:NLOdynqq2qreal}}
\end{figure}

The calculation of the twist-3 contributions in the $qq\to q$ channel is similar to the one we described in Sect.~\ref{sub:qg2q} for the $qg\to q$ channel. The
relevant diagrams are shown in Figs.~\ref{fig:NLOkinqq2qreal} and~\ref{fig:NLOdynqq2qreal}. This channel is less complicated because there are neither kinematical twist-3 contributions nor soft-gluon pole contributions. Soft divergences in the limit $w\to 1$ can be handled in the usual way by means of Eq.~\eqref{eq:epsExpandsion}. 

Curiously, since this channel probes the quark-antiquark content of the nucleon's wave function there are several symmetries between quark-antiquark exchanges in the initial state and in the fragmentation of quarks and antiquarks in the final state. As it turns out, because of these symmetries, it is sufficient to calculate the interference of the diagrams in Fig.~\ref{fig:NLOdynqq2qreal} with those in Fig.~\ref{fig:NLOkinqq2qreal} only. Strictly speaking, this interference would point towards a channel of a quark-antiquark-gluon correlation followed by a quark fragmentation in the final state, hence labeled $q\bar{q}\to q$. However, all other channels such as $q\bar{q}\to \bar{q}$, $\bar{q}q\to q$, and $\bar{q}q\to \bar{q}$ can be linked to the $q\bar{q}\to q$ channel. Therefore, we generically refer to it as ``$qq\to q$ channel''.

Field-theoretically, the hadronic matrix element that enters the factorized description of the $qq\to q$ channel is the quark-gluon-quark correlator \eqref{eq:DefPhiFmain}. However, this correlator is evaluated at different light-cone fractions, that is, it enters as $\Phi^{q,\rho}_{F,ij}(x^\prime-x,x^\prime)$. 
The reason for this is that an antiquark and a gluon exchange their role in comparison with the $qg\to q$ channel. This flip of antiquark and gluon legs is reflected in a change of the variable $x\to x^\prime - x$. The schematic form of the factorization formula for the channel $qq\to q$ reads (cf. Eq.~\eqref{eq:dynInt} for the $qg\to q$ channel)
\be
\int\tfrac{\dd z}{z^2}\int \dd x \int_{x-1}^1 \dd x^\prime \frac{i}{x}\, \left[\,\hat{\sigma}^{qq\to q,1}(x,x^\prime,z)\,F^q(x^\prime-x,x^\prime)+\,\hat{\sigma}^{qq\to q,5}(x,x^\prime,z)\,G^q(x^\prime - x,x^\prime)\,\right]D_1^q(z)+\mathrm{c.c.}.\label{eq:dynIntqqq}
\ee
The partonic factors $\hat{\sigma}^{qq\to q,1,5}(x,x^\prime,z)$ are, as before, constructed as $(d-1)$-dimensional phase space integrals of interfering diagrams of Figs.~\ref{fig:NLOkinqq2qreal} and \ref{fig:NLOdynqq2qreal}. Interestingly, it turns out that the upper and lower diagrams of Fig.~\ref{fig:NLOdynqq2qreal}, i.e., those with a coupling of the exchanged virtual photon to either the quark or the antiquark, form two independent classes of interference effects that may be treated independently. Both of these classes carry a color factor $C_F-N_c/2=-\frac{1}{2N_c}$ that leads to a $1/N_c^2$ suppression compared to the channel $qg\to q$. To be specific, we may split the partonic functions in \eqref{eq:dynIntqqq} into two parts,
\bea
\hat{\sigma}^{q\bar{q}\to q,1,5}(x,x^\prime,z) &=& -\frac{1}{2N_c}\,\left[\hat{\sigma}_1^{q\bar{q}\to q,1,5}(x,x^\prime,z)+\hat{\sigma}^{q\bar{q}\to q,1,5}_2(x,x^\prime,z) \right]\,,\label{eq:qqqsplit}
\eea
where $\hat{\sigma}_{1,2}$ in \eqref{eq:qqqsplit} refer to the interferences of the upper/lower two diagrams in Fig.~\ref{fig:NLOdynqq2qreal} with those in Fig.~\ref{fig:NLOkinqq2qreal}, respectively. The symmetries between quarks/antiquarks in the initial and final states mentioned above occur via a change of integration variables $x^\prime\to x-x^\prime$ in \eqref{eq:dynIntqqq}. Specifically, we find the following symmetries which allow us to restrict ourselves to the configuration $q\bar{q}\to q$ shown in Fig.~\ref{fig:NLOdynqq2qreal} only:
\bea
\hat{\sigma}_{1,2}^{q\bar{q}\to q,1}(x,x^\prime,z) & = & \hat{\sigma}_{2,1}^{q\bar{q}\to \bar{q},1}(x,x-x^\prime,z) = \hat{\sigma}_{2,1}^{\bar{q}q\to q,1}(x,x-x^\prime,z)\,,\nn\\
\hat{\sigma}_{1,2}^{q\bar{q}\to q,5}(x,x^\prime,z) & = & -\hat{\sigma}_{2,1}^{q\bar{q}\to \bar{q},5}(x,x-x^\prime,z)=- \hat{\sigma}_{2,1}^{\bar{q}q\to q,5}(x,x-x^\prime,z)\,.\label{eq:qqqsymmetries}
\eea
It turns out that the computational procedure for the two interference effects $\hat{\sigma}_{1}^{q\bar{q}\to q}$ and $\hat{\sigma}_{2}^{q\bar{q}\to q}$ differs. In order to demonstrate this feature, we start the NLO calculation of the $qq\to q$ channel with the integral contributions as described in Sect.~\ref{sub:qg2q}. We find that the corresponding 
partonic functions carry different denominators (cf. discussion above \eqref{eq:dynIntmodF}) and modify them as follows:
\bea
\hat{\sigma}_{1,\mathrm{Int}}^{qq\to q,1,5}(v,w,\zeta) & \equiv & \zeta\,(w-\zeta)^2\,\hat{\sigma}_{1}^{qq\to q,1,5}(v,w,\zeta)\,,\nn\\
\hat{\sigma}_{2,\mathrm{Int}}^{qq\to q,1,5}(v,w,\zeta) & \equiv & (1-\zeta)\,\hat{\sigma}_{2}^{qq\to q,1,5}(v,w,\zeta)\,.\label{eq:qqqIntmodpart}
\eea
As it was the case for the $qg\to q$ channel, the partonic functions of the integral contributions $\hat{\sigma}_{1,2,\mathrm{Int}}^{qq\to q,1,5}(v,w,\zeta)$ are well-behaved and perfectly integrable. However, the inverses of the prefactors in~(\ref{eq:qqqIntmodpart}) need to be 
shifted to the correlation functions $F,G(-(1-\zeta)\,\tfrac{x_0}{w},\zeta\,\tfrac{x_0}{w})$ which in turn requires new subtraction terms (cf. Eqs.~\eqref{eq:dynIntmodF},\eqref{eq:dynIntmodG}) of the following types:
\bea
F_{\mathrm{Int},1}^{qq\to q}(x_0,w,\zeta) & \equiv & \frac{1}{\zeta\,(w-\zeta)^2}\,\left[ F^q\left(-(1-\zeta)\,\tfrac{x_0}{w},\zeta\,\tfrac{x_0}{w}\right) \right. \nn\\
 &&\hspace{-3cm}\left.- \frac{\zeta\,( 2w-\zeta)}{w^2}\,F^q\left(-\tfrac{1-w}{w}x_0,x_0\right) - \frac{\zeta\,( \zeta-w)}{w^2}\,x_0\,[\partial_1F^q+\partial_2F^q]\left(-\tfrac{1-w}{w}x_0,x_0\right) -  \frac{(w-\zeta)^2}{w^2}\,F^q\left(-\tfrac{x_0}{w},0\right)\right]\,,\label{eq:qqqIntmodF1}
\eea
\bea
G_{\mathrm{Int},1}^{qq\to q}(x_0,w,\zeta) & \equiv & \frac{1}{\zeta\,(w-\zeta)^2}\,\left[ G^q\left(-(1-\zeta)\,\tfrac{x_0}{w},\zeta\,\tfrac{x_0}{w}\right) \right. \nn\\
 &&\hspace{-3cm}\left.- \frac{\zeta\,( 2w-\zeta)}{w^2}\,G^q\left(-\tfrac{1-w}{w}x_0,x_0\right) - \frac{\zeta\,( \zeta-w)}{w^2}\,x_0\,[\partial_1G^q+\partial_2G^q]\left(-\tfrac{1-w}{w}x_0,x_0\right)-  \frac{(w-\zeta)^2}{w^2}\,G^q\left(-\tfrac{x_0}{w},0\right)\right]\,,\label{eq:qqqIntmodG1}
\eea
and
\bea
F_{\mathrm{Int},2}^{qq\to q}(x_0,w,\zeta) & \equiv & \frac{1}{1-\zeta}\,\left[ F^q\left(-(1-\zeta)\,\tfrac{x_0}{w},\zeta\,\tfrac{x_0}{w}\right)-  \,F^q\left(0,\tfrac{x_0}{w}\right)\right] \,,\label{eq:qqqIntmodF2}
\eea
\bea
G_{\mathrm{Int},2}^{qq\to q}(x_0,w,\zeta) & \equiv & \frac{1}{1-\zeta}\,\left[ G^q\left(-(1-\zeta)\,\tfrac{x_0}{w},\zeta\,\tfrac{x_0}{w}\right)-  \,G^q\left(0,\tfrac{x_0}{w}\right)\right] \,.\label{eq:qqqIntmodG2}
\eea
Applying the subtractions, we end up with integrable functions $F_{\mathrm{Int},1/2}^{qq\to q}$ and $G_{\mathrm{Int},1/2}^{qq\to q}$. 

Next, we focus on \emph{hard pole} contributions. The only source of such contributions are the subtraction terms in the second lines of Eqs.~\eqref{eq:qqqIntmodF1},\eqref{eq:qqqIntmodG1}. They can be calculated as described in Sect.~\ref{sub:qg2q}. An infrared singularity of the type $1/(1-w)^{1+2\varepsilon}$ emerges after phase space integration, and we can readily regularize it via Eq.~\eqref{eq:epsExpandsion}. We note that we require neither photon-in-lepton contributions (see Sect.~\ref{sub:PiL}) nor $\overline{\mathrm{MS}}$-renormalization terms for the fragmentation function in \eqref{eq:renFF} to cancel the collinear $1/\varepsilon$-pole for the hard-pole contribution. 
However, we do need to add the NLO $\overline{\mathrm{MS}}$-renormalization terms of the soft-gluon pole matrix element in the last four lines of Eq.~\eqref{eq:LOsplittingfunctionTw3deriv} for a proper removal of collinear divergences of the hard-pole contributions and their derivative terms (see the second line of \eqref{eq:qqqIntmodF1},\eqref{eq:qqqIntmodG1}). Interestingly, there is yet another term originating from the imaginary part of the virtual contributions that is essential for the cancellation of the divergences of the hard-pole contribution. This term has already been discussed in Sect.~\ref{sub:virt} and presented in the last two lines of Eq.~\eqref{eq:NLOvirtpartCS}. After adding all terms, we end up with a well-defined and finite NLO result for the hard-pole contribution for the $qq\to q$ channel.

There are also \emph{soft-fermion poles} that appear as subtraction terms in the second lines of Eqs.~\eqref{eq:qqqIntmodF1},\eqref{eq:qqqIntmodG1}. These SFPs $F,G(-\tfrac{x_0}{w},0)$ correspond to soft-fermion poles for \emph{antiquarks} by virtue of charge conjugation; see the discussion below Eq.~\eqref{eq:DefPhiF}. Interestingly, these antiquark SFPs need to be combined with the \emph{direct} antiquark SFPs, that is, SFPs generated by a propagator hitting its pole similar to \eqref{eq:SFPPole}. However, these direct SFPs appear in the partonic function $\hat{\sigma}_{2}^{qq\to q,1,5}(v,w,\zeta)$ rather than in $\hat{\sigma}_{1}^{qq\to q,1,5}(v,w,\zeta)$. This feature again displays the complicated interplay between various, seemingly different, dynamical twist-3 terms at NLO. Eventually, the collinear divergence cancels once the SFPs $F,G(-\tfrac{x_0}{w},0)$ from photon-in-lepton contributions are added.

In contrast, the SFP contributions $F,G(0,\tfrac{x_0}{w})$ in the subtraction terms of Eqs.~\eqref{eq:qqqIntmodF2},\eqref{eq:qqqIntmodG2} should be combined with direct SFP contributions that are found in $\hat{\sigma}_{1}^{qq\to q,1,5}(v,w,\zeta)$ rather than in $\hat{\sigma}_{2}^{qq\to q,1,5}(v,w,\zeta)$. However, even after adding the corresponding SFP contribution from the photon-in-lepton contribution, a collinear $1/\varepsilon$ pole remains. In fact, this is a feature that we have already described for the SFP contribution in the $qg\to q$ channel (see discussion in \ref{subsub: qgq SFP}). Only if we add the SFP contributions generated by the functions $F,G(0,\tfrac{x_0}{w})$ for the $qg\to q$ and the $qq\to q$ channels does the collinear $1/\varepsilon$ cancel. As mentioned above, we assign the final well-defined NLO SFP result to the $qg\to q$ channel rather than to the $qq\to q$ channel.

The analytical results for the $qq\to q$ channel are shown explicitly in Eq.~\eqref{eq:Channel2}.

\subsection{Channel \texorpdfstring{$qq\to q^\prime$}{qq to q'}}

\begin{figure}[tbp]
\centering
\includegraphics[width=0.5\textwidth]{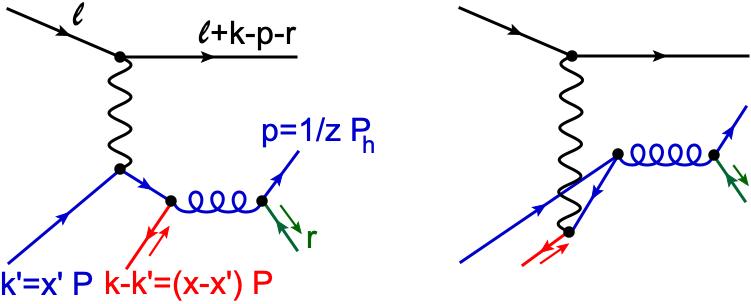}
\caption{NLO diagrams for the $qq\to q^\prime$ channel with a quark-antiquark pair in the initial state. These diagrams are relevant for the dynamical twist-3 contribution. We need to compute
the interference of the sum of the diagrams with the sum of the diagrams in Fig.~\ref{fig:NLOkinqq2qreal}, with $k_T=0$.\label{fig:NLOdynqq2qpreal}}
\end{figure}

The calculation for the $qq\to q^\prime$ channel is quite similar to that for $qq\to q$. The main difference is that the flavor of the fragmenting final-state quark can be different from
that of the initial quark-antiquark pair. In addition, the color factor for this channel is $T_R=\frac{1}{2}$. Therefore, the $qq\to q^\prime$ channel is suppressed only by $\mathcal{O}(1/N_c)$ compared to the $qg\to q$ channel, rather than by $\mathcal{O}(1/N_c^2)$. The relevant partonic functions are generated by an interference of the two diagrams in Fig.~\ref{fig:NLOdynqq2qpreal} with those in Fig.~\ref{fig:NLOkinqq2qreal}. The only reason why this channel leads to a nonvanishing contribution is the difference between quark- and antiquark-fragmentation into the observed hadron. In contrast, this channel vanishes for jet production. However, it may become more important in comparison to the other partonic channels if heavy quarks are studied, for example, in the production of $D$-mesons.

The $qq\to q^\prime$ channel is a standalone channel in the sense that it is finite and well-behaved on its own. It
does not require any $\overline{\mathrm{MS}}$-factorization terms from either the LO SGP matrix elements or the quark fragmentation function to cancel collinear singularities. However, the channel does require the implementation of photon-in-lepton contributions generating logarithms $\ln\left(\tfrac{s\,u}{m^2_\ell\,t}\right)$.

From a technical point of view, the calculation can be performed just as described in the previous section for the $qq\to q$ channel. We observe an antisymmetry of the partonic functions if the fragmenting final state quark is replaced by an antiquark,
\bea
\hat{\sigma}^{q\bar{q}\to q^\prime,1,5}(x,x^\prime,z)& = &-\hat{\sigma}^{q\bar{q}\to \bar{q}^\prime,1,5}(x,x^\prime,z)\,.\label{eq:qq2qpFSantisym}
\eea
As a result, the $qq\to q^\prime$ contribution will enter with the difference of quark- and antiquark fragmentation functions in our final NLO formulas.
Furthermore, we find additional symmetry properties that reflect the behavior under an exchange of quark and antiquark in the initial state:
\bea
\hat{\sigma}^{q\bar{q}\to q^\prime,1}(x,x^\prime,z)& = &-\hat{\sigma}^{q\bar{q}\to q^\prime,1}(x,x-x^\prime,z)\,,\nn\\
\hat{\sigma}^{q\bar{q}\to q^\prime,5}(x,x^\prime,z)& = &+\hat{\sigma}^{q\bar{q}\to q^\prime,5}(x,x-x^\prime,z)\,.\label{eq:qq2qpISantisym}
\eea 
This feature ensures that, like the fragmentation functions, also the $qgq$ correlation functions enter as \emph{differences} of their quark/antiquark content.

Following the same computational procedure, we find that the partonic cross sections of the integral contributions carry a potentially divergent denominator $\frac{1}{\zeta(1-\zeta)}$ that is extracted and combined with the $qgq$-functions with a subsequent subtraction of SFP terms. We define,
\bea
\hspace{-0.8cm}F_{\mathrm{Int}}^{qq\to q^\prime}(x_0,w,\zeta) & \equiv & \frac{F^q\left(-(1-\zeta)\,\tfrac{x_0}{w},\zeta\,\tfrac{x_0}{w}\right)-  \zeta\,F^q\left(0,\tfrac{x_0}{w}\right)-  (1-\zeta)\,F^q\left(-\tfrac{x_0}{w},0\right)}{\zeta(1-\zeta)} \,,\label{eq:qqqpIntmodF}\\
\hspace{-0.8cm}G_{\mathrm{Int}}^{qq\to q^\prime}(x_0,w,\zeta) & \equiv & \frac{G^q\left(-(1-\zeta)\,\tfrac{x_0}{w},\zeta\,\tfrac{x_0}{w}\right)-  \zeta\,G^q\left(0,\tfrac{x_0}{w}\right)-  (1-\zeta)\,G^q\left(-\tfrac{x_0}{w},0\right)}{\zeta(1-\zeta)} \,.\label{eq:qqqpIntmodG}
\eea
These two combinations generate well-behaved and finite integral contributions.

We next turn our attention to the \emph{soft-fermion pole} contributions. From Eqs. \eqref{eq:qqqpIntmodF},\eqref{eq:qqqpIntmodG} we see that two types of SFPs appear as subtraction terms. Together with photon-in-lepton SFPs, these are the only sources of such SFPs. In contrast to the $qq\to q$ channel, there are no direct SFP contributions. It turns out that both types of SFPs, $(F,G)(0,x)$ and $(F,G)(-x,0)$, can be related to each other via charge conjugation and can be interpreted as their corresponding antiquark distributions. As mentioned before, together with the symmetry properties \eqref{eq:qq2qpFSantisym},\eqref{eq:qq2qpISantisym}, this feature causes the contributions for this channel to be driven by the difference of quark- and antiquark correlations.

We present our analytical results below in Eq.~\eqref{eq:Channel3}.

\subsection{Channel \texorpdfstring{$qg\to g$}{qg to g}}

The computation for this channel is quite similar to that for $qg\to q$ discussed previously. Indeed, the partonic factors originate from an interference of the diagrams in Fig.~\ref{fig:NLOkinqg2qreal} among themselves (for the kinematical twist-3 contribution) and from an interference of the diagrams in Figs.~\ref{fig:NLOkinqg2qreal} and~\ref{fig:NLOdynqg2qreal} (for the dynamical twist-3 contributions). However, for $qg\to g$ the observed hadron arises from fragmentation of the final-state gluon in Figs.~\ref{fig:NLOkinqg2qreal}, \ref{fig:NLOdynqg2qreal} rather than from quark fragmentation. Technically, this means that we have to exchange the labels of the external quark and gluon momenta, i.e. $r^\mu \leftrightarrow p^\mu$ in the diagrams in Figs.~\ref{fig:NLOkinqg2qreal}, \ref{fig:NLOdynqg2qreal}. Of course, this exchange modifies the momentum flow within these diagrams, which will lead to different analytical results. In addition, our factorization formula for this channel will incorporate the gluon fragmentation function $D_1^g$ in Eq.~\eqref{eq:DefFFg} rather than the quark fragmentation function. Since the gluon fragmentation function makes its first appearance at NLO level, $\overline{\mathrm{MS}}$-renormalization of the quark fragmentation function via Eq.~\eqref{eq:renFF} is the only ingredient needed for the cancellation of collinear singularities in the hard partonic part for $qg\to g$.

The technical procedure is similar to the one we described for the $qg\to q$-channel. Since the behavior in the $w\to 1$ limit is simpler here, the calculation is less involved. In particular, we do not encounter any hard-pole contributions. Remarkably, one can readily show that in this channel the $C_F$-part of the ``direct'' SGP contributions cancels against that for the kinematical twist-3 contributions. Consequently, the ``direct'' SGP contributions are proportional to the color factor $N_c$, and are thus a result of the non-abelian nature of QCD.

However, there are also SGP contributions with both color factors $C_F$ and $N_c$ that appear through subtraction terms in the integral contributions. 
We find that the partonic functions for the integral contributions, calculated as described below Eq.~\eqref{eq:dynInt}, carry a common singular denominator $\frac{1}{\zeta(1-\zeta)^2}$. As before, we move this denominator from the partonic factors to the $qgq$-correlation functions. In the same way as described earlier, we then need to subtract SGP and SFP terms to ensure integrability of the integral contributions. This procedure requires the introduction of the following integrable combinations of correlation functions:
\be
F_{\mathrm{Int}}^{qg\to g}(x_0,w,\zeta) \equiv  \frac{1}{\zeta\,(1-\zeta)^2}\Bigg[F^q(\tfrac{x_0}{w},\zeta\,\tfrac{x_0}{w})-\zeta(2-\zeta)\,F^q(\tfrac{x_0}{w},\tfrac{x_0}{w})+\tfrac{\zeta\,(1-\zeta)}{2\,w}\,x_0\,(F^q)^\prime(\tfrac{x_0}{w},\tfrac{x_0}{w})-(1-\zeta)^2\,F^q(\tfrac{x_0}{w},0)\Bigg]\,,\label{eq:qg2gmodIntF}
\ee
\be
G_{\mathrm{Int}}^{qg\to g}(x_0,w,\zeta) \equiv  \frac{1}{\zeta\,(1-\zeta)^2}\Bigg[G^q(\tfrac{x_0}{w},\zeta\,\tfrac{x_0}{w})+\tfrac{\zeta\,(1-\zeta)}{w}\,x_0\,(\partial_2 G^q)(\tfrac{x_0}{w},\tfrac{x_0}{w})-(1-\zeta)^2\,G^q(\tfrac{x_0}{w},0)\Bigg]\,.\label{eq:qg2gmodIntG}
\ee
Adding the corresponding SGP subtraction terms in these equations to the ``direct'' SGP contribution (proportional to $N_c$ only), then adding the corresponding photon-in-lepton contributions and the $\overline{\mathrm{MS}}$-renormalization term in Eq.~\eqref{eq:renFF} (proportional to $C_F$) renders the resulting SGP partonic cross section finite. In obtaining this result, integration by parts of the derivative terms is useful. This is entirely possible for this channel, as no plus-distributions or delta functions contribute to those partonic factors accompanying derivative terms.

Lastly, we also need to add those parts of the subtraction terms in Eqs.~\eqref{eq:qg2gmodIntF},\eqref{eq:qg2gmodIntG} that generate soft-fermion pole contributions to the ``direct'' SFP contributions, along with the photon-in-lepton SFP contributions. It is straightforward to see that the collinear poles cancel, and we obtain, also in this case, a finite result.

The explicit analytical results for the $qg\to g$ channel are given in Eq.~\eqref{eq:Channel4} below.

\subsection{Channel \texorpdfstring{$gg\to q^\prime$}{gg to q'}}

Finally, we highlight the most important features of the remaining channel, $gg\to q^\prime$. This channel features gluon-gluon correlations inside the transversely polarized nucleon and quark fragmentation in the final state. In contrast to the $qg\to q$ and $qg\to g$ channels, where the kinematical twist-3 effects were generated by the first moment of the Sivers function alone, we now have the two kinematical distributions $G_T^{(1)}$ and $\Delta H_T^{(1)}$, both contributing at NLO (for their definition see Eq.~\eqref{eq:DefPhidGlu}). For the kinematical twist-3 effects in this channel, one needs to compute the sum of the diagrams shown in Fig.~\ref{fig:NLOkinqq2qreal} in interference among themselves, and with non-zero $k_T$. 

\begin{figure}[tbp]
\centering
\includegraphics[width=0.7\textwidth]{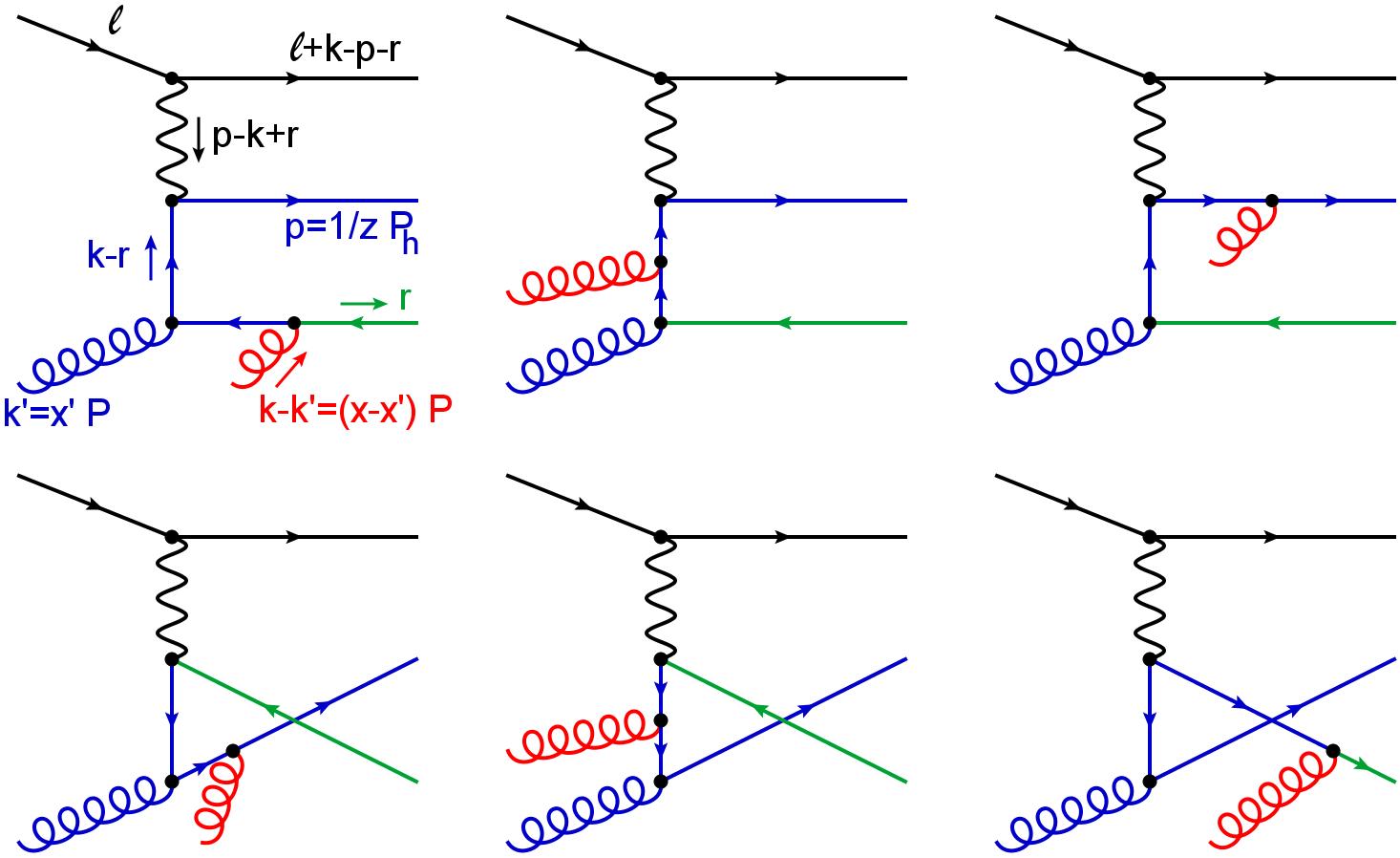}
\caption{NLO diagrams for the $gg\to q^\prime$ channel with a gluon pair in the initial state. These diagrams are relevant for the dynamical twist-3 contribution. 
The sum of the diagrams comes in interference with the sum of the diagrams in Fig.~\ref{fig:NLOkinqq2qreal}, with $k_T=0$.\label{fig:NLOdyngg2qreal}}
\end{figure}

To calculate the dynamical twist-3 effects for the $gg\to q^\prime$ channel, one needs to consider the diagrams in Fig.~\ref{fig:NLOdyngg2qreal}. As far as the color structure of these diagrams is concerned,
we find that both the symmetric $d^{\alpha\beta\gamma}$ and antisymmetric $f^{\alpha\beta\gamma}$ SU(N) structure constants 
appear, and thus both triple-gluon correlators defined in Eqs.~\eqref{eq:DefNF} and \eqref{eq:DefOF} enter the factorized description of the $gg\to q^\prime$ channel. Hence, we obtain the following schematic form of the dynamical part for this channel:
\be
\int\tfrac{\dd z}{z^2}\int \dd x \int \dd x^\prime \frac{i}{xx^\prime(x-x^\prime)} \big[\hat{\sigma}^f_{\sigma\tau\rho}(x,x^\prime,z)\,N_F^{\sigma\tau\rho}(x,x^\prime) + \hat{\sigma}^d_{\sigma\tau\rho}(x,x^\prime,z)\,O_F^{\sigma\tau\rho}(x,x^\prime)\,\big]D_1^q(z)+\mathrm{c.c.}\,.\label{eq:dynIntggqp}
\ee
Remarkably, the Feynman diagrams in Fig.~\ref{fig:NLOdyngg2qreal} that generate the partonic functions $\hat{\sigma}^{f,d}_{\sigma\tau\rho}$ in \eqref{eq:dynIntggqp} do not contain propagators of the form \eqref{eq:PropHP}. Since such propagators \eqref{eq:PropHP} are the only sources of logarithms with negative arguments (such as \eqref{eq:ImLog}) that produce imaginary parts, their absence indicates that there are no \emph{integral} contributions in the $gg\to q^\prime$ channel. This immediately implies that subtraction terms as in Eq.~\eqref{eq:dynIntmodF} $\--$ generating hard-pole and/or other types of pole contributions $\--$ are absent as well. The $gg\to q^\prime$ channel turns out to be the only partonic channel discussed in this paper where a non-zero transverse nucleon SSA originates from propagator poles alone. This feature greatly simplifies the technical procedure to calculate the partonic 
hard-scattering factors for this channel.

In fact, we find that $\hat{\sigma}^{f,d}$ have propagators with a pole at $x^\prime=x$, which we previously called a soft-gluon pole, and at $x^\prime=0$, previously a soft-fermion pole. However, since we are now considering a situation where only gluons in the nucleon are probed by the hard scattering, both cases actually correspond to the scenario where one of the gluons carries a vanishing longitudinal momentum. In addition, the triple-gluon correlators $N_F(x,x^\prime)$ and $O_F(x,x^\prime)$ in Eqs.~\eqref{eq:DefNF},\eqref{eq:DefOF} are parameterized in terms of a combination of the triple-gluon correlation functions $\left(N,O\right)(x,x^\prime),\,\left(N,O\right)(x,x-x^\prime)$ and $\left(N,O\right)(x^\prime,x^\prime-x)$. Therefore, no direct connection can be made between the propagator poles and specific regions of the support of $N,O$. Decomposing the relevant propagators in the hard scattering into a principal value and a delta function (as, e.g., in Eq.~\eqref{eq:LOSGP}), and taking into account the explicit factor $\tfrac{i}{xx^\prime(x-x^\prime)}$ in \eqref{eq:dynIntggqp}, one finds terms proportional to $\tfrac{\delta(x^\prime-x)}{x^\prime-x}$ and $\tfrac{\delta(x^\prime)}{x^\prime}$. This is similar to what happened in Eq.~\eqref{eq:SGPNc1}, and as before these fractions of a $\delta$-function over its argument can be turned into derivatives of the delta functions, i.e. $\tfrac{\delta(x^\prime-x)}{x^\prime-x}=-\tfrac{\dd}{\dd x^\prime}\delta(x^\prime-x)$ and $\tfrac{\delta(x^\prime)}{x^\prime}=-\tfrac{\dd}{\dd x^\prime}\delta(x^\prime)$. Then, after an integration by parts,  terms are obtained that are proportional to $N$ and $O$ and their derivatives at several different arguments. Using the symmetry relations $N(x,x^\prime)=N(x^\prime,x)=-N(-x,-x^\prime)$ and $O(x,x^\prime)=O(x^\prime,x)=O(-x,-x^\prime)$, the only functions remaining in the final expression are $\left(N,O\right)(x,x), \left(N,O\right)(x,0)$ and derivatives $\tfrac{\dd}{\dd x}\left(N,O\right)(x,x),\tfrac{\dd}{\dd x}\left(N,O\right)(x,0)$.

We mention at this point that we performed the calculations for this channel with a \emph{general} light-like vector $n^\mu$ instead of using a specific choice \eqref{eq:choiceLCvectormain}. In fact, we parameterized $n^\mu$ as in Eq.~(59) of  Ref.~\cite{Kanazawa:2015ajw}, with two undetermined parameters indicating the arbitrariness of the choice of the light-cone vector $n^\mu$. As suggested in~\cite{Kanazawa:2015ajw}, it is necessary to relate kinematical and dynamical twist-3 effects to cancel the parameters. Such relations for the gluonic kinematical twist-3 functions $G_T^{(1)}$ and $\Delta H_T^{(1)}$ were presented in~\cite{Koike:2019zxc} in $d=4$ space-time dimensions. We found that the following extensions of the definitions given in Eqs.~(13),(14) of Ref.~\cite{Koike:2019zxc} are required in $d=4-2\varepsilon$ dimensions to ensure that dependence on the light-cone vector $n^\mu$ cancels in the partonic cross sections:
\bea
& G_T^{(1)}(x)\,= \, 4\pi\left(N(x,x)\,-\,(1+\varepsilon)N(x,0)\right)\,, & \nn\\
& \Delta H_T^{(1)}(x) \,= \, -8\pi(1-\varepsilon)N(x,0) .&\label{GandH2N}
\eea
Eventually, we also find for the $gg\to q^\prime$ channel that all collinear $1/\varepsilon$ poles in the partonic cross sections drop out once the 
term of the renormalized SGP function $F^q(x,x)$ 
in the last two lines of Eq.~\eqref{eq:LOsplittingfunctionTw3deriv} is included, along with the corresponding photon-in-lepton contributions to this channel. To facilitate the explicit cancellation of the collinear poles, integration by parts identities \eqref{eq:pISGPidentities} turn out to be useful. 
An intriguing feature of the final result, which can be found in Eq.~\eqref{eq:Channel5} below, is that the partonic hard-scattering factors accompanying the triple-gluon correlation functions $\left(N,O\right)(x,x)$ coincide, while those accompanying $\left(N,O\right)(x,0)$ differ by a sign. 

\section{Analytical Results\label{sec:AnalyticResults}}
In this section we collect the results of the previous sections and present the explicit analytical NLO pQCD formula for the transverse nucleon spin-dependent cross section of the polarized single-inclusive process $\ell+N^\uparrow\to h+X$.

\subsection{Hadron Production}
We first present our NLO results for the production of a hadron with large transverse momentum $P_{hT}$. As elaborated on in the previous section, there are several partonic channels that contribute to the total spin-dependent cross section. We labeled these channels as 1) $qg\to q$, 2) $qq\to q$, 3), $qq\to q^\prime$, 4) $qg\to g$, and 5) $gg\to q^\prime$. In each of the channels, we found that all collinear $1/\varepsilon$ poles eventually canceled, demonstrating that collinear factorization for our twist-3 observable holds at the one-loop level for all contributions
involving multiparton correlations in the nucleon. Our results are therefore valid in the physical $d=4$-dimensional world.

We split the spin-dependent cross section into the contributions of the various partonic channels,
\bea
E_h\frac{\dd \sigma^{\ell N^\uparrow\to hX}_{\mathrm{NLO}}}{\dd^3\bf{P}_h}& = & E_h\frac{\dd \sigma^{qg\to q}_{\mathrm{NLO}}}{\dd^3\bf{P}_h}+E_h\frac{\dd \sigma^{qq\to q}_{\mathrm{NLO}}}{\dd^3\bf{P}_h}+E_h\frac{\dd \sigma^{qq\to q^\prime}_{\mathrm{NLO}}}{\dd^3\bf{P}_h}
+E_h\frac{\dd \sigma^{qg\to g}_{\mathrm{NLO}}}{\dd^3\bf{P}_h}+E_h\frac{\dd \sigma^{gg\to q^\prime}_{\mathrm{NLO}}}{\dd^3\bf{P}_h}\,.\label{eq:Channels}
\eea
As discussed before, each of the partonic channels receives various individual contributions related to the configurations of momentum
fractions of the incoming partons described by the twist-3 correlation functions. We hence further decompose the cross sections for
the various channels.

\paragraph{Channel $qg\to q$:\label{par:qg2q}}
As discussed above, the $qg\to q$ channel can be split up in the following way:
\be
E_h\frac{\dd \sigma^{qg\to q}_{\mathrm{NLO}}}{\dd^3\bf{P}_h}  = E_h\frac{\dd \sigma^{qg\to q}_{\mathrm{Int}}}{\dd^3\bf{P}_h}+E_h\frac{\dd \sigma^{qg\to q}_{\mathrm{HP}}}{\dd^3\bf{P}_h}+E_h\frac{\dd \sigma^{qg\to q}_{\mathrm{SGP}}}{\dd^3\bf{P}_h}+E_h\frac{\dd \sigma^{qg\to q}_{\mathrm{SFP}}}{\dd^3\bf{P}_h}\,,\label{eq:Channel1}
\ee
where we sum the \textit{integral} (Int), \textit{hard pole} (HP), \textit{soft-gluon pole} (SGP) and \textit{soft-fermion pole} (SFP) contributions.\\

The \textbf{integral contributions} have the following analytic form (cf. Eq.~\eqref{eq:dynInt3}), 
\bea
E_h\frac{\dd \sigma^{qg\to q}_{\mathrm{Int}}}{\dd^3\bf{P}_h} &=& \sigma_0(S)\,\frac{\alpha_s(\mu)}{2\pi} \int_{v_0}^{v_1}\dd v\int_{x_0}^1\tfrac{\dd w}{w}\int_0^1\dd \zeta\,\sum_{q}e_q^2\,D_1^q\left(\tfrac{1-v_1}{1-v},\mu\right)\times \nn\\
&&\hspace{-1.cm}\left[\hat{\sigma}_{\mathrm{Int}}^{qg\to q,1}(v,w,\zeta)\,F_{\mathrm{Int}}^{qg\to q}(x_0,w,\zeta,\mu)+\hat{\sigma}_{\mathrm{Int}}^{qg\to q,5}(v,w,\zeta)\,G_{\mathrm{Int}}^{qg\to q}(x_0,w,\zeta,\mu)\right]\,,\label{eq:Channel1Int}
\eea
where the spin-dependent prefactor $\sigma_0(S)$ is given in \eqref{eq:prefactor} and the non-trivial but integrable combinations of quark-gluon-quark correlation functions $F_{\mathrm{Int}}^{qg\to q}$, $G_{\mathrm{Int}}^{qg\to q}$ in Eqs.~\eqref{eq:dynIntmodF} and \eqref{eq:dynIntmodG}. The flavor sum $\sum_q$ is to be understood as a summation over both quarks $q$ and antiquarks $\bar{q}$. The accompanying fragmentation function $D_1^q$ distinguishes between the fragmentation of a quark and an antiquark into a given hadron. All partonic functions $F_{\mathrm{Int}}^{qg\to q}$, $G_{\mathrm{Int}}^{qg\to q}$ and $D_1^q$ are to be understood as $\overline{\mathrm{MS}}$-renormalized that depend on the renormalization/factorization scale $\mu$. Note that the partonic functions $\hat{\sigma}_{\mathrm{Int}}^1$, $\hat{\sigma}_{\mathrm{Int}}^5$ do not depend on $\mu$ for the integral contributions. The explicit analytic forms of the partonic functions $\hat{\sigma}_{\mathrm{Int}}^1$, $\hat{\sigma}_{\mathrm{Int}}^5$ are provided in Eqs.~\eqref{eq:Int1qg2q},\eqref{eq:Int5qg2q}.\\

As described in section \ref{sub:qg2q}), the \textbf{hard pole} contributions in \eqref{eq:Channel1} can be cast into the following analytic form, 
\bea
E_h\frac{\dd \sigma^{qg\to q}_{\mathrm{HP}}}{\dd^3\bf{P}_h} &=& \sigma_0(S)\,\frac{\alpha_s(\mu)}{2\pi} \int_{v_0}^{v_1}\dd v\int_{x_0}^1\tfrac{\dd w}{w}\,\sum_q e_q^2\,D_1^q\left(\tfrac{1-v_1}{1-v},\mu\right)\times \nn\\
&&\hspace{-0cm}\Bigg[\hat{\sigma}_{\mathrm{HP},F}^{qg\to q,1}(v,w,\chi_\mu)\,F^q(\tfrac{x_0}{w},x_0,\mu)\nn\\
&&\hspace{-3cm} -\hat{\sigma}_{\mathrm{HP},\partial_1F}^{qg\to q,1}(v,w,\chi_\mu)\,x_0\,(\partial_1F^q)(\tfrac{x_0}{w},x_0,\mu) -\hat{\sigma}_{\mathrm{HP},\partial_2F}^{qg\to q,1}(v,w,\chi_\mu)\,x_0\,(\partial_2F^q)(\tfrac{x_0}{w},x_0,\mu)\nn\\
&&\hspace{-3cm} +\hat{\sigma}_{\mathrm{HP},\partial^2_1F}^{qg\to q,1}(v,w)\,x^2_0\,(\partial^2_1F^q)(\tfrac{x_0}{w},x_0,\mu) +\hat{\sigma}_{\mathrm{HP},\partial_1\partial_2F}^{qg\to q,1}(v,w)\,x^2_0\,(\partial_1\partial_2F^q)(\tfrac{x_0}{w},x_0,\mu)\nn\\
&&\hspace{-0cm}+\hat{\sigma}_{\mathrm{HP},G}^{qg\to q,5}(v,w,\chi_\mu)\,G^q(\tfrac{x_0}{w},x_0,\mu)\nn\\
&&\hspace{-3cm} -\hat{\sigma}_{\mathrm{HP},\partial_1G}^{qg\to q,5}(v,w)\,x_0\,(\partial_1G^q)(\tfrac{x_0}{w},x_0,\mu) -\hat{\sigma}_{\mathrm{HP},\partial_2G}^{qg\to q,5}(v,w,\chi_\mu)\,x_0\,(\partial_2G^q)(\tfrac{x_0}{w},x_0,\mu)\nn\\
&&\hspace{-3cm} +\hat{\sigma}_{\mathrm{HP},\partial^2_1G}^{qg\to q,5}(v,w)\,x^2_0\,(\partial^2_1G^q)(\tfrac{x_0}{w},x_0,\mu) +\hat{\sigma}_{\mathrm{HP},\partial_1\partial_2G}^{qg\to q,5}(v,w)\,x^2_0\,(\partial_1\partial_2G^q)(\tfrac{x_0}{w},x_0,\mu)\Bigg].\label{eq:Channel1HP}
\eea
There are ten different partonic hard-pole functions $\hat{\sigma}^{qg\to q}_{\mathrm{HP},...}$ whose explicit analytical expressions are given in Eqs.~\eqref{eq:HP1qg2q}--\eqref{eq:HP8qg2q}. Notice that an explicit dependence on the renormalization/factorization scale $\mu$, entering through $\chi_\mu\equiv\tfrac{s\,u}{t\mu^2}$ in the above expression,
only appears in five out of ten partonic factors. Because of the absence of photon-in-lepton contributions (discussed in Sec. \ref{sub:PiL}) for hard poles, the lepton mass $m_{\ell}$ does not enter the partonic cross sections in Eq.~\eqref{eq:Channel1HP}.\\

The \textbf{soft-gluon pole} contributions in \eqref{eq:Channel1} have the following analytic form:
\bea
E_h\frac{\dd \sigma^{qg\to q}_{\mathrm{SGP}}}{\dd^3\bf{P}_h} &=& \sigma_0(S)\,\frac{\alpha_s(\mu)}{2\pi} \int_{v_0}^{v_1}\dd v\int_{x_0}^1\tfrac{\dd w}{w}\,\sum_q e_q^2\,D_1^q\left(
\tfrac{1-v_1}{1-v},\mu\right)\times \nn\\
&&\hspace{-3cm}\Bigg[\hat{\sigma}_{\mathrm{SGP},F}^{qg\to q,1}(v,w,\chi_\mu,\chi_m)\,F^q(\tfrac{x_0}{w},\tfrac{x_0}{w},\mu)+\hat{\sigma}_{\mathrm{SGP},F^\prime}^{qg\to q,1}(v,w,\chi_\mu)\,(-x_0\,(F^q)^\prime(\tfrac{x_0}{w},\tfrac{x_0}{w},\mu))\nn\\
&&\hspace{-2.5cm}+\hat{\sigma}_{\mathrm{SGP},F^{\prime\prime}}^{qg\to q,1}(v,w)\,(x_0^2\,(F^q)^{\prime\prime}(\tfrac{x_0}{w},\tfrac{x_0}{w},\mu))+\hat{\sigma}_{\mathrm{SGP},\partial_1^2F}^{qg\to q,1}(v,w)\,(x_0^2\,(\partial_1^2F^q)(\tfrac{x_0}{w},\tfrac{x_0}{w},\mu))\nn\\
&&\hspace{-2.5cm}+  \hat{\sigma}_{\mathrm{SGP},\partial_1G}^{qg\to q,5}(v,w)\,(-x_0\,(\partial_1G^q)(\tfrac{x_0}{w},\tfrac{x_0}{w},\mu))+  \hat{\sigma}_{\mathrm{SGP},\partial_1^2G}^{qg\to q,5}(v,w)\,(x^2_0\,(\partial_1^2G^q)(\tfrac{x_0}{w},\tfrac{x_0}{w},\mu))\Bigg]\,.\label{eq:Channel1SGP}
\eea
In addition to $\chi_\mu\equiv \frac{s\,u}{t\mu^2}$ we also encounter a dependence on the lepton mass via 
$\chi_m\equiv \frac{s\,u}{tm_\ell^2}$.
The explicit form of the six partonic cross sections in Eq.~\eqref{eq:Channel1SGP} are presented in Appendix \ref{appsub:SGPqg2q}.\\

Finally, we obtain the following result for the \textbf{soft-fermion pole} contributions:
\bea
E_h\frac{\dd \sigma^{qg\to q}_{\mathrm{SFP}}}{\dd^3\bf{P}_h} &=& \sigma_0(S)\,\frac{\alpha_s(\mu)}{2\pi} \int_{v_0}^{v_1}\dd v\int_{x_0}^1\tfrac{\dd w}{w}\,\sum_q e_q^2\,D_1^q
\left(\tfrac{1-v_1}{1-v},\mu\right)\times \nn\\
&&\hspace{-3cm}\Bigg[\hat{\sigma}_{\mathrm{SFP},F}^{qg\to q,1}(v,w,\chi_m)\,F^q(\tfrac{x_0}{w},0,\mu)+\hat{\sigma}_{\mathrm{SFP},\partial_2F}^{qg\to q,1}(v,w)\,(-x_0\,(\partial_2 F^q)(\tfrac{x_0}{w},0,\mu))\nn\\
&&\hspace{-2.5cm}+  \hat{\sigma}_{\mathrm{SFP},G}^{qg\to q,5}(v,w,\chi_m)\,G^q(\tfrac{x_0}{w},0,\mu)+  \hat{\sigma}_{\mathrm{SFP},\partial_2G}^{qg\to q,5}(v,w)\,(-x_0\,(\partial_2G^q)(\tfrac{x_0}{w},0,\mu))\Bigg]\,.\label{eq:Channel1SFP}
\eea
The four partonic cross sections of Eq.~\eqref{eq:Channel1SFP} are given in Appendix \ref{appsub:SFPqg2q}.\\

\paragraph{Channel $qq\to q$:\label{par:qq2q}}
The results for the $qq\to q$ channel can be split up in the following way:
\be
E_h\frac{\dd \sigma^{qq\to q}_{\mathrm{NLO}}}{\dd^3\bf{P}_h}  = E_h\frac{\dd \sigma^{qq\to q}_{\mathrm{Int}}}{\dd^3\bf{P}_h}+E_h\frac{\dd \sigma^{qq\to q}_{\mathrm{HP}}}{\dd^3\bf{P}_h}+E_h\frac{\dd \sigma^{qq\to q}_{\mathrm{SFP}}}{\dd^3\bf{P}_h}\,.\label{eq:Channel2}
\ee

Here, the \textbf{integral contribution} has the following analytic form (cf. Eq.~\eqref{eq:dynIntqqq} and following equations):
\bea
E_h\frac{\dd \sigma^{qq\to q}_{\mathrm{Int}}}{\dd^3\bf{P}_h} &=& \sigma_0(S)\,\frac{\alpha_s(\mu)}{2\pi} \int_{v_0}^{v_1}\dd v\int_{x_0}^1\tfrac{\dd w}{w}\int_0^1\dd \zeta\,\sum_{q}e_q^2\,D_1^q\left(\tfrac{1-v_1}{1-v},\mu\right)\times \nn\\
&&\hspace{-1.5cm}\left[\hat{\sigma}_{\mathrm{Int},1}^{qq\to q,1}(v,w,\zeta)\,F_{\mathrm{Int},1}^{qq\to q}(x_0,w,\zeta,\mu)+\hat{\sigma}_{\mathrm{Int},1}^{qq\to q,5}(v,w,\zeta)\,G_{\mathrm{Int},1}^{qq\to q}(x_0,w,\zeta,\mu)\right.\nn\\
&&\hspace{-1cm}\left.+\hat{\sigma}_{\mathrm{Int},2}^{qq\to q,1}(v,w,\zeta)\,F_{\mathrm{Int},2}^{qq\to q}(x_0,w,\zeta,\mu)+\hat{\sigma}_{\mathrm{Int},2}^{qq\to q,5}(v,w,\zeta)\,G_{\mathrm{Int},2}^{qq\to q}(x_0,w,\zeta,\mu)\right]\,,\label{eq:Channel2Int}
\eea
where the explicit form of the correlation functions $F,G_{\mathrm{Int},1,2}$ is given in Eqs.~\eqref{eq:qqqIntmodF1}--\eqref{eq:qqqIntmodG2}. The partonic functions $\hat{\sigma}_{\mathrm{Int},1,2}^{qq\to q,1,5}$ depend neither on the renormalization scale $\mu$, nor on the lepton mass $m_\ell$. Their explicit analytical forms are presented in Eqs.~\eqref{eq:Int1qq2q}--\eqref{eq:Int4qq2q}.\\

The \textbf{hard pole} contribution in \eqref{eq:Channel2} can be cast into the following analytic form:
\bea
E_h\frac{\dd \sigma^{qq\to q}_{\mathrm{HP}}}{\dd^3\bf{P}_h} &=& \sigma_0(S)\,\frac{\alpha_s(\mu)}{2\pi} \int_{v_0}^{v_1}\dd v\int_{x_0}^1\tfrac{\dd w}{w}\,\sum_q e_q^2\,D_1^q\left(\tfrac{1-v_1}{1-v},\mu\right)\times \nn\\
&&\hspace{-1cm}\Bigg[\hat{\sigma}_{\mathrm{HP},F}^{qq\to q,1}(v,w,\chi_\mu)\,F^q(-\tfrac{1-w}{w}x_0,x_0,\mu) -\hat{\sigma}_{\mathrm{HP},\partial F}^{qq\to q,1}(v,w,\chi_\mu)\,x_0\,(\partial_1F^q+\partial_2F^q)(-\tfrac{1-w}{w}x_0,x_0,\mu)\nn\\
&&\hspace{-1cm}+\hat{\sigma}_{\mathrm{HP},G}^{qq\to q,5}(v,w,\chi_\mu)\,G^q(-\tfrac{1-w}{w}x_0,x_0,\mu) -\hat{\sigma}_{\mathrm{HP},\partial G}^{qq\to q,5}(v,w,\chi_\mu)\,x_0\,(\partial_1G^q+\partial_2G^q)(-\tfrac{1-w}{w}x_0,x_0,\mu)\Bigg].\label{eq:Channel2HP}
\eea
The explicit results for the partonic hard functions $\hat{\sigma}_{\mathrm{HP},...}^{qq\to q}$ can be found in Appendix~\ref{appsub:HPqq2q} 
in Eqs.~\eqref{eq:HP1qq2q}--\eqref{eq:HP4qq2q}.\\

The \textbf{soft-fermion pole} contributions read
\bea
E_h\frac{\dd \sigma^{qq\to q}_{\mathrm{SFP}}}{\dd^3\bf{P}_h} &=& \sigma_0(S)\,\frac{\alpha_s(\mu)}{2\pi} \int_{v_0}^{v_1}\dd v\int_{x_0}^1\tfrac{\dd w}{w}\,\sum_q e_q^2\,D_1^q\left(\tfrac{1-v_1}{1-v},\mu\right)\times \nn\\
&&\hspace{-3cm}\Bigg[\hat{\sigma}_{\mathrm{SFP},F}^{qq\to q,1}(v,w,\chi_m)\,F^q(-\tfrac{x_0}{w},0,\mu)+  \hat{\sigma}_{\mathrm{SFP},G}^{qq\to q,5}(v,w,\chi_m)\,G^q(-\tfrac{x_0}{w},0,\mu)\Bigg]\,,\label{eq:Channel2SFP}
\eea
where the two partonic cross sections are presented in Eqs.~\eqref{eq:SFP1qq2q},\eqref{eq:SFP2qq2q}.\\

\paragraph{Channel $qq\to q^\prime$:\label{par:qq2qp}}
The contributions for this channel may be split up in the following way:
\be
E_h\frac{\dd \sigma^{qq\to q^\prime}_{\mathrm{NLO}}}{\dd^3\bf{P}_h}  = E_h\frac{\dd \sigma^{qq\to q^\prime}_{\mathrm{Int}}}{\dd^3\bf{P}_h}+E_h\frac{\dd \sigma^{qq\to q^\prime}_{\mathrm{SFP}}}{\dd^3\bf{P}_h}\,.\label{eq:Channel3}
\ee

The \textbf{integral contribution} is given as (cf. Eq.~\eqref{eq:dynIntqqq} and below)
\bea
E_h\frac{\dd \sigma^{qq\to q^\prime}_{\mathrm{Int}}}{\dd^3\bf{P}_h} &=& \sigma_0(S)\,\frac{\alpha_s(\mu)}{2\pi} \int_{v_0}^{v_1}\dd v\int_{x_0}^1\tfrac{\dd w}{w}\int_0^1\dd \zeta\,\sum_{q^\prime}(D_1^{q^\prime}-D_1^{\bar{q}^\prime})\left(\tfrac{1-v_1}{1-v},\mu\right)\times \nn\\
&&\hspace{-2.5cm}\sum_q e_q^2\,\left[\hat{\sigma}_{\mathrm{Int}}^{qq\to q^\prime,1}(v,w,\zeta)\,F_{\mathrm{Int}}^{qq\to q^\prime}(x_0,w,\zeta,\mu)+\hat{\sigma}_{\mathrm{Int}}^{qq\to q^\prime,5}(v,w,\zeta)\,G_{\mathrm{Int}}^{qq\to q^\prime}(x_0,w,\zeta,\mu)\right]\,,\label{eq:Channel3Int}
\eea
where the explicit form of the correlation functions $F,G_{\mathrm{Int}}$ is given in Eqs.~\eqref{eq:qqqpIntmodF},\eqref{eq:qqqpIntmodG}. The partonic functions $\hat{\sigma}_{\mathrm{Int}}^{qq\to q^\prime,1,5}$ do not depend on $\mu$ and $m_\ell$. Their explicit analytical forms are presented in Eqs.~\eqref{eq:Int1qq2qp},\eqref{eq:Int2qq2qp}. Note that, unlike the other channels, the flavor sum $\sum_q=\sum_{u,d,s,...}$ does not include anti-flavors $\bar{q}$ in this case. Because of the symmetries in Eqs.~\eqref{eq:qq2qpISantisym} the antiquark contributions are already included in the partonic cross sections.\\

We obtain the following form for the \textbf{soft-fermion pole} contribution:
\bea
E_h\frac{\dd \sigma^{qq\to q^\prime}_{\mathrm{SFP}}}{\dd^3\bf{P}_h} &=& \sigma_0(S)\,\frac{\alpha_s(\mu)}{2\pi} \int_{v_0}^{v_1}\dd v\int_{x_0}^1\tfrac{\dd w}{w}\,\sum_{q^\prime}\,(D_1^{q^\prime}-D_1^{\bar{q}^\prime})\left(\tfrac{1-v_1}{1-v},\mu\right)\times \nn\\
&&\hspace{-1.5cm}\sum_q e_q^2\Bigg[\hat{\sigma}_{\mathrm{SFP},F}^{qq\to q^\prime,1}(v,w,\chi_m)\,\left(F^q(-\tfrac{x_0}{w},0,\mu)-F^q(0,\tfrac{x_0}{w},\mu)\right)\nn\\
&&\hspace{-1cm}+  \hat{\sigma}_{\mathrm{SFP},G}^{qq\to q^\prime,5}(v,w,\chi_m)\,\left(G^q(-\tfrac{x_0}{w},0,\mu)+G^q(0,\tfrac{x_0}{w},\mu)\right)\Bigg]\,.\label{eq:Channel3SFP}
\eea
The explicit form of the two partonic cross sections is presented in Eqs~\eqref{eq:SFP1qq2qp},\eqref{eq:SFP2qq2qp}. Again, because of  Eqs.~\eqref{eq:qq2qpISantisym}, the flavor sum 
only runs over quarks but not antiquarks. \\

\paragraph{Channel $qg\to g$:\label{par:qg2g}}
Our results for the $qg\to g$ channel can be split up in the following way:
\be
E_h\frac{\dd \sigma^{qg\to g}_{\mathrm{NLO}}}{\dd^3\bf{P}_h}  = E_h\frac{\dd \sigma^{qg\to g}_{\mathrm{Int}}}{\dd^3\bf{P}_h}+E_h\frac{\dd \sigma^{qg\to g}_{\mathrm{SGP}}}{\dd^3\bf{P}_h}+E_h\frac{\dd \sigma^{qg\to g}_{\mathrm{SFP}}}{\dd^3\bf{P}_h}\,.\label{eq:Channel4}
\ee

The \textbf{integral contribution} reads
\bea
E_h\frac{\dd \sigma^{qg\to g}_{\mathrm{Int}}}{\dd^3\bf{P}_h} &=& \sigma_0(S)\,\frac{\alpha_s(\mu)}{2\pi} \int_{v_0}^{v_1}\dd v\int_{x_0}^1\tfrac{\dd w}{w}\int_0^1\dd \zeta\,D_1^g\left(\tfrac{1-v_1}{1-v},\mu\right)\times \nn\\
&&\hspace{-1.5cm}\sum_{q}e_q^2\,\left[\hat{\sigma}_{\mathrm{Int}}^{qg\to g,1}(v,w,\zeta)\,F_{\mathrm{Int}}^{qg\to g}(x_0,w,\zeta,\mu)+\hat{\sigma}_{\mathrm{Int}}^{qg\to g,5}(v,w,\zeta)\,G_{\mathrm{Int}}^{qg\to g}(x_0,w,\zeta,\mu)\right]\,,\label{eq:Channel4Int}
\eea
where the explicit form of the correlation functions $F,G_{\mathrm{Int}}$ is given in Eqs.~\eqref{eq:qg2gmodIntF},\eqref{eq:qg2gmodIntG}. The partonic functions $\hat{\sigma}_{\mathrm{Int}}^{qg\to g,1,5}$ do not depend on $\mu$ and $m_\ell$. Their explicit analytical form is presented in Eqs.~\eqref{eq:Int1qg2g}, \eqref{eq:Int2qg2g}.\\

The \textbf{soft-gluon pole} contribution in \eqref{eq:Channel4} can be cast into the following analytic form:
\bea
E_h\frac{\dd \sigma^{qg\to g}_{\mathrm{SGP}}}{\dd^3\bf{P}_h} &=& \sigma_0(S)\,\frac{\alpha_s(\mu)}{2\pi} \int_{v_0}^{v_1}\dd v\int_{x_0}^1\tfrac{\dd w}{w}\,D_1^g\left(\tfrac{1-v_1}{1-v},\mu\right)\times \nn\\
&&\hspace{-3cm}\sum_q e_q^2\,\Bigg[\hat{\sigma}_{\mathrm{SGP},F}^{qg\to g,1}(v,w,\chi_\mu,\chi_m)\,F^q(\tfrac{x_0}{w},\tfrac{x_0}{w},\mu)-\hat{\sigma}_{\mathrm{SGP},\partial_2 G}^{qg\to g,5}(v,w)\,x_0\,(\partial_2G^q)(\tfrac{x_0}{w},\tfrac{x_0}{w},\mu)\Bigg],\label{eq:Channel4SGP}
\eea
where as before $\chi_\mu\equiv\tfrac{s\,u}{t\mu^2}$ and $\chi_m\equiv\tfrac{s\,u}{tm_\ell^2}$.
The explicit analytical form of the partonic hard functions $\hat{\sigma}_{\mathrm{SGP},...}^{qg\to g}$ can be found in Appendix~\ref{appsub:Integralqg2g} 
in Eqs.~\eqref{eq:SGP1qg2g},\eqref{eq:SGP2qg2g}.\\

Finally, we obtain the following form for the \textbf{soft-fermion pole} contributions:
\bea
E_h\frac{\dd \sigma^{qg\to g}_{\mathrm{SFP}}}{\dd^3\bf{P}_h} &=& \sigma_0(S)\,\frac{\alpha_s(\mu)}{2\pi} \int_{v_0}^{v_1}\dd v\int_{x_0}^1\tfrac{\dd w}{w}\,D_1^g\left(\tfrac{1-v_1}{1-v},\mu\right)\times \nn\\
&&\hspace{-3cm}\sum_q e_q^2\,\Bigg[\hat{\sigma}_{\mathrm{SFP},F}^{qg\to g,1}(v,w,\chi_m)\,F^q(\tfrac{x_0}{w},0,\mu)+  \hat{\sigma}_{\mathrm{SFP},G}^{qg\to g,5}(v,w,\chi_m)\,G^q(\tfrac{x_0}{w},0,\mu)\Bigg]\,.\label{eq:Channel4SFP}
\eea
The explicit form of the two partonic cross sections is presented in Eqs.~\eqref{eq:SFP1qg2g},\eqref{eq:SFP2qg2g}.\\

\paragraph{Channel $gg\to q^\prime$:} Here we find the following form of the cross section:
\bea
E_h\frac{\dd \sigma^{gg\to q^\prime}_{\mathrm{NLO}}}{\dd^3\bf{P}_h} &=& \sigma_0(S)\,\frac{\alpha_s(\mu)}{2\pi} \int_{v_0}^{v_1}\dd v\int_{x_0}^1\tfrac{\dd w}{w}\,\sum_{q^\prime} e_{q^\prime}^2\,D_1^{q^\prime}\left(\tfrac{1-v_1}{1-v},\mu\right)\times \nn\\
&&\hspace{-3cm}\left[-\hat{\sigma}_{xx}^{gg\to q^\prime}\left(v,w,\chi_m,\chi_\mu\right)\frac{N\left(\tfrac{x_0}{w},\tfrac{x_0}{w},\mu\right)+O\left(\tfrac{x_0}{w},\tfrac{x_0}{w},\mu\right)}{\tfrac{x_0}{w}}\,+\,\hat{\sigma}_{x0}^{gg\to q^\prime}\left(v,w,\chi_m,\chi_\mu\right)\frac{O\left(\tfrac{x_0}{w},0,\mu\right)-N\left(\tfrac{x_0}{w},0,\mu\right)}{\tfrac{x_0}{w}}\right]\,,\label{eq:Channel5}
\eea
where we again have a dependence of the partonic cross sections on $\chi_\mu\equiv\tfrac{s\,u}{t\mu^2}$ and $\chi_m\equiv\tfrac{s\,u}{tm_\ell^2}$. As discussed below Eq.~\eqref{eq:dynIntggqp}, integral and hard-pole contributions are absent for this channel. The explicit form of the two partonic cross sections 
$\hat{\sigma}_{xx}^{gg\to q^\prime},\hat{\sigma}_{x0}^{gg\to q^\prime}$ is given in Eqs.~\eqref{eq:tripleGxx},\eqref{eq:tripleGx0}.

\subsection{Jet Production\label{sub:Jets}}

Now that we have established the NLO analytic formulas for single-inclusive hadron production in Eqs.~\eqref{eq:LO},\eqref{eq:Channels}, we can extend them without too much effort to
single-inclusive jet production, $\ell(l)+N^\uparrow(P)\to \mathrm{jet}(P_j)+X$. For jets, no fragmentation functions are involved, 
and the first step is to replace the fragmentation functions in Eqs.~\eqref{eq:LO},\eqref{eq:Channels} by $\delta$ functions:
\be
D_1^{h/q,g}\left(\tfrac{1-v_1}{1-v},\mu\right)\to \delta\left(1-\tfrac{1-v_1}{1-v}\right)=(1-v_1)\,\delta(v-v_1).\label{eq:ReplFjet2}
\ee 
This simple procedure works for the LO contribution \eqref{eq:LO}, but is not sufficient at NLO. The reason for this is that final-state divergences 
cancel for a jet cross section, whereas for the single-hadron cross section we had to subtract collinear singularities from the hard single-parton
cross sections, canceling them by renormalization of the fragmentation functions. As was shown in Refs.~\cite{Jager:2002xm,Jager:2003vy,Jager:2004jh,Jager:2005uf,Jager:2008qm,Mukherjee:2012uz}, at NLO it is relatively straightforward
to account for this mismatch, at least when the produced jet is relatively narrow. 

For the purposes of the present paper, we define jets by the anti-$k_T$ algorithm~\cite{Cacciari:2008gp}. If the jet radius $R$ is small, $R\ll 1$,
one can systematically derive analytical matching terms that translate from the collinear-subtracted single-parton cross sections to a jet cross 
section~\cite{Jager:2002xm,Jager:2003vy,Jager:2004jh,Jager:2005uf,Jager:2008qm}. This approach is sometimes called the \emph{small-cone approximation (SCA)}
and actually turns out to be rather accurate also at larger values of $R$, even out to $R\sim 0.7$. We note that we used the SCA already  
for the unpolarized and longitudinally polarized NLO single-hadron cross sections in Refs.~\cite{Hinderer:2015hra,Hinderer:2017ntk}.
It results in additional finite and jet-specific contributions to the NLO partonic cross sections that depend on the jet definition and the jet radius $R$. 
Applying the SCA to our present calculation, we find that the jet-specific terms for the transversely polarized cross section are different from the ones obtained for the unpolarized and longitudinally polarized cross sections in~\cite{Hinderer:2015hra,Hinderer:2017ntk}. 

Our analytical result for the partonic hard-scattering functions relevant for jet production read, in the SCA,
\bea
E_j\,\frac{\dd \sigma_{\mathrm{NLO}}^{\ell N^\uparrow\to \mathrm{jet}X}}{\dd ^3\bm{P_{j}}}(S,R)& = & \left(E_h\,\frac{\dd \sigma_{\mathrm{NLO}}^{\ell N^\uparrow\to h X}}{\dd ^3\bm{P_{h}}}(S)\right)\Bigg|_{D_1^{h/q,g}\left(\tfrac{1-v_1}{1-v},\mu\right)\to (1-v_1)\,\delta(v-v_1)}+ E_j\,\frac{\dd \sigma_{\mathrm{NLO,\,jet\,def}}^{\ell N^\uparrow\to \mathrm{jet}X}}{\dd ^3\bm{P_{j}}}(S,R)\,,\label{eq:h2jet}
\eea
where the latter term is given by
\bea
E_j\,\frac{\dd \sigma_{\mathrm{NLO,\,jet\,def}}^{\ell N^\uparrow\to \mathrm{jet}X}}{\dd ^3\bm{P_{j}}}(S,R) & = & \sigma_0(S)\,\frac{\alpha_s(\mu)}{2\pi} \int_{x_0}^1\tfrac{\dd w}{w}\times \nn\\
&&\hspace{-3cm}\Bigg[\hat{\sigma}_{\mathrm{SGP,\,jet},F}^{qg\to \mathrm{jet}(q+g)}(w,R,\mu)\,\,\sum_q e_q^2\,F^q(\tfrac{x_0}{w},\tfrac{x_0}{w},\mu)+\hat{\sigma}_{\mathrm{SGP,\,jet},F^\prime}^{qg\to \mathrm{jet}(q)}(w,R,\mu)\,\left(-\tfrac{x_0}{w}\,\left(\sum_q e_q^2\,F^q\right)^\prime(\tfrac{x_0}{w},\tfrac{x_0}{w},\mu)\right)\Bigg]\,.\label{eq:jetsub1}
\eea
We note that the additional jet-specific terms only appear as SGP contributions, including a derivative term. The term $\hat{\sigma}_{\mathrm{SGP,\,jet},F}^{qg\to \mathrm{jet}(q+g)}$ receives contributions from both the previous quark and gluon fragmentation channels, while the derivative term $\hat{\sigma}_{\mathrm{SGP,\,jet},F^\prime}^{qg\to \mathrm{jet}(q)}(w,R,\mu)$ 
is generated only by the previous quark channel. The only relevant color factor for both partonic factors is $C_F$. Similarly to the case of hadron production, we moved all regular terms originating from the partonic factors for the derivative term to the SGP term via integration by parts. We present the explicit analytic forms of the partonic factors in Appendix \ref{Appsub:jet}.

\section{Numerical Study for the SSA at the EIC\label{sec:Numerics}}

In this section, we explore the numerical impact of the NLO corrections for the transverse SSA \eqref{eq:SSA} we have derived, both for
single-inclusive hadron and for jet production. We perform our study for $\sqrt{s}=100\,\mathrm{GeV}$, which corresponds to the collision energy expected for the EIC. We work in the lepton-nucleon c.m.s. frame rather than in the asymmetric laboratory frame. We consider the doubly differential cross section and the corresponding SSA in the single-inclusive hadron's (or jet's) pseudorapidity $\eta_h$ ($\eta_j$) and transverse momentum $P_{h,T}$ ($P_{j,T}$). The direction of the lepton defines the positive $z$-axis, that is, we count pseudorapidity to be positive in the forward lepton direction. To be specific, the three relevant momenta have the explicit form
\bea
l^\mu & = & \tfrac{1}{2}\sqrt{s}\,\left(1,\,0,\,0\,,+1\right)\,,\nn\\
P^\mu & = & \tfrac{1}{2}\sqrt{s}\,\left(1,\,0,\,0\,,-1\right)\,,\nn\\
P^\mu_{h} & = & P_{T,h}\,\left(\cosh(\eta_{h}),\,\cos(\phi_{h}),\,\sin(\phi_{h}),\,\sinh(\eta_{h})\right)\,,\label{eq:cmframemomenta}
\eea
in our frame of reference, and analogously for the jet momentum $P^\mu_j$. 
The Mandelstam variables are expressed in terms of $\eta_{h}$ and $P_{T,h}$ as
\bea
t & = & -P_{h,T}\,\sqrt{s}\,\mathrm{e}^{+\eta_{h}}\,,\nn\\
u & = & -P_{h,T}\,\sqrt{s}\,\mathrm{e}^{-\eta_{h}}\,,\label{eq:Mandelstamcm}
\eea
again likewise for jet production. The unpolarized cross section depends only on pseudorapidity and transverse momentum
so that we can integrate over the hadron's (or jet's) azimuthal angle:
\be
\frac{\dd \sigma}{\dd \eta_{h}\,\dd P_{h,T}}=2\pi\,P_{h,T}\,\left(E_{h}\frac{\dd \sigma}{\dd^3\bm{P}_{h}}\right)\,.\label{eq:cmframeunpCS}
\ee
The situation is different for the spin-dependent cross section because of the prefactor $\sigma_0(S)$ defined in Eq.~\eqref{eq:prefactor}. 
The spin four-vector $S$ is constrained by the condition $P\cdot S=0$ and normalization $S^2=-1$. 
In the frame specified by \eqref{eq:cmframemomenta} the following expression for $S$ is consistent with these two conditions:
\be
S^\mu = \left(S^0,\,\cos\phi_s,\,\sin\phi_s,\,-S^0\right)\,.\label{eq:cmspinvector}
\ee
The azimuthal angle $\phi_s$ of the transverse components determines the location of the spin vector in a plane transverse to the beam direction.
The spin vector appears in $\sigma_0(S)$ in combination with a totally antisymmetric tensor as $\epsilon^{lPP_{h} S}$. One can easily work out this term in the c.m.s. frame \eqref{eq:cmframemomenta}, along with \eqref{eq:cmspinvector} to find
\be
\epsilon^{lPP_{h} S}=\tfrac{1}{2}\sqrt{stu}\,\sin(\phi_s-\phi_{h})\,.\label{eq:cmeps}
\ee
Modulation with $\sin(\phi_s-\phi_{h})$ is of course a hallmark of single transverse spin asymmetries. 
There are several equivalent ways of dealing with this dependence on the azimuthal angle. For the present study, we consider the so-called \emph{right-left} asymmetry. For a given fixed angle $\phi_s$ one counts numbers of events to the right
and the left of a plane spanned by the beam (positive direction determined by the lepton beam) and the transverse spin vector
and takes their difference:
\be
A_{RL}=\frac{\int_{\phi_s+\pi}^{\phi_s+2\pi}\dd \phi_{h}\,\frac{\dd \sigma(S)}{\dd \eta_{h}\,\dd P_{h,T}\,\dd \phi_{h}}-\int_{\phi_s}^{\phi_s+\pi}\dd \phi_{h}\,\frac{\dd \sigma(S)}{\dd \eta_{h}\,\dd P_{h,T}\,\dd \phi_{h}}}{\int_{0}^{2\pi}\dd \phi_{h}\,\frac{\dd \sigma}{\dd \eta_{h}\,\dd P_{h,T}\,\dd \phi_{h}}}\,.\label{eq:cmrightleft}
\ee
Effectively, this means that we replace the Levi-Civita tensor in $\sigma_0(S)$ in the numerator of $A_{RL}$ by 
\be
\epsilon^{lPP_{h}S}\to 2\sqrt{stu}\,.\label{eq:repleps}
\ee
As an alternative to $A_{RL}$, one could also take a $\sin(\phi_s-\phi_{h})$-weight of the spin-dependent cross section. \\

In the following, we will study the right-left asymmetry $A_{RL}$ for $\pi^+$ production, as well as for jet production at the EIC. 
For the numerator of the right-left asymmetry we need input distributions for the quark-gluon-quark correlation functions $F^q$ and $G^q$ for all flavors $q$, as well as for the tri-gluon correlation functions $N(x,x^\prime)$ and $O(x,x^\prime)$. For single-inclusive hadron production one would also need to include multiparton correlations in the fragmentation process (see the discussion below \eqref{eq:LO}) for a realistic prediction of transverse spin effects. Since none of these functions are currently known to a satisfactory extent, a realistic theoretical prediction is
not feasible. However, we can study the generic effect of the NLO corrections compared to the LO approximation for $A_{RL}$. For this purpose, we adopt three scenarios (labeled
\emph{Scenario 0,1,2}) for the quark-gluon-quark correlation functions $F^q$ and $G^q$ for $u$ and $d$-quarks, assuming all other multiparton correlation functions to vanish. Details of the three scenarios are presented in Appendix \ref{appsec:Model}.

For our numerical studies, we will focus on the pseudo-rapidity dependence of this observable at fixed transverse 
momentum $P_{\pi,T}$ or $P_{j,T}$, which we will set to $5\,\mathrm{GeV}$ throughout this work. Our default choice for the 
renormalization/factorization scale is the pion or jet transverse momentum, that is, $\mu = P_{\pi,T}$ (or $\mu = P_{j,T}$). 
Although the numerical implementation of QCD evolution is of course not a problem for the leading-twist PDFs and FFs, it is more involved for the multiparton correlation functions
for which only few numerical codes have been released so far~\cite{Braun:2009mi,Rodini:2024usc}, and valid only at LO.
The main purpose of our present numerical study is to see how much the observable $A_{RL}$ is affected by the NLO corrections to the partonic hard-scattering
functions that we have derived, and less so to provide a realistic phenomenological prediction for $A_{RL}$. For this reason we do not implement the proper scale evolution of the multiparton correlation functions into our numerical code. Instead, keeping in mind that all our models for the twist-3 correlation functions $F^q$ and $G^q$ are formulated
in terms of the unpolarized $f_1^q$ quark PDFs (see Appendix \ref{appsec:Model}), we simply use the evolution of the $f_1^q$ to get a first estimate of the scale dependence of the
$F^q$ and $G^q$ and hence of the spin asymmetries. In other words, we assume that the evolution of the twist-3 functions closely follows that of the twist-2 PDFs. In view of the similarity
of the respective evolution kernels (see~\cite{Braun:2009mi,Kang:2008ey}), this approximation is expected to be quite realistic. 
We vary the scale between $\mu = P_{\pi,T}/2$ and $\mu = 2P_{\pi,T}$ for pion production and analogously for jets. The theoretical uncertainty bands obtained by this oversimplified approach are only shown for the asymmetries. That said, the proper implementation of the true scale dependence 
and the study of the dependence of the asymmetry on transverse momentum remain important tasks for the future, once NLO
corrections to evolution of the twist-3 matrix elements have also been derived.

\begin{figure}[t!]
\centering
\includegraphics[width=0.49\textwidth]{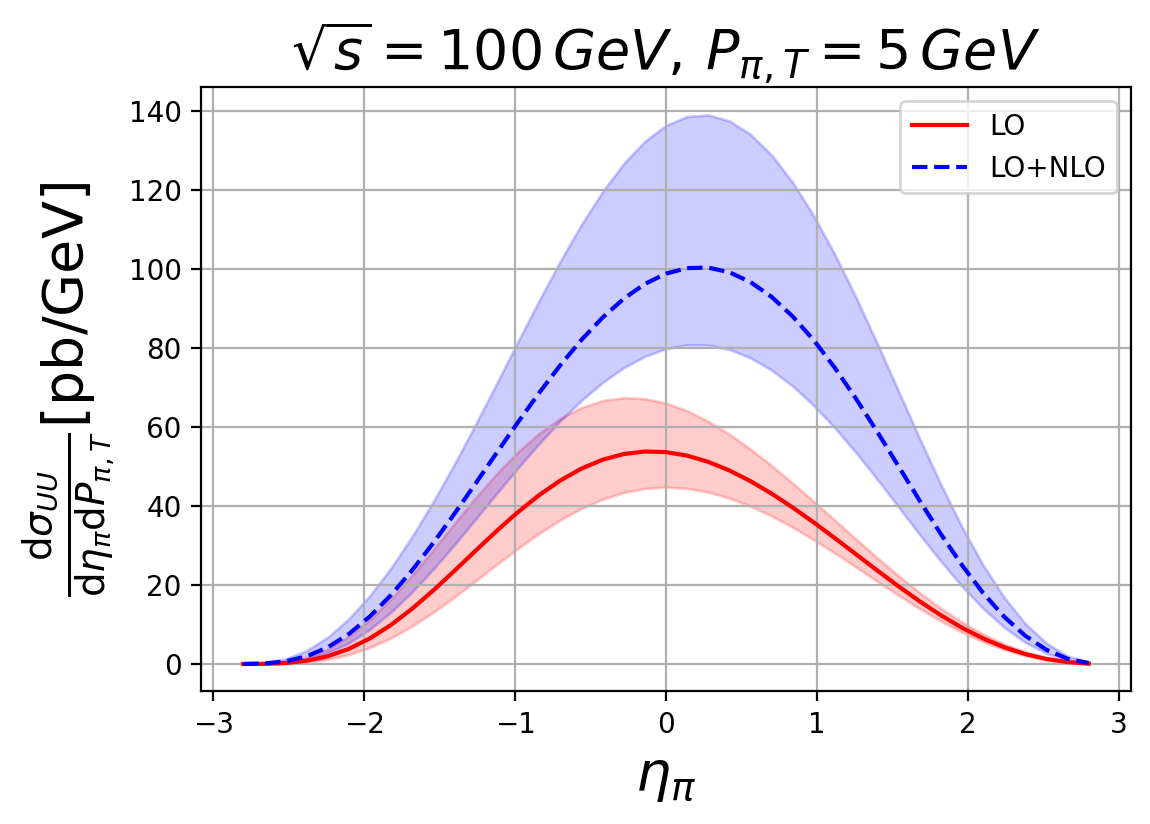}
\includegraphics[width=.49\textwidth]{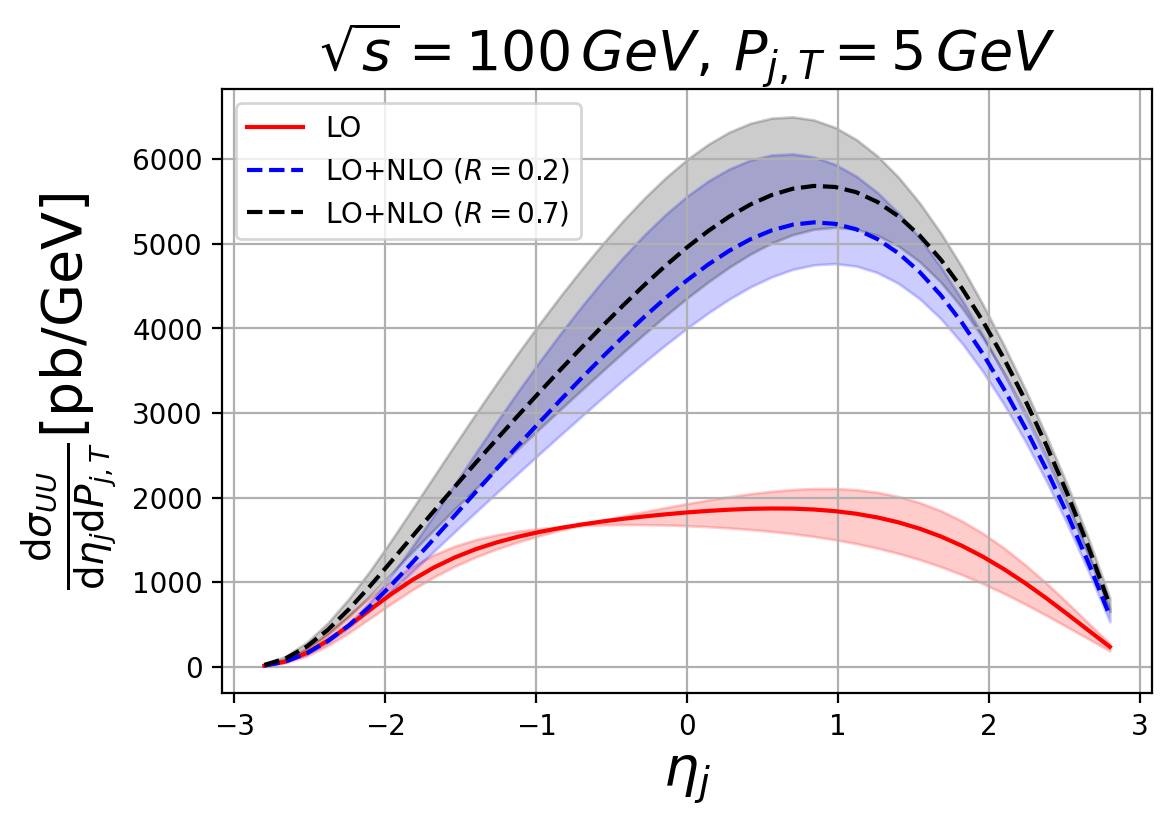}
\caption{Unpolarized differential cross section \eqref{eq:cmframeunpCS} plotted vs. the pseudorapidity $\eta$ of a $\pi^+$ \textbf{(left)} or a jet \textbf{(right)}. The transverse momentum 
is fixed to $5\,\mathrm{GeV}$ in both cases. We plot the NLO jet cross section for two jet radii, $R_1=0.2$ and $R_2=0.7$.\label{fig:PlotUU}}
\end{figure}

We start the discussion of our numerical results with the denominator of the right-left asymmetry $A_{RL}$ in \eqref{eq:cmrightleft}, the unpolarized NLO cross section.
Here we can simply rely on the results of Ref.~\cite{Hinderer:2015hra} and recalculate the cross section for the kinematics of interest for our present study. 
We use the MSTW2008 parameterization of~\cite{Martin:2009iq} for the proton's PDFs and the DSS parameterization \cite{deFlorian:2014xna} for the pion fragmentation functions.
The results are presented in Fig.~\ref{fig:PlotUU}. Theoretical uncertainty bands were generated by varying the scale between $\mu =P_T/2$ and $\mu = 2P_T$. We find that the LO cross sections for both single-inclusive pions and jets receive sizable NLO corrections, of about a factor two to three.
This is in line with observations made previously in Refs.~\cite{Hinderer:2015hra,Schlegel:2015hga}.

As a final comment, we mention that in the phenomenological plots presented below 
we will refrain from displaying separately the contributions to the polarized cross section caused by the \textit{photon-in-lepton} distributions $f_1^{\gamma/\ell,\overline{\mathrm{MS}}}(x,\mu)$ in \eqref{eq:f1WW} (see discussion in Sec.~\ref{sub:PiL}). The reason is that,
as we will see in the following section, $A_{RL}$ tends to become largest in the backward rapidity region where the quasi-real photon contributions are expected to 
be small according to the study of the unpolarized cross section in Ref.~\cite{Hinderer:2015hra}.

\subsection{Single-Inclusive \texorpdfstring{$\pi^+$}{pi+}-Production\label{sub:NumericsPi+}}

\subsubsection{Scenario 0\label{subsub:Scenario0}}

\begin{figure}[t]
\centering
\includegraphics[width=0.49\textwidth]{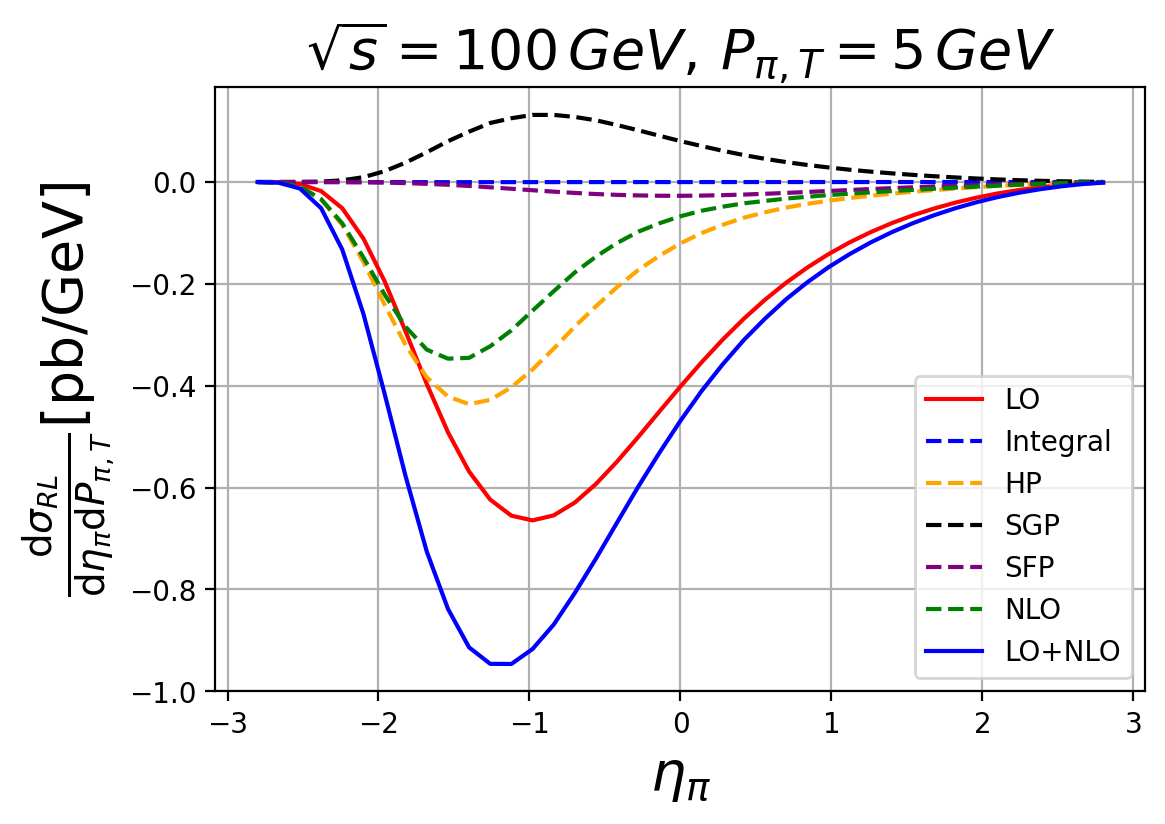}
\includegraphics[width=.49\textwidth]{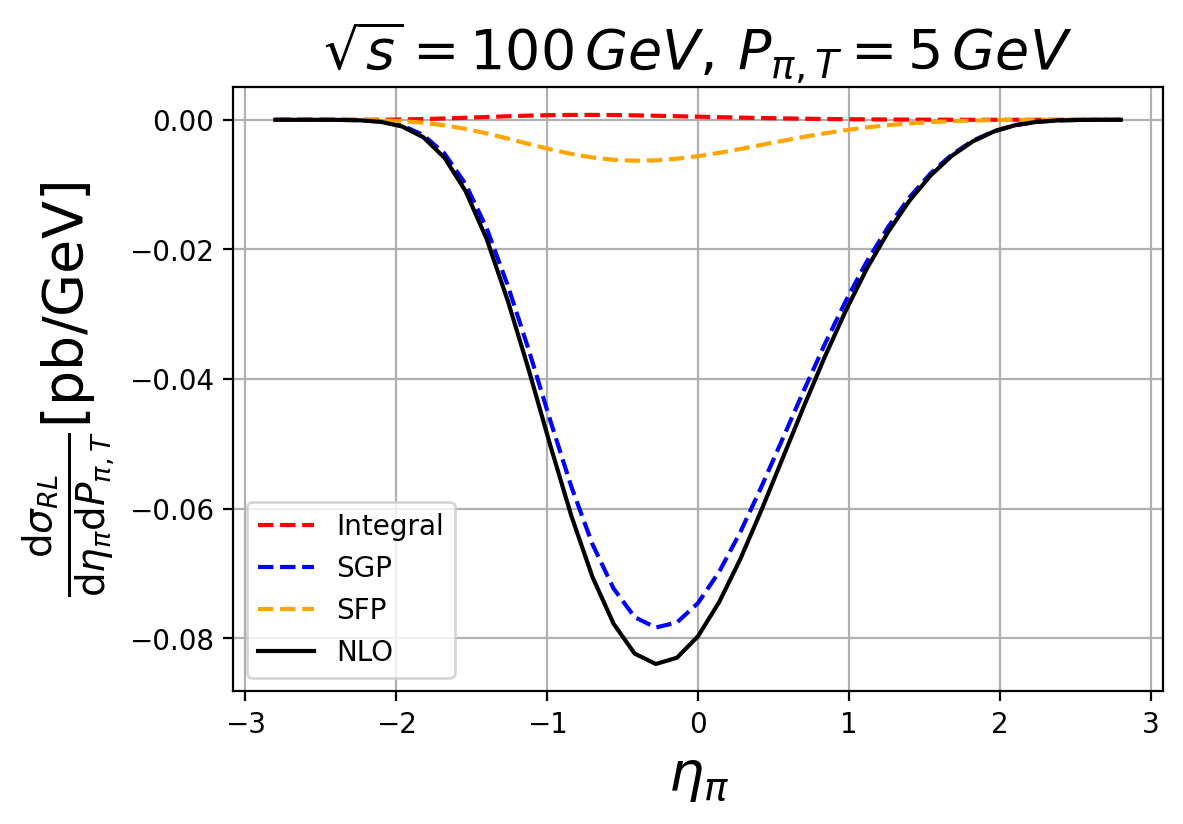}\\\vspace{0.5cm}
\includegraphics[width=0.49\textwidth]{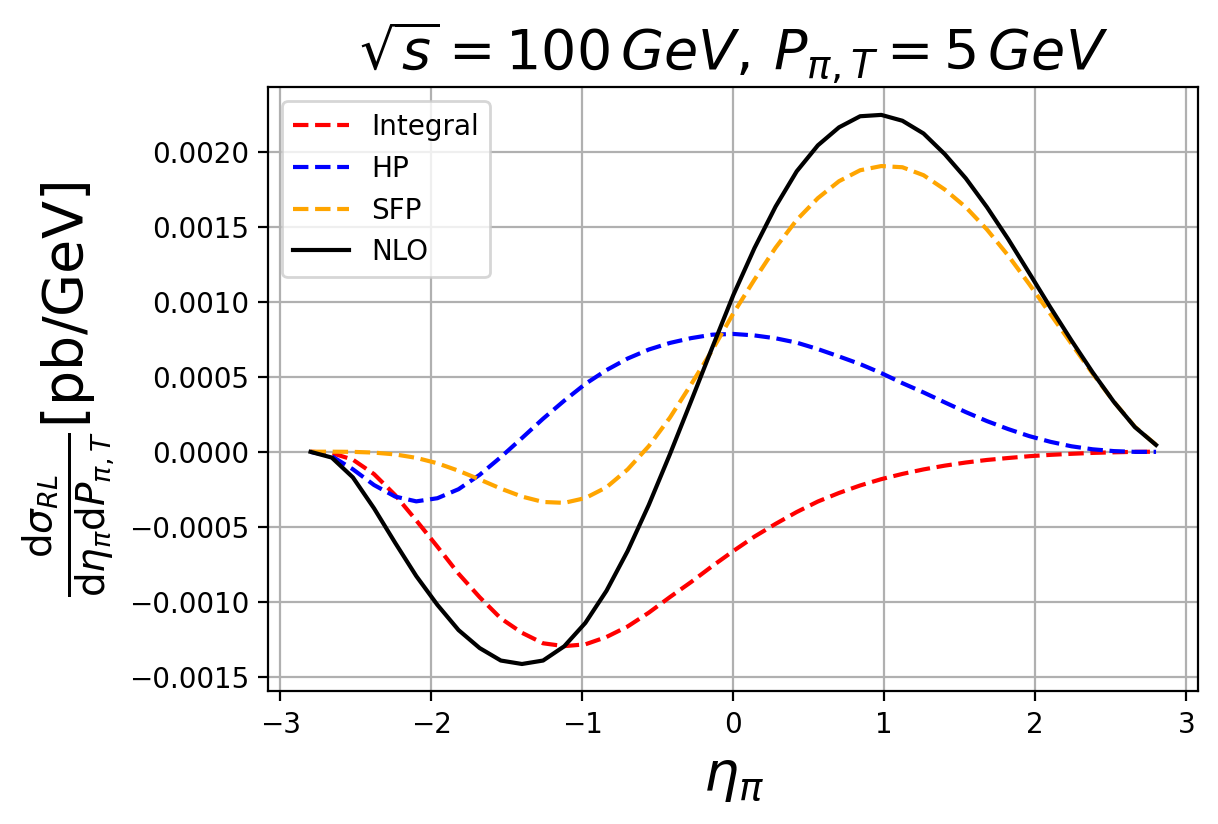}
\includegraphics[width=.49\textwidth]{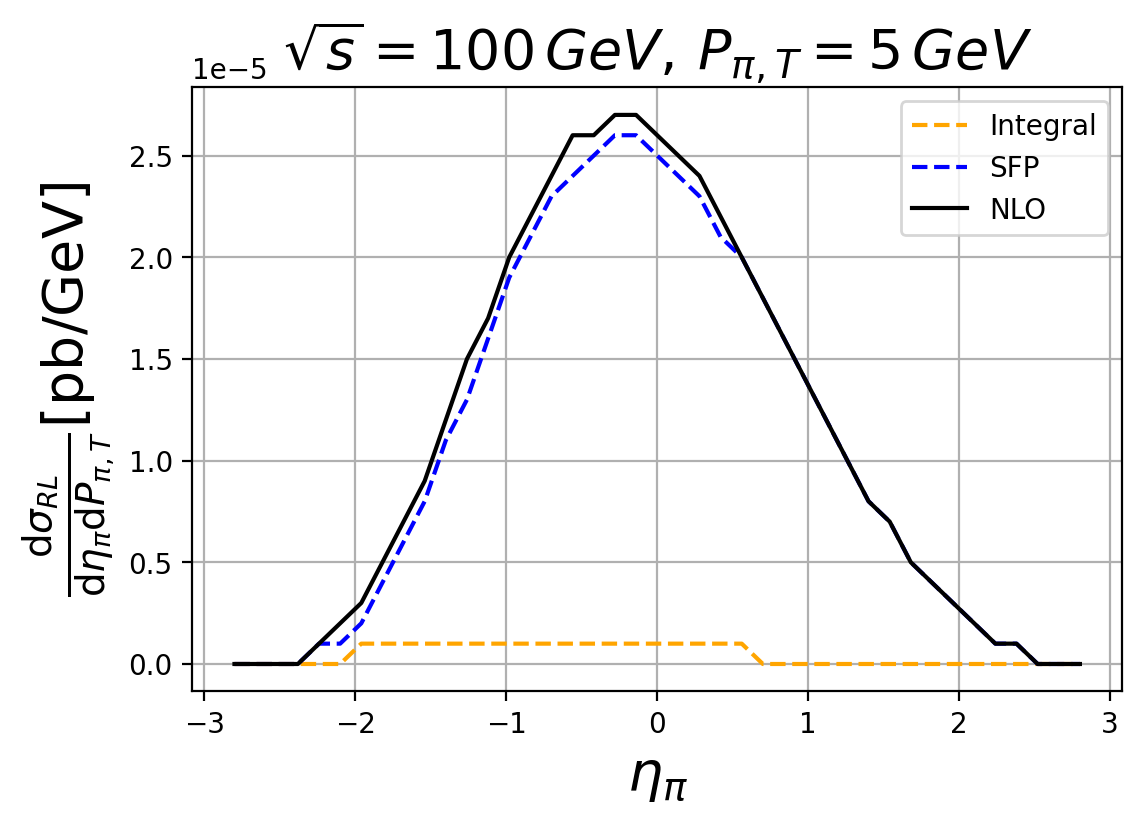}
\caption{Cross sections for the various NLO channels of Eq.~\eqref{eq:Channels} plotted vs. the pion's pseudorapidity $\eta_\pi$ at fixed transverse momentum $P_{\pi,T}=5\,\mathrm{GeV}$. The curves are produced for Scenario 0 for the correlation functions $F$ and $G$, see \eqref{eq:Scenario0}. The channels are: $qg\to q$ (upper left), $qg\to g$ (upper right), 
$qq\to q$ (lower left), $qq\to q^\prime$ (lower right).\label{fig:ChannelsUTPiScen0}}
\end{figure}

We first present our LO and NLO results for the numerator of $A_{RL}$. Scenario 0, discussed in Appendix \ref{appsub:Scen0}, may be regarded as ``minimal''
in the sense that most model parameters (see Eq.~\eqref{eq:Scenario0}) vanish, and $F$ shows the least variation on its support in $x,x'$ (see
contour plot Fig.~\ref{fig:Scen0F}). In addition, in this scenario it is assumed that $G$ vanishes identically.

Figure~\ref{fig:ChannelsUTPiScen0} shows the NLO right-left cross section for the various partonic channels in \eqref{eq:Channels}, also distinguished by their
pole contributions. We observe that the largest NLO corrections originate from $qg\to q$. Interestingly, the hard-pole contribution dominates this channel in Scenario 0 and is only partially canceled by the soft-gluon pole contributions. The soft-fermion pole and integral contributions appear to be negligible in this scenario. The second largest channel is the gluon fragmentation channel $qg\to g$, which is dominated by the soft-gluon pole contribution. Again, SFP and integral contributions are negligible for this channel. The other channels generated by quark-antiquark-gluon correlations, $qq\to q$ and $qq\to q^\prime$, are non-zero, but irrelevant (note that the y-axis is scaled by a factor $10^{-5}$ for the $qq\to q^\prime$ channel!).

\begin{figure}[t]
\centering
\includegraphics[width=0.49\textwidth]{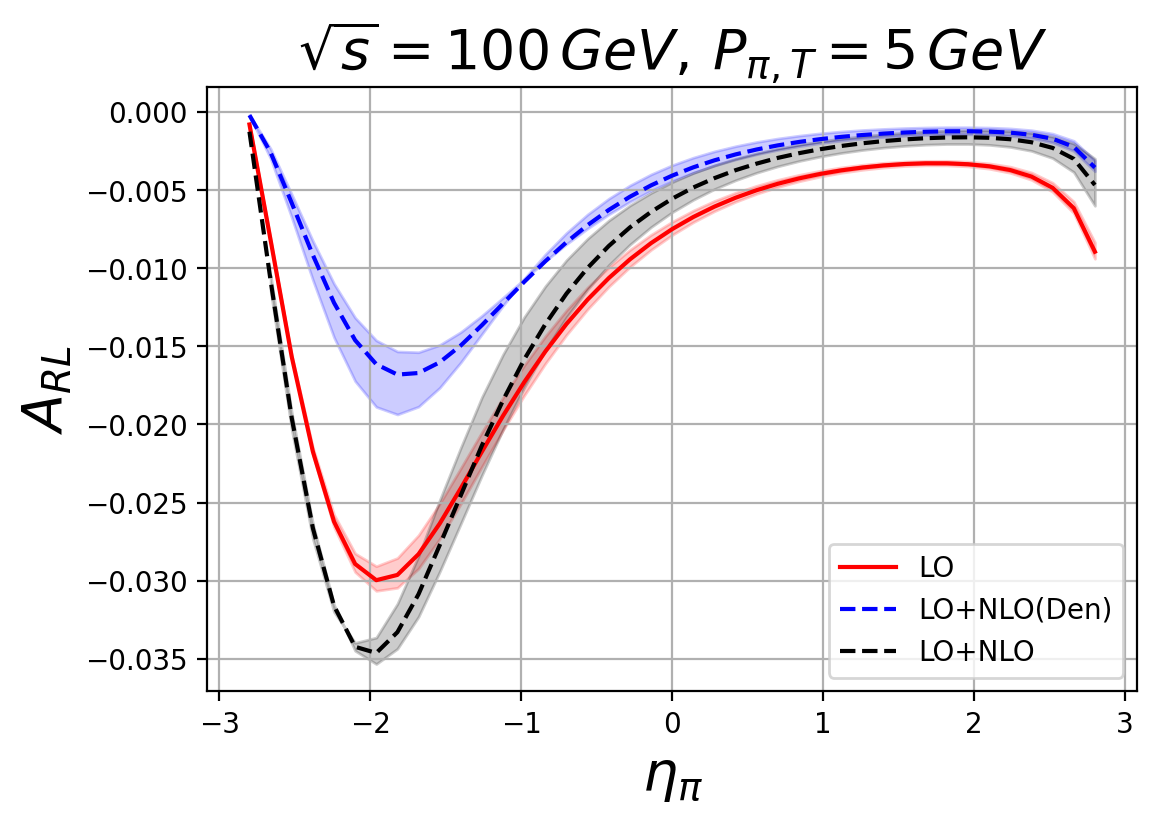}
\includegraphics[width=.49\textwidth]{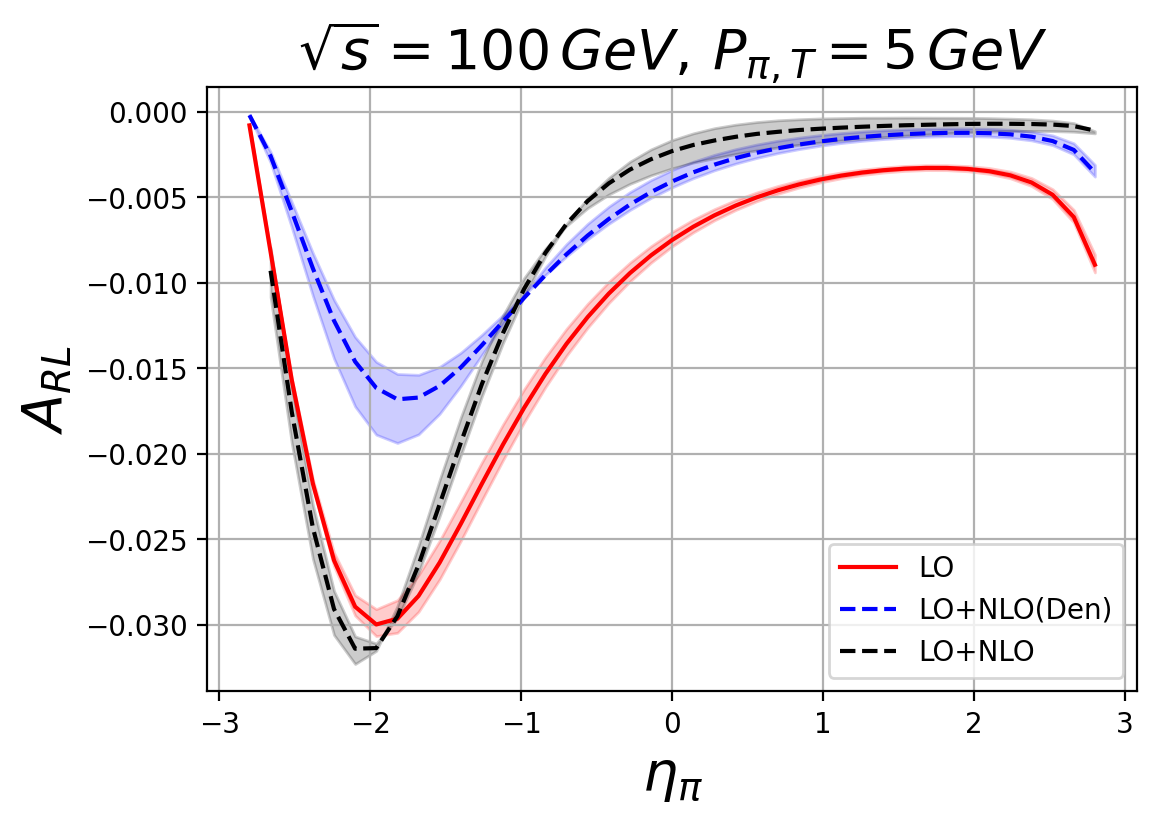}
\caption{Right-left asymmetry $A_{RL}$ of \eqref{eq:cmrightleft} plotted vs. the pion's pseudorapidity $\eta_\pi$ at fixed transverse momentum $P_{\pi,T}=5\,\mathrm{GeV}$ in Scenarios 0 (\textbf{left}) and 1 (\textbf{right}). The red curve shows the asymmetry at LO, the dashed black curve at NLO. The dashed blue curve shows the asymmetry that one obtains if NLO corrections are only included in the denominator of the asymmetry \eqref{eq:cmrightleft}.\label{fig:ChannelsAUTPiScen0and1}}
\end{figure}

We proceed by adding all partonic channels and calculating the right-left asymmetry $A_{RL}$ to NLO accuracy. The result is shown in the left panel of Fig.~\ref{fig:ChannelsAUTPiScen0and1} versus pseudorapidity $\eta_\pi$, again at fixed transverse momentum of the pion. We observe that the asymmetry peaks at far backward pseudorapidities of around $-2$. This backward direction corresponds to the nucleon direction in the frame \eqref{eq:cmframemomenta} we have adopted. The asymmetry at LO is predicted to be about $-3.5\,\%$ at the peak, but drops to about $-0.2\,\%$ in the forward region. If one takes into account NLO corrections only for the unpolarized cross section, i.e., in the denominator of the asymmetry $A_{RL}$, the asymmetry is roughly reduced by a factor of two. This was already found in Ref.~\cite{Hinderer:2015hra} and recently quantified in Ref.~\cite{Fitzgibbons:2024zsh}. However, Figure~\ref{fig:ChannelsAUTPiScen0and1}
demonstrates that if the full NLO corrections are taken into account in \emph{both} the numerator and the denominator, 
the NLO corrections for the numerator overcompensate those in the denominator, at least near the peak, thus increasing the asymmetry.  In contrast,
in the forward direction, the situation is reversed, and the NLO corrections of the denominator dominate. Using our simple estimate for the scale dependence of the asymmetry we observe a moderate uncertainty for the NLO corrections and an even smaller effect at LO.

\subsubsection{Scenario 1\label{subsub:Scenario1}}

\begin{figure}[t]
\centering
\includegraphics[width=0.49\textwidth]{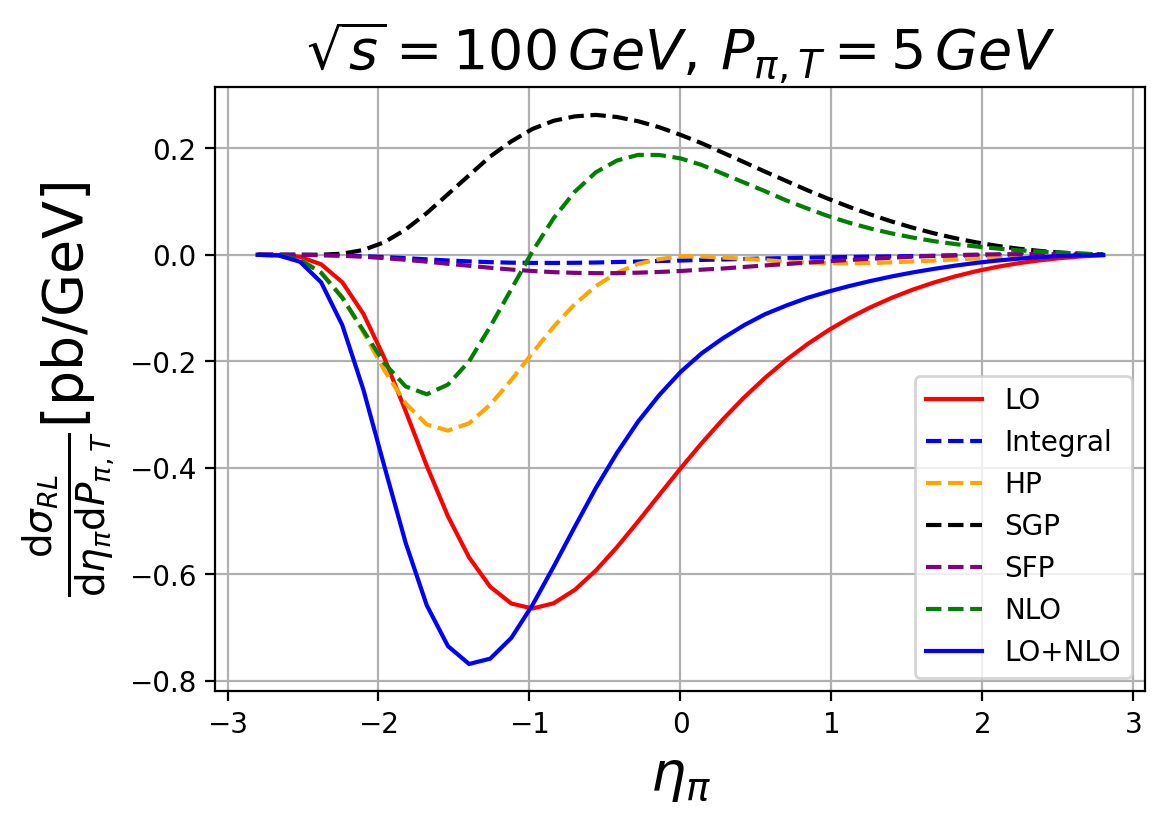}
\includegraphics[width=.49\textwidth]{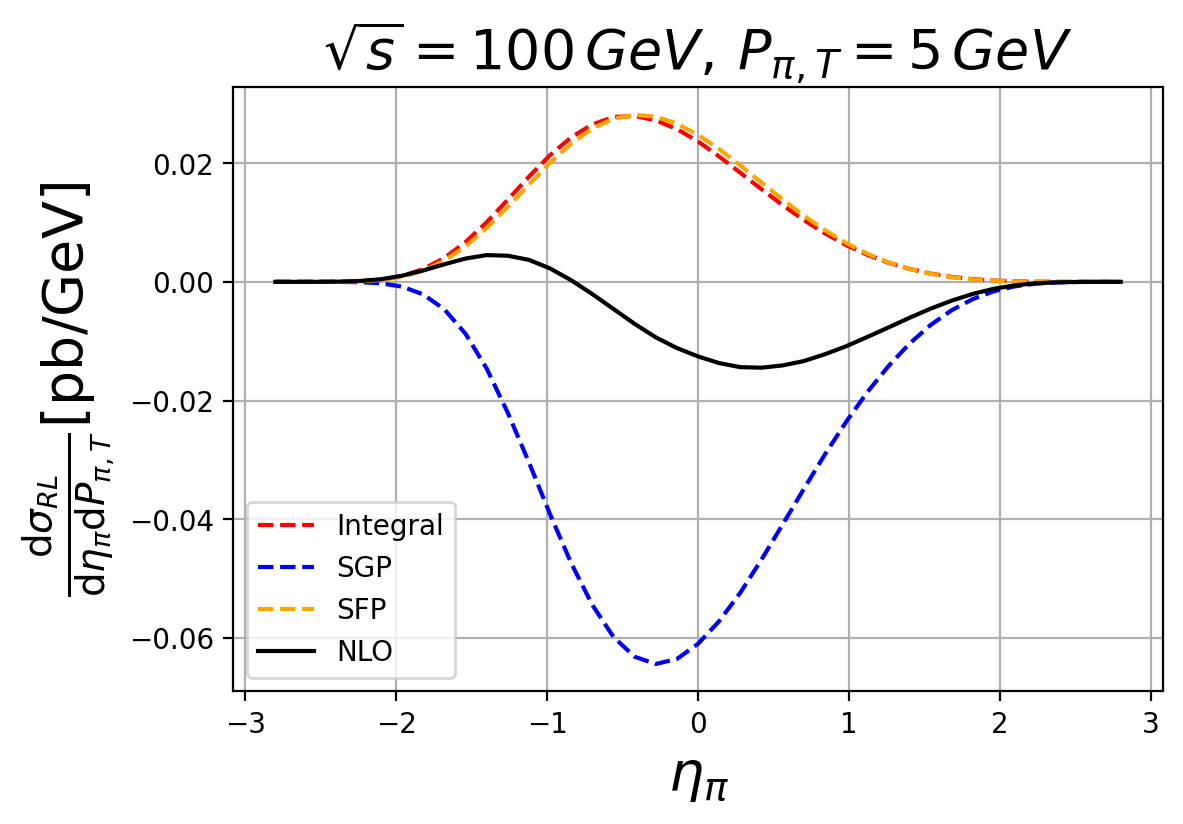}\\\vspace{0.5cm}
\includegraphics[width=0.49\textwidth]{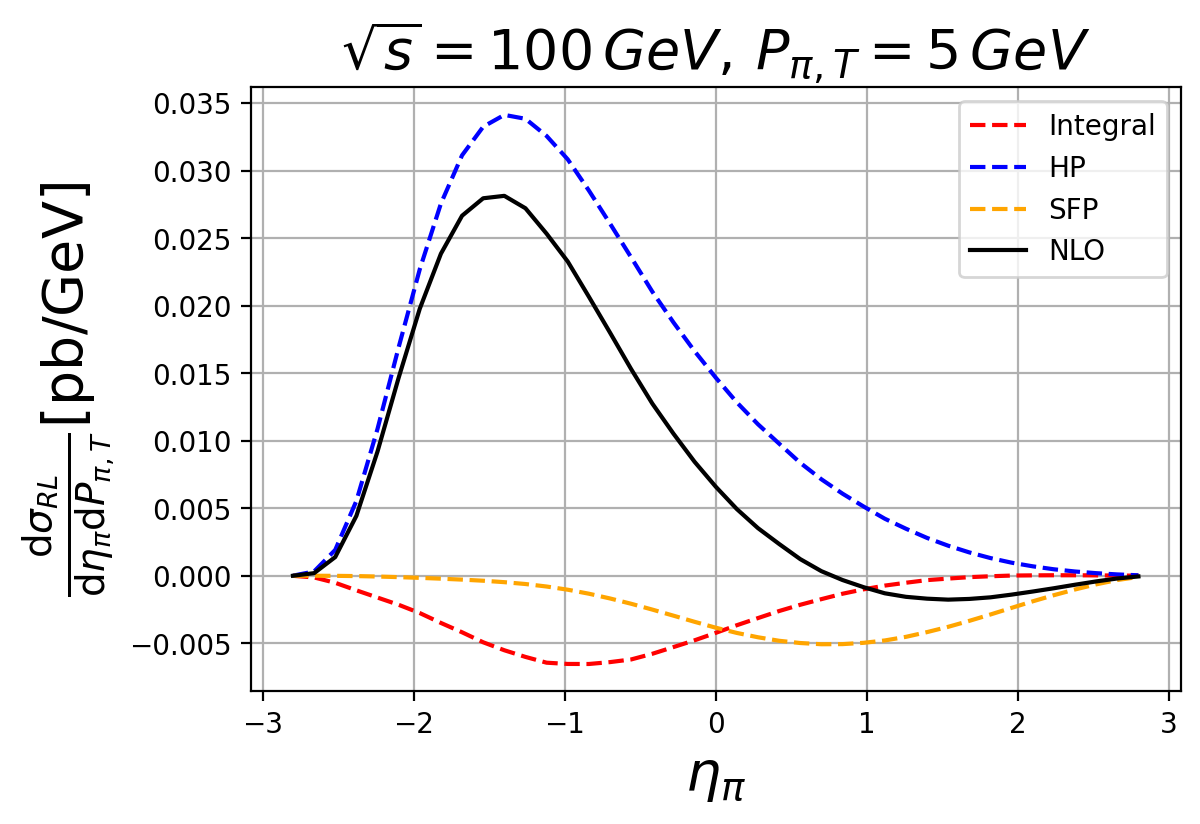}
\includegraphics[width=.49\textwidth]{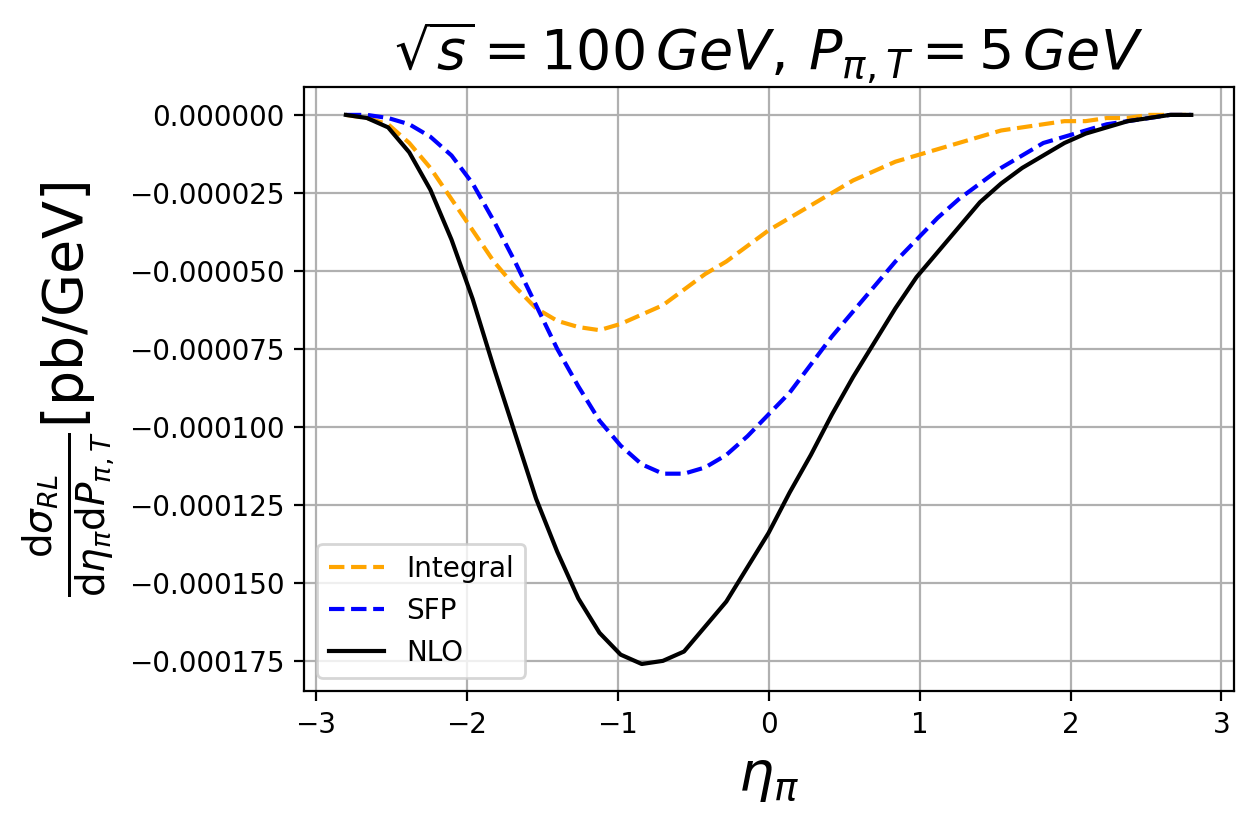}
\caption{Same as Fig.~\ref{fig:ChannelsUTPiScen0}, but for Scenario 1.\label{fig:ChannelsUTPiScen1}}
\end{figure}

We now turn to Scenario 1, discussed in Appendix \ref{appsub:Scen1}. In this scenario, the contour plots for the correlation functions $F$ and $G$ show 
moderately more ``structure'' (see Figs.~\ref{fig:Scen1F}, \ref{fig:Scen1G}), as compared to those for Scenario 0. In particular, we now have a nonvanishing function $G$.

The NLO results for the spin-dependent cross section are shown for the various channels in Fig.~\ref{fig:ChannelsUTPiScen1}. By comparison of the dominant $qg\to q$ channel in Scenario 0 and Scenario 1 we observe that, in particular, the hard-pole contribution differs quite a lot. 
For the gluon fragmentation channel, we observe that the SFP and integral contributions roughly cancel against the SGP contribution. Hence, NLO corrections from this channel almost drop out. The NLO contributions for the $qq\to q$ channel in particular become larger in Scenario 1 compared to 
those in Scenario 0, but still remain small. The same statement holds for the $qq\to q^\prime$ channel.

We again use the results to compute the right-left asymmetry, see Fig.~\ref{fig:ChannelsAUTPiScen0and1} (right). We find that the asymmetries in Scenarios 0 and 1 do not differ too much, and one can readily draw the same conclusions as in the previous subsection. Whether or not the precision of EIC measurements of $A_{RL}$ will be good enough to 
distinguish between both scenarios and possibly rule out one of them, remains to be seen. Our simplified approach for the scale variation again predicts only moderately sized uncertainty bands for Scenario 1, the smallest out of all three scenarios.

\subsubsection{Scenario 2\label{subsub:Scenario2}}

Scenario 2 is described in Appendix \ref{appsub:Scen2}, and the corresponding contour plots for the correlation functions $F$ and $G$ are shown in Figs.~\ref{fig:Scen2F}, \ref{fig:Scen2G}. These plots display more complex ``structures'' of the correlation functions on their support in $x,x'$. 
The reason is that the Fourier coefficients in Eq.~\eqref{eq:Scenario2} parameterizing Scenario 2 have been inflated by a factor of three relative to those for
Scenario 1, Eq.~\eqref{eq:Scenario1}. As a result, the correlation functions increase for $x\neq x^\prime$. It is interesting to note that a model calculation in Ref.~\cite{Braun:2011aw} based on an overlap representation of the functions $F$ and $G$ in terms of light-cone wave functions supports such a behavior. In any case, we emphasize that all three scenarios agree with all known constraints.

\begin{figure}[t]
\centering
\includegraphics[width=0.49\textwidth]{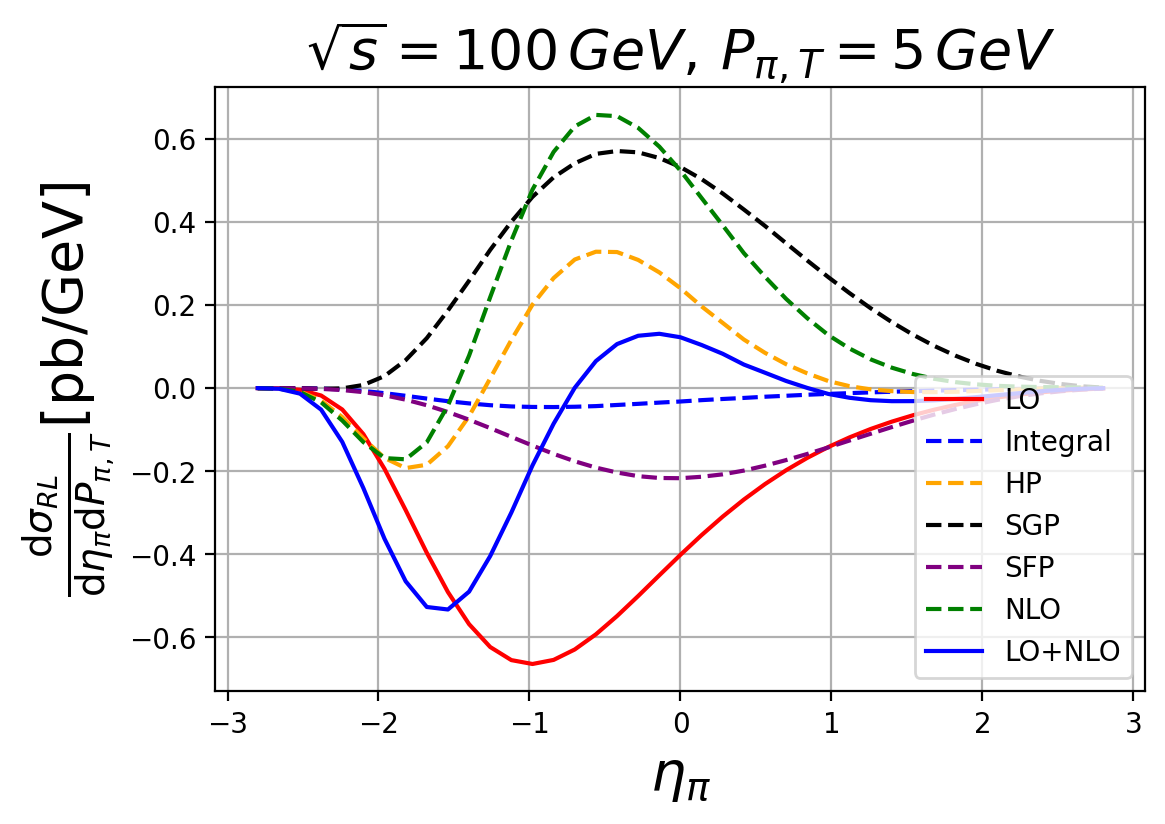}
\includegraphics[width=.49\textwidth]{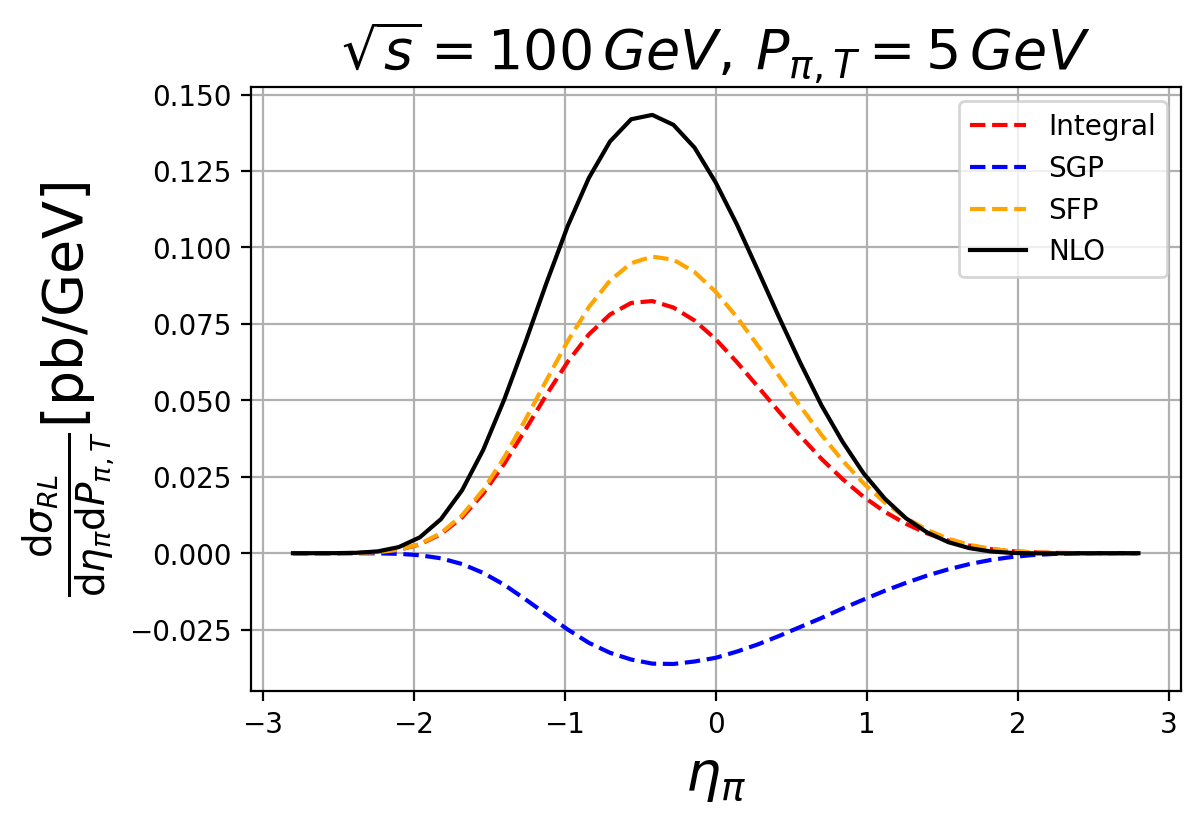}\\\vspace{0.5cm}
\includegraphics[width=0.49\textwidth]{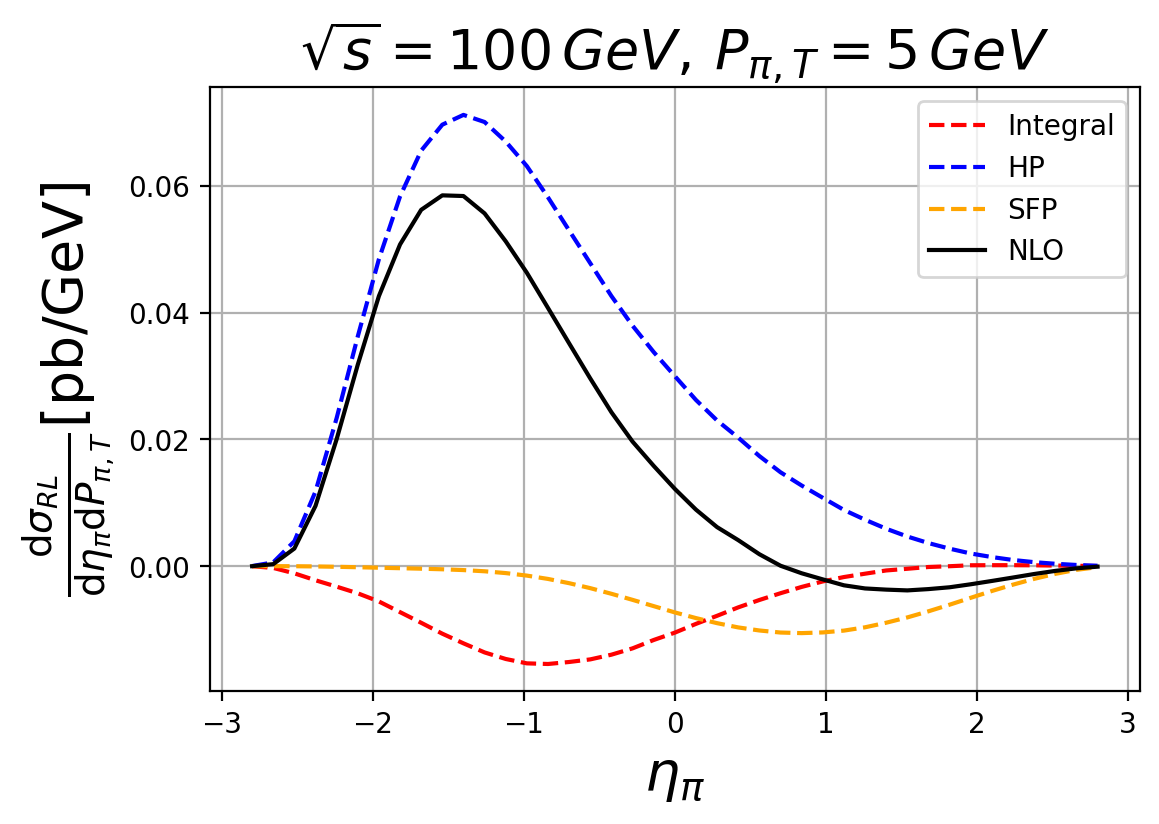}
\includegraphics[width=.49\textwidth]{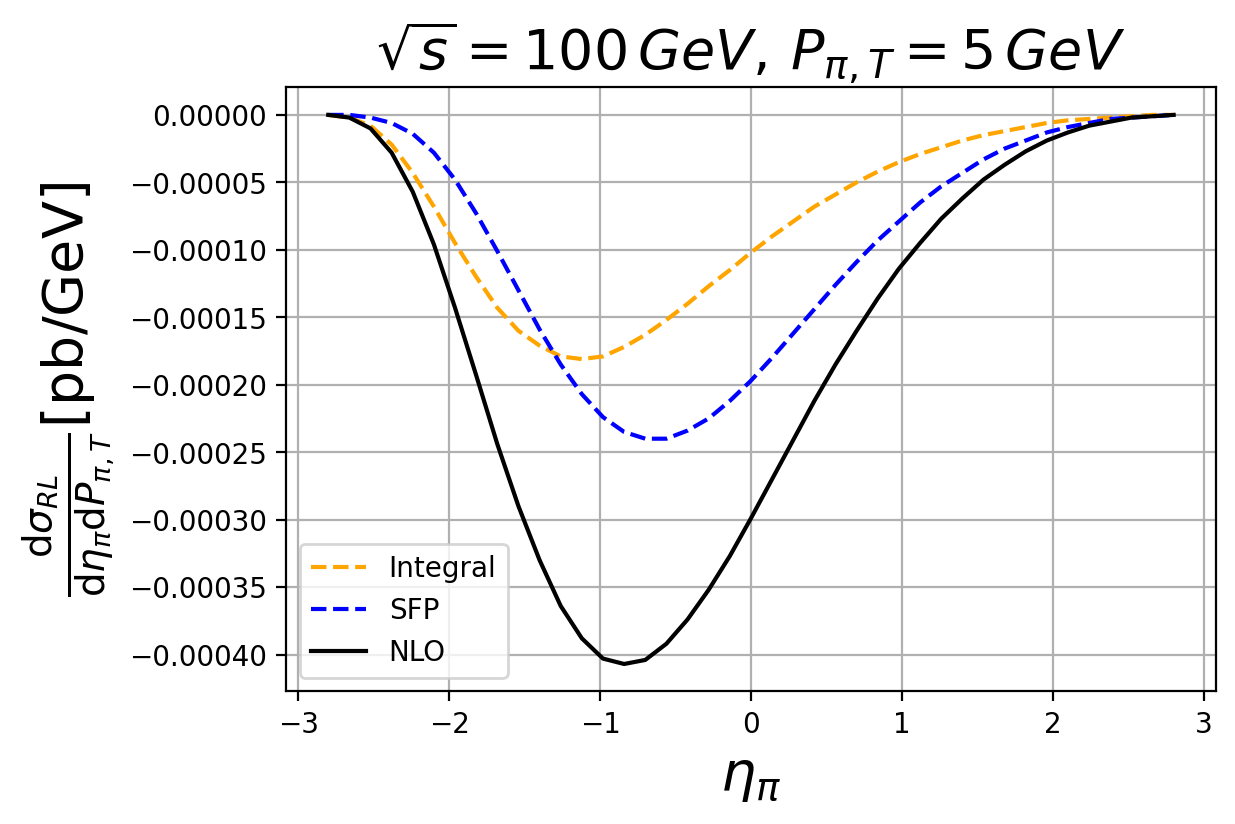}
\caption{Same as Fig.~\ref{fig:ChannelsUTPiScen0}, but for Scenario 2.\label{fig:ChannelsUTPiScen2}}
\end{figure}

\begin{figure}[t]
\centering
\includegraphics[width=0.7\textwidth]{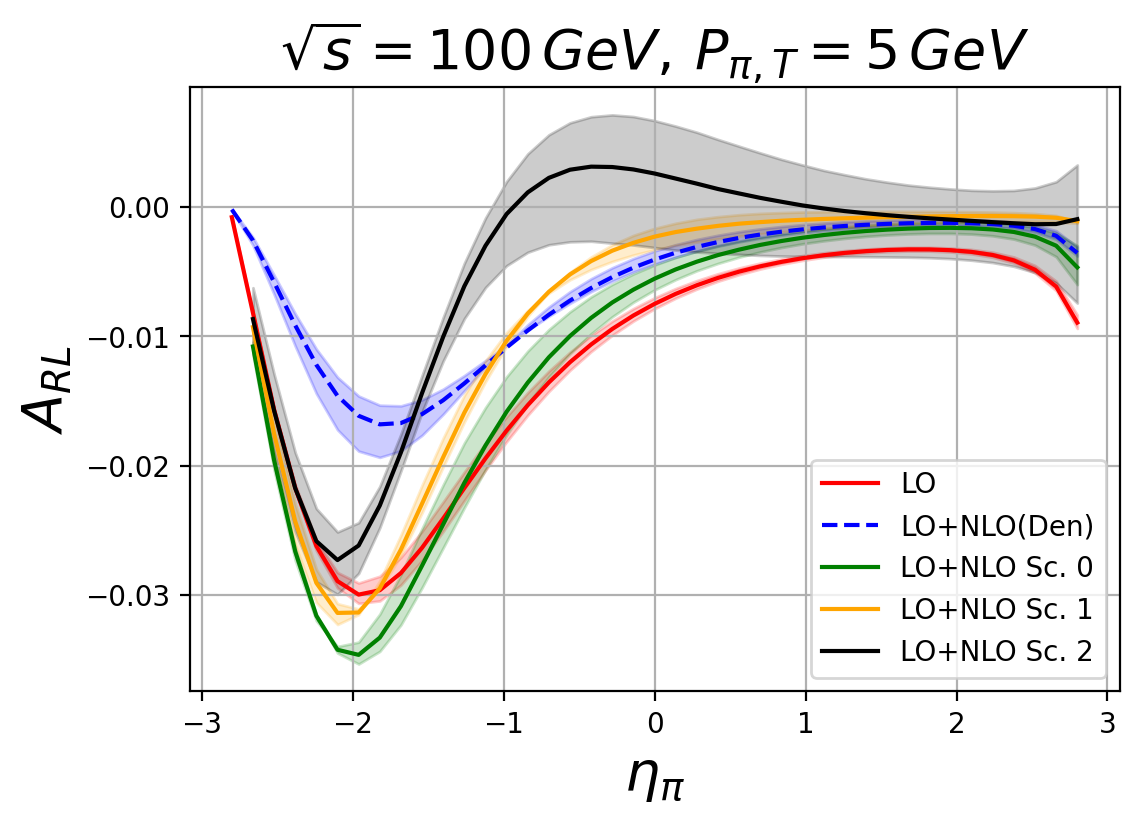}
\caption{Same as Fig.~\ref{fig:ChannelsAUTPiScen0and1}, but now showing all Scenarios 0, 1 and 2 in a combined plot for easier comparison.\label{fig:ChannelsAUTPiScen2}}
\end{figure}

In Fig.~\ref{fig:ChannelsUTPiScen2} we plot the numerator of the right-left asymmetry for Scenario 2 for all contributing partonic channels at NLO. As for the other scenarios, $qg\to q$ 
dominates over all other NLO channels. In contrast to the previous scenarios, the NLO corrections drastically change the LO result. In particular, the hard-pole contribution to the $qg\to q$ channel is completely reversed compared to Scenario 0. All other contributions, i.e., SGP, SFP and integral contributions, are larger as well. Note that the LO result (which is 
sensitive only to $x=x'$) is the same in all the plots in Figs.~\ref{fig:ChannelsUTPiScen0},\ref{fig:ChannelsUTPiScen1},\ref{fig:ChannelsUTPiScen2}. When comparing the results
for the second-most relevant channel, gluon fragmentation $qg\to g$, for Scenarios 0 and 2 we observe that because of non-trivial cancellation of SGP, SFP and integral contributions the NLO corrections flip sign. As before the other channels $qq\to q$ and $qq\to q^\prime$ remain irrelevant.

In Fig.~\ref{fig:ChannelsAUTPiScen2} we turn again to the asymmetry. Comparing $A_{RL}$ for all three Scenarios, we observe that the NLO effects are largest for Scenario 2 where the unknown support ``away from the SGP diagonal'' of the correlation functions $F$ and $G$ causes the asymmetry to even flip sign around 
mid pseudorapidity. One may hope that experimental data, once available from the EIC, would be able to resolve such large differences,
and thus help to distinguish between different scenarios. That being said, our simple implementation of scale variation produces theoretical uncertainty bands that do not show a clean sign change anymore.

\subsection{Single-Inclusive Jet Production\label{sub:NumericsJet}}

In the following we use our analytical result \eqref{eq:h2jet} for the NLO spin-dependent cross section, along with the NLO result for the unpolarized cross section taken from Refs.~\cite{Hinderer:2015hra,Hinderer:2017ntk}, to study the numerical NLO effects on the right-left asymmetry $A_{RL}$ in jet production. As input for the quark-gluon-quark correlation functions $F$ and $G$ we use the same model Scenarios 0, 1, 2 as above. We provide exemplary numerical studies for two jet radii, $R_1=0.2$ and $R_2=0.7$. 

\begin{figure}[t]
\centering
\includegraphics[width=0.49\textwidth]{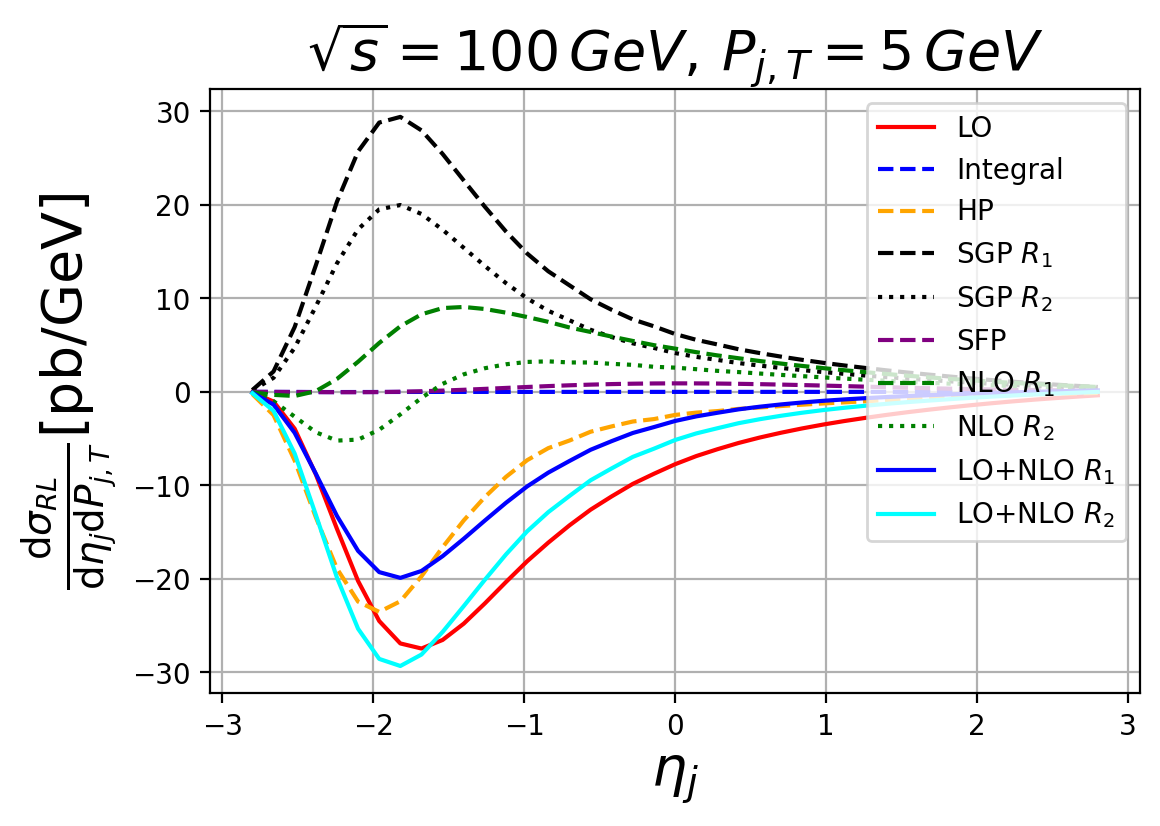}
\includegraphics[width=.49\textwidth]{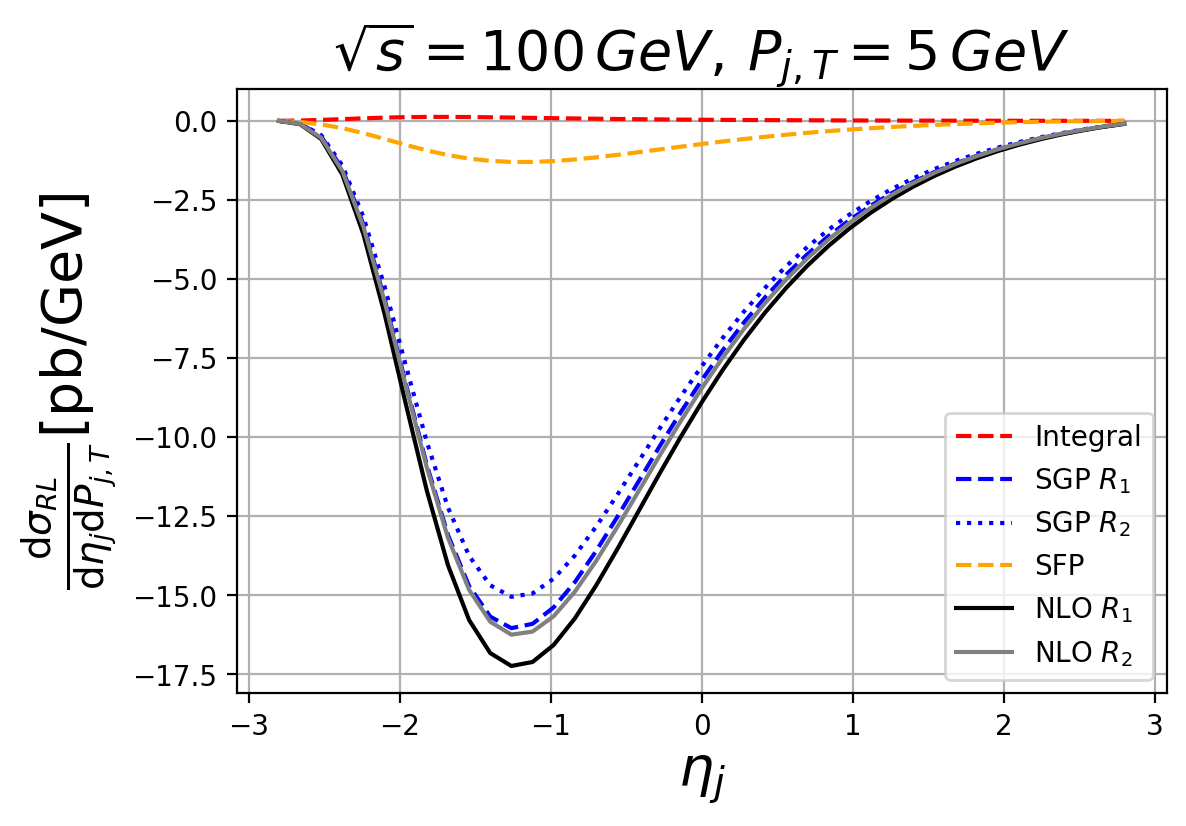}\\\vspace{0.5cm}
\includegraphics[width=0.49\textwidth]{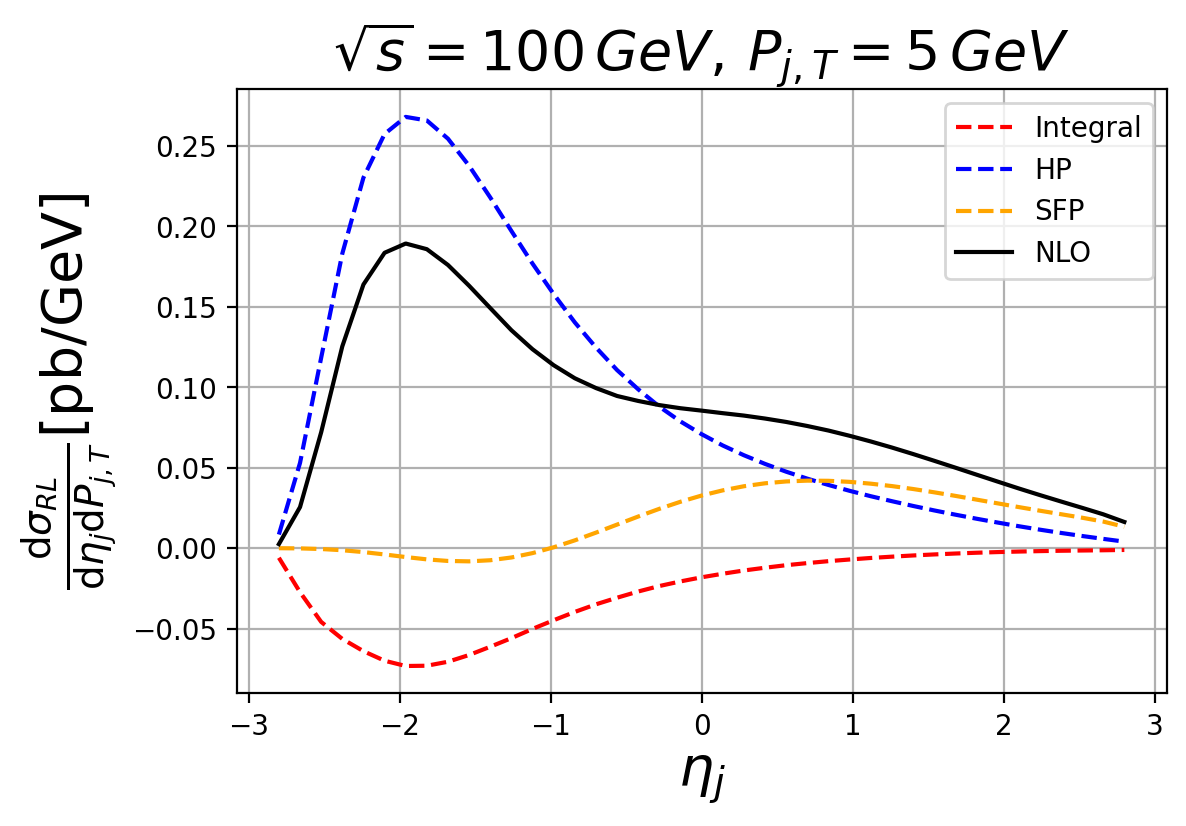}
\includegraphics[width=.49\textwidth]{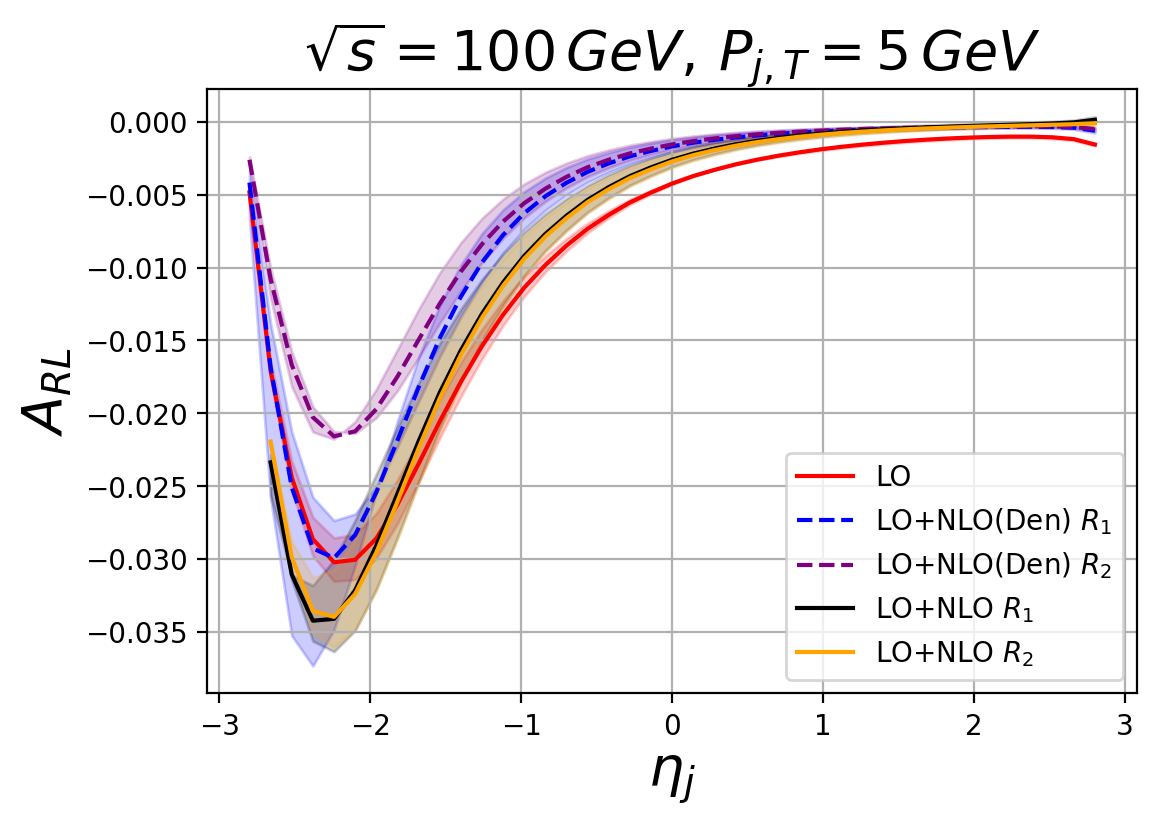}
\caption{Contributions by the various NLO channels plotted versus the jet's pseudorapidity $\eta_j$ at fixed transverse momentum $P_{j,T}=5\,\mathrm{GeV}$ at the EIC. The curves are obtained for Scenario 0 for the non-perturbative correlation functions $F$ and $G$, see \eqref{eq:Scenario0}. The channels are: $qg\to \mathrm{jet}(q)$ (upper left), 
$qg\to \mathrm{jet}(g)$ (upper right), $qq\to \mathrm{jet}(q)$ (lower left). In the lower right we show the full NLO asymmetry $A_{RL}$. Results are shown for two jet radii, $R_1=0.2$ and $R_2=0.7$.\label{fig:ChannelsUTJetScen0}}
\end{figure}

\subsubsection{Scenario 0\label{subsub:Scen0Jet}}
Our numerical results for Scenario 0 are shown in Fig.~\ref{fig:ChannelsUTJetScen0}. As for single-inclusive pion production we show plots for the NLO contributions by the 
various partonic channels. The dominant channels are those originating from quark-gluon-quark correlations ($x^\prime\ge 0$),  where either a radiated quark or gluon generates a jet in the partonic cross section. We note that for Scenario 0 large soft-gluon pole contributions as well as large hard-pole contributions appear in the 
$qg\to \mathrm{jet}(q)$ channel, along with negligible soft-fermion and integral contributions. However, a partial cancellation of SGP and HP contributions leads to rather small NLO corrections. Since HP contributions do not appear in the $qg\to\mathrm{jet}(g)$ channel, the whole NLO correction for that channel is generated by the SGP contribution and consequently is larger. The $qq\to \mathrm{jet}(q)$ channel remains irrelevant, while the $qq\to \mathrm{jet}(q^\prime)$ channel vanishes for jets.

The right-left asymmetry $A_{RL}$ for single-inclusive jet production is shown in the lower-right plot of Fig.~\ref{fig:ChannelsUTJetScen0} as function of 
the jet's pseudo-rapidity. We obtain a similar behavior for this observable in Scenario 0 as for $\pi^+$-production: the sum of all NLO corrections to the numerator
is even a little larger than the corrections to the unpolarized cross section in the denominator of $A_{RL}$. Overall, however, the asymmetry in Scenario 0 does not turn out to be too sensitive to NLO corrections, in contrast to the individual numerator and denominator. The NLO results show only little dependence on the jet radius $R$. Moreover, they are remarkably similar to their $\pi^+$ counterpart in the left panel of Fig.~\ref{fig:ChannelsAUTPiScen0and1}. 

One would regard the jet asymmetry as an even cleaner observable than the pion asymmetry. From a theoretical point of view,
the main reason is that jet production is independent of fragmentation. As discussed earlier, this is especially important in the polarized case since
no twist-3 fragmentation correlation functions 
are present for jets.  In this sense, the plots in Fig.~\ref{fig:ChannelsUTJetScen0} represent a full NLO prediction 
for the right-left asymmetry, based on Scenario 0.
We also note that the event rate for jets in the backward pseudo-rapidity region (where the asymmetry is largest) seems to be larger roughly by a factor of 100 compared to that for pion production.

\subsubsection{Scenario 1\label{subsub:Scen1Jet}}

\begin{figure}[t]
\centering
\includegraphics[width=0.49\textwidth]{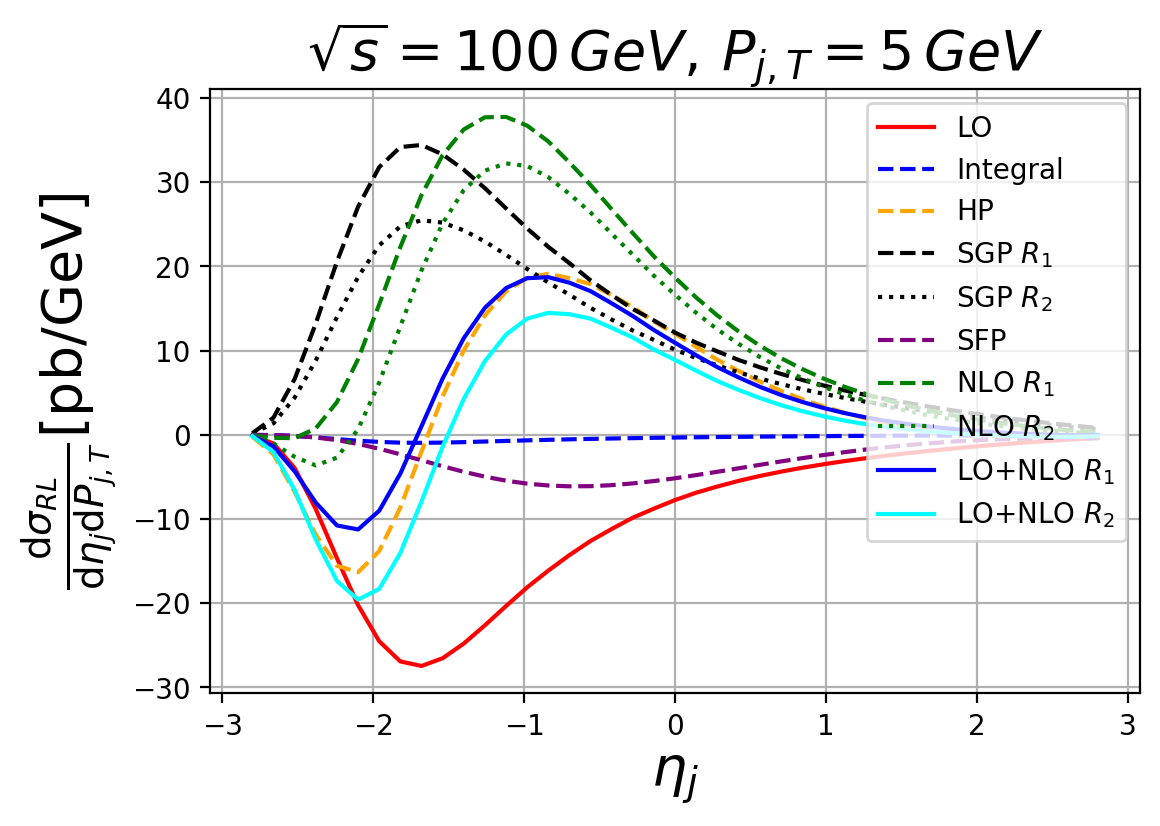}
\includegraphics[width=.49\textwidth]{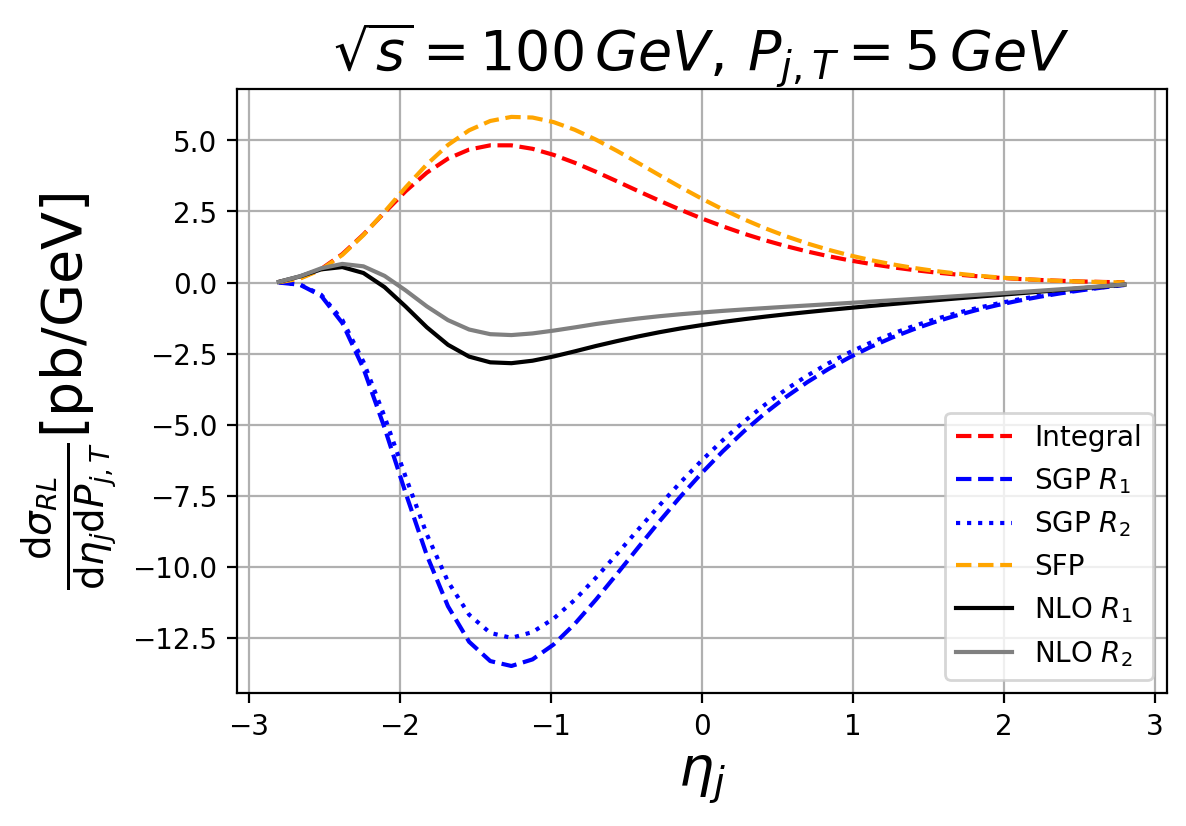}\\\vspace{0.5cm}
\includegraphics[width=0.49\textwidth]{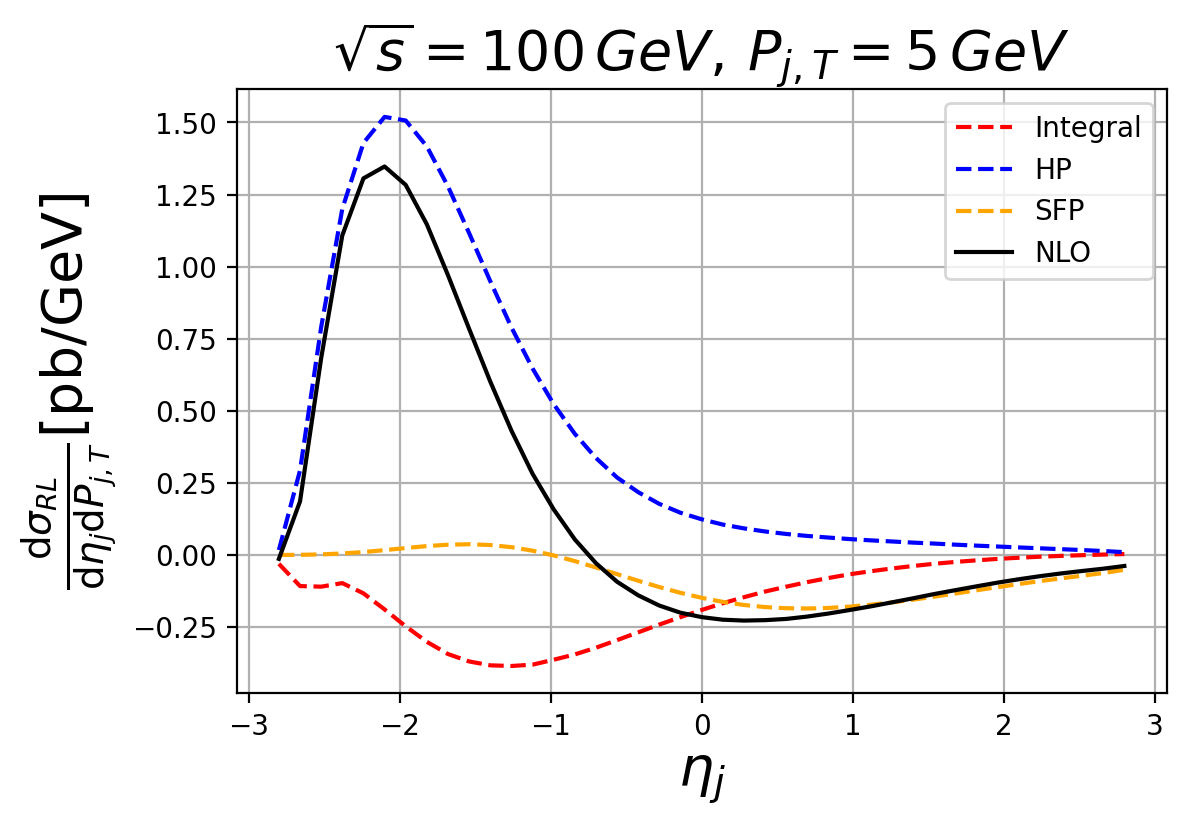}
\includegraphics[width=.49\textwidth]{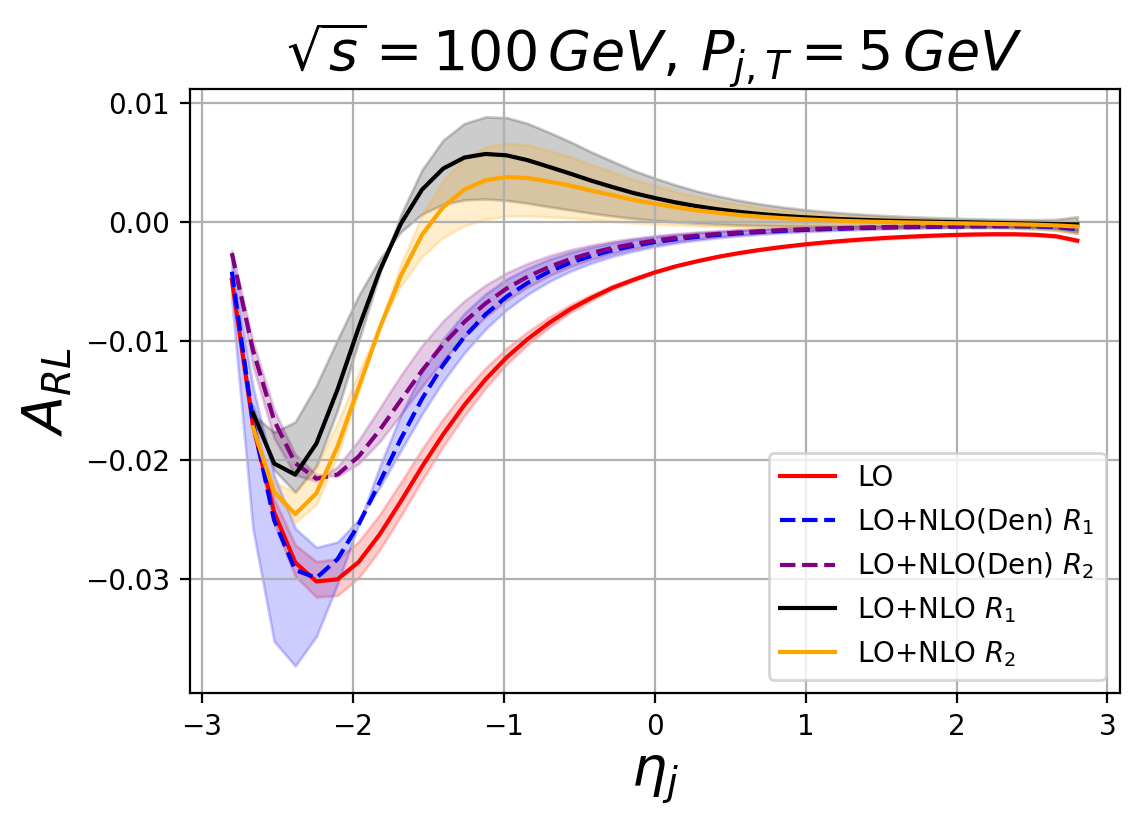}
\caption{Same as Fig.~\ref{fig:ChannelsUTJetScen0}, but for Scenario 1.\label{fig:ChannelsUTJetScen1}}
\end{figure}

Next, we consider jet production in Scenario 1 (see \eqref{eq:Scenario1}). Our numerical results for this scenario are shown in Fig.~\ref{fig:ChannelsUTJetScen1}. Comparing the curves with the corresponding ones for pion production in Fig.~\ref{fig:ChannelsUTPiScen1} we observe a similar behavior for all channels, except that the HP contribution of the $qg\to\mathrm{jet}(q)$ channel and the SFP contribution in the $qg\to\mathrm{jet}(g)$ channel are modified compared to Scenario 0. This leads to a sign change for the asymmetry, indicating that in Scenario 1 the NLO corrections affect the asymmetry in jets to a somewhat greater extent than for pions.

\subsubsection{Scenario 2\label{subsub:Scen2Jet}}

\begin{figure}[t]
\centering
\includegraphics[width=0.49\textwidth]{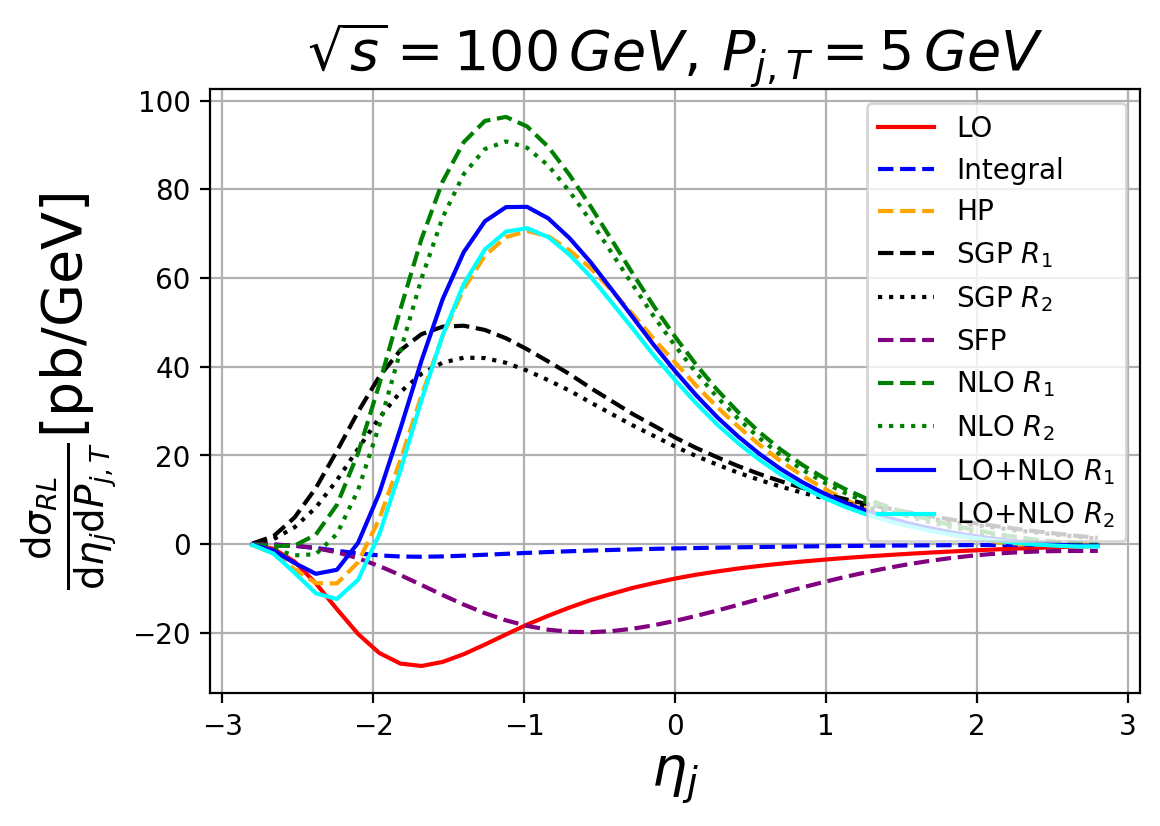}
\includegraphics[width=.49\textwidth]{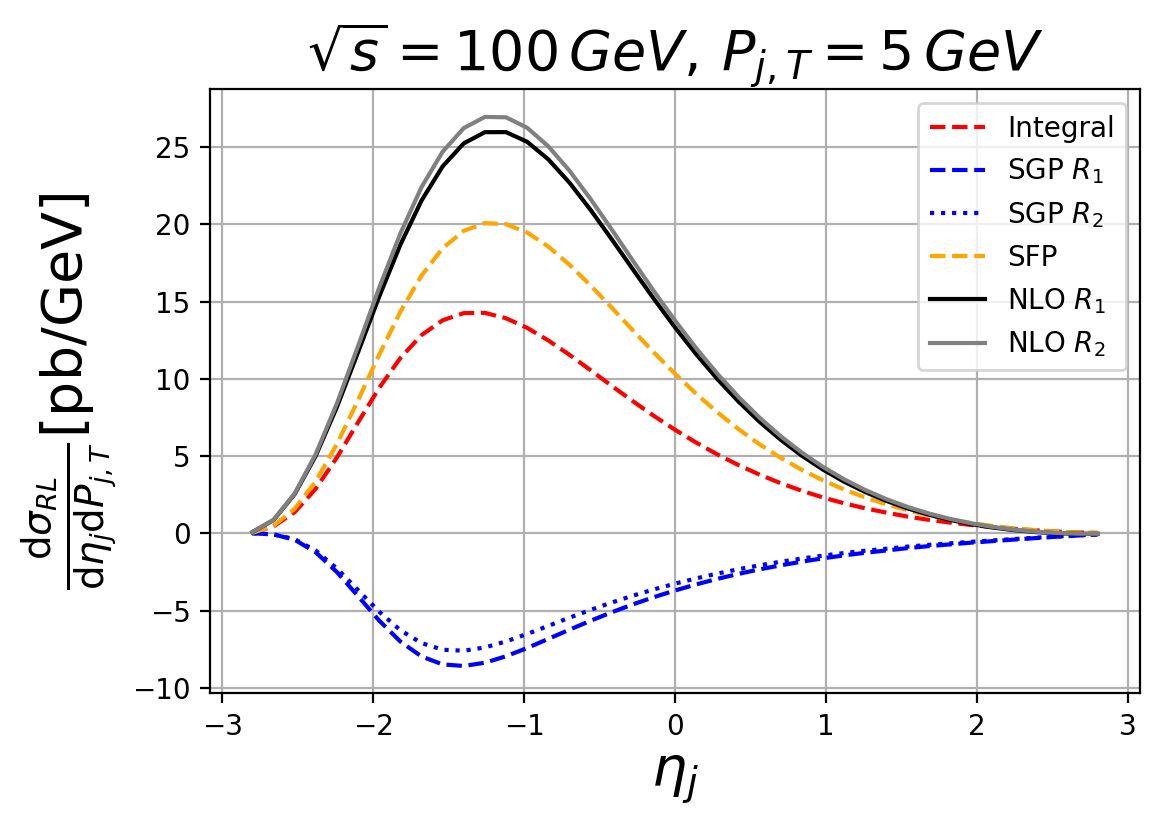}\\\vspace{0.5cm}
\includegraphics[width=0.49\textwidth]{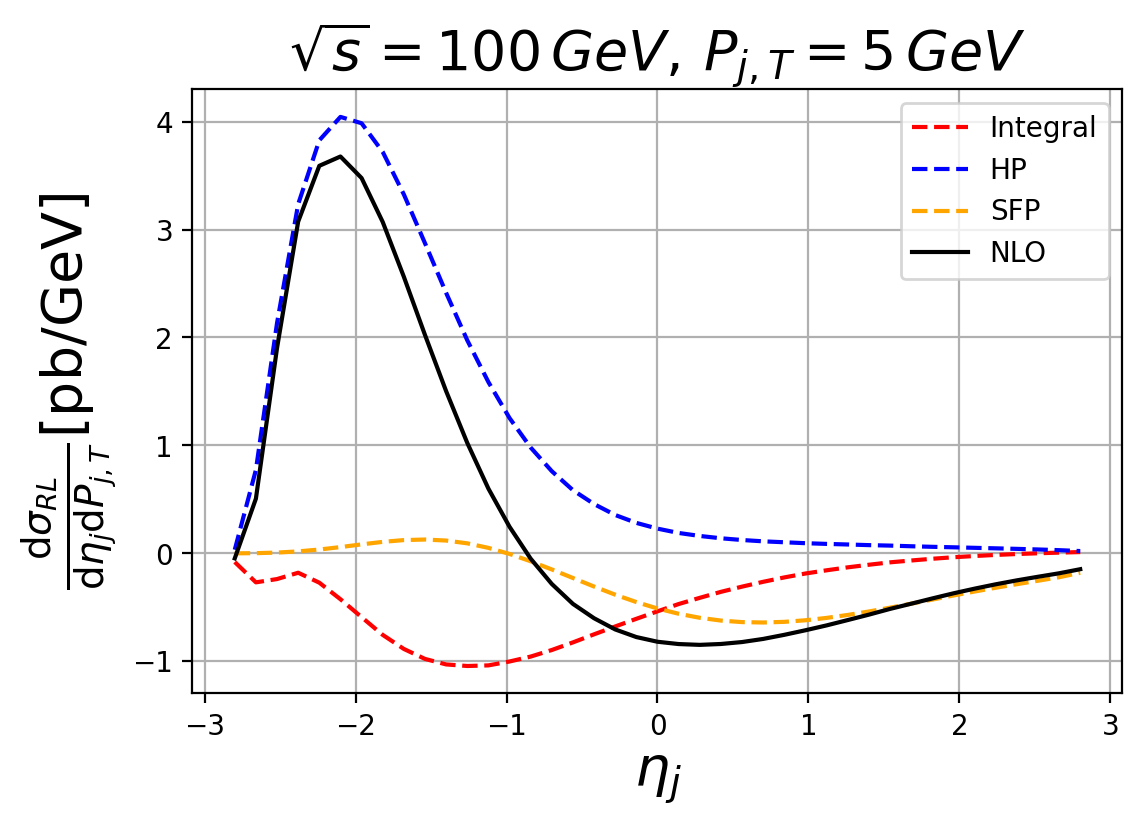}
\includegraphics[width=.49\textwidth]{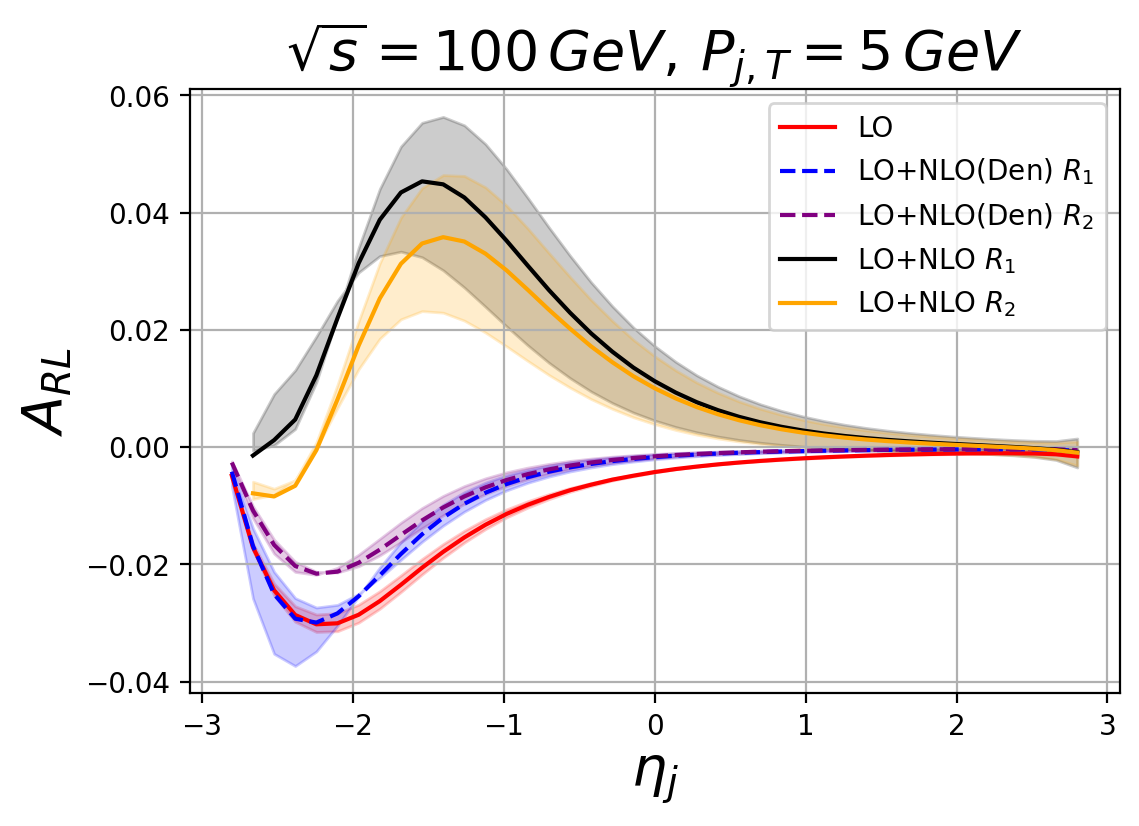}
\caption{Same as Fig.~\ref{fig:ChannelsUTJetScen0}, but for Scenario 2.\label{fig:ChannelsUTJetScen2}}
\end{figure}

\begin{figure}[t]
\centering
\includegraphics[width=0.49\textwidth]{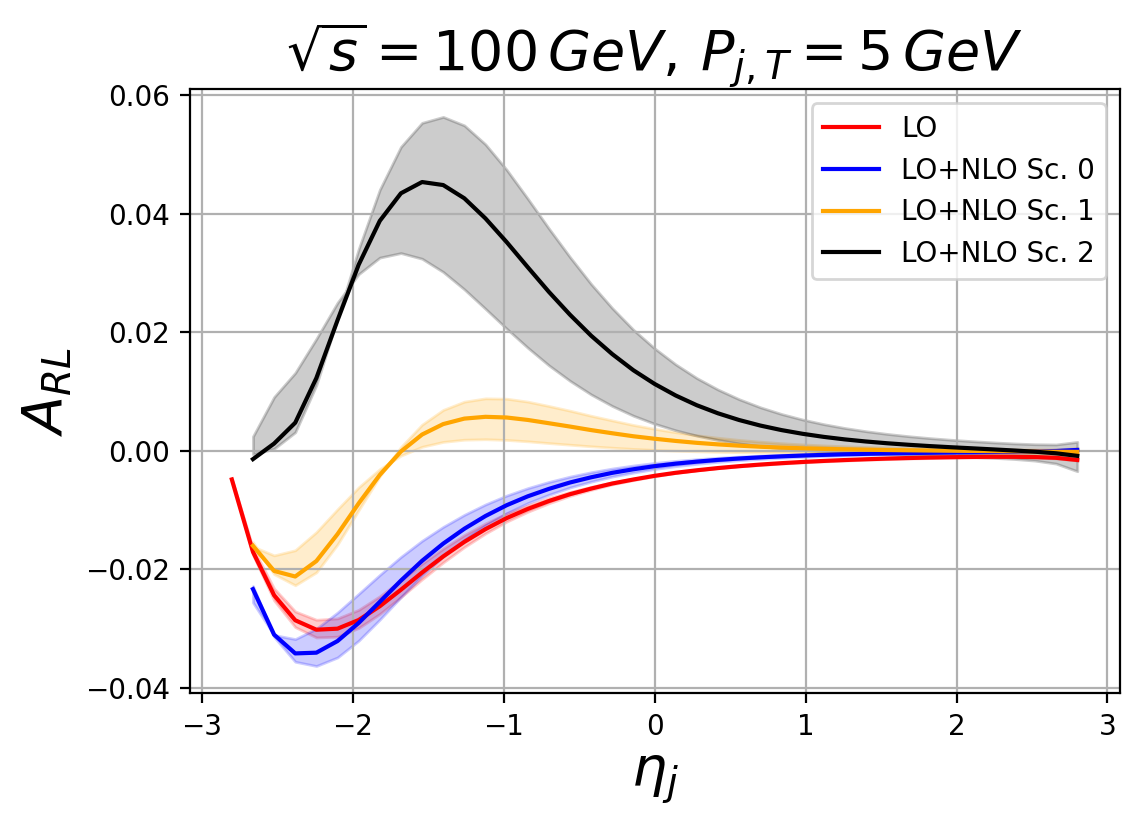}
\includegraphics[width=.49\textwidth]{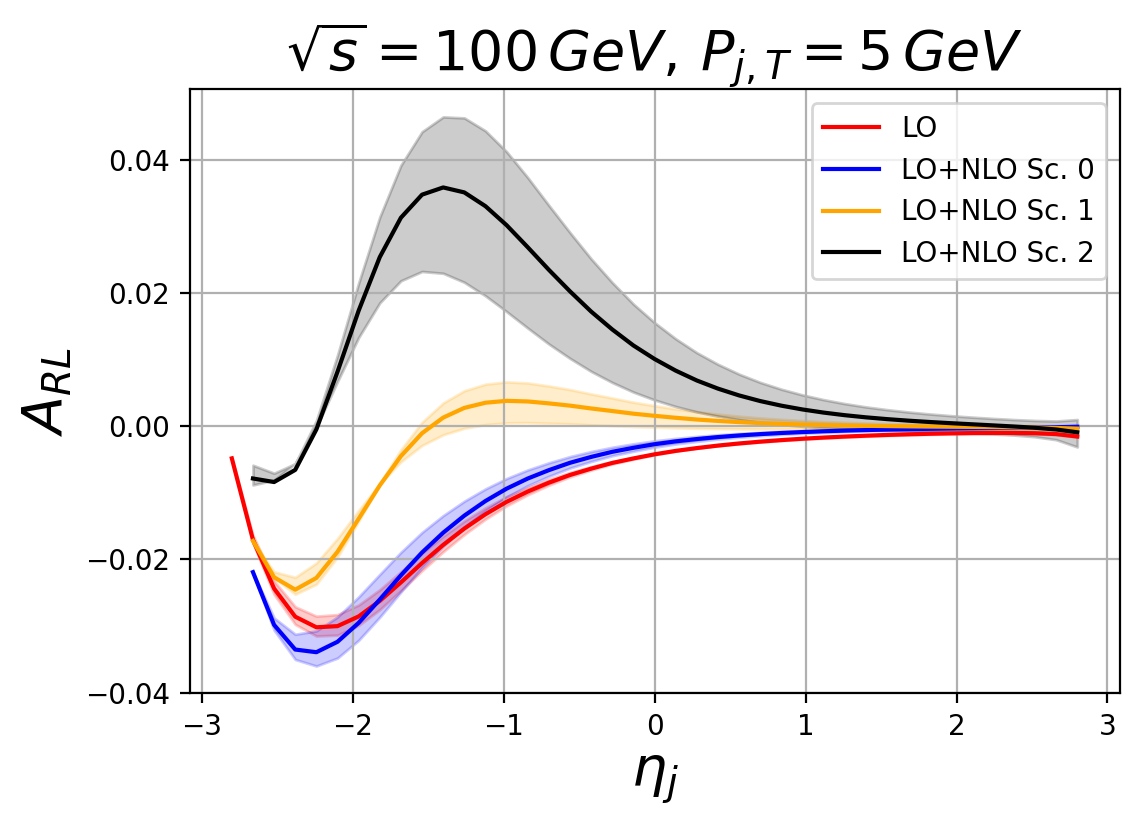}
\caption{Comparison of the NLO asymmetries $A_{RL}$ computed for Scenarios 0,1,2. The asymmetries are again plotted vs. the jet's pseudorapidity $\eta_j$ for jet cone radii $R_1=0.2$ \textbf{(left)} and $R_2=0.7$ \textbf{(right)}.\label{fig:ComparisonJet}}
\end{figure}

This effect becomes even more pronounced for Scenario 2, as can be seen from Fig.~\ref{fig:ChannelsUTJetScen2}. Again, it is the HP contribution of the $qg\to\mathrm{jet}(q)$ channel as well as the SFP contribution in the $qg\to\mathrm{jet}(g)$ channel that drastically change the behavior of the right-left asymmetry. As we can see from the lower-right plot in Fig.~\ref{fig:ChannelsUTJetScen2} the NLO asymmetry peaks somewhere in the backward pseudo-rapidity region between $-2<\eta_j<-1$ with a value 
slightly above $4\,\%$, but with different sign compared to the LO curve. This means that at NLO the right-left asymmetry for single-inclusive jet production shows high sensitivity to the precise 
form of the ``off-diagonal'' support of the $qgq$ correlation functions $F$ and $G$. This again indicates that future EIC data for the right-left (or the corresponding single-spin) asymmetry might be in the position to confirm or rule out scenarios for these functions.
In order to better illustrate this interesting result, we plot all three NLO asymmetries for jet production within Scenarios 0,1,2 
in Fig.~\ref{fig:ComparisonJet}, again for both jet radii. 

\section{Conclusions \label{sec:Conclusions}}

We have performed a next-to-leading order pQCD calculation 
of the transverse single-spin asymmetry 
for single-inclusive high-$P_T$ hadron or jet production 
in polarized lepton-nucleon scattering, $\ell N^\uparrow\to h\,\mathrm{or\,jet}\,X$. 
We have taken into account all contributions to the spin-dependent cross section that, in collinear
factorization, enter with twist-3 parton correlation functions in the incoming nucleon. 
As far as jet production is concerned, our calculation thus constitutes the first 
complete NLO calculation of a single-inclusive single-transverse spin observable. 
For hadron production, our calculation will need to be extended to account also for
twist-3 fragmentation contributions at NLO. 
We have explicitly demonstrated that collinear twist-3 factorization holds at the one-loop level for the 
contributions associated with twist-3 nucleon matrix elements. This finding provides a great boost
for theoretical confidence in the collinear twist-3 approach. 

Furthermore, we have presented estimates of the size of the NLO corrections. To this end, we have 
adopted three models for the relevant quark-gluon-quark correlation functions $F(x,x^\prime)$ and $G(x,x^\prime)$
and numerically computed the spin-dependent cross section based on our analytical NLO results, as well
as the transverse nucleon single-spin asymmetry for typical kinematics relevant at the EIC. 
We found that a large part of the NLO corrections originate from hard-pole configurations where the
correlation functions are probed on their ``off-diagonal'' support $x\neq x^\prime$, implying that 
future EIC data may have the potential to shed light on this so far completely unexplored region. 
That said, the fact that numerous separate contributions influence the size of the spin asymmetry may
make it ultimately difficult to unravel all the individual contributions experimentally. Perhaps new 
jet physics techniques such as the concept of ``jet charge'' \cite{Kang:2020fka} may be of value here at the EIC.
Possible identification of quark- and gluon-induced jets could help further to study individual partonic channels.

Our work can and will be extended in various ways. First, as mentioned above, we have
not yet addressed the twist-3 multiparton fragmentation contribution to the hadron production cross section. 
We plan to work out the NLO corrections for these contributions in the future. We also note
that so far we have only properly included the evolution of the twist-2 functions in our phenomenological studies. For the twist-3 distribution
functions, on the other hand, we merely used the scale dependence of the $f_1^q$ PDF, which is the only input of our model. The correct implementation of the true twist-3 scale evolution will be an important next step, e.g. for
studying the transverse-momentum dependence of the spin asymmetry, which may
offer additional insights into $F(x,x^\prime)$ and $G(x,x^\prime)$. Of course,
for a fully consistent next-to-leading order study, evolution at next-to-leading order will
be required, which currently is not on the horizon. 

A salient result of our paper is the sheer complexity of the calculation, even for the 
relatively simple single-inclusive hadron or jet observables in lepton-nucleon scattering that we
have considered. In a way, our calculation serves as a ``proof of principle'' that full
NLO twist-3 calculations for transverse spin asymmetries are possible. 
We expect that the methods presented in this paper will also be relevant for other, even more complicated processes, 
especially for $p^\uparrow p\to hX$. A long-term goal would be to achieve a global QCD analysis at the NLO level 
combining all possible data sets for transverse-spin effects from medium-energy fixed-target experiments all the way to RHIC and the EIC.
The results of this paper are one step towards the completion of this long-term goal.

\acknowledgments

W.V. is grateful to Patriz Hinderer and Yuji Koike for their collaboration at initial stages of this work. We thank Jianwei Qiu for helpful discussions. This work has been supported by Deutsche Forschungsgemeinschaft (DFG) through the Research Unit FOR 2926 (Project No. 409651613).

\newpage

\appendix

\section{Definition of (Multiparton) Correlation Functions in the Nucleon and Fragmentation Functions \label{app:Def}}

In this appendix, we briefly recall the definitions of the multiparton correlation functions and fragmentation functions that are relevant for our work. These definitions and their properties have been discussed in Refs.~\cite{Kanazawa:2015ajw,Koike:2019zxc,Koike:2011nx}, and we refer the reader to these references for more details. 

\subsection{Quark-Quark Correlator}

The first relevant field theoretical definition we provide is for the bare, unrenormalized quark-quark correlator,
\bea
\Phi_{\partial,ij}^{q,\rho}(x) & = & \int_{-\infty}^\infty \tfrac{\dd \lambda}{2\pi}\,\e^{i\lambda x}\langle P,S |\,\bar{q}_j(0)\,[0,\infty n]\,\times\nn\\
&&\hspace{2cm}\lim_{z_T\to 0}i\partial_{z_T}^\rho\left([\infty n+z_T,\lambda n+z_T]\,q_i(\lambda n+z_T)\right)\,|P,S\rangle\nn\\
&=&  \frac{1}{2}\,M\,\epsilon^{Pn\rho S}\,\slash{P}_{ij}\,f_{1T}^{\perp (1),q}(x)+...\,.\label{eq:DefPhid}
\eea
The diagonal matrix element between nucleon states of given momentum $P$ and spin $S$ in \eqref{eq:DefPhid} contains the Heisenberg quark fields $q$ (of flavor $q=u,d,...$) and Wilson lines $[\lambda n,\xi n]$ from light-cone position $\xi$ to $\lambda$ along an additional light-like four-vector $n^\mu$ (see Ref.~\cite{Kanazawa:2015ajw}). This vector, unlike the nucleon momentum $P$, is unphysical and can be arbitrarily chosen as long as $n^2=0$ and $P\cdot n=1$ are satisfied. However, $n$ is actually needed to characterize what we mean by the term \emph{transverse}. In fact, we may introduce a \emph{transverse} projector,
\be
g_T^{\mu \nu}\equiv g^{\mu\nu}-P^\mu n^\nu-P^\nu n^\mu\,,\, \mathrm{with}\quad a_T^\mu\equiv g_T^{\mu\nu}a_\nu\,,\label{eq:TransProjector}
\ee
that separates out the transverse components $a_T^\mu$ of an arbitrary four-vector $a^\mu$. So, if we say that the nucleon is transversely polarized, we mean $S^\mu=S_T^\mu$, or $n\cdot S=0$.

In the parameterization \eqref{eq:DefPhid} we encounter the collinear function $f_{1T}^{\perp (1),q}(x)$ which is commonly referred to as the \emph{first moment of the Sivers function} (see Refs.~\cite{Sivers:1989cc,Sivers:1990fh}). It is accompanied by a factor $\epsilon^{\mu\nu\rho\sigma}P_\mu n_\nu S_\sigma\equiv \epsilon^{Pn\rho S}$, where $\epsilon$ is the totally antisymmetric tensor with $\epsilon^{0123}=+1$. The ellipsis $...$ in \eqref{eq:DefPhid} indicates that there are several more terms that appear in the parameterization of the quark-quark correlator $\Phi_\partial^q$. However, these terms are irrelevant for the purpose of this paper. 

The quark-quark correlator $\Phi_\partial$ in \eqref{eq:DefPhid} generates contributions to a transverse nucleon spin-dependent observable where $\Phi_\partial$ is accompanied by partonic cross sections $\hat{\sigma}_\partial$ in the collinear twist-3 formalism. These partonic cross sections $\hat{\sigma}_\partial$ are calculated so that a kinematical approximation is applied to the four-momentum $k^\mu$ of the initial state quarks (cf. \eqref{eq:kinApproxktintw2}),
\[k^\mu \simeq x\,P^\mu-\tfrac{k_T^2}{2x}n^\mu+k_T^\mu\,,\]
with a subsequent Taylor expansion of the partonic cross section in the quark's transverse momentum $k_T$. It is the first-order coefficient $\mathcal{O}(k_T^1)$ in this expansion that then constitutes the final collinear partonic cross section $\hat{\sigma}_\partial(k=x\,P)$. Since $\hat{\sigma}_\partial(k=x\,P)$ is an indirect result of a non-zero transverse parton momentum, the contributions generated by $\Phi_\partial$ in the collinear twist-3 formalism were called \emph{kinematical} twist-3 contributions in Ref.~\cite{Kanazawa:2015ajw}.

\subsection{Gluon-Gluon Correlator}

Similarly to the quark-quark correlator one can define the following analog for gluons (see Ref.~\cite{Koike:2019zxc}, note that this reference uses a different sign convention for the $\epsilon$-tensor):

\bea
\Phi_{\partial}^{g, \sigma \tau \rho}(x) & = & \int_{-\infty}^\infty \tfrac{\dd \lambda}{2\pi}e^{i\lambda x}\langle P,S|G^{n \sigma}(0)\,[0,\infty n]\,\times\nn\\
&&\hspace{2cm}\lim_{z_T\to 0}i\partial_{z_T}^{\rho}\left([\infty n+z_T,\lambda n+z_T]\,G^{n \tau}(\lambda n+z_T)\right)\,|P,S\rangle\nn\\
& = & -\frac{M}{2}g_T^{\sigma\tau}\epsilon^{PnS\rho}G_T^{(1)}(x)\,-\,\frac{M}{8}\left(\epsilon^{PnS\{\sigma}g_T^{\tau\}\rho}\,+\,\epsilon^{Pn\rho\{\sigma}S_T^{\tau\}}\right)\Delta H_T^{(1)}(x)\,+\,\dots,\label{eq:DefPhidGlu}
\eea
where $G^{n\sigma}\,\equiv\,n_{\mu}G^{\mu\sigma}$ and $G^{n\tau}$ are gluonic field strength tensors and an implicit sum over their color indices of the adjoint representation is understood. Again, the ellipsis $...$ in \eqref{eq:DefPhidGlu} indicates terms in the parameterization that do not contribute to our calculation. Like the collinear function $f_{1T}^{\perp (1),q}(x)$ from \eqref{eq:DefPhid}, the kinematical distributions $G_T^{(1)}(x)$ and $\Delta H_T^{(1)}(x)$ in \eqref{eq:DefPhidGlu} are the first moments of the corresponding transverse momentum dependent distributions (TMDs). For the same reason as for the quark-quark correlator, the contribution to an observable generated by the gluon-gluon correlator is counted as a \emph{kinematical} twist-3 contribution.

\subsection{Quark-Gluon-Quark Correlator}

Another non-perturbative matrix element that is of relevance for this work is the \emph{quark-gluon-quark correlator},
\bea
\Phi_{F,ij}^{q,\rho}(x,x^\prime)&=& \int_{-\infty}^\infty \tfrac{\dd\lambda}{2\pi} \int_{-\infty}^\infty \tfrac{\dd \mu}{2\pi} \,\e^{i\lambda x^\prime}\e^{i\mu (x-x^\prime)}\,\times\nn\\
&&\hspace{2cm}\langle P,S|\,\bar{q}_j(0)\,[0,\mu n]\,ig\,G^{n\rho}(\mu n)\,[\mu n,\lambda n]\,q_i(\lambda n)\,|P,S\rangle\nn\\
&=&\frac{1}{2}M\,i\epsilon^{Pn\rho S}\,\slash{P}_{ij}\,F_{FT}^{q}(x,x^\prime) - \frac{1}{2}M\,S_T^\rho\,(\slash{P}\gamma_5)_{ij}\,G_{FT}^q(x,x^\prime)+...\,.\label{eq:DefPhiF}
\eea
The matrix element in \eqref{eq:DefPhiF} contains, in addition to two quark fields $q$, the gluonic field strength tensor $G^{n\rho}\equiv n_\mu G^{\mu\rho}$. Because of the appearance of this third dynamical quantum field in \eqref{eq:DefPhiF} contributions to spin observables generated by $\Phi_F$ are called \emph{dynamical} twist-3 contributions (see Ref.~\cite{Kanazawa:2015ajw}). The \emph{quark-gluon-quark} correlation functions $F_{FT}^q$ and $G_{FT}^q$, as introduced, for example, in Ref.~\cite{Kanazawa:2015ajw}, are key to this paper whereas other structures in the second line of Eq.~\eqref{eq:DefPhiF}, denoted by $...$, are irrelevant for our purposes. In order to ease the notation, from now on and throughout this paper we shall drop the subscript $FT$ for the quark-gluon-quark correlation functions and simply denote them as $F,G$ instead of $F_{FT},G_{FT}$.

The \emph{quark-gluon-quark} correlation functions $F$ and $G$ 
depend on two light-cone momentum fractions $x$ and $x^\prime$. Their support 
is constrained by the conditions $-1\le x,x^\prime \le 1$ and $|x-x^\prime|\le 1$. Most importantly, the vector-type function $F$ is symmetric under exchange $x\leftrightarrow x^\prime$, while the axial vector-type function $G$ is antisymmetric:
\bea
F(x,x^\prime)&=&+F(x^\prime,x)\,,\nn\\
G(x,x^\prime)&=&-G(x^\prime,x)\,.\label{eq:Symmetry}
\eea
Consequently, $G(x,x)=0$. The functions $F$, $G$ also contain information on their corresponding antiquark content in the region of negative $x$ and $x^\prime$, that is, $F^{\bar{q}}(x,x^\prime)=F^q(-x,-x^\prime)$ and $G^{\bar{q}}(x,x^\prime)=G^q(-x,-x^\prime)$ \cite{Kanazawa:2015ajw}.

At higher orders, not only the quark-gluon-quark correlation functions but also their derivatives become important. To be precise, in the following we consider the two functions $F(x_1,x_2)$ and $G(x_1,x_2)$ as generic functions of two variables $x_1$, $x_2$. We perform the partial derivatives with respect to $x_1$ or $x_2$, and then replace $(x_1,x_2)\to (x,x^\prime)$. In this way, we can easily examine the (anti)symmetry properties of the functions $F$ and $G$ in \eqref{eq:Symmetry} and extend them to their derivatives.
As a short-hand notation, we may introduce ($n_1$, $n_2$ integers)
\bea
(\partial_{1}^{n_1}\partial_{2}^{n_2} F)(x,x^\prime) &\equiv & \left(\frac{\partial^{n_1+n_2} F}{\partial x_1^{n_1}\,\partial x_2^{n_2}}\right)(x_1=x,x_2=x^\prime)\,,\nn\\
(\partial_{1}^{n_1}\partial_{2}^{n_2} G)(x,x^\prime) &\equiv & \left(\frac{\partial^{n_1+n_2} G}{\partial x_1^{n_1}\,\partial x_2^{n_2}}\right)(x_1=x,x_2=x^\prime)\,.\label{eq:Derivatives}
\eea
If the quark-gluon-quark correlation functions are assumed to be smooth, then the order of the partial derivatives is arbitrary.

With these notations it is easy to see that the (anti-)symmetry properties of the functions $F$, $G$ are inherited by their partial derivatives,
\bea
(\partial_1^{n_1}\partial_2^{n_2}F)(x,x^\prime) & = & (\partial_2^{n_1}\partial_1^{n_2}F)(x^\prime,x)\,,\nn\\
(\partial_1^{n_1}\partial_2^{n_2}G)(x,x^\prime) & = & -(\partial_2^{n_1}\partial_1^{n_2}G)(x^\prime,x)\,.\label{eq:DerivProp}
\eea
Of particular interest is the so-called \emph{soft-gluon pole}, i.e., the region of the support of the functions $F$, $G$ where $x^\prime = x$. From \eqref{eq:DerivProp} the obvious relations follow:
\begin{align}
(\partial_1 F)(x,x)&=(\partial_2 F)(x,x),&
(\partial^2_1 F)(x,x)&=(\partial^2_2 F)(x,x),&
(\partial_1\partial_2 F)(x,x)& = (\partial_2 \partial_1 F)(x,x),\nn\\
(\partial_1 G)(x,x)&  =  -(\partial_2 G)(x,x) ,&
(\partial^2_1 G)(x,x) &= -(\partial^2_2 G)(x,x),&
(\partial_1\partial_2 G)(x,x)&=0\,.\label{eq:DerivProp2}
\end{align}
We introduce an even shorter notation for derivative terms along the
\emph{diagonal} support,
\bea
F^\prime(x,x)\equiv\tfrac{\dd}{\dd x}F(x,x)&=& (\partial_1F+\partial_2F)(x,x)=2(\partial_1F)(x,x)=2(\partial_2F)(x,x),\nn\\
F^{\prime\prime}(x,x)\equiv\tfrac{\dd^2}{\dd x^2}F(x,x)&=& (\partial^2_1F+\partial^2_2F+\partial_1\partial_2F+\partial_2\partial_1F)(x,x).\label{eq:Dev1SGP}
\eea
Lastly, we discuss the derivatives of antiquark correlation functions $F^{\bar{q}}$, $G^{\bar{q}}$. It is easy to see that the first partial derivatives differ by a sign between quarks and antiquarks, while the second derivatives keep their sign, ($i,j=1,2$),
\begin{align}
(\partial_i F^{\bar{q}})(x,x^\prime) & =  - (\partial_i F^q)(-x,-x^\prime),& (\partial_i G^{\bar{q}})(x,x^\prime)  &=  - (\partial_i G^q)(-x,-x^\prime)\,,\nn\\
(\partial_i\partial_j F^{\bar{q}})(x,x^\prime) & =  + (\partial_i \partial_j F^q)(-x,-x^\prime),& (\partial_i \partial_j G^{\bar{q}})(x,x^\prime) & =  + (\partial_i \partial_j G^q)(-x,-x^\prime)\,.\label{eq:DerivAntiQuark}
\end{align}

\subsection{Triple-Gluon Correlator}
Yet another type of correlator of the nucleon that we need in our calculation is the triple-gluon correlator. It appears in the following two varieties (cf. Refs.~\cite{Koike:2011nx,Koike:2019zxc}; note that these references use a different convention for the $\epsilon$-tensor):
\bea
N_F^{\sigma\tau\rho}(x,x^\prime) & = & \int_{-\infty}^{\infty}\tfrac{\dd \lambda}{2\pi}\int_{-\infty}^{\infty}\tfrac{\dd \mu}{2\pi} e^{i\lambda x}e^{i\mu(x^\prime -x)}\,\times\nn\\
&&\hspace{1.5cm}\langle P,S|i\,f^{\alpha\beta\gamma}G^{n\sigma ,\alpha}(0)\,[0,\mu\,n]\,ig\,G^{n\rho , \beta}(\mu n)\,[\mu\,n,\lambda\,n]\,G^{n\tau , \gamma}(\lambda n)|P,S\rangle \nn\\
& = &2iM\left[g_T^{\sigma\tau}\epsilon^{\rho PnS}N(x,x^\prime)-g_T^{\tau\rho}\epsilon^{\sigma PnS}N(x,x-x^\prime)-g_T^{\rho\sigma}\epsilon^{\tau PnS}N(x^\prime,x^\prime -x)\right],\label{eq:DefNF}
\eea
\bea
O_F^{\sigma\tau\rho}(x,x^\prime) & = &  \int_{-\infty}^{\infty}\tfrac{\dd \lambda}{2\pi}\int_{-\infty}^{\infty}\tfrac{\dd \mu}{2\pi} e^{i\lambda x}e^{i\mu(x^\prime -x)}\,\times\nn\\
&&\hspace{1.5cm}\langle P,S|\,d^{\alpha\beta\gamma}G^{n\sigma ,\alpha}(0)\,[0,\mu\,n]\,ig\,G^{n\rho , \beta}(\mu n)\,[\mu\,n,\lambda\,n]\,G^{n\tau , \gamma}(\lambda n)|P,S\rangle \nn\\
& = &2iM\left[g_T^{\sigma\tau}\epsilon^{\rho PnS}O(x,x^\prime)+g_T^{\tau\rho}\epsilon^{\sigma PnS}O(x,x-x^\prime)+g_T^{\rho\sigma}\epsilon^{\tau PnS}O(x^\prime,x^\prime -x)\right],\label{eq:DefOF}
\eea
where $d^{\alpha\gamma\beta}$ and $f^{\alpha\gamma\beta}$ are the totally symmetric/antisymmetric SU(N) structure constants. We encounter the real-valued functions $N(x,x^\prime)$ and $O(x,x^\prime)$. They have the same support properties as $F^q(x,x^\prime)$ and $G^q(x,x^\prime)$, introduced in the previous subsection, $-1\le x,x^\prime \le 1$ and $|x-x^\prime|\le 1$, and they also have certain symmetry relations. To be specific, $N(x,x^\prime)=N(x^\prime,x)=-N(-x,-x^\prime)$ and $O(x,x^\prime)=O(x^\prime,x)=O(-x,-x^\prime)$. Naturally, the contributions to a spin observable generated by the triple-gluon correlators $N_F$ and $O_F$ are counted as \emph{dynamical} twist-3 contributions.

\subsection{Fragmentation Functions \label{app:fragmentationFunctions}}

We further introduce the more familiar parton-to-hadron fragmentation functions which describe the hadronization of a quark or a gluon into a specific hadron. Both fragmentation functions may be provided by the following well-known collinear fragmentation correlators (cf. the review article \cite{Metz:2016swz} on fragmentation functions):
\bea
\Delta_{ij}^{h/q}(z) & = & \tfrac{1}{N_c}\sumint \int_{-\infty}^{\infty} \tfrac{\dd \lambda}{2\pi}\,\e^{-i\lambda/z}\,\langle \Omega |\,[ \infty\,m,0]\,q_i(0)\,|P_h;X\rangle \langle P_h;X|\,\bar{q}_j(\lambda m)\,[\lambda\, m,  \infty\,m]\,|\Omega\rangle\nn\\
&=& z^{-1+2\varepsilon}\,\slash{P}_{h,ij}\,D_1^{h/q}(z) + ...\,,\label{eq:DefFFq}\\
\Delta^{h/g,\mu\nu}(z) & = & \tfrac{1}{N_c^2-1} \sumint \int_{-\infty}^\infty \tfrac{\dd \lambda}{2\pi}\,\e^{-i\lambda/z}\,
\langle \Omega|\,G^{m\mu}(0)\,[0,\infty m]\,|P_h;X\rangle\langle P_h;X|\,[\infty m,\lambda m]\,G^{m\nu}(\lambda m)\,|\Omega\rangle \nn\\
&=& -z^{-2+2\varepsilon}\,g_\perp^{\mu \nu}\,D_1^{h/g}(z)+...\,.\label{eq:DefFFg}
\eea
In these definitions another light-cone vector $m^\mu$ with $m^2=0$, $m\cdot P_h=1$ is introduced. This light-cone vector may differ from $n^\mu$, included in the above definition of nucleon correlators. $P_h^\mu$ and $m^\mu$ provide a different transverse projector,
\be
g_\perp^{\mu \nu}\equiv g^{\mu \nu}-P_h^\mu m^\nu-P_h^\nu m^\mu\,.\label{eq:TransProjector2}
\ee
We note that since a hadron is assumed to be observed with momentum $P_h^\mu$, the summation $\sumint\,$ over intermediate states $|P_h;X\rangle$ is incomplete, which is always the case  for a fragmentation process. Furthermore, $|\Omega\rangle$ represents the full QCD vacuum state. The definitions \eqref{eq:DefFFq},\eqref{eq:DefFFg} are given in $d=4-2\varepsilon$ dimensions. The ellipsis in \eqref{eq:DefFFq},\eqref{eq:DefFFg} indicates other spin-dependent fragmentation functions that are irrelevant for the purpose of this paper. 

As explained in Sec.~\ref{sub:Renormalization}, an $\overline{\mathrm{MS}}$-subtraction of UV-divergences emerging in our perturbative calculation is required. It is well understood (see, e.g., the discussion in Refs.~\cite{Hinderer:2015hra,Rein:2024fns}) how this subtraction works for the leading-twist fragmentation functions, like, e.g., the quark-to-hadron fragmentation function $D_{1,\mathrm{bare}}^{h/q}$ that appears in the LO formula \eqref{eq:LO}, which is a priori a \emph{bare}, unsubtracted quantity. For the SSA \eqref{eq:SSA} at NLO one needs to replace $D_{1,\mathrm{bare}}^{h/q}$ by the subtracted, $\overline{\mathrm{MS}}$-renormalized fragmentation function in the following way:
\bea
D_{1,\mathrm{bare}}^{h/q}(z,\mu) & = & D_{1}^{h/q,\overline{\mathrm{MS}}}(z,\mu)+ \frac{\alpha_s(\mu)}{2\pi}\frac{S_\varepsilon}{\varepsilon}(P_{qq}\otimes D_{1}^{h/q,\overline{\mathrm{MS}}})(z,\mu)\nn\\
&& \hspace{2.3cm}+ \frac{\alpha_s(\mu)}{2\pi}\frac{S_\varepsilon}{\varepsilon}(P_{gq}\otimes D_{1}^{h/g,\overline{\mathrm{MS}}})(z,\mu)\,+\mathcal{O}(\alpha_s^2)\,,\label{eq:renFF}
\eea
where $\mu$ denotes the renormalization/factorization scale, $S_\varepsilon=(4\pi)^\varepsilon/\Gamma(1-\varepsilon)$ is a convenient prefactor consistent with the $\overline{\mathrm{MS}}$-scheme at NLO, and $P_{qq}$, $P_{gq}$ are the well-known LO splitting functions
\bea
P_{qq}(w) & = & C_F\left[\frac{1+w^2}{(1-w)_+}+\frac{3}{2}\delta(1-w)\right]\,,\nn\\
P_{gq}(w) & = & C_F\,\left[\frac{1+(1-w)^2}{w}\right]\,.\label{eq:LOsplittingfunctionTw2}
\eea
The convolution integral in Eq. \eqref{eq:renFF} is defined as usual as
\be
(P\otimes D)(z,\mu) = \int_z^1 \tfrac{\dd w}{w}\,P(w)\,D\left(\tfrac{z}{w},\mu\right)\,.\label{eq:convolution}
\ee

\subsection{Photon-in-Lepton Distribution}

Here, we introduce a photon-in-lepton distribution in terms of the following matrix element (cf. Refs.~\cite{Hinderer:2015hra,Hinderer:2017ntk,Rein:2024fns}):
\bea
\Phi^{\gamma/\ell,\mu\nu}(x) & = & \int_{-\infty}^\infty \tfrac{\dd \lambda}{2\pi}\,\e^{i\lambda x}\,\langle \ell(l)|\,F^{o\nu}(0)\,F^{o\mu}(\lambda\,o)|\ell(l)\rangle \nn\\
& = & - \frac{x}{2}\,\frac{\tilde{g}_T^{\mu\nu}}{1-\varepsilon}\,f^{\gamma/\ell}(x)+...\,.\label{eq:Defgine}
\eea
The two photon field-strength tensors $F^{o\sigma}(x)=o_\rho F^{\rho\sigma}(x)=o_\rho (\partial^\rho A^\sigma(x)-\partial^\sigma A^\rho(x))$, $A$ being the photon field, are to be evaluated between lepton states of momentum $l^\mu$. The definition \eqref{eq:Defgine} utilizes yet another light-cone vector $o^\mu$ with $o^2=0$, $o\cdot l=1$, and the corresponding transverse projector $\tilde{g}_T$ reads
\be
\tilde{g}_T^{\mu\nu}\equiv g^{\mu\nu}-l^\mu o^\nu-l^\nu o^\mu\,.\label{eq:TransProjector3}
\ee
The special role of the photon-in-lepton distribution $f^{\gamma/\ell}_1(x)$ is discussed in Sec.~\ref{sub:PiL} and a discussion of the various light-cone vectors $n^\mu$, $m^\mu$, $o^\mu$ can be found below in the next appendix.

We also mention the renormalization of the bare photon-in-lepton distribution $f_{1,\mathrm{bare}}^{\gamma/\ell}(x)$. The distribution
includes a UV-divergence that needs to be $\overline{\mathrm{MS}}$-subtracted. Following Ref.~\cite{Hinderer:2015hra} the $\overline{\mathrm{MS}}$-subtraction term reads
\bea
f_{1,\mathrm{bare}}^{\gamma/\ell}(x,\mu) & = & f_{1}^{\gamma/\ell,\overline{\mathrm{MS}}}(x,\mu) + \frac{\alpha_{\mathrm{em}}}{2\pi}\frac{S_\varepsilon}{\varepsilon}\,(P_{\gamma \ell}\otimes f_{1}^{\ell/\ell,\overline{\mathrm{MS}}})(x,\mu)+\mathcal{O}(\alpha_{\mathrm{em}}^2)\,,\label{eq:renWW}
\eea
where $P_{\gamma \ell}(w)=P_{gq}(w)/C_F$ is the lepton-to-photon QED splitting function and $f^{\ell/\ell}(x)$ the lepton-in-lepton distribution which, at LO in QED, reduces to $\delta(1-x)$.
In contrast to the QCD parton distributions, the renormalized photon-in-lepton distribution $f_{1}^{\gamma/\ell,\overline{\mathrm{MS}}}(x,\mu)$ can be calculated in QED perturbation theory, and one obtains \cite{Hinderer:2015hra},
\be
f_{1}^{\gamma/\ell,\overline{\mathrm{MS}}}(x,\mu) = \frac{\alpha_{\mathrm{em}}}{2\pi}\,P_{\gamma \ell}(x)\,\left[\ln\left(\frac{\mu^2}{x^2\,m_\ell^2}\right)-1\right]+\mathcal{O}(\alpha_{\mathrm{em}}^2)\,.\label{eq:f1WW}
\ee
Note the explicit dependence on the lepton mass $m_\ell$ in the logarithm in \eqref{eq:f1WW}. Since the renormalization/factorization scale $\mu$ is typically of the order of the hard scale of the underlying process, that is, of the order of several GeV at least, the logarithm $\ln\left(\frac{\mu^2}{x^2\,m_\ell^2}\right)$ can potentially become large and its resummation to all orders may be necessary. More discussion on the treatment of the lepton mass $m_{\ell}$ in \eqref{eq:f1WW} is included in Sec.~\ref{sub:PiL}.

\section{Discussion on the Choice of Gauge and Related Subjects\label{app:gauge}}

In this appendix we discuss the gauge conditions which we impose throughout this paper. In addition, we address the special role of the light-cone vectors $n^{\mu}, m^{\mu}, o^{\mu}$,
which were featured in several of the definitions in the previous Appendix~\ref{app:Def} for the (multiparton) correlation functions and fragmentation functions 
(see Eqs.~\eqref{eq:DefPhid},\eqref{eq:DefPhidGlu},\eqref{eq:DefPhiF},\eqref{eq:DefNF},\eqref{eq:DefOF},\eqref{eq:DefFFg}, and\eqref{eq:Defgine}).
\subsection{Light-Cone Gauge}
First, we point out that we organize our calculation of the partonic hard factors in the collinear twist-3 factorization approach using a specific light-cone gauge for 
the gluonic field $G^{\mu,\alpha}(x)$ ($\alpha$ being a color index in the adjoint representation) with asymmetric boundary conditions,
\be
n_\mu G^{\mu,\alpha}(x) = 0 \quad\mathrm{and}\quad G^\alpha_T(P\cdot x=+\infty)+G^\alpha_T(P\cdot x=-\infty)=0\,.\label{eq:LCgauge}
\ee
This gauge is a pragmatic choice already adopted in a similar, yet simpler NLO calculation for the production of polarized $\Lambda$ particles in electron-positron collisions \cite{Gamberg:2018fwy}.  It leads to simplifications since the gluonic field-strength tensors $G^{n\rho}$ in the definitions \eqref{eq:DefPhidGlu},\eqref{eq:DefPhiF},\eqref{eq:DefNF}, and \eqref{eq:DefOF} reduce to $(n\cdot \partial) G_T^\rho$, which can be easily inverted. In addition, the Wilson lines in these definitions reduce to unity. All of this simplifies the handling of the quark-gluon-quark and triple-gluon correlations.

In addition, the matrix element \eqref{eq:DefPhid} simplifies in the light-cone gauge \eqref{eq:LCgauge}. However, we must be careful when applying this gauge to the Sivers function $f_{1T}^\perp$. As described in Ref.~\cite{Boer:2003cm}, in light-cone gauge the Sivers function is generated by \emph{transverse} Wilson lines at light-cone infinity, and, the boundary conditions are important. According to Ref.~\cite{Belitsky:2002sm}, the asymmetric boundary condition in \eqref{eq:LCgauge} guarantees a non-zero Sivers function for both a DIS-like process (like the single-inclusive hadron production $\ell N\to hX$) and a Drell-Yan (DY) like process. However, the Sivers functions in DIS and DY 
differ by a sign \cite{Collins:2002kn}. In turn, the identity \eqref{eq:SiversSGP}, derived for a DIS-type process in Ref.~\cite{Boer:2003cm}, 
requires a change of sign on the r.h.s. for a DY-type process.

However, there is a price to pay when using the light-cone gauge \eqref{eq:LCgauge}. Gluonic polarization sums as well as the numerator of gluonic Feynman propagators $\--$ both objects that copiously appear in the calculation of partonic hard factors at NLO $\--$ are more complicated in this gauge compared to covariant Feynman gauge. It is well known that both must be modified in light-cone gauge \eqref{eq:LCgauge} according to
\be
-g^{\mu\nu}\to -d^{\mu\nu}(k,n)\equiv -\left(g^{\mu \nu}-\kappa \,\mathcal{P}\frac{k^\mu n^\nu+k^\nu n^\mu}{k\cdot n}\right)\,.\label{eq:LCpolsum}
\ee
The parameter $\kappa$ allows us to switch between Feynman gauge ($\kappa=0$) and light-cone gauge ($\kappa =1$). It turns out that all terms proportional to $\kappa$ in the final result for the partonic hard factors cancel upon application of the identity \eqref{eq:SiversSGP}. This means that our results for partonic hard factors coincide in Feynman gauge and in light-cone gauge. We take this as an indication that all partonic hard factors in the collinear twist-3 approach to the SSA \eqref{eq:SSA} are color gauge invariant. Hence, in turn, this means that we can use the gauge-invariant extensions of the twist-3 matrix elements \eqref{eq:DefPhid},\eqref{eq:DefPhiF} in our collinear twist-3 approach.

However, we emphasize that the choice of a light-cone gauge \eqref{eq:LCgauge} is certainly not a necessity for an NLO calculation; one may also choose the Feynman gauge right away. 

\subsection{Choice of the Light-Cone Vector}

Secondly, as discussed in Sect.~\ref{sub:LO}, there is an additional potential dependence on the light-cone vector $n^\mu$ of physical observables through the parameterizations \eqref{eq:DefPhid},\eqref{eq:DefPhidGlu},\eqref{eq:DefPhiF},\eqref{eq:DefNF},\eqref{eq:DefOF}. In Ref.~\cite{Kanazawa:2015ajw} it was shown for the LO contribution to the SSA \eqref{eq:SSA} that an arbitrary light-cone vector $n^\mu$, parameterized by the three external physical vectors of the process, can be adopted. Applying the identity \eqref{eq:SiversSGP}, one finds that
all dependences on $n^\mu$ cancel in the final LO result \eqref{eq:LO}. However, at NLO, working with an arbitrary light-cone vector $n^\mu$ would require the computation of numerous redundant terms that are eventually expected to disappear. In order to avoid this, we take the LO finding of Ref.~\cite{Kanazawa:2015ajw} for granted in our NLO calculation, and assume a specific light-cone vector right away from the beginning,
\be
n^\mu = -\tfrac{2}{t}P_h^\mu \,.\label{eq:choiceLCvector}
\ee
This would correspond to a natural choice in a specific frame in which the nucleon momentum $P^\mu$ and the momentum $P_h^\mu$ of the produced hadron are collinear along one axis, e.g. the $z$-axis. We assume that any arbitrariness of the choice \eqref{eq:choiceLCvector} drops out at NLO once the constraint \eqref{eq:SiversSGP} is applied.

\section{Analytic Continuation and Useful Integration by Parts Identities\label{sub:analContinue}}

In the calculation of several of the contributions, we find terms of the form $1/(1-w)^{n+\varepsilon}$ that require careful treatment with respect to the soft limit $w\to 1$. \textit{A priori} such terms are only integrable on an interval $(x_0,1)$ under the condition that $\varepsilon<-(n-1)$. However, integration by parts can be applied to find an analytic continuation of such terms that extends also to the region $-(n-1)<\varepsilon <0$. Although we only need the specific cases $n=2$ and $n=3$ in our calculation, we note that a closed form can also be found for a general integer $n$. Let $f$ be a sufficiently differentiable function satisfying the boundary conditions $f(1)=f^\prime(1)=\dots =f^{(n-2)}(1)=0$ (where $f^{(n)}$ denotes the $n$-th derivative) and let $x_0$ be a real constant with $0<x_0<1$. Then the following general analytic continuation holds
\be
\int_{x_0}^1\dd w \frac{f\left(\tfrac{x_0}{w}\right)}{(1-w)^{n+\varepsilon}}\,=\,\int_{x_0}^1\dd w \frac{(-1)^{n-1}}{(1-w)^{1+\varepsilon}}\frac{\Gamma(1+\varepsilon)}{\Gamma(n+\varepsilon)}\frac{\dd^{n-1}}{\dd w^{n-1}}f\left(\tfrac{x_0}{w}\right)\,.\label{app:AnalyticContinuation}
\ee
This can be shown in a straightforward way by induction. The induction step from $n$ to $n+1$ looks like this:
\bea
\int_{x_0}^1\dd w \frac{f\left(\tfrac{x_0}{w}\right)}{(1-w)^{n+1+\varepsilon}} & = & -\frac{(1-w)^{-n-\varepsilon}}{-n-\varepsilon}f\left(\tfrac{x_0}{w}\right)\bigg\vert_{x_0}^1+\int_{x_0}^1\dd w \frac{(1-w)^{-n-\varepsilon}}{-n-\varepsilon}\frac{\dd}{\dd w}f\left(\tfrac{x_0}{w}\right)\nn\\
&&\hspace{-0cm}=\int_{x_0}^1\dd w\frac{-1}{n+\varepsilon}\frac{(-1)^{n-1}}{(1-w)^{1+\varepsilon}}
\frac{\Gamma(1+\varepsilon)}{\Gamma(n+\varepsilon)}\frac{\dd^{n-1}}{\dd w^{n-1}}\left(\frac{\dd }{\dd w}f\left(\tfrac{x_0}{w}\right)\right)\nn\\
&&\hspace{-0cm}=\int_{x_0}^1\dd w \frac{(-1)^n}{(1-w)^{1+\varepsilon}}
\frac{\Gamma(1+\varepsilon)}{\Gamma(n+1+\varepsilon)}\frac{\dd^n}{\dd w^n}f\left(\tfrac{x_0}{w}\right)\,,
\eea
where the boundary term at $w=1$ vanishes because $\varepsilon<-n$ and the boundary term at $w=x_0$ because $f(1)=0$. We emphasize that only after the analytic continuation \eqref{app:AnalyticContinuation} has been applied can one proceed with the small-$\varepsilon$ expansion and make use of the familiar relation between $(1-w)^{-1-\varepsilon}$ and the $\delta$- and plus-distributions given by Eq.~\eqref{eq:epsExpandsion}.

For completeness we also give the explicit formulas for the specific cases $n=2$ and $n=3$ that were encountered in our calculation, with the generic function $f$ replaced by other relevant
objects.

\subsection{Kinematical Contributions} 
For the analytic continuation of the term $1/(1-w)^{2+2\varepsilon}$ appearing in the computation of kinematical contributions, for example in the $qg\to q$-channel, it is convenient to use the following identity:
\bea
\int_{x_0}^1\tfrac{\dd w}{w}\,\frac{\sigma_1(v,w,\varepsilon)\,f_{1T}^{\perp (1)}(\frac{x_0}{w})}{(1-w)^{2+2\varepsilon}}&=& -\frac{1}{1+2\varepsilon}\int_{x_0}^1\tfrac{\dd w}{w}\frac{[(\partial_w\sigma_1)-\frac{1}{w}\sigma_1](v,w,\varepsilon)\,f_{1T}^{\perp (1)}(\frac{x_0}{w})}{(1-w)^{1+2\varepsilon}}\nn\\
&& +\frac{1}{1+2\varepsilon}\int_{x_0}^1\tfrac{\dd w}{w}\frac{\frac{1}{w}\sigma_1(v,w,\varepsilon)\,\frac{x_0}{w}(f_{1T}^{\perp (1)})^\prime(\frac{x_0}{w})}{(1-w)^{1+2\varepsilon}}\,.\label{eq:pIkintw3}
\eea

\subsection{Hard Poles} 
For the calculation of the hard-pole contributions in the $qg\to q$-channel, that is, the partonic factors accompanying the functions $F\left(\tfrac{x_0}{w},x_0\right)$ and $G\left(\tfrac{x_0}{w},x_0\right)$, we apply the following identities:
\bea
\int_{x_0}^1\dd w\,\frac{\sigma_3(v,\varepsilon)}{(1-w)^{3+\varepsilon}}F(\tfrac{x_0}{w},x_0) &=& \frac{\sigma_3(v,\varepsilon)}{(1+\varepsilon)(2+\varepsilon)}\times\nn\\
&&\hspace{-2cm}\int_{x_0}^1\tfrac{\dd w}{w}\,\frac{\frac{1}{w^3}\left(x_0^2\,\partial_1^2F\right)(\tfrac{x_0}{w},x_0)-\frac{2}{w^2}\left(-x_0\,\partial_1F\right)(\tfrac{x_0}{w},x_0)}{(1-w)^{1+\varepsilon}}\,,\label{eq:pIHP3pe}\\
\int_{x_0}^1\dd w\,\frac{\sigma_2(v,\varepsilon)}{(1-w)^{2+\varepsilon}}F(\tfrac{x_0}{w},x_0) &=& -\frac{\sigma_2(v,\varepsilon)}{1+\varepsilon}\int_{x_0}^1\tfrac{\dd w}{w}\,\frac{\frac{1}{w}\left(-x_0\,\partial_1F\right)(\tfrac{x_0}{w},x_0)}{(1-w)^{1+\varepsilon}}\,.\label{eq:pIHP2pe}
\eea
The right-hand-sides of Eqs.~\eqref{eq:pIHP3pe},\eqref{eq:pIHP2pe} can then be expanded in $\varepsilon$ using \eqref{eq:epsExpandsion}. 
Note that the above procedure works for both correlation functions $F$ and $G$.
The hard-pole subtraction terms in \eqref{eq:dynIntmodF},\eqref{eq:dynIntmodG} also contain contributions of other derivative terms $-x_0\,(\partial_2 F)(\tfrac{x_0}{w},x_0)$ and $-x_0\,(\partial_2 G)(\tfrac{x_0}{w},x_0)$. These terms need to be dealt with in the same way as described in Eq.~\eqref{eq:pIHP2pe}, followed by application of \eqref{eq:epsExpandsion}, and (only afterwards) expansion in $\varepsilon$.
We note that all our results are organized in such a way that derivative terms are integrated by parts as much as possible. Integration by parts identities are applied to all regular, non-distributional terms of the corresponding partonic cross sections. As a consequence, only delta- and plus distributions remain in our results for the partonic factors that accompany derivatives of the $qgq$ functions. This procedure is performed using the following list of useful formulas:
\bea
\int_{x_0}^1\tfrac{\dd w}{w}\,\sigma(w)\,x_0^2\,(\partial_1^2 F)(\tfrac{x_0}{w},x_0) & = & -(\sigma(1)+\sigma^\prime(1))\,F(x_0,x_0)+\tfrac{1}{2}\sigma(1)\,(-x_0\,F^\prime(x_0,x_0))\nn\\
&&\hspace{0cm}+\int_{x_0}^1\tfrac{\dd w}{w}\,\left[2w^2\,\sigma(w)+4w^3\,\sigma^\prime(w)+w^4\,\sigma^{\prime\prime}(w)\right]\,F(\tfrac{x_0}{w},x_0)\,,\nn\\
\int_{x_0}^1\tfrac{\dd w}{w}\,\sigma(w)\,(-x_0)\,(\partial_1 F)(\tfrac{x_0}{w},x_0) & = & \sigma(1)\,F(x_0,x_0) -\int_{x_0}^1\tfrac{\dd w}{w}\left[w\,\sigma(w)+w^2\,\sigma^\prime(w)\right]\,F(\tfrac{x_0}{w},x_0)\,,\nn\\
\int_{x_0}^1\tfrac{\dd w}{w}\,\tfrac{\sigma(v)\,\ln(1-w)}{w}\,(-x_0)\,(\partial_1 F)(\tfrac{x_0}{w},x_0) & = &\sigma(v)\int_{x_0}^1\tfrac{\dd w}{w}\,\tfrac{F(\tfrac{x_0}{w},x_0)}{(1-w)_+}-\sigma(v)\int_{x_0}^1\tfrac{\dd w}{w}\,F(\tfrac{x_0}{w},x_0)\,.\label{eq:pIidentitiesHP1}
\eea

\subsection{Soft-Gluon Poles} 

Additional complications arise for the NLO computation of the SGP contributions (in the $qg\to q$ channel). Let us focus on the SGP subtraction terms in the third lines of Eqs.~\eqref{eq:dynIntmodF},\eqref{eq:dynIntmodG}. As for the hard poles, the all-order (in $\varepsilon$) results for the partonic cross sections involve terms $1/(1-w)^{3+\varepsilon}$, $1/(1-w)^{2+\varepsilon}$ that we need to deal with via Eq.~\eqref{app:AnalyticContinuation}. This implies the following two identities:
\bea
\int_{x_0}^1\dd w\,\frac{\sigma_3(v,\varepsilon)\,F(\tfrac{x_0}{w},\tfrac{x_0}{w})}{(1-w)^{3+\varepsilon}} & = & \frac{\sigma_3(v,\varepsilon)}{(2+\varepsilon)(1+\varepsilon)}\,\int_{x_0}^1\dd w\,\frac{\frac{1}{w^4}\,(x_0^2\,F^{\prime\prime}(\frac{x_0}{w},\frac{x_0}{w}))-\frac{2}{w^3}\,(-x_0\,F^\prime(\frac{x_0}{w},\frac{x_0}{w}))}{(1-w)^{1+\varepsilon}}\,,\nn\\
\int_{x_0}^1\dd w\,\frac{\sigma_2(v,\varepsilon)\,F(\tfrac{x_0}{w},\tfrac{x_0}{w})}{(1-w)^{2+\varepsilon}} & = & -\frac{\sigma_2(v,\varepsilon)}{1+\varepsilon}\int_{x_0}^1\dd w\,\frac{\frac{1}{w^2}(-x_0\,F^\prime(\frac{x_0}{w},\frac{x_0}{w}))}{(1-w)^{1+\varepsilon}}\,.\label{eq:pISGP3pe}
\eea
However, the all-$\varepsilon$-order results for the partonic cross sections also contain terms that are proportional to a hypergeometric function which, interestingly, comes with a term $1/(1-w)^{2+2\varepsilon}$,
\[ \frac{_2 F_1(-\varepsilon,-\varepsilon;1-\varepsilon;\tfrac{w}{1-v+v\,w})}{(1-w)^{2+2\varepsilon}}.\]
This term needs extra care. We again perform an integration by parts, which in turn requires a derivative of the hypergeometric function. We use the following identity for this derivative:
\[\tfrac{\dd}{\dd w}{}_2F_1(-\varepsilon,-\varepsilon;1-\varepsilon;\tfrac{w}{1-v+v\,w})=\varepsilon\,\tfrac{1-v}{w(1-v+v\,w)}\,_2F_1(-\varepsilon,-\varepsilon;1-\varepsilon;\tfrac{w}{1-v+v\,w})-\varepsilon\,\tfrac{(1-v)^{1+\varepsilon}(1-w)^\varepsilon}{w\,(1-v+v\, w)^{1+\varepsilon}}.\]
By means of this identity we derive the following integral:
\bea
\int_{x_0}^1\dd w\,\frac{\tilde{\sigma}_2(v,\varepsilon)\,_2F_1(-\varepsilon,-\varepsilon;1-\varepsilon;\tfrac{w}{1-v+vw})\,F(\tfrac{x_0}{w},\tfrac{x_0}{w})}{(1-w)^{2+2\varepsilon}} & = & -\frac{\tilde{\sigma}_2(v,\varepsilon)}{1+2\varepsilon}\int_{x_0}^1\dd w\times\nn\\
&&\hspace{-10cm}\Bigg[\frac{\frac{1}{w^2}\,_2F_1(-\varepsilon,-\varepsilon;1-\varepsilon;\tfrac{w}{1-v+vw})\,(-x_0\,F^\prime(\tfrac{x_0}{w},\tfrac{x_0}{w}))}{(1-w)^{1+2\varepsilon}}-\frac{\varepsilon\,(1-v)^{1+\varepsilon}\,F(\tfrac{x_0}{w},\tfrac{x_0}{w})}{(1-w)^{1+\varepsilon}\,w\,(1-v+vw)^{1+\varepsilon}}\nn\\
&&\hspace{-6cm}+\frac{\frac{1-v}{w\,(1-v+vw)}\,\varepsilon\,_2F_1(-\varepsilon,-\varepsilon;1-\varepsilon;\tfrac{w}{1-v+vw})\,F(\tfrac{x_0}{w},\tfrac{x_0}{w})}{(1-w)^{1+2\varepsilon}}\Bigg]\,.\label{eq:pISGP2p2eHGF}
\eea
These expressions can now safely be expanded in $\varepsilon$ by virtue of Eq.~\eqref{eq:epsExpandsion} (with a replacement $\varepsilon\to2\varepsilon$ where needed), and an expansion of the hypergeometric function
\[_2F_1(-\varepsilon,-\varepsilon;1-\varepsilon;\tfrac{w}{1-v+vw})=1+\varepsilon^2\,\mathrm{Li}_2(\tfrac{w}{1-v+vw})+\mathcal{O}(\varepsilon^3)\,.\]
Another apparent complication in the calculation of the all-$\varepsilon$-order results for the partonic cross sections is that another special function emerges, the Appell hypergeometric function of two arguments,
\[F_1(-2\varepsilon;-\varepsilon,-\varepsilon;1-2\varepsilon;\tfrac{vw}{1-v+vw};-\tfrac{v(1-w)}{1-v+vw})=1+\mathcal{O}(\varepsilon^3)\,.\]
Since this function is typically accompanied by a factor $1/(1-w)^\varepsilon$ no dangerous behavior as $w\to 1$ is observed, and one can readily replace the Appell function by unity.

The partonic cross sections for the derivative parts $F^\prime(\tfrac{x_0}{w},\tfrac{x_0}{w})$, $(\partial_2 G)(\tfrac{x_0}{w},\tfrac{x_0}{w})$ of the SGP subtraction terms in Eq.~\eqref{eq:dynIntmodF},\eqref{eq:dynIntmodG} are accompanied only by a factor $1/(1-w)^{2+\varepsilon}$ and can be handled using \eqref{eq:pISGP3pe},\eqref{eq:pISGP2p2eHGF} in a similar way.

As for the HP contributions, we list some useful formulas that facilitate the integration by parts of all regular, non-distributional terms in the partonic factors accompanying the derivatives of $qgq$ functions. For the SGP contributions, we make use of the following identities,
\bea
\int^1_{x_0} \tfrac{\dd w}{w}\,\sigma(w)\,\left(-x_0\,F^\prime(\tfrac{x_0}{w},\tfrac{x_0}{w})\right) & = & \sigma(1)\,F(x_0,x_0)-\int^1_{x_0}\tfrac{\dd w}{w}\,w\left(w\,\sigma(w)\right)^\prime\,F(\tfrac{x_0}{w},\tfrac{x_0}{w})\,,\nn\\
\int^1_{x_0} \tfrac{\dd w}{w}\,\sigma(w)\,\ln(1-w)\,\left(-x_0\,F^\prime(\tfrac{x_0}{w},\tfrac{x_0}{w})\right) & = &  \int^1_{x_0} \tfrac{\dd w}{w}\,\frac{w^2\,\sigma(w)}{(1-w)_+}\,F(\tfrac{x_0}{w},\tfrac{x_0}{w})
-\int^1_{x_0}\tfrac{\dd w}{w}\,w\left(w\,\sigma(w)\right)^\prime\,\ln(1-w)\,F(\tfrac{x_0}{w},\tfrac{x_0}{w})\,,\nn
\eea
\bea
\hspace{-0.5cm}\int^1_{x_0} \tfrac{\dd w}{w}\,\sigma(w)\,\left(x^2_0\,F^{\prime\prime}(\tfrac{x_0}{w},\tfrac{x_0}{w})\right) & = & \sigma(1)\,\left(-x_0\,F^\prime(x_0,x_0)\right)-(w\,\sigma(w))^\prime\Big|_{w=1}\,F(x_0,x_0)
+\int^1_{x_0} \tfrac{\dd w}{w}\,w\,[w^2\,(w\,\sigma(w))^\prime]^\prime\,F(\tfrac{x_0}{w},\tfrac{x_0}{w})\,,\nn\\
\int^1_{x_0} \tfrac{\dd w}{w}\,\sigma(w)\,\left(x^2_0\,(\partial_1^2G)(\tfrac{x_0}{w},\tfrac{x_0}{w})\right) & = & \sigma(1)\,\left(-x_0\,(\partial_1G)(x_0,x_0)\right)
-\int^1_{x_0}\tfrac{\dd w}{w}\,w\left(w\,\sigma(w)\right)^\prime\,\left(-x_0\,(\partial_1G)(\tfrac{x_0}{w},\tfrac{x_0}{w})\right)\,.\label{eq:pISGPidentities}
\eea

\newpage

\section{Analytical Results for the Partonic Hard-Scattering Factors at NLO \label{app:Analytics}}

In this appendix, we present the explicit analytical results for the various partonic factors.
\subsection{Channel \texorpdfstring{$qg\to q$}{qg to q}} 
\subsubsection{Integral Contribution\label{appsub:Integralqg2q}}
The partonic factor accompanying the vector-type correlation function $F$ in our result for the integral contribution to the $qg\to q$-channel in Eq.~\eqref{eq:Channel1Int} reads explicitly,
\bea
\hat{\sigma}_{\mathrm{Int}}^{qg\to q,1}(v,w,\zeta) & = & \frac{w}{4(1-v)^4}\Bigg(\frac{  (N_c-2\,C_F) \,\zeta\,(1-v)^2 v (w-\zeta )^2 (1+v (w-\zeta ))}{\sqrt{1-2\zeta v+2(2 \zeta -1) v w+v^2 (w-\zeta )^2}}\nn\\
& & -2 C_F\, v (w-\zeta )^2\,\Big( \zeta ^2+2 (\zeta -1) \zeta ^2 v^3 w \left(6 w^2-8 w+3\right)\nn\\
&& + 2 \zeta  v \left(-\zeta +\left(\zeta ^2+3 \zeta -3\right) w+1\right)\nn\\
& & + v^2 \left(\zeta ^2+\left(8 \zeta ^3-4 \zeta ^2-2\right) w^2+2 \zeta  \left(-4 \zeta ^2+2 \zeta +1\right) w\right)\nn\\
& & + \mathrm{sgn}(w-\zeta ) \left(\zeta  (\zeta +1)+v^2 \left(\zeta ^2+\zeta +2 w^2-2 \zeta  w\right)+2 \zeta  v (w-\zeta )\right)\Big)\nn\\
& & -N_c\,\Big(v (w-\zeta )^2 \left(\zeta  (\zeta +2)+v^2 \left(\zeta ^2+\left(4 \zeta ^2-2\right) w^2+2 (1-2 \zeta ) \zeta  w\right)\right.\nn\\
&&\left.+v \left(\left(4 \zeta
   ^2-2\right) w-2 \zeta ^2\right)\right)\nn\\
&& + \mathrm{sgn}(w-\zeta ) \left(v^3 (w-\zeta )^2 \left((\zeta -1) \zeta +2 w^2-2 \zeta  w\right)\right.\nn\\
&&+2 v^2 \left(-(\zeta -1) \zeta ^3+w^3-\zeta  (\zeta +1) w^2+\zeta  \left(3 \zeta ^2-\zeta -1\right) w\right)\nn\\
 &&\left.  +\zeta  (\zeta +1) v (w-\zeta )^2+2 \zeta  \left(\zeta ^2-1\right) w\right)\Big)\Bigg)\,,\label{eq:Int1qg2q}   
\eea
where $\mathrm{sgn}(x)$ is the sign function. Furthermore, the partonic factor of the axial-vector type correlation function $G$ in Eq.~\ref{eq:Channel1Int} reads, 
\bea
\hat{\sigma}_{\mathrm{Int}}^{qg\to q,5}(v,w,\zeta) & = & \frac{w}{4(1-v)^4}\Bigg( \frac{(N_c-2\,C_F)\,\zeta (1-v)^2 v (w-\zeta )^2 (1+v (w-\zeta ))}{\sqrt{1-2\zeta v+2(2 \zeta -1) v w+v^2 (w-\zeta )^2}}\nn\\
&& + 2 C_F\, v (w-\zeta )^2 \Big(\zeta ^2+2 (\zeta -1) \zeta ^2 v^3 w \left(6 w^2-8 w+3\right)\nn\\
&& + 2 \zeta  v \left(-\zeta +\left(\zeta ^2-\zeta +1\right) w-1\right)\nn\\
&& + v^2 \left(\zeta ^2+\left(8 \zeta ^3-4 \zeta ^2-2\right) w^2+2 \zeta  \left(-4 \zeta ^2+2 \zeta +1\right) w\right)\nn\\
&& +\mathrm{sgn}(w-\zeta ) \left((\zeta -1) \zeta +v^2 \left((\zeta -1) \zeta +2 w^2-2 \zeta  w\right)+2 \zeta  v (w-\zeta )\right)\Big)\nn\\
&& +N_c\,\Big(v (w-\zeta )^2 \left((\zeta -2) \zeta +v^2 \left(\zeta ^2+\left(4 \zeta ^2-2\right) w^2+2 (1-2 \zeta ) \zeta  w\right)\right.\nn\\
&&\left.+v \left(\left(4 \zeta  ^2-2\right) w-2 \zeta ^2\right)\right)\nn\\
&& +\mathrm{sgn}(w-\zeta ) \left(v^3 (w-\zeta )^2 \left(\zeta ^2+\zeta +2 w^2-2 \zeta  w\right)\right.\nn\\
&&\left. +2 v^2 \left(-\zeta ^3 (\zeta +1)+w^3-\zeta  (\zeta +3)  w^2+\zeta  \left(3 \zeta ^2+\zeta +1\right) w\right)\right.\nn\\
&&\left.   +(\zeta -1) \zeta  v (w-\zeta )^2+2 (\zeta -1)^2 \zeta  w\right)  \Big)\Bigg)\,.\label{eq:Int5qg2q}
\eea

\subsubsection{Hard-Pole Contribution\label{appsub:HPqg2q}}
Here we list the explicit analytic results for the partonic factors that appear in our formula for the hard-pole contribution to the $qg\to q$-channel in Eq.~\eqref{eq:Channel1HP},
\bea
\hat{\sigma}_{\mathrm{HP},F}^{qg\to q,1}(v,w,\chi_\mu) & = & C_F \Bigg[\frac{ \left(v^2-14 v+18\right) \ln (1-v)}{v^2\, (1-v)^2\, (1-w)_+}\nn\\
&&\hspace{-2cm}-\frac{75 v^5-100 v^4+151 v^3-1720 v^2+2460 v-1080}{60\,v\, (1-v)^4\,(1-w)_+}+\frac{2 (2 v-3)\, \ln (1-v)}{v^2\,w\, (1-v)^2}\nn\\
&&\hspace{-2cm}+\frac{v^5 (75 w-9)+v^4 (22-130 w)+v^3 (7-30 w)-20 v^2 (w+12)+390 v-180}{30\,v\,w\, (1-v)^4}\Bigg]\nn\\
&&\hspace{-2.5cm}+N_c\,\Bigg[\frac{1+v^2}{(1-v)^4}\left(\frac{\ln(1-w)}{1-w}\right)_+  +\frac{ \left(3 v^4+16 v^3-43 v^2+50 v-18\right) \ln (1-v)}{2v^2(1-v)^4\,(1-w)_+}\nn\\
&&\hspace{-2cm} +\frac{  (1+v^2) \ln (\chi_\mu)}{(1-v)^4\,(1-w)_+}+\frac{  16 v^4-85 v^3-176 v^2+133 v-108}{12v(1-v)^4\,(1-w)_+}\nn\\
&&\hspace{-2cm}-\frac{\left(v^4 w+2 v^3+v^2 (w-7)+8 v-3\right) \ln (1-v)}{v^2\,(1-v)^4 w}-\frac{\left(1+v^2\right) (\ln (1-w)+\ln(\chi_\mu))}{2\,(1-v)^4}\nn\\
&&\hspace{-2cm}+\frac{18+4v^2(6+w)-v^4(13w-1)}{6 v\,(1-v)^4 w}+\frac{3 \left(1+v^2\right) w \left(w^2+2 w-1\right) \ln (w)}{6 \,(1-v)^4 w\,(1-w)^2}\nn\\
&&\hspace{-2cm}-\frac{\left(35 w^2-76 w+35+v^2 \left(45 w^2-53 w+2\right)\right)}{6 \,(1-v)^4 w\,(1-w)}\Bigg].\label{eq:HP1qg2q}
\eea
Note that the last two terms combined $\--$ although individually divergent $\--$ are finite for $w\to 1$.
\bea
\hat{\sigma}_{\mathrm{HP},\partial_1 F}^{qg\to q,1}(v,w,\chi_\mu) & = & C_F\Bigg[-\frac{27 v^5-44 v^4+291 v^3-1600 v^2+2040 v-840}{60 \, v\,(1-v)^4\,(1-w)_+}\nn\\
   && \hspace{1cm}+\frac{ \left(v^2-13 v+14\right) \ln (1-v)}{v^2\, (1-v)^2\,(1-w)_+}\Bigg]\nn\\
 && \hspace{-2.5cm}+ N_c\,\Bigg[\frac{1+v^2}{(1-v)^4}\left(\frac{\ln(1-w)}{1-w}\right)_+ + \frac{\left(3 v^4+15 v^3-37 v^2+41
   v-14\right) \ln (1-v)}{2\,v^2\, (1-v)^4\, (1-w)_+}\nn\\
   && \hspace{-2cm} + \frac{ \left(1+v^2\right) \ln (\chi_\mu) }{ (1-v)^4 \, (1-w)_+}+ \frac{6 v^4-51 v^3-142 v^2+115 v-84}{12\,v\, (1-v)^4 \, (1-w)_+}\Bigg].\label{eq:HP2qg2q}
\eea
\bea
\hat{\sigma}_{\mathrm{HP},\partial^2_1 F}^{qg\to q,1}(v,w) & = & \hat{\sigma}_{\mathrm{HP},\partial_1\partial_2 F}^{qg\to q,1}(v,w)\nn\\
&=&C_F\Bigg[\frac{ \ln (1-v)}{ v^2\,(1-v)\, (1-w)_+}-\frac{  2 v^5-5 v^4+22 v^3-125 v^2+150 v-60}{60\,v\, (1-v)^4\,(1-w)_+}\Bigg]\nn\\
 && \hspace{0cm}+ N_c\,\Bigg[-\frac{\ln (1-v)}{2\,v^2\,(1-v)\,(1-w)_+}+\frac{v^4-4 v^3-19 v^2+20 v-12}{24\,v\,(1-v)^4\,(1-w)_+}\Bigg].\label{eq:HP4qg2q}
\eea
\bea
\hat{\sigma}_{\mathrm{HP},\partial_2 F}^{qg\to q,1}(v,w,\chi_\mu) & = & C_F \,\Bigg[\frac{\left(v^2-6 v+6\right) \ln (1-v)}{v^2\, (1-v)^2\,(1-w)_+}\nn\\
&&\hspace{-2cm}-\frac{ 9 v^5-7 v^4+171 v^3-725 v^2+900 v-360}{60\,v\, (1-v)^4\,(1-w)_+}+\frac{(2 v-3) \ln(1-v)}{w\,v^2\,(1-v)^2}\nn\\
&&\hspace{-2cm}+\frac{v^5 (30 w-9)+v^4 (22-45 w)+v^3 (7-10 w)-5 v^2 (w+48)+390 v-180}{60\,w\,v\, (1-v)^4}\Bigg]\nn\\
&&\hspace{-2cm}+N_c\,\Bigg[\frac{1+v^2}{(1-v)^4}\left(\frac{\ln(1-w)}{1-w}\right)_+ +\frac{\left(3 v^4+8 v^3-15 v^2+18
   v-6\right) \ln (1-v)}{2\,v^2\,(1-v)^4\,(1-w)_+}\nn\\
&& +\frac{v \left(24\, v\,\left(1+v^2\right) \ln \left(\chi_\mu\right)+3 v^4-68 v^3-143 v^2+74 v-72\right)}{24\,v^2\, (1-v)^4\, (1-w)_+}   \nn\\
&&\hspace{-2cm}-\frac{\left(2 v^4 w+2 v^3+v^2 (2 w-7)+8 v-3\right) \ln (1-v)}{2\,v^2\, (1-v)^4\, w}-\frac{\left(1+v^2\right) \ln (1-w)}{2 (1-v)^4}\nn\\
&& -\frac{\left(1+v^2\right)\,  \ln (\chi_\mu)}{2\,(1-v)^4 }-\frac{\left(1+v^2\right)\, (1+w)\, \ln (w)}{2\, (1-v)^4\, (1-w)}\nn\\
&&\hspace{-1.5cm} +\frac{v^4 (2-9 w)+v^3 (70 w-4)-v^2 (w-48)+v (68 w-70)+36}{24\,v\,w\, (1-v)^4}\Bigg]\,.\label{eq:HP3qg2q}
\eea
\bea
\hat{\sigma}_{\mathrm{HP},G}^{qg\to q,1}(v,w,\chi_\mu) & = &C_F\,\Bigg[ \frac{\left(v^2-14 v+18\right) \ln (1-v)}{v^2\,(1-v)^2\,(1-w)_+}+\frac{2 (2 v-3) \ln (1-v)}{v^2\,w\,(1-v)^2}\nn\\
&& +\frac{75 v^5-100 v^4-369 v^3+1640 v^2-2460 v+1080}{60\,v\,(1-v)^4\,(1-w)_+} \nn\\
&&\hspace{-2.5cm} + \frac{v^5 (9-75 w)+2 v^4 (65 w-11)+v^3 (30 w+33)+20 v^2 (w-12)+390 v-180}{30\,v\,w\, (1-v)^4}\Bigg]\nn\\
&&\hspace{-2cm}+N_c\,\Bigg[-  \frac{\left(v^2-14 v+18\right)\, \ln (1-v)}{2\,v^2\,(1-v)^2\, (1-w)_+}  - \frac{16 v^4-73 v^3+160 v^2-287 v+108}{12\,v\, (1-v)^4\,(1-w)_+}\nn\\
&& \hspace{-1.5cm} + \frac{\left(v^4 w-2 v^3+v^2 (w+7)-8 v+3\right) \ln (1-v)}{v^2\,w\,(1-v)^4} + \frac{1+v^2 }{2 (1-v)^4}\ln\left(\tfrac{1-w}{w}\chi_\mu\right)\nn\\
&&\hspace{-1.5cm}+\frac{v^4 (13 w-1)+v^3 (2-51 w)-4 v^2 (w-6)-v (41 w+37)+18}{6\,v\,w\, (1-v)^4}\Bigg]\,.\label{eq:HP6qg2q}
\eea
Since $G^q(x_0,x_0)=0$, we may replace the plus-distribution $1/(1-w)_+$ just by $1/(1-w)$.
\bea
\hat{\sigma}_{\mathrm{HP},\partial_1G}^{qg\to q,1}(v,w) & = & C_F\,\Bigg[ \frac{ \left(v^2-13 v+14\right) \ln (1-v)}{v^2\,(1-v)^2\,
   (1-w)_+} 
+ \frac{27 v^5-44 v^4-309 v^3+1520 v^2-2040 v+840}{60\,v\, (1-v)^4
   (1-w)_+}\Bigg]\nn   \\
&&\hspace{0cm} + N_c\,\Bigg[-\frac{\left(v^2-13 v+14\right)\, \ln (1-v)}{2\,v^2\, (1-v)^2\, (1-w)_+} - \frac{ 6 v^4-39 v^3+150 v^2-209 v+84}{12\,v\, (1-v)^4\,(1-w)_+} \Bigg]\,.\label{eq:HP7qg2q}
\eea
\bea
\hat{\sigma}_{\mathrm{HP},\partial_1^2G}^{qg\to q,1}(v,w) & = & \hat{\sigma}_{\mathrm{HP},\partial_1\partial_2G}^{qg\to q,1}(v,w) \nn\\
&=&\hspace{0cm}C_F\,\Bigg[\frac{\ln (1-v)}{v^2\, (1-v)\, (1-w)_+}   +\frac{ 2 v^5-5 v^4-18 v^3+115 v^2-150 v+60}{60\,v\, (1-v)^4\,(1-w)_+}\Bigg]\nn\\
&&\hspace{0cm}+N_c\,\Bigg[ -\frac{\ln (1-v)}{2\,v^2\, (1-v)\, (1-w)_+}  - \frac{v^4-4 v^3+21 v^2-28 v+12}{24\,v\, (1-v)^4\, (1-w)_+}\Bigg]\,.\label{eq:HP9qg2q}
\eea
\bea
\hat{\sigma}_{\mathrm{HP},\partial_2G}^{qg\to q,1}(v,w,\chi_\mu) & = & C_F\,\Bigg[ \frac{ \left(v^2-6 v+6\right) \,\ln (1-v)}{ v^2\,(1-v)^2\,(1-w)_+}   \nn\\
&& \hspace{-1.5cm}+\frac{9 v^5-7 v^4-189 v^3+715 v^2-900 v+360}{60\,v\,(1-v)^4\,(1-w)_+} + \frac{(2 v-3)\, \ln (1-v)}{v^2\,w\, (1-v)^2}\nn\\
&& \hspace{-2cm}+ \frac{v^5 (9-30 w)+v^4 (45 w-22)+v^3 (10 w+33)+5 v^2 (w-48)+390 v-180}{60 \,v\,w\,(1-v)^4}\Bigg] \nn\\
&& \hspace{-2cm}+ N_c\,\Bigg[ -\frac{ \left(v^2-6 v+6\right) \,\ln (1-v)}{2\,v^2\, (1-v)^2\, (1-w)_+}  - \frac{3 v^4-44 v^3+145 v^2-190 v+72}{24\,v\, (1-v)^4\, (1-w)_+} \nn\\
&& \hspace{-1.5cm}+\frac{\left(2 v^4 w-2 v^3+v^2 (2 w+7)-8 v+3\right) \ln (1-v)}{2\,v^2\,w\, (1-v)^4}+\frac{1+v^2}{2 (1-v)^4}\ln\left(\tfrac{1-w}{w}\chi_\mu\right)\nn\\
&& \hspace{-1cm}+\frac{v^4 (9 w-2)+v^3 (4-70 w)+v^2 (w+48)-2 v (34 w+37)+36}{24 \,v\,w\,(1-v)^4}\Bigg]\,. \label{eq:HP8qg2q}
\eea

\subsubsection{Soft-Gluon Pole Contribution\label{appsub:SGPqg2q}}
In the following passage, the partonic factors of Eq.~\eqref{eq:Channel1SGP} are listed, which constitutes our result for the soft-gluon pole contribution to the $qg\to q$-channel,
\bea
\hat{\sigma}_{\mathrm{SGP},F}^{qg\to q,1}(v,w,\chi_\mu,\chi_m) & = & N_c\,\hat{\sigma}_{\mathrm{SGP},F,N_c}^{qg\to q,1}(v,w,\chi_\mu,\chi_m) \label{eq:SGP1qg2q}\\
&&+ C_F\,\hat{\sigma}_{\mathrm{SGP},F,C_F}^{qg\to q,1}(v,w,\chi_\mu,\chi_m)\,,\nn
\eea
where we divided the partonic cross section into its contributions entering with various color factors. The $N_c$ contribution reads,
\bea
\hat{\sigma}_{\mathrm{SGP},F,N_c}^{qg\to q,1}(v,w,\chi_\mu,\chi_m) & = &A_{1,\mathrm{SGP},F,N_c}^{qg\to q,1}(v,\chi_\mu,\chi_m)\,\delta(1-w)\,\nn\\
&&\hspace{0cm}+A_{2,\mathrm{SGP},F,N_c}^{qg\to q,1}(v)\,\left(\frac{\ln(1-w)}{1-w}\right)_+ +A_{3,\mathrm{SGP},F,N_c}^{qg\to q,1}(v,\chi_\mu)\,\frac{1}{(1-w)_+}\nn\\
&&\hspace{0cm}+A_{4,\mathrm{SGP},F,N_c}^{qg\to q,1}(v,w)\,\ln(\chi_m) +A_{5,\mathrm{SGP},F,N_c}^{qg\to q,1}(v,w)\,\ln(1-w)\nn\\
&&\hspace{0cm}+A_{6,\mathrm{SGP},F,N_c}^{qg\to q,1}(v)\,\ln(\chi_\mu) +A_{7,\mathrm{SGP},F,N_c}^{qg\to q,1}(v,w)\,\ln(1-v\,w)\nn\\
&&\hspace{0cm}+A_{8,\mathrm{SGP},F,N_c}^{qg\to q,1}(v,w)\,\ln(1-v) +A_{9,\mathrm{SGP},F,N_c}^{qg\to q,1}(v,w)\,\ln(1-v+v\,w)\,\nn\\
&&\hspace{0cm}+ A_{10,\mathrm{SGP},F,N_c}^{qg\to q,1}(v,w)\,.\label{eq:SGP1qg2qNc}
\eea
with ten coefficient functions of the following explicit form:
\bea
A_{1,\mathrm{SGP},F,N_c}^{qg\to q,1}(v,\chi_\mu,\chi_m) & = & \frac{v^5-4 v^4-12 v^3+38 v^2-49 v+18}{4\,v^2\, (1-v)^4}\ln (1-v)\nn\\
&&\hspace{0cm}-\frac{ 1+v^2  }{2\, (1-v)^4}\ln (\chi_\mu) -\frac{ 1+v^2  }{2\, (1-v)^3}\ln (\chi_m) -\frac{(1+v) v }{(1-v)^4}\ln(v)\nn\\
&&-\frac{9 v^4-58 v^3-150 v^2+150 v-108}{24\,v\, (1-v)^4} \,,    \nn\\
A_{2,\mathrm{SGP},F,N_c}^{qg\to q,1}(v) & = &  -\frac{1+v^2}{(1-v)^4} \,, \nn\\
A_{3,\mathrm{SGP},F,N_c}^{qg\to q,1}(v,\chi_\mu) & = & -\frac{1+v^2  }{(1-v)^4}\,\ln(\chi_\mu)-\frac{3 v^4+16 v^3-43 v^2+50 v-18  }{2\,v^2\, (1-v)^4}\ln (1-v)\nn\\
&& \hspace{0cm}-\frac{10 v^4-61 v^3-176 v^2+151 v-108}{12\,v\, (1-v)^4},\nn
\eea
\bea
A_{4,\mathrm{SGP},F,N_c}^{qg\to q,1}(v,w) & = & \frac{2 v^5 (1-2 w) w^4+v^4 w^2 \left(8 w^2-4 w+1\right)}{2\, (1-v)^4\, (1-v w)^2}\nn\\
&& \hspace{0cm}- \frac{2 v^3 w \left(2 w^2+w-1\right)+v^2 \left(1-w^2\right)-2 v w+1}{2\, (1-v)^4\, (1-v w)^2}, \nn \\
A_{5,\mathrm{SGP},F,N_c}^{qg\to q,1}(v,w) & = & -\frac{v^2 w \left(v^3 w^2 \left(3 w^2-2 w+1\right)+2 v^2 w \left(-2 w^2+w-2\right)\right)}{2 \,(1-v)^4\, (1-v w)^2}\nn\\
&&  -\frac{v^2 w \left(v \left(-w^2+6 w+1\right)-2\right)}{2 \,(1-v)^4\, (1-v w)^2}\,,\nn\\
A_{6,\mathrm{SGP},F,N_c}^{qg\to q,1}(v) & = & \frac{1+v^2}{2 \,(1-v)^4}\,, \nn\\
A_{7,\mathrm{SGP},F,N_c}^{qg\to q,1}(v,w) & = &  \frac{2 v^2 w^2}{(1-v)^2\, (1-v w)^2}\,, \nn
\eea
\bea
A_{8,\mathrm{SGP},F,N_c}^{qg\to q,1}(v,w) & = &  -\frac{v^4-20 v^3+47 v^2-50 v+18}{2\,v^2\, (1-v)^4}-\frac{\left(v^3-9 v+16\right) w}{4\,v^2\, (1-v)^2}\nn\\
& & -\frac{(3 v-5) w^2}{v^2\,(1-v)^2}+\frac{1-2 v w}{(1-v)^2\, (1-v\, w)^2}\,,\nn\\
 A_{9,\mathrm{SGP},F,N_c}^{qg\to q,1}(v,w) & = & \frac{v w}{4 \,(1-v)^2}\,,\nn
\eea
\bea
A_{10,\mathrm{SGP},F,N_c}^{qg\to q,1}(v,w) & = &  \frac{w^2\,\ln(w)}{(1-w)^2}\times\,\nn\\
&& \hspace{-3cm}\left(\frac{v^3 (w-1)^2 (2 w-1)+2 v^2 (w-2) w+2 v-2}{(1-v)^4}+\frac{1-v (2 w-1)}{(1-v)\, (1-v\, w)^2}\right)  \nn \\
&& \hspace{-3cm}-\frac{1+v^2}{(1-v)^4 (1-w)}+\frac{13 v^4-76 v^3-95 v^2+130 v-108}{12\,v\, (1-v)^4}\nn\\
&&\hspace{-3cm}+\frac{\left(37 v^4-122 v^3-313 v^2+228 v-192\right) w}{48\,v\, (1-v)^4}+\frac{\left(5 v^4+38 v^2-63 v+30\right) w^2}{6\,v\, (1-v)^4}\nn\\
&&\hspace{-3cm}+\frac{\left(17 v^3+6 v^2+v+8\right) w^3}{48\, (1-v)^4}+\frac{4 v^2 w-v (5 w+2)+3}{(1-v)^3 (1-v\, w)^2}+\frac{1}{4(1-v+v\,w)}\,.\label{eq:SGPNcFA1}  
\eea
Notice that the last coefficient function $A_{10,\mathrm{SGP},F,N_c}^{qg\to q,1}(v,w)$ is regular as $w\to 1$, despite the factors $\ln(w)/(1-w)^2$ and $1/(1-w)$. 
Its limit for $w\to 1$ is 
\[\lim_{w\to 1}A_{10,\mathrm{SGP},F,N_c}^{qg\to q,1}(v,w)=\frac{6 v^5+49 v^4-210 v^3-314 v^2+168 v-192}{24\,v\, (1-v)^4}\,.\]

The $C_F$ contribution in \eqref{eq:SGP1qg2q} takes a similar form as the $N_c$ one,
\bea
\hat{\sigma}_{\mathrm{SGP},F,C_F}^{qg\to q,1}(v,w,\chi_\mu,\chi_m) & = &A_{1,\mathrm{SGP},F,C_F}^{qg\to q,1}(v,\chi_\mu,\chi_m)\,\delta(1-w)\,\nn\\
&&\hspace{0cm}+A_{2,\mathrm{SGP},F,C_F}^{qg\to q,1}(v)\,\left(\frac{\ln(1-w)}{1-w}\right)_+ +A_{3,\mathrm{SGP},F,C_F}^{qg\to q,1}(v,\chi_\mu)\,\frac{1}{(1-w)_+}\nn\\
&&\hspace{0cm}+A_{4,\mathrm{SGP},F,C_F}^{qg\to q,1}(v,w)\,\ln(\chi_m) +A_{5,\mathrm{SGP},F,C_F}^{qg\to q,1}(v,w)\,\ln(1-w)\nn\\
&&\hspace{0cm}+A_{6,\mathrm{SGP},F,C_F}^{qg\to q,1}(v,w)\,\ln(\chi_\mu) +A_{7,\mathrm{SGP},F,C_F}^{qg\to q,1}(v,w)\,\ln(1-v\,w)\nn\\
&&\hspace{0cm}+A_{8,\mathrm{SGP},F,C_F}^{qg\to q,1}(v,w)\,\ln(1-v) +A_{9,\mathrm{SGP},F,C_F}^{qg\to q,1}(v,w)\ln(1-v+v\,w)\nn\\
&&\hspace{0cm}+A_{10,\mathrm{SGP},F,C_F}^{qg\to q,1}(v,w)\label{eq:SGP1qg2qCF}\,.
\eea
with ten coefficient functions of the following explicit form:
\bea
A_{1,\mathrm{SGP},F,C_F}^{qg\to q,1}(v,\chi_\mu,\chi_m) & = & \frac{1+v^2}{(1-v)^4}\ln(v)\,(4\ln(1-v)+2\ln(\chi_\mu)-\ln(v))\nn\\
&&\hspace{-4cm}
-\frac{9 v^5+4 v^4+12 v^3+30 v^2-49 v+18}{2 (v-1)^4 v^2}\ln(1-v)-\frac{2 v^3+v^2+6 v-3}{(1-v)^4}\ln(\chi_\mu)\nn\\
&&\hspace{-4cm}+\frac{1+v^2}{(1-v)^3}\ln(\chi_m)-5\frac{1+v^2}{(1-v)^4}\ln(v)+\frac{45 v^5-21 v^4-365 v^3-840 v^2+780 v-540}{60\,v\, (1-v)^4},\nn
\eea
\bea
A_{2,\mathrm{SGP},F,C_F}^{qg\to q,1}(v) & = & 8\frac{1+v^2}{(1-v)^4}\,,\nn\\
A_{3,\mathrm{SGP},F,C_F}^{qg\to q,1}(v,\chi_\mu) & = &  4\frac{1+v^2}{(1-v)^4}\ln(\chi_\mu) +\frac{7 v^4+16 v^3-39 v^2+50 v-18}{v^2\,(1-v)^4}\ln(1-v) \nn\\
&&\hspace{0cm} +\frac{75 v^5-340 v^4-509 v^3-2440 v^2+2280 v-1080}{60\,v \,(1-v)^4}\,,\nn\\
A_{4,\mathrm{SGP},F,C_F}^{qg\to q,1}(v,w) & = & \frac{2 v^5 w^4 (2 w-1)-v^4 w^2 \left(8 w^2-4 w+1\right)}{(1-v)^4\, (1-v\, w)^2}\,\nn\\
&&+\frac{2 v^3 w \left(2 w^2+w-1\right)-v^2 \left(w^2-1\right)-2 v w+1}{(1-v)^4\, (1-v\, w)^2}\,,\nn\\
A_{5,\mathrm{SGP},F,C_F}^{qg\to q,1}(v,w) & = & \frac{6 v^4 w^4}{(1-v)^4}-\frac{v^3 (8 v-17) w^3}{(1-v)^4}+\frac{v^2 \left(3 v^2-10 v+13\right) w^2}{(1-v)^4}+\frac{v^2 w}{(1-v)^4}\nn\\
&&-\frac{4 \left(2
   v^2-v+2\right)}{(1-v)^4}+\frac{2-4 v w}{(1-v)^2 (1-v\, w)^2}
\,, \nn  
\eea
\bea
A_{6,\mathrm{SGP},F,C_F}^{qg\to q,1}(v,w) & = &\frac{v^4 w^2 \left(6 w^2-8 w+3\right)+4 v^3 w^2 (3 w-2)+v^2 \left(9 w^2-3\right)-3}{(1-v)^4}\,,\nn\\
A_{7,\mathrm{SGP},F,C_F}^{qg\to q,1}(v,w) & = &  -\frac{4 v^2 w^2}{(1-v)^2\, (1-v\, w)^2}\,,\nn\\
A_{8,\mathrm{SGP},F,C_F}^{qg\to q,1}(v,w) & = & \frac{-3 v^4-20 v^3+43 v^2-50 v+18}{v^2\,(1-v)^4} + \frac{\left(v^3-9 v+16\right) w}{2\,v^2\, (1-v)^2}\nn\\
&& +\frac{2 \left(3 v^6-8 v^5+9 v^4+3 v^3-11 v^2+13 v-5\right) w^2}{v^2\,(1-v)^4}\nn\\
&&-\frac{8 v^3 (2 v-3) w^3}{(1-v)^4}+\frac{12 v^4 w^4}{(1-v)^4}+\frac{4 v w-2}{(1-v)^2\, (1-v \,w)^2}\,,\nn\\
A_{9,\mathrm{SGP},F,C_F}^{qg\to q,1}(v,w) & = & -\frac{v w}{2(1-v)^2}\,,
\nn    
\eea
\bea
A_{10,\mathrm{SGP},F,C_F}^{qg\to q,1}(v,w) & = &\,\ln(w)\left(-\frac{6 v^4 w^4}{(1-v)^4}+\frac{8 (v-2) v^3 w^3}{(1-v)^4}-\frac{v^2 \left(3 v^2-10 v+13\right) w^2}{(1-v)^4}\right.\nn\\
&&\hspace{-0cm}\left.+\frac{2 \left(3 v^2-2 v+3\right)}{(1-v)^4}+\frac{4 \left(1+v^2\right) (2 w-1)}{(1-v)^4\, (1-w)^2}+\frac{4 v w-2}{(1-v)^2\, (1-v\, w)^2}\right)\nn\\
&&\hspace{-0cm}-\frac{3 v^4 w^4}{(1-v)^4}
+\frac{v \left(76 v^3-15 v^2+6 v+1\right) w^3}{24 (1-v)^4}\nn\\
&&\hspace{-0cm}
-\frac{\left(64 v^5-210 v^4-195 v^3+360 v^2-630 v+300\right) w^2}{30\,v\,(1-v)^4}\nn\\
&&\hspace{-0cm}
-\frac{\left(114 v^5-569 v^4-1120 v^3-2735 v^2+1860 v-960\right)
   w}{120\,v\, (1-v)^4}\nn\\
&&\hspace{-0cm}
-\frac{75 v^5-310 v^4-539 v^3-1630 v^2+2190 v-1080}{60\,v\,(1-v)^4}\nn\\
&&\hspace{-0cm}+\frac{4 \left(1+v^2\right)}{(1-v)^4 (1-w)}
-\frac{8 v^2 w-2 v (5 w+2)+6}{(1-v)^3\, (1-v\, w)^2}
-\frac{1}{2 (1-v+v\,w)}\,.\label{eq:SGP1qg2qCF1}
\eea
As for the $N_c$-part the last coefficient function appears to diverge for $w\to 1$, but is actually regular and integrable since
\[\lim_{w\to 1}A_{10,\mathrm{SGP},F,C_F}^{qg\to q,1}(v,w) = 
\frac{-280 v^5+1097 v^4+1684 v^3+2880 v^2-1890 v+960}{60\,v\, (1-v)^4 }\,.\]

The partonic cross section for the term with the first derivative can be decomposed as well, but after integration by parts only distributions appear in the analytic forms, and the results are significantly less complex. We obtain
\bea
\hat{\sigma}_{\mathrm{SGP},F^\prime}^{qg\to q,1}(v,w,\chi_\mu) & = & N_c\,\hat{\sigma}_{\mathrm{SGP},F^\prime,N_c}^{qg\to q,1}(v,w,\chi_\mu) + C_F\,\hat{\sigma}_{\mathrm{SGP},F^\prime,C_F}^{qg\to q,1}(v,w,\chi_\mu)\,,\label{eq:SGP2qg2q}
\eea
where the $N_c$ part reads,
\bea
\hat{\sigma}_{\mathrm{SGP},F^\prime,N_c}^{qg\to q,1}(v,w,\chi_\mu) & = &A_{1,\mathrm{SGP},F^\prime,N_c}^{qg\to q,1}(v,\chi_\mu)\,\delta(1-w)\,\nn\\
&&\hspace{0cm}+A_{2,\mathrm{SGP},F^\prime,N_c}^{qg\to q,1}(v)\,\left(\frac{\ln(1-w)}{1-w}\right)_+ +A_{3,\mathrm{SGP},F^\prime,N_c}^{qg\to q,1}(v,\chi_\mu)\,\frac{1}{(1-w)_+}\,,\label{eq:SGP2qg2qNc}
\eea
with the coefficient functions
\bea
A_{1,\mathrm{SGP},F^\prime,N_c}^{qg\to q,1}(v,\chi_\mu) & = & -\frac{1+v^2}{(1-v)^4}\ln (\chi_\mu) -\frac{7 v^4+15 v^3-33 v^2+41 v-14}{4\,v^2\, (1-v)^4}\ln (1-v)\nn\\
&&\hspace{0cm} -\frac{6 v^4-51 v^3-142 v^2+115 v-84}{24\,v\, (1-v)^4} \,,   \nn\\
A_{2,\mathrm{SGP},F^\prime,N_c}^{qg\to q,1}(v) & = & -\frac{1+v^2}{(1-v)^4}\,,\nn
\eea
\bea
A_{3,\mathrm{SGP},F^\prime,N_c}^{qg\to q,1}(v,\chi_\mu) & = &  -\frac{1+v^2}{(1-v)^4}\ln(\chi_\mu)-\frac{6 v^4+23 v^3-52 v^2+59 v-20}{4\,v^2\, (1-v)^4}\ln(1-v)\nn\\
&& -\frac{15 v^4-98 v^3-427 v^2+376 v-240}{48\,v\, (1-v)^4}\,.\label{eq:SGPNcdFA1}
\eea   
Furthermore, the $C_F$ part reads,
\bea
\hat{\sigma}_{\mathrm{SGP},F^\prime,C_F}^{qg\to q,1}(v,w,\chi_\mu) & = &A_{1,\mathrm{SGP},F^\prime,C_F}^{qg\to q,1}(v,\chi_\mu)\,\delta(1-w)\,\nn\\
&&\hspace{0cm}+A_{2,\mathrm{SGP},F^\prime,C_F}^{qg\to q,1}(v)\,\left(\frac{\ln(1-w)}{1-w}\right)_+ +A_{3,\mathrm{SGP},F^\prime,C_F}^{qg\to q,1}(v,\chi_\mu)\,\frac{1}{(1-w)_+}\,,\label{eq:SGP2qg2qCF}
\eea
with the coefficient functions
\bea
A_{1,\mathrm{SGP},F^\prime,C_F}^{qg\to q,1}(v,\chi_\mu) & = & \frac{1+v^2}{(1-v)^4} \ln(v)\,\left(4\ln(1-v)+2\ln(\chi_\mu)-\ln(v)\right)\nn\\
&& \hspace{0cm}+\frac{11 v^4+15 v^3-29 v^2+41 v-14}{2\,v^2\, (1-v)^4}\ln(1-v) +3\frac{1+v^2}{(1-v)^4} \,\left(\ln(\chi_\mu)-\ln(v)\right)  \nn\\
&&\hspace{0cm}+\frac{27 v^5-44 v^4-669 v^3-1600 v^2+1080 v-840}{120\,v\, (1-v)^4} \,,   \nn\\
A_{2,\mathrm{SGP},F^\prime,C_F}^{qg\to q,1}(v) & = & 8\frac{1+v^2}{(1-v)^4}\,,\nn\\
A_{3,\mathrm{SGP},F^\prime,C_F}^{qg\to q,1}(v,\chi_\mu) & = &  4\frac{1+v^2}{(1-v)^4}\ln(\chi_\mu)+\frac{14 v^4+23 v^3-44 v^2+59 v-20}{2\,v^2\, (1-v)^4}\ln(1-v)\nn\\
&& +\frac{12 v^5-17 v^4+154 v^3-775 v^2+980 v-400}{40\,v \,(1-v)^4}\,.\label{eq:SGPCFdFA1}
\eea   
\\
The partonic cross sections for the terms with the second derivative have similar forms. Again we perform a separation into color structures,
\bea
\hat{\sigma}_{\mathrm{SGP},F^{\prime\prime}}^{qg\to q,1}(v,w) & = & N_c\,\hat{\sigma}_{\mathrm{SGP},F^{\prime\prime},N_c}^{qg\to q,1}(v,w) + C_F\,\hat{\sigma}_{\mathrm{SGP},F^{\prime\prime},C_F}^{qg\to q,1}(v,w)\,,\label{eq:SGP3qg2q}
\eea
where the $N_c$ part reads,
\bea
\hat{\sigma}_{\mathrm{SGP},F^{\prime\prime},N_c}^{qg\to q,1}(v,w) & = &A_{1,\mathrm{SGP},F^{\prime\prime},N_c}^{qg\to q,1}(v)\,\left(\delta(1-w)
+\frac{1}{(1-w)_+}\right)\,,\label{eq:SGP3qg2qNc}
\eea
with the coefficient function
\bea
A_{1,\mathrm{SGP},F^{\prime\prime},N_c}^{qg\to q,1}(v) & = &\frac{1}{4\,v^2\,(1-v)}\ln (1-v)-\frac{v^4-4 v^3-19 v^2+20 v-12}{48\,v\, (1-v)^4}\,.   \label{eq:SGPNcddFA1}
\eea   
The $C_F$ part reads
\bea
\hat{\sigma}_{\mathrm{SGP},F^{\prime\prime},C_F}^{qg\to q,1}(v,w) & = &A_{1,\mathrm{SGP},F^{\prime\prime},C_F}^{qg\to q,1}(v)\,\left(\delta(1-w)
+\frac{1}{(1-w)_+}\right)\,,\label{eq:SGP3qg2qCF}
\eea
with the coefficient function
\bea
A_{1,\mathrm{SGP},F^{\prime\prime},C_F}^{qg\to q,1}(v) & = &-\frac{\ln (1-v)}{2\,v^2\,(1-v)}+\frac{2 v^5-5 v^4+22 v^3-125 v^2+150 v-60}{120\,v\, (1-v)^4}\,.   \label{eq:SGPCFddFA1}
\eea  
The other partonic cross section related to a second derivative term in Eq.~\eqref{eq:Channel1SGP} is $\hat{\sigma}_{\mathrm{SGP},\partial_1^2 F}^{qg\to q,1}(v,w)$. It shows the least analytical complexity and reads
\bea
\hat{\sigma}_{\mathrm{SGP},\partial_1^2 F}^{qg\to q,1}(v,w) & = & \delta(1-w)\,\Bigg[N_c\,\left(-\frac{\ln(1-v)}{4\,v^2\,(1-v)}+\frac{v^4-4 v^3-19 v^2+20 v-12}{48\,v\, (1-v)^4}\right)\nn\\
&&\hspace{-1cm}+C_F\,\left(\frac{\ln (1-v)}{2\,v^2\,(1-v)}-\frac{2 v^5-5 v^4+22 v^3-125 v^2+150 v-60}{120\,v\, (1-v)^4}\right)\Bigg].\label{eq:SGP4qg2q}
\eea

We next turn to the terms in Eq.~\eqref{eq:Channel1SGP} that are generated by the quark-gluon-quark correlation function $G$. Only derivative terms contribute here,
\bea
\hat{\sigma}_{\mathrm{SGP},\partial_1 G}^{qg\to q,5}(v,w) & = & N_c\,\hat{\sigma}_{\mathrm{SGP},\partial_1 G,N_c}^{qg\to q,5}(v,w) + C_F\,\hat{\sigma}_{\mathrm{SGP},\partial_1 G,C_F}^{qg\to q,5}(v,w)\,,\label{eq:SGP5qg2q}
\eea
where we again divided the partonic cross section into its color factor contributions. The $N_c$ contribution reads,
\bea
\hat{\sigma}_{\mathrm{SGP},\partial_1 G,N_c}^{qg\to q,5}(v,w) & = &A_{1,\mathrm{SGP},\partial_1 G,N_c}^{qg\to q,5}(v)\,\delta(1-w)+A_{2,\mathrm{SGP},\partial_1 G,N_c}^{qg\to q,5}(v)\,\frac{1}{(1-w)_+}\nn\\
&&\hspace{0cm} +A_{3,\mathrm{SGP},\partial_1 G,N_c}^{qg\to q,5}(v,w)\,\ln(1-w) +A_{4,\mathrm{SGP},\partial_1 G,N_c}^{qg\to q,5}(v)\,\ln(1-v+v\,w)\nn\\
&&\hspace{0cm}+A_{5,\mathrm{SGP},\partial_1 G,N_c}^{qg\to q,5}(v,w)\,\ln(1-v) + A_{6,\mathrm{SGP},\partial_1 G,N_c}^{qg\to q,5}(v,w)\,,\label{eq:SGP5qg2qNc}
\eea
with six coefficient functions of the following explicit form:
\bea
A_{1,\mathrm{SGP},\partial_1 G,N_c}^{qg\to q,5}(v) & = & -\frac{v^2-13 v+14}{2\,v^2\,(1-v)^2}\ln(1-v)-\frac{6 v^4-39 v^3+150 v^2-209 v+84}{12\,v\,(1-v)^4}\,,\nn\\
A_{2,\mathrm{SGP},\partial_1 G,N_c}^{qg\to q,5}(v) & = & \frac{8-7 v}{2 \,v^2\,(1-v)^2}\ln(1-v)+\frac{9 v^4-34 v^3+155 v^2-228 v+96}{24\,v\, (1-v)^4}\,,\nn\\
A_{3,\mathrm{SGP},\partial_1 G,N_c}^{qg\to q,5}(v,w) & = & \frac{v^2 \left(v \left(w^2-w+1\right)+w-2\right)}{(1-v)^4}\,,\nn\\
A_{4,\mathrm{SGP},\partial_1 G,N_c}^{qg\to q,5}(v) & = &-\frac{v}{2 (1-v)^2}\,,\nn\\
A_{5,\mathrm{SGP},\partial_1 G,N_c}^{qg\to q,5}(v,w) & = & \frac{v^3+6 v w+7 v-10 w-8}{2\,v^2\, (1-v)^2}\,,\nn\\
A_{6,\mathrm{SGP},\partial_1 G,N_c}^{qg\to q,5}(v,w) & = & -\frac{9 v^4-34 v^3+155 v^2-228 v+96}{24\,v\, (1-v)^4}\nn\\
&& \hspace{0cm}-\frac{w\,(v^4-2 v^3+73 v^2-128 v+60)}{12\,v\,(1-v)^4}-\frac{w^2\,(7 v^3-6 v^2-v-8)}{24 \,(1-v)^4}\,.\label{eq:SGPNcD2GA1}
\eea
The $C_F$ contribution reads,
\bea
\hat{\sigma}_{\mathrm{SGP},\partial_1 G,C_F}^{qg\to q,5}(v,w) & = &A_{1,\mathrm{SGP},\partial_1 G,C_F}^{qg\to q,5}(v)\,\delta(1-w)+A_{2,\mathrm{SGP},\partial_1 G,C_F}^{qg\to q,5}(v)\,\frac{1}{(1-w)_+}\nn\\
&&\hspace{0cm} +A_{3,\mathrm{SGP},\partial_1 G,C_F}^{qg\to q,5}(v,w)\,\ln(1-w) +A_{4,\mathrm{SGP},\partial_1 G,C_F}^{qg\to q,5}(v)\,\ln(1-v+v\,w)\nn\\
&&\hspace{0cm}+A_{5,\mathrm{SGP},\partial_1 G,C_F}^{qg\to q,5}(v,w)\,\ln(1-v) + A_{6,\mathrm{SGP},\partial_1 G,C_F}^{qg\to q,5}(v,w)\,,\label{eq:SGP5qg2qCF}
\eea
with six coefficient functions of the following explicit form:
\bea
A_{1,\mathrm{SGP},\partial_1 G,C_F}^{qg\to q,5}(v) & = & \frac{v^2-13 v+14}{v^2\,(1-v)^2}\ln(1-v)\nn\\
&&+\frac{27 v^5-44 v^4-309 v^3+1520 v^2-2040 v+840}{60\,v\,(1-v)^4}\,,\nn\\
A_{2,\mathrm{SGP},\partial_1 G,C_F}^{qg\to q,5}(v) & = & -\frac{8-7 v}{v^2\,(1-v)^2}\ln(1-v)\nn\\
&&-\frac{18 v^5-37 v^4-120 v^3+805 v^2-1140 v+480}{60\,v\,(1-v)^4}\,,\nn
\eea
\bea
A_{3,\mathrm{SGP},\partial_1 G,C_F}^{qg\to q,5}(v,w) & = & -\frac{2 v^2 (1-w) (1+v w)}{(1-v)^4}\,,\nn\\
A_{4,\mathrm{SGP},\partial_1 G,C_F}^{qg\to q,5}(v) & = &\frac{v}{(1-v)^2}\,,\nn
\eea
\bea
A_{5,\mathrm{SGP},\partial_1 G,C_F}^{qg\to q,5}(v,w) & = & -\frac{v^3+6 v w+7 v-10 w-8}{v^2\,(1-v)^2}\,,\nn\\
A_{6,\mathrm{SGP},\partial_1 G,C_F}^{qg\to q,5}(v,w) & = & \frac{18 v^5-37 v^4-120 v^3+805 v^2-1140 v+480}{60\,v\,(1-v)^4}\nn\\
&& \hspace{0cm}+\frac{w\,(26 v^5-25 v^4-120 v^3+375 v^2-630 v+300)}{30\,v\,(1-v)^4}\nn\\
&& -\frac{w^2\,(20 v^4-9 v^3-6 v^2-v)}{12\,(1-v)^4}+\frac{2\,w^3\,v^4}{(1-v)^4}\,.\label{eq:SGPCFD2GA1}
\eea
Finally, the cross section entering with the second derivative of $G$ reads
\bea
\hat{\sigma}_{\mathrm{SGP},\partial_1^2 G}^{qg\to q,5}(v,w) & = & \left(\delta(1-w)-\frac{2}{(1-w)_+}\right)\,\times\nn\\
&&\hspace{0cm}\Bigg[N_c\,\left(-\frac{\ln(1-v)}{4\,v^2\,(1-v)}-\frac{v^4-4 v^3+21 v^2-28 v+12}{48\,v\, (1-v)^4}\right)\nn\\
&&\hspace{0cm}+C_F\,\left(\frac{\ln (1-v)}{2\,v^2\,(1-v)}+\frac{2 v^5-5 v^4-18 v^3+115 v^2-150 v+60}{120\,v\,(1-v)^4}\right)\Bigg].\label{eq:SGP6qg2q}
\eea

\subsubsection{Soft-Fermion Pole Contribution\label{appsub:SFPqg2q}}

The soft-fermion pole contribution is the final part of the $qg\to q$-channel and below we present the explicit expressions for the partonic factors found in Eq.~\eqref{eq:Channel1SFP}. We separate the SFP partonic factors by means of their color factors:
\bea
\hat{\sigma}_{\mathrm{SFP},F}^{qg\to q,1}(v,w,\chi_m) & = & N_c\,\hat{\sigma}_{\mathrm{SFP},F,N_c}^{qg\to q,1}(v,w,\chi_m)+C_F\,\hat{\sigma}_{\mathrm{SFP},F,C_F}^{qg\to q,1}(v,w,\chi_m)\,,\label{eq:SFP1qg2q}
\eea
with the $N_c$ part given by,
\bea
\hat{\sigma}_{\mathrm{SFP},F,N_c}^{qg\to q,1}(v,w,\chi_m) & = & A_{1,\mathrm{SFP},F,N_c}^{qg\to q,1}(v,w)\,\ln(\chi_m) +A_{2,\mathrm{SFP},F,N_c}^{qg\to q,1}(v,w)\,\ln(1-w)\nn\\
&&\hspace{0cm} +A_{3,\mathrm{SFP},F,N_c}^{qg\to q,1}(v,w)\,\ln(1-v\,w)+A_{4,\mathrm{SFP},F,N_c}^{qg\to q,1}(v,w)\,\ln(1-v)\nn\\
&&\hspace{0cm} +A_{5,\mathrm{SFP},F,N_c}^{qg\to q,1}(v,w)\,\ln(1-v+v\,w)+A_{6,\mathrm{SFP},F,N_c}^{qg\to q,1}(v,w)\,\ln(w)\,\nn\\
&&\hspace{0cm}+ A_{7,\mathrm{SFP},F,N_c}^{qg\to q,1}(v,w)\,,\label{eq:SFP1qg2qNc}
\eea
with seven coefficient functions of the following explicit form:
\bea
A_{1,\mathrm{SFP},F,N_c}^{qg\to q,1}(v,w) & = & -\frac{v^3 w^3}{(1-v)^4}-\frac{v^2 w^2}{(1-v)^3}-\frac{v \left(v^2-4 v+5\right) w}{2 (1-v)^4}\nn\\
&&-\frac{v^2-4 v+2}{(1-v)^4}+\frac{3-2 v}{(1-v)^3 (1-v\, w)}-\frac{1}{2
   (1-v)^2 (1-v\, w)^2}\,,\nn\\
A_{2,\mathrm{SFP},F,N_c}^{qg\to q,1}(v,w) & = & -\frac{2 v^3 w^3}{(1-v)^4}+\frac{v^2 (3 v-1) w^2}{(1-v)^4}-\frac{v \left(3 v^2-4 v+5\right) w}{2\, (1-v)^4}\nn\\
&&-\frac{v^2-4 v+2}{(1-v)^4}+\frac{3-2 v}{(1-v)^3 (1-v\,
   w)}-\frac{1}{2 (1-v)^2 (1-v\, w)^2},\nn\\
A_{3,\mathrm{SFP},F,N_c}^{qg\to q,1}(v,w) & = & -\frac{v w \left(v^3 (w-1) w^2+v^2 (w-1) w+v-1\right)}{(1-v)^3\, (1-v \,w)^2},\nn 
\eea
\bea
A_{4,\mathrm{SFP},F,N_c}^{qg\to q,1}(v,w) & = & \frac{2 v-3}{(1-v)^2 \,v^2 \,w}-\frac{2 v^3-10 v^2+16 v-9}{(1-v)^3\, v^2}-\frac{w\,(v^4-5 v^3-v^2+5 v-4)}{2 \,(1-v)^3 \,v^2}\nn\\
&&+\frac{2 (3 v-5) w^2}{(1-v)^2\, v^2}-\frac{4 v^2 w-3 v (2 w+1)+5}{2 \,(1-v)^3\, (1-v\, w)^2}\,,\nn\\
A_{5,\mathrm{SFP},F,N_c}^{qg\to q,1}(v,w)&=& \frac{v w (2 v w-v-1)}{2\, (1-v)^3}\,,\nn
\eea
\bea
A_{6,\mathrm{SFP},F,N_c}^{qg\to q,1}(v,w) & = & \frac{2 v^2-4 v+3}{(1-v)^4}+\frac{(2-v) v w}{(1-v)^3}-\frac{v^2 (2 v+1) w^2}{(1-v)^4}-\frac{4 v^2 w-3 v (2 w+1)+5}{2\, (1-v)^3\, (1-v\, w)^2},\nn\\
A_{7,\mathrm{SFP},F,N_c}^{qg\to q,1}(v,w) & = & -\frac{v^4-2 v^3+24 v^2-35 v+18}{6\, (1-v)^4 \,v \,w}+\frac{10 v^4-23 v^3+126 v^2-203 v+108}{12\, (1-v)^4\, v}\nn\\
&&\hspace{-0cm}-\frac{\left(5 v^4-5 v^3-29 v^2+9 v-12\right) w}{6\, (1-v)^4\, v}\nn\\
&&\hspace{0cm}-\frac{\left(33 v^4-16 v^3+155 v^2-238 v+120\right) w^2}{12\, (1-v)^4\, v}+\frac{\left(31 v^3-6 v^2-v-8\right) w^3}{12\, (1-v)^4}\nn\\
&&\hspace{-0cm}-\frac{4 v^2 w-v (7 w+2)+5}{2\, (1-v)^3\, (1-v\, w)^2}+\frac{1}{2\, (1-v)\, (1-v+v\,w)}\,.\label{eq:SFPNcFA1}
\eea
The $C_F$ part has a similar form,
\bea
\hat{\sigma}_{\mathrm{SFP},F,C_F}^{qg\to q,1}(v,w,\chi_m) & = & A_{1,\mathrm{SFP},F,C_F}^{qg\to q,1}(v,w)\,\ln(\chi_m) +A_{2,\mathrm{SFP},F,C_F}^{qg\to q,1}(v,w)\,\ln(1-w)\nn\\
&&\hspace{0cm} +A_{3,\mathrm{SFP},F,C_F}^{qg\to q,1}(v,w)\,\ln(1-v\,w)+A_{4,\mathrm{SFP},F,C_F}^{qg\to q,1}(v,w)\,\ln(1-v)\nn\\
&&\hspace{0cm} +A_{5,\mathrm{SFP},F,C_F}^{qg\to q,1}(v,w)\,\ln(1-v+v\,w)+A_{6,\mathrm{SFP},F,C_F}^{qg\to q,1}(v,w)\,\ln(w)\,\nn\\
&&\hspace{0cm}+ A_{7,\mathrm{SFP},F,C_F}^{qg\to q,1}(v,w)\,,\label{eq:SFP1qg2qCF}
\eea
as well with seven coefficient functions of the following explicit form:
\bea
A_{1,\mathrm{SFP},F,C_F}^{qg\to q,1}(v,w) & = & \frac{5-3 v}{(1-v)^3}+\frac{w\,v\, \left(v^2-6 v+3\right)}{(1-v)^4}+\frac{2\,w^2\,v^2\, (2-v)}{(1-v)^4}\nn\\
&&+\frac{2\, w^3\, v^3}{(1-v)^4}-\frac{4 v^2 w-3 v (2 w+1)+5}{(1-v)^3\, (1-v\, w)^2}\,,\nn\\
A_{2,\mathrm{SFP},F,C_F}^{qg\to q,1}(v,w) & = & \frac{5-3 v}{(1-v)^3} +\frac{3\,w\, v}{(1-v)^2}+\frac{2\,w^2\, v^2 \,(2-3 v)}{(1-v)^4}+\frac{4\,w^3\,v^3}{(1-v)^4}\nn\\
&&-\frac{4 v^2 w-3 v (2 w+1)+5}{(1-v)^3\, (1-v\, w)^2}\,,\nn\\
A_{3,\mathrm{SFP},F,C_F}^{qg\to q,1}(v,w) & = &\frac{2 v w \left(v^3 (w-1) w^2+v^2 (w-1) w+v-1\right)}{(1-v)^3\, (1-v\, w)^2} ,\,\nn
\eea
\bea
A_{4,\mathrm{SFP},F,C_F}^{qg\to q,1}(v,w)&=& \frac{6-4 v}{(1-v)^2 v^2 w} +\frac{4 v^3-20 v^2+32 v-18}{v^2\,(1-v)^3}\nn\\
&&\hspace{-2cm}+\frac{w\,\left(v^4-5 v^3-v^2+5 v-4\right)}{v^2\,(1-v)^3}-\frac{4\,w^2\, (3 v-5)}{v^2\,(1-v)^2}+\frac{4 v^2 w-v (6 w+3)+5}{(1-v)^3\, (1-v\, w)^2},\nn\\
A_{5,\mathrm{SFP},F,C_F}^{qg\to q,1}(v,w) &=& \frac{v\, w\, (v (1-2 w)+1)}{(1-v)^3}, \nn\\
A_{6,\mathrm{SFP},F,C_F}^{qg\to q,1}(v,w) &=& -\frac{5-3 v}{(1-v)^3} -\frac{2\,w\, (2-v) v}{(1-v)^3} + \frac{2\,w^2\, v^2 (4 v-3)}{(1-v)^4}-\frac{8\,w^3\, v^3}{(1-v)^4}\nn\\
&&+ \frac{4 v^2 w-v (6 w+3)+5}{(1-v)^3\, (1-v\, w)^2}\,, \nn
\eea
\bea
A_{7,\mathrm{SFP},F,C_F}^{qg\to q,1}(v,w) & = &\frac{9 v^5-22 v^4-7 v^3+240 v^2-390 v+180}{30\,v\, (1-v)^4\, w} \nn\\
&&\hspace{0cm}- \frac{75 v^5-160 v^4-151 v^3+1380 v^2-2220 v+1080}{60\,v\, (1-v)^4}\nn\\
&& +\frac{w\,\left(12 v^5-14 v^4+55 v^3-90 v^2+105 v-60\right)}{15\,v \,(1-v)^4}\nn\\
&& + \frac{w^2\,\left(128 v^5-65 v^4-115 v^3+705 v^2-1260 v+600\right)}{30\,v\, (1-v)^4}\nn\\
&& -\frac{w^3\,v \left(52 v^3-3 v^2+6 v+1\right)}{6\, (1-v)^4}+\frac{5\,w^4\,v^4}{(1-v)^4}\nn\\
&&+\frac{4 v^2 w-v (7 w+2)+5}{(1-v)^3\, (1-v \,w)^2}-\frac{1}{(1-v)\, (1-v+v\,w)}\,.\label{eq:SFPCFFA1}
\eea
The partonic cross section for the corresponding SFP derivative term reads,
\bea
\hat{\sigma}_{\mathrm{SFP},\partial_2F}^{qg\to q,1}(v,w,\chi_m) & = & N_c\,\hat{\sigma}_{\mathrm{SFP},\partial_2F,N_c}^{qg\to q,1}(v,w,\chi_m)+C_F\,\hat{\sigma}_{\mathrm{SFP},\partial_2F,C_F}^{qg\to q,1}(v,w,\chi_m)\,,\label{eq:SFP2qg2q}
\eea
with the $N_c$ part given by,
\bea
\hat{\sigma}_{\mathrm{SFP},\partial_2F,N_c}^{qg\to q,1}(v,w) & = &A_{1,\mathrm{SFP},\partial_2F,N_c}^{qg\to q,1}(v,w)\,\ln(1-v)+A_{2,\mathrm{SFP},\partial_2F,N_c}^{qg\to q,1}(v,w)\,\ln(w)
\hspace{0cm}+ A_{3,\mathrm{SFP},\partial_2F,N_c}^{qg\to q,1}(v,w)\,,\label{eq:SFP2qg2qNc}
\eea
with three coefficient functions of the following explicit form:
\bea
A_{1,\mathrm{SFP},\partial_2F,N_c}^{qg\to q,1}(v,w) & = & \frac{v \left(-6 w^2+8 w-2\right)+10 w^2-12 w+3}{2 \,w\,v^2\,(1-v)^2}\,,\nn\\
A_{2,\mathrm{SFP},\partial_2F,N_c}^{qg\to q,1}(v,w) & = & \frac{w\,v^2\, (1+v\, w)}{(1-v)^4}\,,\nn\\
A_{3,\mathrm{SFP},\partial_2F,N_c}^{qg\to q,1}(v,w)&=& \frac{v^4-2 v^3+24 v^2-35 v+18}{12\,w\,v\, (1-v)^4}-\frac{7 v^4-16 v^3+96 v^2-144 v+72}{12\,v\, (1-v)^4} \nn\\
&&\hspace{-1cm}+\frac{\left(11 v^4-9 v^3+38 v^2-63 v+30\right) w}{6\,v\, (1-v)^4}-\frac{\left(31 v^3-6 v^2-v-8\right)\,w^2}{24\, (1-v)^4}.\label{eq:SFPNcD2FA1}
\eea
The $C_F$ part reads
\bea
\hat{\sigma}_{\mathrm{SFP},\partial_2F,C_F}^{qg\to q,1}(v,w) & = &A_{1,\mathrm{SFP},\partial_2F,C_F}^{qg\to q,1}(v,w)\,\ln(1-v)+A_{2,\mathrm{SFP},\partial_2F,C_F}^{qg\to q,1}(v,w)\,\ln(w)
\hspace{0cm}+ A_{3,\mathrm{SFP},\partial_2F,NC_F}^{qg\to q,1}(v,w)\,,\label{eq:SFP2qg2qCF}
\eea
with three coefficient functions of the following explicit form:
\bea
A_{1,\mathrm{SFP},\partial_2F,C_F}^{qg\to q,1}(v,w) & = & \frac{v \left(6 w^2-8 w+2\right)-10 w^2+12 w-3}{w\,v^2\,(1-v)^2}\,,\nn\\
A_{2,\mathrm{SFP},\partial_2F,C_F}^{qg\to q,1}(v,w) & = & \frac{2 v^3 w^2}{(1-v)^4}\,,\nn\\
A_{3,\mathrm{SFP},\partial_2F,C_F}^{qg\to q,1}(v,w)&=& \frac{-9 v^5+22 v^4+7 v^3-240 v^2+390 v-180}{60\,w\,v\, (1-v)^4} \nn\\
&&\hspace{0cm}+\frac{27 v^5-61 v^4-15 v^3+480 v^2-780 v+360}{30\,v\, (1-v)^4}\nn\\
&&\hspace{0cm}-\frac{w\,\left(64 v^5-150 v^4+15 v^3+360 v^2-630 v+300\right)}{30\,v\, (1-v)^4}\nn\\
&&+\frac{w^2\,v\, \left(28 v^3-39 v^2+6 v+1\right)}{12 \,(1-v)^4}-\frac{v^4 w^3}{(1-v)^4}.\label{eq:SFPCFD2FA1}
\eea\\
We now turn to the SFP contributions by the quark-gluon-quark correlation function $G$. The corresponding partonic factor reads,
\bea
\hat{\sigma}_{\mathrm{SFP},G}^{qg\to q,5}(v,w,\chi_m) & = & N_c\,\hat{\sigma}_{\mathrm{SFP},G,N_c}^{qg\to q,5}(v,w,\chi_m)+C_F\,\hat{\sigma}_{\mathrm{SFP},G,C_F}^{qg\to q,5}(v,w,\chi_m)\,,\label{eq:SFP3qg2q}
\eea
with the $N_c$ part given by,
\bea
\hat{\sigma}_{\mathrm{SFP},G,N_c}^{qg\to q,5}(v,w,\chi_m) & = & A_{1,\mathrm{SFP},G,N_c}^{qg\to q,5}(v,w)\,\ln(\chi_m) +A_{2,\mathrm{SFP},G,N_c}^{qg\to q,5}(v,w)\,\ln(1-w)\nn\\
&&\hspace{0cm} +A_{3,\mathrm{SFP},G,N_c}^{qg\to q,5}(v,w)\,\ln(1-v\,w)+A_{4,\mathrm{SFP},G,N_c}^{qg\to q,5}(v,w)\,\ln(1-v)\nn\\
&&\hspace{0cm} +A_{5,\mathrm{SFP},G,N_c}^{qg\to q,5}(v,w)\,\ln(1-v+v\,w)+A_{6,\mathrm{SFP},G,N_c}^{qg\to q,5}(v,w)\,\ln(w)\,\nn\\
&&\hspace{0cm}+ A_{7,\mathrm{SFP},G,N_c}^{qg\to q,5}(v,w)\,,\label{eq:SFP3qg2qNc}
\eea
with seven coefficient functions of the following explicit form:
\bea
A_{1,\mathrm{SFP},G,N_c}^{qg\to q,5}(v,w) & = &-\frac{v^3 w^3}{(1-v)^4}-\frac{v^2 w^2}{(1-v)^3}-\frac{v \left(v^2-4 v+5\right) w}{2 (1-v)^4}-\frac{v^2-4 v+2}{(1-v)^4} \nn\\
&&+\frac{4 v^2 w-v (6 w+3)+5}{2\, (1-v)^3\, (1-v\, w)^2}\,,\nn\\
A_{2,\mathrm{SFP},G,N_c}^{qg\to q,5}(v,w) & = & -\frac{v^2 (1+v) w^2}{(1-v)^4}-\frac{v^2-4 v+2}{(1-v)^4}-\frac{v (5+v) w}{2 \,(1-v)^3}\nn\\
&&+\frac{4 v^2 w-v (6 w+3)+5}{2\, (1-v)^3\, (1-v\, w)^2}\,,\nn\\
A_{3,\mathrm{SFP},G,N_c}^{qg\to q,5}(v,w) &=& \frac{v^2 w^2}{(1-v)^3}-\frac{(3-v)\, v\, w}{(1-v)^3}+\frac{3 v-5}{(1-v)^3}+\frac{4 v^2 w-v (6 w+3)+5}{(1-v)^3\, (1-v \,w)^2}\,,\nn\\
A_{4,\mathrm{SFP},G,N_c}^{qg\to q,5}(v,w) &=& \frac{2 v-3}{v^2\,(1-v)^2\, w}-\frac{2 v^3-10 v^2+16 v-9}{v^2\,(1-v)^3}+\frac{\left(3 v^3-v+4\right) w}{2\,v^2\, (1-v)^2} \nn\\
&& + \frac{2 (3 v-5) w^2}{v^2\,(1-v)^2} - \frac{4 v^2 w-3 v (2 w+1)+5}{2\, (1-v)^3\, (1-v\, w)^2}\,, \nn\\
A_{5,\mathrm{SFP},G,N_c}^{qg\to q,5}(v,w) &=& -\frac{v^2 w^2}{(1-v)^3}+\frac{v (1+v)\, w}{2\, (1-v)^3}\,,\nn\\
A_{6,\mathrm{SFP},G,N_c}^{qg\to q,5}(v,w) &=& \frac{2 v^3 w^3}{(1-v)^4}+\frac{3 v^2 w^2}{(1-v)^4}+\frac{(2-v) v\, w}{(1-v)^3}+\frac{2 v^2-4 v+3}{(1-v)^4} \nn\\
&& - \frac{4 v^2 w-3 v (2 w+1)+5}{2\, (1-v)^3\, (1-v\, w)^2}\,,\nn
\eea
\bea
A_{7,\mathrm{SFP},G,N_c}^{qg\to q,5}(v,w) &=& \frac{v^4-2 v^3-24 v^2+37 v-18}{6\, (1-v)^4\, v\, w} - \frac{10 v^4-35 v^3-102 v^2+181 v-108}{12 (1-v)^4 v}\nn\\
&& \hspace{0cm}+ \frac{\left(5 v^4-17 v^3+23 v^2-33 v+12\right) w}{6 (1-v)^4 v} + \frac{\left(9 v^4+8 v^3-149 v^2+242 v-120\right) w^2}{12 (1-v)^4 v}\nn\\
&& \hspace{0cm}-\frac{\left(7 v^3-6 v^2-v-8\right) w^3}{12 (1-v)^4}- \frac{4 v^2 w-v (7 w+2)+5}{2 \,(1-v)^3\, (1-v\, w)^2} -\frac{1}{2 \,(1-v)\, (1-v+v\,w)}\,.\label{eq:SFPNcGA1}
\eea
On the other hand, the $C_F$ part is given by,
\bea
\hat{\sigma}_{\mathrm{SFP},G,C_F}^{qg\to q,5}(v,w,\chi_m) & = & A_{1,\mathrm{SFP},G,C_F}^{qg\to q,5}(v,w)\,\ln(\chi_m) +A_{2,\mathrm{SFP},G,C_F}^{qg\to q,5}(v,w)\,\ln(1-w)\nn\\
&&\hspace{0cm} +A_{3,\mathrm{SFP},G,C_F}^{qg\to q,5}(v,w)\,\ln(1-v\,w)+A_{4,\mathrm{SFP},G,C_F}^{qg\to q,5}(v,w)\,\ln(1-v)\nn\\
&&\hspace{0cm} +A_{5,\mathrm{SFP},G,C_F}^{qg\to q,5}(v,w)\,\ln(1-v+v\,w)+A_{6,\mathrm{SFP},G,C_F}^{qg\to q,5}(v,w)\,\ln(w)\,\nn\\
&&\hspace{0cm}+ A_{7,\mathrm{SFP},G,C_F}^{qg\to q,5}(v,w)\,,\label{eq:SFP3qg2qCF}
\eea
with seven coefficient functions of the following explicit form:
\bea
A_{1,\mathrm{SFP},G,C_F}^{qg\to q,5}(v,w) & = & \frac{2 v^3 w^3}{(1-v)^4}-\frac{2 (v-2) v^2 w^2}{(1-v)^4}+\frac{v \left(v^2-6 v+3\right) w}{(1-v)^4}+\frac{5-3 v}{(1-v)^3} 
- \frac{4 v^2 w-3 v (2 w+1)+5}{(1-v)^3\, (1-v\, w)^2}\,,\nn\\
A_{2,\mathrm{SFP},G,C_F}^{qg\to q,5}(v,w) & = & \frac{2 v^2 (v+2) w^2}{(1-v)^4}-\frac{v \left(v^2+6 v-3\right) w}{(1-v)^4}+\frac{5-3 v}{(1-v)^3}  - \frac{4 v^2 w-3 v (2 w+1)+5}{(1-v)^3 \,(1-v\, w)^2}\,,\nn\\
A_{3,\mathrm{SFP},G,C_F}^{qg\to q,5}(v,w) & = & -\frac{2 v^2 w^2}{(1-v)^3}+\frac{2 (3-v) v\, w}{(1-v)^3}+\frac{2 (5-3 v)}{(1-v)^3}  - \frac{2 \left(4 v^2 w-3 v (2 w+1)+5\right)}{(1-v)^3\, (1-v\, w)^2}\,, \nn
\eea
\bea
A_{4,\mathrm{SFP},G,C_F}^{qg\to q,5}(v,w) & = & \frac{6-4 v}{(1-v)^2 v^2\, w} + \frac{4 v^3-20 v^2+32 v-18}{(1-v)^3 v^2} -\frac{\left(3 v^3-v+4\right)\, w}{(1-v)^2 v^2} \nn\\
&& +\frac{4 (5-3 v)\, w^2}{(1-v)^2 v^2}  + \frac{4 v^2 w-v (6 w+3)+5}{(1-v)^3\, (1-v\, w)^2} \,,\nn \\
A_{5,\mathrm{SFP},G,C_F}^{qg\to q,5}(v,w) & = & -\frac{v (1+v) w}{(1-v)^3}+\frac{2 v^2 w^2}{(1-v)^3}\,,\nn\\
A_{6,\mathrm{SFP},G,C_F}^{qg\to q,5}(v,w) & = & \frac{4 v^3 w^3}{(1-v)^4}-\frac{2 v^2 (2 v+1) w^2}{(1-v)^4}+\frac{2 (v-2) v w}{(1-v)^3}+\frac{3 v-5}{(1-v)^3} \nn\\
&& + \frac{4 v^2 w-v (6 w+3)+5}{(1-v)^3\, (1-v\, w)^2}\,, \nn\\
A_{7,\mathrm{SFP},G,C_F}^{qg\to q,5}(v,w) & = & \frac{-9 v^5+22 v^4-33 v^3+240 v^2-390 v+180}{30 (1-v)^4 v w}\nn\\
&&\hspace{0cm}+ \frac{75 v^5-160 v^4+129 v^3-1140 v^2+2100 v-1080}{60 \,v\,(1-v)^4} \nn\\
&& -\frac{\left(12 v^5-14 v^4-75 v^3+60 v^2-105 v+60\right) w}{15\,v\, (1-v)^4}\nn\\
&& -\frac{\left(128 v^5-185 v^4+325 v^3-735 v^2+1260 v-600\right) w^2}{30\,v\, (1-v)^4}\nn\\
&& + \frac{v \left(52 v^3-27 v^2+6 v+1\right) w^3}{6 (1-v)^4}-\frac{5 v^4 w^4}{(1-v)^4}\nn\\
&& + \frac{4 v^2 w-v (7 w+2)+5}{(1-v)^3\, (1-v\, w)^2} +\frac{1}{(1-v)\, (1-v+v\,w)}\,.\label{eq:SFPCFGA1}
\eea

The partonic cross section for the corresponding SFP derivative term for $G$ reads,
\bea
\hat{\sigma}_{\mathrm{SFP},\partial_2G}^{qg\to q,5}(v,w,\chi_m) & = & N_c\,\hat{\sigma}_{\mathrm{SFP},\partial_2G,N_c}^{qg\to q,5}(v,w,\chi_m)+C_F\,\hat{\sigma}_{\mathrm{SFP},\partial_2G,C_F}^{qg\to q,5}(v,w,\chi_m)\,,\label{eq:SFP4qg2q}
\eea
with the $N_c$ part given by,
\bea
\hat{\sigma}_{\mathrm{SFP},\partial_2G,N_c}^{qg\to q,5}(v,w) & = &A_{1,\mathrm{SFP},\partial_2G,N_c}^{qg\to q,5}(v,w)\,\ln(1-v)+A_{2,\mathrm{SFP},\partial_2G,N_c}^{qg\to q,5}(v,w)\,\ln(w)\nn\\
&&\hspace{0cm}+ A_{3,\mathrm{SFP},\partial_2G,N_c}^{qg\to q,5}(v,w)\,,\label{eq:SFP4qg2qNc}
\eea
with three coefficient functions of the following explicit form:
\bea
A_{1,\mathrm{SFP},\partial_2G,N_c}^{qg\to q,5}(v,w) & = & \frac{3-2 v}{2\,v^2\, (1-v)^2\, w}+\frac{2 (2 v-3)}{v^2\,(1-v)^2}+\frac{(5-3 v) w}{v^2\,(1-v)^2}  \,,\nn\\
A_{2,\mathrm{SFP},\partial_2G,N_c}^{qg\to q,5}(v,w) & = & -\frac{v^2 w \,(1+v\, w)}{(1-v)^4} \,,\nn\\
A_{3,\mathrm{SFP},\partial_2G,N_c}^{qg\to q,5}(v,w)&=&  \frac{-v^4+2 v^3+24 v^2-37 v+18}{12\,v\, (1-v)^4\, w}+\frac{7 v^4-16 v^3-96 v^2+144 v-72}{12\,v\, (1-v)^4}\nn\\
&&\hspace{0cm}-\frac{\left(11 v^4-9 v^3-38 v^2+57 v-30\right) w}{6\,v\, (1-v)^4}+\frac{\left(31 v^3-6 v^2-v-8\right) w^2}{24 \,(1-v)^4}.\label{eq:SFPNcD2GA1}
\eea

The $C_F$ part reads
\bea
\hat{\sigma}_{\mathrm{SFP},\partial_2G,C_F}^{qg\to q,5}(v,w) & = &A_{1,\mathrm{SFP},\partial_2G,C_F}^{qg\to q,5}(v,w)\,\ln(1-v)+A_{2,\mathrm{SFP},\partial_2G,C_F}^{qg\to q,5}(v,w)\,\ln(w)\nn\\
&&\hspace{0cm}+ A_{3,\mathrm{SFP},\partial_2G,NC_F}^{qg\to q,5}(v,w)\,,\label{eq:SFP4qg2qCF}
\eea
with three coefficient functions of the following explicit form:
\bea
A_{1,\mathrm{SFP},\partial_2G,C_F}^{qg\to q,5}(v,w) & = & \frac{2 v-3}{v^2\,(1-v)^2\, w}-\frac{4 (2 v-3)}{v^2\,(v-1)^2}+\frac{2 (3 v-5) w}{v^2\,(1-v)^2}\,,\nn\\
A_{2,\mathrm{SFP},\partial_2G,C_F}^{qg\to q,5}(v,w) & = & -\frac{2 v^3 w^2}{(1-v)^4}\,,\nn
\eea
\bea
A_{3,\mathrm{SFP},\partial_2G,C_F}^{qg\to q,5}(v,w)&=& \frac{9 v^5-22 v^4+33 v^3-240 v^2+390 v-180}{60\,v\, (1-v)^4\, w}\nn\\
&&\hspace{0cm}-\frac{27 v^5-61 v^4+65 v^3-480 v^2+780 v-360}{30\,v\, (1-v)^4}\nn\\
&&+\frac{\left(64 v^5-150 v^4+55 v^3-360 v^2+630 v-300\right) w}{30\,v\, (1-v)^4}\nn\\
&&-\frac{v \left(28 v^3-39 v^2+6 v+1\right) w^2}{12\, (1-v)^4} + \frac{v^4 w^3}{(1-v)^4}\,.\label{eq:SFPCFD2GA1}
\eea

\subsection{Channel \texorpdfstring{$qq\to q$}{qq to q}} 
\subsubsection{Integral Contribution\label{appsub:Integralqq2q}}
The partonic factors that appear in our result for the integral contribution to the $qq\to q$-channel in Eq.~\ref{eq:Channel2Int} read explicitly,
\bea
\hat{\sigma}_{\mathrm{Int},1}^{qq\to q,1}(v,w,\zeta) & = & \frac{1}{N_c}\frac{w  }{4 (1-\zeta) \zeta  (1-v)^4}\left[v (w-\zeta )^2 \left(\zeta  (2 \zeta -1)\right.\right.\nn\\
&&\hspace{0cm}\left.+v^2 \left(\zeta  (2 \zeta -1)+\left(8 \zeta ^2-4 \zeta -2\right) w^2+2 (3-4
   \zeta ) \zeta  w\right)-2 v \left(\zeta ^2-3 \zeta  w+w\right)\right)\nn\\
 &&\hspace{0cm}+\text{sgn}(w-\zeta ) \left(v^3 (w-\zeta )^2 \left(\zeta  (2 \zeta -1)+2 w^2-2 \zeta  w\right)\right.\nn\\
 &&\hspace{0cm}+2 v^2 \left(-\zeta ^4+(\zeta +1)
   w^3-(\zeta +4) \zeta  w^2+\left(3 \zeta ^3+\zeta \right) w\right)\nn\\
 &&\hspace{0cm} \left.\left. +\zeta  v \left(\zeta ^2 (2 \zeta -1)+(6 \zeta -5) w^2+2 (3-4 \zeta ) \zeta  w\right)+2 \zeta 
   \left(2 \zeta ^2-3 \zeta +1\right) w\right)\right]\,,\label{eq:Int1qq2q}
\eea
\bea
\hat{\sigma}_{\mathrm{Int},1}^{qq\to q,5}(v,w,\zeta) & = & -\frac{1}{N_c}\frac{w  }{4 (1-\zeta) \zeta
    (1-v)^4}\left[v (w-\zeta )^2
   \left(\zeta +v^2 \left(\zeta +(4 \zeta -2) w^2-2 \zeta  w\right)\right.\right.\nn\\
   &&\hspace{0cm}\left.+v \left(\left(8 \zeta ^2-6 \zeta -2\right) w-2 \zeta ^2\right)\right) +\text{sgn}(w-\zeta ) \left(v^3 (w-\zeta )^2 \left(\zeta +2 w^2-2 \zeta  w\right)\right.\nn\\
   &&\hspace{0cm}-2 v^2 \left(\zeta ^4+(\zeta -1) w^3+\zeta ^2
   w^2+\left(\zeta -3 \zeta ^3\right) w\right)\nn\\
 &&\hspace{0cm} \left.\left. +\zeta  v \left(\zeta ^2+(5-4 \zeta ) w^2+2 \zeta  (2 \zeta -3) w\right)+2 (\zeta -1) \zeta  w\right)\right]\,,\label{eq:Int2qq2q}
\eea
\bea
\hat{\sigma}_{\mathrm{Int},2}^{qq\to q,1}(v,w,\zeta) & = & \frac{1}{N_c}\frac{v \,w}{4\, \zeta \,
   (1-v)^4} \Bigg(\frac{2 \zeta ^2-3 \zeta -2 v^2 (1-w-\zeta)^2 \text{sgn}(1-w-\zeta)}{1-\zeta}\nn\\
  &&\hspace{-3cm} +\frac{v^2 \left(2 \zeta ^2-3 \zeta +2 \left(4 \zeta ^2-6
   \zeta +1\right) w^2-2 \left(4 \zeta ^2-5 \zeta +1\right) w+1\right)+2 (\zeta -1) v (w-\zeta )+1}{1-\zeta}\nn\\
 &&  \hspace{-3cm}+\frac{1-v}{\left[1-2 v (1-w-\zeta +2 \zeta  w)+v^2 (1-w-\zeta)^2\right]^{\tfrac{3}{2}}} \Big(2 \zeta +v^4 (1-2 w) (1-w-\zeta)^3\nn\\
 &&\hspace{-2cm}+v^3 \left((1-\zeta )^2 \left(2 \zeta ^2-3 \zeta -2\right)+(5-6 \zeta ) w^3+\left(-6 \zeta ^2+17 \zeta -12\right)w^2\right.\nn\\
 &&\hspace{-2cm}\left.+\left(2 \zeta ^3+\zeta ^2-12 \zeta
   +9\right) w\right)+v^2 \left(3 \zeta  \left(2 \zeta ^2-5 \zeta +3\right)+\left(3-4 \zeta ^2\right) w^2\right.\nn\\
  &&\hspace{-2cm}\left. -\left(4 \zeta ^3-22 \zeta ^2+12 \zeta +3\right) w\right)-v
   \left(-6 \zeta ^2+9 \zeta +12 \zeta ^2 w-12 \zeta  w+w-2\right)-1\Big)\Bigg)\,,\label{eq:Int3qq2q}
\eea
\bea
\hat{\sigma}_{\mathrm{Int},2}^{qq\to q,5}(v,w,\zeta) & = & \frac{1}{N_c}\frac{v\, w}{4 \,\zeta\,  (1-v)^4}  \Bigg(\frac{-\zeta -2 v^2 (1-\zeta -w)^2\, \text{sgn}(1-w-\zeta)}{1-\zeta }\nn\\
&&\hspace{-3cm}+\frac{v^2 \left(-\zeta +(2-4 \zeta ) w^2-2 (1-\zeta) w+1\right)+2 (1-\zeta) v ((4 \zeta -1) w-\zeta )+1}{1-\zeta }\nn\\
 && \hspace{-3cm} +\frac{1-v}{\left[1-2 v (1-w-\zeta +2 \zeta 
   w)+v^2 (1-w-\zeta)^2\right]^{\tfrac{3}{2}}} \Big(v^4 (1-w-\zeta)^3 (1-2w-2 \zeta)\nn\\
 &&\hspace{-2cm}  +v^3 \big[(1-\zeta)^2 (5 \zeta -2)+(5-16 \zeta ) w^3+\left(-32
   \zeta ^2+45 \zeta -12\right) w^2\nn\\
 &&\hspace{-2cm}  +\left(-16 \zeta ^3+45 \zeta ^2-38 \zeta +9\right) w\big]+v^2 \big[-3 (1-\zeta ) \zeta +\left(32 \zeta ^2-22 \zeta +3\right)
   w^2\nn\\
  &&\hspace{-2cm} +\left(-22 \zeta ^2+22 \zeta -3\right) w\big]-v (\zeta +2 \zeta  w+w-2)-1\Big)\Bigg)\,.\label{eq:Int4qq2q}
\eea

\subsubsection{Hard-Pole Contribution\label{appsub:HPqq2q}}
Moving on with the hard-pole contribution to the $qq\to q$-channel found in Eq.~\eqref{eq:Channel2HP} we have the following partonic factors,
\bea
\hat{\sigma}_{\mathrm{HP},F}^{qq\to q,1}(v,w,\chi_\mu) & = & -\frac{1}{2N_c} \Bigg[\delta(1-w)\,\frac{1+v^2}{(1-v)^4} \left(\ln \chi_\mu+2 \ln (1-v)\right)\nn\\
&& \hspace{0cm}+\frac{2(1+v^2)}{(1-v)^4\, (1-w)_+} + \frac{v^3 \left(8 w^2-7 w+1\right)-9 v^2 w+v (7 w-3)-8 w+2}{(1-v)^4}\nn\\
&&\hspace{0cm} -\frac{1+v^2}{(1-v)^4}\, (\ln \chi_\mu+2\,\ln(1-v))+\frac{2\,v\, (1+v)\, w }{(1-v)^4}\ln v + \frac{1-4 v^3 w^2+v^2 (1-4 w)}{(1-v)^4}\,\ln w \nn\\
&&\hspace{0cm}+ \frac{v^3 \left(4 w^3-6 w^2+4 w-1\right)+v^2 \left(8 w^2-9 w+2\right)+v (2 w-1)-w }{(1-v)^4\, w}\ln(1-w)\Bigg]\,,\label{eq:HP1qq2q}
\eea
\bea
\hat{\sigma}_{\mathrm{HP},\partial F}^{qq\to q,1}(v,w,\chi_\mu) & = & \frac{1}{2N_c} \Bigg[\frac{(1+v^2)\,(1-2w)}{(1-v)^4}\, (\ln \chi_\mu+2\,\ln(1-v))\nn\\
&& \hspace{0cm}+ \frac{v^3 \left(-4 w^2+5 w-1\right)+v^2 (7 w-2)+v (3-5 w)+8 w-4}{(1-v)^4}\nn\\
&&\hspace{0cm}+ \frac{2 v^3 w^2+v^2 (4 w-1)+2 w-1 }{(1-v)^4} \,\ln w\nn\\
&&\hspace{0cm}+ \frac{v^3 \left(-2 w^3+4 w^2-3 w+1\right)+v^2 \left(-6 w^2+7 w-2\right)-v w+v-2 w^2+w }{(1-v)^4\, w}\,\ln(1-w)\Bigg]\,,\label{eq:HP2qq2q}
\eea
\bea
\hat{\sigma}_{\mathrm{HP},G}^{qq\to q,5}(v,w,\chi_\mu) & = & -\frac{1}{2N_c} \Bigg[-\delta(1-w)\,\frac{1+v^2}{(1-v)^4} \left(\ln \chi_\mu+2 \ln (1-v)\right)\nn\\
&& \hspace{0cm}-\frac{2(1+v^2)}{(1-v)^4\, (1-w)_+} + \frac{v^3+v^2 (w+4)+v (8 w-1)+4}{(1-v)^4}\nn\\
&&\hspace{0cm} -\frac{1+v^2}{(1-v)^4}\, (\ln \chi_\mu+2\,\ln(1-v))+\frac{2\,v\, (1+v)\, w }{(1-v)^4}\ln v + \frac{1+4 v^3 w^2+v^2 (1+4 w)}{(1-v)^4}\,\ln w \nn\\
&&\hspace{0cm}+ \frac{v^3 \left(-4 w^3+6 w^2-4 w+1\right)+v^2 (3 w-2)-2 v w+v-w }{(1-v)^4\, w}\,\ln(1-w)\Bigg]\,,\label{eq:HP3qq2q}
\eea
\bea
\hat{\sigma}_{\mathrm{HP},\partial G}^{qq\to q,5}(v,w,\chi_\mu) & = & \frac{1}{2N_c} \Bigg[\frac{1+v^2}{(1-v)^4}\, (\ln \chi_\mu+2\,\ln(1-v))\nn\\
&& \hspace{0cm}+ \frac{-v^3+v^2 (w-2)-4 v w+v-2}{(1-v)^4}-\frac{2 v^3 w^2+v^2 (2 w+1)+1}{(1-v)^4}\,\ln w\nn\\
&&\hspace{0cm}+ \frac{v^3 \left(2 w^3-4 w^2+3 w-1\right)-v^2 (w-2)+v (w-1)+w }{(1-v)^4\, w}\,\ln(1-w)\Bigg]\,,\label{eq:HP4qq2q}
\eea

\subsubsection{Soft-Fermion Pole Contribution\label{appsub:SFPqq2q}}
At last, our formula Eq.~\eqref{eq:Channel2SFP} for the soft-fermion pole contribution concludes our results for the $qq\to q$-channel and the partonic factors therein are given as,
\bea
\hat{\sigma}_{\mathrm{SFP},F}^{qq\to q,1}(v,w,\chi_m) & = & \frac{-1}{2N_c}\Bigg[ \frac{2 v^3 w \left(2 w^2-2 w+1\right)+v^2 \left(-6 w^2+2 w-1\right)+4 v w-1}{(1-v)^4}\,\ln\chi_m\nn \\
&& \hspace{-0cm}+ \frac{v w (1+v (1-4 w))}{(1-v)^3}\,\ln(1-v)+\frac{2 v^2 w^2 }{(1-v)^3}\,\ln(1-v\,w) - \frac{v w (1+v-2 v w)}{(1-v)^3}\,\ln(1-v+v\,w)\nn\\
&&\hspace{-0cm}+\Bigg(\frac{v}{(1-v)^2\, w}-\frac{1+4 v}{(1-v)^2}+\frac{v \left(7 v^2-8 v+5\right) w}{(1-v)^4}-\frac{2 v^2 (3 v+1) w^2}{(1-v)^4}+\frac{4 v^3 w^3}{(1-v)^4}\Bigg)\,\ln(1-w)\nn\\
&&\hspace{-0cm} +\frac{2 v^3 (5-2 w) w^2+v^2 \left(-2 w^2+6 w-1\right)+2 v w-1}{(1-v)^4}\ln w -\frac{8 v^3 w^3}{(1-v)^4}+\frac{\left(7 v^3-8 v+4\right) w}{(1-v)^4}\nn\\
&&-\frac{v^3+v^2-7 v+3}{(1-v)^4}+\frac{2 v (3 v-1) w^2}{(1-v)^4}+\frac{1}{(1-v)\, (1-v+v\, w)}\Bigg]\,,\label{eq:SFP1qq2q}
\eea
\bea
\hat{\sigma}_{\mathrm{SFP},G}^{qq\to q,5}(v,w,\chi_m) & = & \frac{-1}{2N_c}\Bigg[  \frac{2 v^3 w \left(2 w^2-2 w+1\right)+v^2 \left(-6 w^2+2 w-1\right)+4 v w-1}{(1-v)^4}\,\,\ln\chi_m\nn \\
&& \hspace{0cm}+\frac{v \,(1+v)\, w}{(1-v)^3}\,\ln(1-v)-\frac{2 v^2 w^2 }{(1-v)^3}\,\ln(1-v\,w) - \frac{v w (1+v-2 v w)}{(1-v)^3}\,\ln(1-v+v\,w)\nn\\
&&\hspace{0cm}+\Bigg(-\frac{v}{(1-v)^2\, w}-\frac{4 v^2-v+1}{(1-v)^3}+\frac{v (7 v+3) w}{(1-v)^3}-\frac{6 v^2 w^2}{(1-v)^3}\Bigg)\,\ln(1-w)\nn\\
&&\hspace{0cm} +\frac{-6 v^3 w^2+v^2 \left(6 w^2-2 w-1\right)+2 v w-1 }{(1-v)^4}\,\ln w +\frac{8 v^2 w^2}{(1-v)^4}\nn\\
&&\hspace{0cm}-\frac{v^3+v^2-5 v+1}{(1-v)^4}-\frac{v (6 v+5) w}{(1-v)^4}+\frac{1}{(1-v)\, (1-v+v\, w)}\Bigg]\,.\label{eq:SFP2qq2q}
\eea

\subsection{Channel \texorpdfstring{$qq\to q^\prime$}{qq to q'}} 
\subsubsection{Integral Contribution\label{appsub:Integralqq2qp}}
For the integral contribution to the $qq\to q^\prime$-channel in Eq.~\eqref{eq:Channel3Int} we found the following partonic factors,
\bea
\hat{\sigma}_{\mathrm{Int},1}^{qq\to q^\prime,1}(v,w,\zeta) & = & \frac{1}{2}\frac{v\, w}{2\, (1-v)^4} \Bigg(\frac{(2 \zeta -1) (1-v)^2 (1+v (w-\zeta ))}{\sqrt{1-2 \zeta  v+2(2 \zeta -1) v w+v^2 (w-\zeta )^2}}\nn\\
&& \hspace{0cm}+\frac{1}{\zeta
   }\text{sgn}(w-\zeta ) \left(\zeta  (2 \zeta -1)+v^2 \left(\zeta  (4 \zeta -1)+4 w^2-6 \zeta  w\right)+2 \zeta  v (w-\zeta
   )\right)\nn\\
 && \hspace{0cm} +\frac{2 v}{\zeta
   } \Big(2 (\zeta -1) \zeta  v^2 w \left(6 \zeta +2 (6 \zeta -1) w^2+(2-16 \zeta ) w-1\right)\nn\\
 &&\hspace{0cm}  +v \left(\zeta ^2+\left(8 \zeta ^2-4 \zeta
   -2\right) w^2+\zeta  \left(-4 \zeta ^2-2 \zeta +3\right) w\right)\nn\\
 && \hspace{0cm} +\zeta  \left(\zeta +\left(4 \zeta ^2-6 \zeta +3\right) w-1\right)\Big)\Bigg)\,,\nn\\
 \hat{\sigma}_{\mathrm{Int},2}^{qq\to q^\prime,1}(v,w,\zeta) & = & -\hat{\sigma}_{\mathrm{Int},1}^{qq\to q^\prime,1}(v,w,1-\zeta)\,,\nn\\
 \hat{\sigma}_{\mathrm{Int}}^{qq\to q^\prime,1}(v,w,\zeta) & \equiv & \hat{\sigma}_{\mathrm{Int},1}^{qq\to q^\prime,1}(v,w,\zeta) + \hat{\sigma}_{\mathrm{Int},2}^{qq\to q^\prime,1}(v,w,\zeta)\,.\label{eq:Int1qq2qp}
\eea
\bea
\hat{\sigma}_{\mathrm{Int},1}^{qq\to q^\prime,5}(v,w,\zeta) & = & \frac{1}{2}\frac{v\, w }{2\, (1-v)^4}\Bigg(-\frac{(1-v)^2 (1+v (w-\zeta ))}{\sqrt{1-2 \zeta  v+2(2 \zeta -1) v w+v^2 (w-\zeta )^2}}\nn\\
&&\hspace{0cm}-\frac{1}{\zeta }\text{sgn}(w-\zeta ) \left(\zeta +v^2 \left(2 \zeta ^2+\zeta +4 w^2-6 \zeta  w\right)+2 \zeta  v (w-\zeta )\right)\nn\\
&&\hspace{0cm}-\frac{2 v }{\zeta }\Big(-2
   (1-\zeta) \zeta  v^2 w \left(2 w^2-2 w+1\right)+v \left(\zeta ^2+(4 \zeta -2) w^2+3 (1-2 \zeta ) \zeta  w\right)\nn\\
 &&\hspace{0cm}  +\zeta  (-\zeta +(6 \zeta -5)
   w+1)\Big)\Bigg)\,,\nn\\
 \hat{\sigma}_{\mathrm{Int},2}^{qq\to q^\prime,5}(v,w,\zeta) & = & +\hat{\sigma}_{\mathrm{Int},1}^{qq\to q^\prime,5}(v,w,1-\zeta)\,,\nn\\
 \hat{\sigma}_{\mathrm{Int}}^{qq\to q^\prime,5}(v,w,\zeta) & \equiv & \hat{\sigma}_{\mathrm{Int},1}^{qq\to q^\prime,5}(v,w,\zeta) + \hat{\sigma}_{\mathrm{Int},2}^{qq\to q^\prime,5}(v,w,\zeta)\,.\label{eq:Int2qq2qp}
\eea

\subsubsection{Soft-Fermion Pole Contribution\label{appsub:SFPqq2qp}}
The two partonic factors in our formula for the soft-fermion contribution to the $qq\to q^\prime$-channel Eq.~\ref{eq:Channel3SFP} read explicitly,  
\bea
\hat{\sigma}_{\mathrm{SFP},F}^{qq\to q^\prime,1}(v,w,\chi_m) & = & \frac{1}{2}\Bigg[ \Big(\frac{4 v^4 w^4}{(1-v)^4}-\frac{4 v^3 (v+1) w^3}{(1-v)^4}+\frac{2 v^2 \left(v^2+2\right) w^2}{(1-v)^4}\nn\\
&&\hspace{0cm}-\frac{2 v^2 w}{(1-v)^4}+\frac{1}{(1-v)^2}-\frac{1}{(1-v)^2 (1-v\,w)}\Big)\,\ln\chi_m +\frac{w \left(4 v w^2+v-4 w\right) }{(1-v)^2\, (1-v\, w)}\,\ln(1-v)\nn \\
&& \hspace{0cm}-\frac{v w (1+v\, w)}{(1-v)^2 \,(1-v\, w)}\,\ln(1-v\,w)-\frac{v\, w }{(1-v)^2}\,\ln(1-v+v\,w)\nn\\
&&\hspace{0cm}+\Big(\frac{4 v^4 w^4}{(1-v)^4}-\frac{4 v^4 w^3}{(1-v)^4}+\frac{2 \left(v^2-3 v+3\right) v^2 w^2}{(1-v)^4}+\frac{(1-3 v) v w}{(1-v)^3}+\frac{1}{(1-v)^2}\nn\\
&&\hspace{0cm}-\frac{1}{(1-v)^2\, (1-v\,w)}\Big)\,\ln(1-w)+\Big(-\frac{4 v^4 w^4}{(1-v)^4}+\frac{4 v^4 w^3}{(1-v)^4}-\frac{2 \left(v^2+3 v+1\right) v^2 w^2}{(1-v)^4}\nn\\
&&\hspace{0cm}-\frac{v w}{(1-v)^2}-\frac{1}{(1-v)^2}+\frac{1}{(1-v)^2 (1-v\,w)}\Big)\,\ln w-\frac{12 v^4 w^4}{(1-v)^4}+\frac{12 v^3 (v+2) w^3}{(1-v)^4}\nn\\
&&\hspace{0cm} -\frac{2 v^2 \left(2 v^2+8 v+1\right) w^2}{(1-v)^4}+\frac{2 v^2 w}{(1-v)^4}\Bigg]\,,\label{eq:SFP1qq2qp}
\eea
\bea
\hat{\sigma}_{\mathrm{SFP},G}^{qq\to q^\prime,5}(v,w,\chi_m) & = & \frac{1}{2}\Bigg[ \Big(\frac{4 v^4 w^4}{(1-v)^4}-\frac{4 v^3 (v+1) w^3}{(1-v)^4}+\frac{2 v^2 \left(v^2+2\right) w^2}{(1-v)^4}\nn\\
&&\hspace{0cm}-\frac{2 v^2 w}{(1-v)^4}+\frac{1}{(1-v)^2}-\frac{1}{(1-v)^2 (1-v\,w)}\Big)\,\ln\chi_m +\frac{v\,w }{(1-v)^2\, (1-v\, w)}\,\ln(1-v)\nn \\
&& \hspace{0cm}-\frac{v w (1+v\, w)}{(1-v)^2 \,(1-v\, w)}\,\ln(1-v\,w)-\frac{v\, w }{(1-v)^2}\,\ln(1-v+v\,w)\nn\\
&&\hspace{0cm}+\Big(\frac{4 v^4 w^4}{(1-v)^4}-\frac{4 v^4 w^3}{(1-v)^4}+\frac{2 \left(v^2-3 v+3\right) v^2 w^2}{(1-v)^4}+\frac{(1-3 v) v w}{(1-v)^3}+\frac{1}{(1-v)^2}\nn\\
&&\hspace{0cm}-\frac{1}{(1-v)^2\, (1-v\,w)}\Big)\,\ln(1-w)+\Big(-\frac{4 v^4 w^4}{(1-v)^4}+\frac{4 v^4 w^3}{(1-v)^4}-\frac{2 \left(v^2-3 v+3\right) v^2 w^2}{(1-v)^4}\nn\\
&&\hspace{0cm}-\frac{v w}{(1-v)^2}-\frac{1}{(1-v)^2}+\frac{1}{(1-v)^2 (1-v\,w)}\Big)\,\ln w-\frac{4 v^4 w^4}{(1-v)^4}+\frac{4 v^4 w^3}{(1-v)^4}\nn\\
&&\hspace{0cm} -\frac{2 v^2 w^2}{(1-v)^4}+\frac{2 v^2 w}{(1-v)^4}\Bigg]\,.\label{eq:SFP2qq2qp}
\eea

\subsection{Channel \texorpdfstring{$qg\to g$}{qg to g}} 

\subsubsection{Integral Contribution\label{appsub:Integralqg2g}}

For the $qg\to g$-channel we again start with the partonic factors of the integral contribution, cf. Eq.~\eqref{eq:Channel4Int}. As for the $qg\to q$-channel we encounter two color factors, $C_F$ and $N_c$. We can split the results according to these two factors. For the integral contributions, we write
\bea
\hat{\sigma}_{\mathrm{Int}}^{qg\to g,1}(v,w,\zeta) & = & C_F\,\hat{\sigma}_{\mathrm{Int},C_F}^{qg\to g,1}(v,w,\zeta) + N_c\,\hat{\sigma}_{\mathrm{Int},N_c}^{qg\to g,1}(v,w,\zeta) \,,\label{eq:Int1qg2g}
\eea
with
\bea
\hat{\sigma}_{\mathrm{Int},C_F}^{qg\to g,1}(v,w,\zeta) & = & \frac{v\, w}{2\, (1-v)^4}  \Bigg(-\zeta(1-2v-v^2) -2 (1-\zeta) \zeta  v^3 w \left(6
   w^2-8 w+3\right)\nn\\
 && \hspace{0cm} -2 \left(4 \zeta ^2-6 \zeta +\tfrac{1}{\zeta}\right)v^2\, w^2+4  \left(\zeta ^2-3 \zeta +1\right) v^2\,w+2 \left(\zeta ^2+\zeta -2\right) v\,w+2v\nn\\
 &&\hspace{0cm}+\frac{(1+\zeta)
   (1-v)^2 (1+v (w-\zeta ))}{\sqrt{1-2 \zeta  v+2(2 \zeta -1) v w+v^2 (w-\zeta )^2}}  +\text{sgn}(w-\zeta )\left(1 + v^2\left(1 + \frac{2(w - \zeta)^2}{\zeta}\right)\right)\Bigg)\,,\label{eq:Int1CFqg2g}
\eea
\bea
\hat{\sigma}_{\mathrm{Int},N_c}^{qg\to g,1}(v,w,\zeta) & = & \frac{v w}{4\, (1-v)^4}\Bigg( - (2+\zeta)-v^2 \left(\zeta+\left(4 \zeta -\tfrac{2}{\zeta}\right) w^2+2 (1-2 \zeta )   w\right)\nn\\
&&-2 (1-2 \zeta )  v w-\text{sgn}(w-\zeta ) \,\left(1 +v^2+v^2\left(2 \,\zeta +2 \tfrac{w^2}{\zeta}-4 \, w\right)\right)\nn\\
&&\hspace{0cm}-\frac{(1-v)}{\left[1-2 \zeta  v+2(2 \zeta -1) v w+v^2 (w-\zeta )^2\right]^{\tfrac{3}{2}}} \Big[-1-\zeta +v^4 (w-\zeta )^3 (-\zeta +2 w-1)\nn\\
&&\hspace{0cm}+v^3 \left(\zeta ^2 \left(\zeta ^2-2 \zeta -3\right)+(11 \zeta -7) w^3+\left(-19 \zeta ^2+4 \zeta +3\right)
   w^2+\zeta ^2 (7 \zeta +5) w\right)\nn\\
&&\hspace{0cm}+(\zeta -1) v^2 \left(-3 \zeta  (\zeta +1)+(14 \zeta -5) w^2+\left(2 \zeta ^2-2 \zeta +1\right) w\right)\nn\\
&&+v \left(3 \zeta ^2+2
   \zeta +\left(-6 \zeta ^2+\zeta +1\right) w-1\right)\Big]\Bigg) \,,\label{eq:Int1Ncqg2g}
\eea
and
\bea
\hat{\sigma}_{\mathrm{Int}}^{qg\to g,5}(v,w,\zeta) & = & C_F\,\hat{\sigma}_{\mathrm{Int},C_F}^{qg\to g,5}(v,w,\zeta) + N_c\,\hat{\sigma}_{\mathrm{Int},N_c}^{qg\to g,5}(v,w,\zeta) \,,\label{eq:Int2qg2g}
\eea
with
\bea
\hat{\sigma}_{\mathrm{Int},C_F}^{qg\to g,5}(v,w,\zeta) & = & \frac{v\, w}{2\, (1-v)^4} \Bigg(\zeta(1-2v-v^2) +2 (1-\zeta) \zeta  v^3 w \left(6 w^2-8
   w+3\right)\label{eq:Int2CFqg2g}\\
   &&\hspace{0cm}+2 \left(4 \zeta ^2-6 \zeta +\tfrac{1}{\zeta}\right) v^2\,w^2-4  \left(\zeta ^2-3 \zeta +1\right)v^2\, w-2 (1-\zeta ) (2-\zeta)v\, w+2  v\nn\\
  &&\hspace{0cm} +\frac{(1-\zeta)
   (1-v)^2 (1+v (w-\zeta))}{\sqrt{1-2 \zeta  v+2(2 \zeta -1) v w+v^2 (w-\zeta )^2}}+\text{sgn}(w-\zeta ) \left(1+v^2-v^2 \left(2 \zeta +\tfrac{2 w^2}{\zeta }-4 w\right)\right)  \Bigg)\,,\nn
\eea
\bea
\hat{\sigma}_{\mathrm{Int},N_c}^{qg\to g,5}(v,w,\zeta) & = & \frac{v\, w}{4 \,(1-v)^4}\Bigg(-(2-\zeta) +v^2 \left(\zeta +\left(4 \zeta -\tfrac{2}{\zeta }\right) w^2+2(1-2 \zeta ) w\right)\nn\\
&&+2(1-2 \zeta ) v w-\text{sgn}(w-\zeta ) \left(1 +v^2-v^2\,\left(2 \,\zeta +\tfrac{2\, w^2}{\zeta}-4\,   w\right) \right)\nn\\
&&\hspace{0cm}+\frac{(1-v)}{\left[1-2 \zeta  v+2(2 \zeta -1) v w+v^2 (w-\zeta )^2\right]^{\tfrac{3}{2}}} \Big[1-\zeta +v^4 (w-\zeta )^3 (-\zeta +2 w+1)\nn\\
&&\hspace{0cm}+v^3 \left(\zeta ^2 \left(\zeta ^2-4 \zeta +3\right)+(11 \zeta -5) w^3+\left(-19 \zeta ^2+10 \zeta
   -3\right) w^2+\zeta ^2 (7 \zeta -1) w\right)\nn\\
 &&\hspace{0cm}  +v^2 \left(-3 (\zeta -1)^2 \zeta +\left(14 \zeta ^2-15 \zeta +7\right) w^2+\left(2 \zeta ^3-8 \zeta ^2-\zeta +1\right)
   w\right)\nn\\
 &&\hspace{0cm}  +v \left(3 \zeta ^2-4 \zeta +\left(-6 \zeta ^2+13 \zeta -5\right) w+1\right)\Big]\Bigg)\,.\label{eq:Int2Ncqg2g}
\eea

\subsubsection{Soft-Gluon Pole Contribution\label{appsub:SGPqg2g}}

Next, we state our results for the partonic factors appearing in Eq.~\eqref{eq:Channel4SGP}, the soft-gluon pole contribution to the $qg\to g$-channel. This contribution can again be separated according to its color structures,
\be
\hat{\sigma}_{\mathrm{SGP},F}^{qg\to g,1}(v,w,\chi_\mu,\chi_m) = C_F\,\hat{\sigma}_{\mathrm{SGP},F,C_F}^{qg\to g,1}(v,w,\chi_\mu) + N_c\,\hat{\sigma}_{\mathrm{SGP},F,N_c}^{qg\to g,1}(v,w,\chi_m)\,,\label{eq:SGP1qg2g}
\ee
with 
\bea
\hat{\sigma}_{\mathrm{SGP},F,C_F}^{qg\to g,1}(v,w,\chi_\mu) & = & A_{1,\mathrm{SGP},F,C_F}^{qg\to g,1}(v,\chi_\mu)\,\delta(1-w) \nn\\
&&\hspace{0cm}+A_{2,\mathrm{SGP},F,C_F}^{qg\to g,1}(v)\frac{1}{(1-w)_+}+A_{3,\mathrm{SGP},F,C_F}^{qg\to g,1}(v,w)\,\ln(\chi_\mu)\nn\\
&&\hspace{0cm}+A_{4,\mathrm{SGP},F,C_F}^{qg\to g,1}(v,w)\,\ln(1-w)+A_{5,\mathrm{SGP},F,C_F}^{qg\to g,1}(v,w)\,\ln(1-v)\nn\\
&&\hspace{0cm}+A_{6,\mathrm{SGP},F,C_F}^{qg\to g,1}(v,w)\,\ln(1-v+v\,w)+A_{7,\mathrm{SGP},F,C_F}^{qg\to g,1}(v,w)\,\ln(w)\nn\\
&&+A_{8,\mathrm{SGP},F,C_F}^{qg\to g,1}(v,w)\,.\label{eq:SGP1qg2gCF}
\eea
The eight coefficients read,
\bea
A_{1,\mathrm{SGP},F,C_F}^{qg\to g,1}(v,\chi_\mu) & = & \frac{v \left(1+v^2\right)}{(1-v)^4}\ln \chi_\mu + \frac{2 v^3+5 v^2+1}{2\, (1-v)^4}\ln(1-v)- \frac{v^4-8 v^3+9 v^2-6v}{4 \,(1-v)^4}\,,\nn\\
A_{2,\mathrm{SGP},F,C_F}^{qg\to g,1}(v) & = & \frac{3\, v\, \left(1+v^2\right)}{2\, (1-v)^4}\,,\nn\\
A_{3,\mathrm{SGP},F,C_F}^{qg\to g,1}(v,w) & = & -\frac{6 v^4 w^4}{(1-v)^4}+\frac{8 v^4 w^3}{(1-v)^4}-\frac{\left(3 v^2-2 v+3\right) v^2 w^2}{(1-v)^4}+\frac{2}{(1-v)^2} + \frac{v (2-4 w)-2}{(1-v)\, (1-v+v\,w)^2}\,,\nn\\
A_{4,\mathrm{SGP},F,C_F}^{qg\to g,1}(v,w) & = & -\frac{6 v^4 w^4}{(1-v)^4}+\frac{v^3 (8 v-1) w^3}{(1-v)^4}-\frac{v^2 \left(3 v^2-2 v+3\right) w^2}{(1-v)^4}-\frac{v (v+1) w}{2 (1-v)^3}\nn\\
&& +\frac{2}{(1-v)^2}+ \frac{v (2-4 w)-2}{(1-v)\, (1-v+v\,w)^2}\,,\nn\\
A_{5,\mathrm{SGP},F,C_F}^{qg\to g,1}(v,w) & = & -\frac{12 v^4 w^4}{(1-v)^4}+\frac{16 v^4 w^3}{(1-v)^4}-\frac{2 \left(3 v^2-2 v+3\right) v^2 w^2}{(1-v)^4}+\frac{2}{(1-v)^2}+\frac{v (2-4 w)-2}{(1-v)\, (1-v+v\,w)^2}\,,\nn\\
A_{6,\mathrm{SGP},F,C_F}^{qg\to g,1}(v,w) & = & \frac{4 v^2 w^2}{(1-v)^2\, (1-v+v\,w)^2}
\,,\nn\\
A_{7,\mathrm{SGP},F,C_F}^{qg\to g,1}(v,w) & = & \frac{6 v^4 w^4}{(1-v)^4}-\frac{v^3 (8 v-1) w^3}{(1-v)^4}+\frac{v^2 \left(3 v^2-2 v+3\right) w^2}{(1-v)^4}-\frac{2}{(1-v)^2}+\frac{v (4 w-2)+2}{(1-v)\, (1-v+v\,w)^2}\,,\nn\\
A_{8,\mathrm{SGP},F,C_F}^{qg\to g,1}(v,w) & = & \frac{3 v^4 w^4}{(1-v)^4}-\frac{2 v^3 (8+v) w^3}{(1-v)^4}+\frac{v^2 \left(11 v-2\right) w^2}{(1-v)^4}+\frac{3v \left(v^2-4 v+1\right) w}{2
   (1-v)^4}\nn\\
&&+\frac{3 v^3-26 v^2+31 v-14}{2 (1-v)^4}+\frac{11-3 v}{(1-v) (1-v+v\, w)}-\frac{4}{(1-v+v\, w)^2}\,.\label{eq:SGP1qg2gCFAs}
\eea
The $N_c$-part reads
\bea
\hat{\sigma}_{\mathrm{SGP},F,N_c}^{qg\to g,1}(v,w,\chi_m) & = & A_{1,\mathrm{SGP},F,N_c}^{qg\to g,1}(v,\chi_m)\,\delta(1-w) \nn\\
&&\hspace{0cm}+A_{2,\mathrm{SGP},F,N_c}^{qg\to g,1}(v)\frac{1}{(1-w)_+}+A_{3,\mathrm{SGP},F,N_c}^{qg\to g,1}(v,w)\,\ln(\chi_m)\nn\\
&&\hspace{0cm}+A_{4,\mathrm{SGP},F,N_c}^{qg\to g,1}(v,w)\,\ln(1-w)+A_{5,\mathrm{SGP},F,N_c}^{qg\to g,1}(v,w)\,\ln(1-v)\nn\\
&&\hspace{0cm}+A_{6,\mathrm{SGP},F,N_c}^{qg\to g,1}(v,w)\,\ln(1-v\,w)+A_{7,\mathrm{SGP},F,N_c}^{qg\to g,1}(v,w)\,\ln(w)\nn\\
&&\hspace{0cm}+A_{8,\mathrm{SGP},F,N_c}^{qg\to g,1}(v,w)\,,\label{eq:SGP1qg2gNc}
\eea
with the eight coefficients
\bea
A_{1,\mathrm{SGP},F,N_c}^{qg\to g,1}(v,\chi_m) & = & \frac{v^3-2 v^2+2v }{2\,(1-v)^4}\ln(\chi_m)+\frac{1-2 v}{2\,(1-v)^4}\ln(1-v)-\frac{(1+v) v^2}{4\, (1-v)^4}\,,\nn\\
A_{2,\mathrm{SGP},F,N_c}^{qg\to g,1}(v) &=& \frac{v\,(3-v)}{4\, (1-v)^3}\,,\nn\\
A_{3,\mathrm{SGP},F,N_c}^{qg\to g,1}(v,w) & = & -\frac{2 v^3 w^3}{(1-v)^4}+\frac{v^2 (v+1) w^2}{(1-v)^4}+\frac{1}{(1-v)^3}-\frac{3-v}{(1-v)^3\, (1-v\, w)}\nn\\
&&+\frac{3-2 v}{(1-v)^3 (1-v\, w)^2}-\frac{1}{(1-v)^2\, (1-v\, w)^3}\,,\nn
\eea
\bea
A_{4,\mathrm{SGP},F,N_c}^{qg\to g,1}(v,w) & = & -\frac{3 v^3 w^3}{2\,(1-v)^4}+\frac{v^2 (v+1) w^2}{(1-v)^4}+\frac{v (1+v) w}{4\, (1-v)^3}-\frac{3-v}{(1-v)^3\, (1-v\, w)}\nn\\
&&+\frac{3-2 v}{(1-v)^3\, (1-v\,
   w)^2}-\frac{1}{(1-v)^2 \,(1-v\, w)^3}+\frac{1}{(1-v)^3}\,,\nn\\
A_{5,\mathrm{SGP},F,N_c}^{qg\to g,1}(v,w) & = & -\frac{v^3 (1-w) w^2}{(1-v)^3\, (1-v \,w)^3}\,,\nn   \\
A_{6,\mathrm{SGP},F,N_c}^{qg\to g,1}(v,w) & = & \frac{2 v^3 (1-w) w^2}{(1-v)^3\, (1-v\, w)^3}\,,\nn   \\
A_{7,\mathrm{SGP},F,N_c}^{qg\to g,1}(v,w) & = & \frac{3 v^3 w^3}{2\,(1-v)^4}-\frac{v^2 (1+v) w^2}{(1-v)^4}+\frac{3-v}{(1-v)^3\, (1-v\, w)}\nn\\
&&-\frac{3-2 v}{(1-v)^3 (1-v\, w)^2}+\frac{1}{(1-v)^2 (1-v\,
   w)^3}-\frac{1}{(1-v)^3}\,,\nn\\
A_{8,\mathrm{SGP},F,N_c}^{qg\to g,1}(v,w) & = & \frac{3 v^3 w^3}{2\,(1-v)^4}-\frac{v^2 (1+2 v) w^2}{2\,(1-v)^4}-\frac{v \left(1+v^2\right) w}{4 (1-v)^4}+\frac{8 v^2-32 v+25}{2\,(1-v)^4\, (1-v\, w)}\nn\\
&&-\frac{v^3+4 v^2-29 v+26}{4
   (1-v)^4}+\frac{2}{(1-v)\, (1-v+v\, w)}\nn\\
 && -\frac{21-13 v}{2\,(1-v)^3\, (1-v\, w)^2} -\frac{1}{2\,(1-v+v\, w)^2}+\frac{3}{(1-v)^2\, (1-v\, w)^3} \,.\label{eq:SGP1qg2gNcAs} 
\eea
The other SGP contribution entering with the axial-vector $qgq$ correlation function has a relatively simple expression:
\bea
\hat{\sigma}_{\mathrm{SGP},\partial_2 G}^{qg\to g,5}(v,w) &=& \frac{v (2 C_F-N_c) \left(v^2 \left(2 w^2-4 w+1\right)-1\right)}{2\, (1-v)^4}\ln(1-w)\nn\\
&& \hspace{0cm}-\frac{w \left(C_F(1-v)+N_c \left(1-v+v^2\right)\right)}{(1-v)^3}\ln(1-v)-\frac{v^3 w^2 (2 C_F-N_c)}{(1-v)^4}\ln(w)\nn\\
&&\hspace{0cm}+\frac{N_c v^2 w}{(1-v)^3}\ln(1-v+v\,w)+C_F\,\frac{v\, w \left(v^3 \left(6 w^2-8 w+3\right)-4 v^2+v-2\right)}{2 (1-v)^4}\nn\\
&&+N_c\,\frac{v\,w\,\left(v^2 (1-2 w)+3 v-2\right)}{2\, (1-v)^4}\,.\label{eq:SGP2qg2g}
\eea

\subsubsection{Soft-Fermion Pole Contribution\label{appsub:SFPqg2g}}

The explicit result for the partonic functions of the soft-fermion pole contribution to the $qg\to g$-channel in Eq.~\eqref{eq:Channel4SFP} can be split up according to its color factors,
\be
\hat{\sigma}_{\mathrm{SFP},F}^{qg\to g,1}(v,w,\chi_m) = C_F\,\hat{\sigma}_{\mathrm{SFP},F,C_F}^{qg\to g,1}(v,w,\chi_m)+N_c\,\hat{\sigma}_{\mathrm{SFP},F,N_c}^{qg\to g,1}(v,w,\chi_m)\,,\label{eq:SFP1qg2g}
\ee
with
\bea
\hat{\sigma}_{\mathrm{SFP},F,C_F}^{qg\to g,1}(v,w,\chi_m) & = &  A_{1,\mathrm{SFP},F,C_F}^{qg\to g,1}(v,w)\,\ln(\chi_m)\nn\\
&&\hspace{0cm}+A_{2,\mathrm{SFP},F,C_F}^{qg\to g,1}(v,w)\,\ln(1-w)+A_{3,\mathrm{SFP},F,C_F}^{qg\to g,1}(v,w)\,\ln(1-v)\nn\\
&&\hspace{0cm}+A_{4,\mathrm{SFP},F,C_F}^{qg\to g,1}(v,w)\,\ln(1-v+v\,w)+A_{5,\mathrm{SFP},F,C_F}^{qg\to g,1}(v,w)\,\ln(1-v\,w)\nn\\
&&\hspace{0cm}+A_{6,\mathrm{SFP},F,C_F}^{qg\to g,1}(v,w)\,\ln(w)+A_{7,\mathrm{SFP},F,C_F}^{qg\to g,1}(v,w)\,,\label{eq:SFP1qg2gCF}
\eea
with the seven coefficients
\bea
A_{1,\mathrm{SFP},F,C_F}^{qg\to g,1}(v,w) & = & \frac{v^2 w^2 \left(v^2 \left(2 w^2-2 w+1\right)-2 v w+1\right)}{(1-v)^4\, (1-v\, w)}\,,\nn\\
A_{2,\mathrm{SFP},F,C_F}^{qg\to g,1}(v,w) & = & \frac{v^2 w \left(v^2 w (2 w-1)+v (2-4 w)+w\right)}{(1-v)^4\, (1-v\, w)}\,,\nn
\eea
\bea
A_{3,\mathrm{SFP},F,C_F}^{qg\to g,1}(v,w) & = & -\frac{w \left(2 v^3 w (2 w-1)+v^2 \left(w^2-4 w+1\right)-v \left(w^2+w-1\right)+w\right)}{(1-v)^3\, (1-v\, w)}\,,\nn\\
A_{4,\mathrm{SFP},F,C_F}^{qg\to g,1}(v,w) & = & \frac{v\, w\, (1+v (1-2 w))}{(1-v)^3}\,,\nn\\
A_{5,\mathrm{SFP},F,C_F}^{qg\to g,1}(v,w) & = & \frac{v\, w\, \left(v^2 w (2 w-1)-v (w+1)+1\right)}{(1-v)^3\, (1-v\, w)}\,,\nn\\
A_{6,\mathrm{SFP},F,C_F}^{qg\to g,1}(v,w) & = & \frac{v^2\, w^2\, \left(v^2 (6 w-1)-4 v (w+1)+3\right)}{(1-v)^4\, (1-v\, w)}\,,\nn\\
A_{7,\mathrm{SFP},F,C_F}^{qg\to g,1}(v,w) & = & -\frac{3 v^4 w^4}{(1-v)^4}+\frac{2 v^3 (2 v+5) w^3}{(1-v)^4}-\frac{v \left(3 v^3+20 v^2-3 v-2\right) w^2}{2 (1-v)^4}\nn\\
&&-\frac{v w}{(1-v)^3}-\frac{1}{(1-v) (1-v+v\,w)}+\frac{1}{(1-v)^2}\,.\label{eq:SFP1qg2gCFAs}
\eea
The $N_c$ part reads,
\bea
\hat{\sigma}_{\mathrm{SFP},F,N_c}^{qg\to g,1}(v,w,\chi_m) & = &  A_{1,\mathrm{SFP},F,N_c}^{qg\to g,1}(v,w)\,\ln(\chi_m)\nn\\
&&\hspace{0cm}+A_{2,\mathrm{SFP},F,N_c}^{qg\to g,1}(v,w)\,\ln(1-w)+A_{3,\mathrm{SFP},F,N_c}^{qg\to g,1}(v,w)\,\ln(1-v)\nn\\
&&\hspace{0cm}+A_{4,\mathrm{SFP},F,N_c}^{qg\to g,1}(v,w)\,\ln(1-v+v\,w)+A_{5,\mathrm{SFP},F,N_c}^{qg\to g,1}(v,w)\,\ln(1-v\,w)\nn\\
&&\hspace{0cm}+A_{6,\mathrm{SFP},F,N_c}^{qg\to g,1}(v,w)\,\ln(w)+A_{7,\mathrm{SFP},F,N_c}^{qg\to g,1}(v,w)\,,\label{eq:SFP1qg2gN_c}
\eea
with the seven coefficients
\bea
A_{1,\mathrm{SFP},F,N_c}^{qg\to g,1}(v,w) & = & \frac{v^2 w^2 \left(v^3 w \left(2 w^2-2 w+1\right)+v^2 \left(-6 w^2+4 w-2\right)+5 v w-2\right)}{2 \,(1-v)^4\, (1-v\, w)^2}\,,\nn\\
A_{2,\mathrm{SFP},F,N_c}^{qg\to g,1}(v,w) & = & \frac{v^2 w \left(v^3 w^2 (2 w-1)-2 v^2 w \left(w^2+2 w-1\right)+v \left(3 w^2+4 w-2\right)-2 w\right)}{2\, (1-v)^4\, (1-v\, w)^2}\,,\nn\\
A_{3,\mathrm{SFP},F,N_c}^{qg\to g,1}(v,w) & = & -\frac{\left(3 v^2-v+1\right) w^2}{(1-v)^3}+\frac{v (1+v) w}{(1-v)^3}-\frac{3-v}{2\, (1-v)^3\, (1-v\, w)}\nn\\
&&+\frac{1}{2\, (1-v)^2\, (1-v \,w)^2}+\frac{1}{(1-v)^3}\,,\nn\\
A_{4,\mathrm{SFP},F,N_c}^{qg\to g,1}(v,w) & = & -\frac{v\, w\, (1+v (1-2 w))}{2 \,(1-v)^3}\,,\nn\\
A_{5,\mathrm{SFP},F,N_c}^{qg\to g,1}(v,w) & = & \frac{v \,w \,\left(v^3 w^2 (4 w-1)+v^2 (2-9 w) w+2 v w+v+1\right)}{2\, (1-v)^3\, (1-v \,w)^2}\,,\nn\\
A_{6,\mathrm{SFP},F,N_c}^{qg\to g,1}(v,w) & = & \frac{v^2 w^2 \left(v^3 w (6 w-1)-2 v^2 \left(w^2+6 w-1\right)+v (5 w+4)-2\right)}{2\, (1-v)^4\, (1-v\, w)^2}\,,\nn\\
A_{7,\mathrm{SFP},F,N_c}^{qg\to g,1}(v,w) & = & -\frac{4 v^3 w^3}{(1-v)^4}+\frac{v \left(7 v^2+7 v-4\right) w^2}{2\, (1-v)^4}+\frac{2 v w}{(1-v)^3}+\frac{4-v}{2\, (1-v)^3}\nn\\
&&-\frac{7-4 v}{2\, (1-v)^3\, (1-v\, w)}+\frac{1}{2\, (1-v)\, (1-v+v\,w)}+\frac{1}{(1-v)^2\, (1-v \,w)^2}\,.\label{eq:SFP1qg2gNcAs}
\eea

The result for the partonic functions for the axial-vector soft-fermion pole contributions can also be split up according to its color factors,
\be
\hat{\sigma}_{\mathrm{SFP},G}^{qg\to g,5}(v,w,\chi_m) = C_F\,\hat{\sigma}_{\mathrm{SFP},G,C_F}^{qg\to g,5}(v,w,\chi_m)+N_c\,\hat{\sigma}_{\mathrm{SFP},G,N_c}^{qg\to g,5}(v,w,\chi_m)\,.\label{eq:SFP2qg2g}
\ee
The $\hat{\sigma}_{\mathrm{SFP},G}^{qg\to g,5}$ are quite similar to the $\hat{\sigma}_{\mathrm{SFP},F}^{qg\to g,1}$. Indeed, we find,
\bea
\hat{\sigma}_{\mathrm{SFP},G,C_F}^{qg\to g,5}(v,w,\chi_m) & = & \hat{\sigma}_{\mathrm{SFP},F,C_F}^{qg\to g,1}(v,w,\chi_m) + \frac{2 \left(1-v-2 v^2\right) w^2}{(1-v)^3}\ln(1-v)\nn\\
&&\hspace{0cm}+\frac{4 v^2 \,w^2}{(1-v)^3}\,\ln(1-v\,w)-\frac{4 v^2 (1-3 v) w^2}{(1-v)^4}\,\ln(w)\nn\\
&& + \frac{v^2 \,w^2\, \left(v^2 \left(6 w^2-8 w+3\right)-16 v (w-1)-1\right)}{(1-v)^4}\,,\label{eq:SFP2qg2gCF}
\eea
and
\bea
\hat{\sigma}_{\mathrm{SFP},G,N_c}^{qg\to g,5}(v,w,\chi_m) & = & \hat{\sigma}_{\mathrm{SFP},F,N_c}^{qg\to g,1}(v,w,\chi_m)+\frac{2 \left(1-v+2 v^2\right) w^2}{(1-v)^3}\ln(1-v)\nn\\
&&\hspace{0cm}-\frac{4 v^2 w^2}{(1-v)^3}\,\ln(1-v\,w)+\frac{2 (1-3 v) v^2 w^2}{(1-v)^4}\,\ln(w)+\frac{v w^2 \left(v^2 (6 w-5)-4 v+3\right)}{(1-v)^4}\,.\label{eq:SFP2qg2gNc}
\eea

\subsection{Channel \texorpdfstring{$gg\to q^\prime$}{gg to q'}\label{appsub:tripleG}}
We conclude our explicit results for hadron production with the two partonic factors of Eq.~\eqref{eq:Channel5}, our formula for the $gg\to q^\prime$-channel. They are given as
\bea
\hat{\sigma}_{xx}^{gg\to q^\prime}(v,w,\chi_m,\chi_\mu) & = & A_{1,xx}^{gg\to q^\prime}(v,\chi_m,\chi_\mu)\delta(1-w)+A_{2,xx}^{gg\to q^\prime}(v)\frac{1}{(1-w)_+}\nn\\
&&+A_{3,xx}^{gg\to q^\prime}(v,w)\ln(\chi_m)+A_{4,xx}^{gg\to q^\prime}(v,w)\ln(1-w)\nn\\
&&+A_{5,xx}^{gg\to q^\prime}(v,w)\ln(\chi_\mu)+A_{6,xx}^{gg\to q^\prime}(v,w)\ln(1-v\,w)\nn\\
&&+A_{7,xx}^{gg\to q^\prime}(v,w)\ln(1-v)+A_{8,xx}^{gg\to q^\prime}(v,w)\ln(w)+A_{9,xx}^{gg\to q^\prime}(v,w)\,\label{eq:tripleGxx}
\eea
with the following nine coefficient functions:
\bea
A_{1,xx}^{gg\to q^\prime}(v,\chi_m,\chi_\mu) & = &\frac{2 (v-1) v+1 }{(1-v)^4}\ln (\chi_m)+\frac{1+v^2}{(1-v)^4}\ln (\chi_\mu)+\frac{v^2-(1-v)^2}{(1-v)^4}\nn\\
&&+\frac{(3 v-2) v+2 }{(1-v)^4}\ln (1-v)\,,\nn\\
A_{2,xx}^{gg\to q^\prime}(v) & = & \frac{2-2v+3v^2}{(1-v)^4}\,,\nn\\
A_{3,xx}^{gg\to q^\prime}(v,w) & = & \frac{2 v \left(-2 v w^2+v-2\right)+4}{(1-v)^4}-\frac{4 (2-v) v^2 w^2+(3-v) (1-3 v w)}{(1-v)^3 (1-v\,w)^3}\,,\nn\\
A_{4,xx}^{gg\to q^\prime}(v,w) & = & \frac{-2 \left(3 v^2+1\right) w^2+3 v^2-4 v+5}{(1-v)^4}-\frac{4 (2-v) v^2 w^2+(3-v) (1-3 v w)}{(1-v)^3 (1-v\,w)^3}\,,\nn\\
A_{5,xx}^{gg\to q^\prime}(v,w) & = & \frac{\left(1+v^2\right) \left(1-2 w^2\right)}{(1-v)^4}\,,\nn\\
A_{6,xx}^{gg\to q^\prime}(v,w) & = & \frac{2 v^2 w^2 ((v-3) v w+v+1)}{(1-v)^3 (1-v\,w)^3} \,,\nn\\
A_{7,xx}^{gg\to q^\prime}(v,w) & = & \frac{v \left(-6 v w^2+v+4\right)-1-2w^2}{(1-v)^4}+\frac{4 (2-v) v^2 w^2+(3-v) (1-3 v w)}{(1-v)^3 (1-v\,w)^3}\,,\nn\\
A_{8,xx}^{gg\to q^\prime}(v,w) & = & \frac{2 (v+1) w^2+v-3}{(1-v)^3}+\frac{4 (2-v) v^2 w^2+(3-v) (1-3 v w)}{(1-v)^3 (1-v\,w)^3}\,,\nn\\
A_{9,xx}^{gg\to q^\prime}(v,w) & = & \frac{8 v^2 w-2 (v+1) (4 v-1) w^2+(19-7 v) v-20}{(1-v)^4}\nn\\
&&+\frac{12 v^2-45 v+34}{(1-v)^4 (1-v\,w)}+\frac{15v-23}{(1-v)^3 (1-v\,w)^2}+\frac{6}{(1-v)^2 (1-v\,w)^3}\,.\label{eq:tripleGxxCoeff}
\eea
\bea
\hat{\sigma}_{x0}^{gg\to q^\prime}(v,w,\chi_m,\chi_\mu) & = & A_{1,x0}^{gg\to q^\prime}(v,\chi_m,\chi_\mu)\delta(1-w)+A_{2,x0}^{gg\to q^\prime}(v)\frac{1}{(1-w)_+}\nn\\
&&+A_{3,x0}^{gg\to q^\prime}(v,w)\ln(\chi_m)+A_{4,x0}^{gg\to q^\prime}(v,w)\ln(1-w)\nn\\
&&+A_{5,x0}^{gg\to q^\prime}(v,w)\ln(\chi_\mu)+A_{6,x0}^{gg\to q^\prime}(v,w)\ln(1-v\,w)\nn\\
&&+A_{7,x0}^{gg\to q^\prime}(v,w)\ln(1-v)+A_{8,x0}^{gg\to q^\prime}(v,w)\ln(w)+A_{9,x0}^{gg\to q^\prime}(v,w)\,\label{eq:tripleGx0}
\eea
with the following nine coefficient functions:
\bea
A_{1,x0}^{gg\to q^\prime}(v,\chi_m,\chi_\mu) & = &-\frac{2 (v-1) v+1 }{(1-v)^4}\ln (\chi_m)-\frac{1+v^2}{(1-v)^4}\ln (\chi_\mu)+\frac{(2-v)^2}{(1-v)^4}+\frac{(2-3 v) v-2 }{(1-v)^4}\ln (1-v)\,,\nn\\
A_{2,x0}^{gg\to q^\prime}(v) & = & \frac{-2+2v-3v^2}{(1-v)^4}\,,\nn\\
A_{3,x0}^{gg\to q^\prime}(v,w) & = & \frac{v^2 \left(4 w^2-2\right)+4 v-4}{(1-v)^4}+\frac{4 (2-v) v^2 w^2+(3-v) (1-3 v w)}{(1-v)^3 (1-v\,w)^3}\,,\nn\\
A_{4}^{gg\to q^\prime}(v,w) & = &\frac{v \left(v \left(6 w^2-3\right)+4\right)+2 w^2-5}{(1-v)^4}+\frac{4 (2-v) v^2 w^2+(3-v) (1-3 v w)}{(1-v)^3 (1-v\,w)^3}\,,\nn\\
A_{5,x0}^{gg\to q^\prime}(v,w) & = &-\frac{\left(1+v^2\right) \left(1-2 w^2\right)}{(1-v)^4} \,,\nn\\
A_{6}^{gg\to q^\prime}(v,w) & = & -\frac{2 v^2 w^2 (-4 v^2 w^2 (3-v w)+v (v+9) w+v-3)}{(1-v)^3 (1-v\,w)^3}\,,\nn\\
A_{7,x0}^{gg\to q^\prime}(v,w) & = & \frac{\left((8 v-2) v^2+2\right) w^2-v^2-4 v+1}{(1-v)^4}-\frac{4 (2-v) v^2 w^2+(3-v) (1-3 v w)}{(1-v)^3 (1-v\,w)^3}\,,\nn\\
A_{8,x0}^{gg\to q^\prime}(v,w) & = & -\frac{2 \left(4 v^2+v+1\right) w^2+v-3}{(1-v)^3}-\frac{4 (2-v) v^2 w^2+(3-v) (1-3 v w)}{(1-v)^3 (1-v\,w)^3}\,,\nn
\eea
\bea
A_{9,x0}^{gg\to q^\prime}(v,w) & = & \frac{2 \left(7 v^2+v-4\right) w^2+8 v^2+8 (1-2 v) v w-23 v+25}{(1-v)^4}\nn\\
&&-\frac{12 v^2-49 v+38}{(1-v)^4 (1-v\,w)}-\frac{15 v-23}{(1-v)^3 (1-v\,w)^2}-\frac{6}{(1-v)^2 (1-v\,w)^3}\,.\label{eq:tripleGx0Coeff}
\eea

\subsection{Specific Contributions for Jet Production \label{Appsub:jet}} 

In this last subsection, we present the jet-specific partonic factors that appear in \eqref{eq:jetsub1}. The variable $v$ in the following is to be understood as $v=v_1=1+\tfrac{t}{s}$. For the SGP term, we divide the partonic factors into two parts,
\be
\hat{\sigma}_{\mathrm{SGP,\,jet},F}^{qg\to \mathrm{jet}(q+g)}(w,R,\mu) = C_F\,\hat{\sigma}_{\mathrm{SGP,\,jet},F}^{qg\to \mathrm{jet}(q)}(w,R,\mu)+C_F\,\hat{\sigma}_{\mathrm{SGP,\,jet},F}^{qg\to \mathrm{jet}(g)}(w,R,\mu)\,,\label{eq:jetspec1}
\ee
where the part corresponding to former quark fragmentation is given as
\bea
\hat{\sigma}_{\mathrm{SGP,\,jet},F}^{qg\to \mathrm{jet}(q)}(w,R,\mu) & = & A_{0,\mathrm{SGP,\,jet},F}^{qg\to \mathrm{jet}(q)}(R,\mu)\,\delta(1-w)+ A_{1,\mathrm{SGP,\,jet},F}^{qg\to \mathrm{jet}(q)}\,\left(\frac{\ln(1-w)}{1-w}\right)_+\nn\\
&&+A_{2,\mathrm{SGP,\,jet},F}^{qg\to \mathrm{jet}(q)}(R,\mu)\,\frac{1}{(1-w)_+}\nn\\
&&+A_{3,\mathrm{SGP,\,jet},F}^{qg\to \mathrm{jet}(q)}\,\ln\left(R^2v^2(1-w)^2\,\tfrac{t\,u}{s\,\mu^2}\right)+A_{4,\mathrm{SGP,\,jet},F}^{qg\to \mathrm{jet}(q)},\label{eq:ASGPqjet}
\eea
with
\bea
A_{0,\mathrm{SGP,\,jet},F}^{qg\to \mathrm{jet}(q)}(R,\mu) & = &\frac{4 v^3+9 v^2+12 v+1-4(1+v^2)\, \ln (v) }{2 (1-v)^3}\ln \left(R^2\,\tfrac{t\, u}{s\,\mu^2}\right)\nn\\
&&-2\frac{1+v^2}{(1-v)^3}\ln ^2(v)+4\frac{(1+v)^3}{(1-v)^3} \ln(v)+\frac{1+v^2}{(1-v)^3}\left(\tfrac{13}{2}-\tfrac{2}{3} \pi^2\right)\,,\nn\\
A_{1,\mathrm{SGP,\,jet},F}^{qg\to \mathrm{jet}(q)} & = & -4\frac{1+v^2}{(1-v)^3}\,,\nn
\eea
\bea
A_{2,\mathrm{SGP,\,jet},F}^{qg\to \mathrm{jet}(q)}(R,\mu) & = & 4\frac{(1+v)^3-(1+v^2)\ln (v)}{(1-v)^3}-2\frac{1+v^2 }{(1-v)^3}\,\ln \left(R^2\tfrac{t\, u}{s\,\mu^2}\right)\,,\nn\\
A_{3,\mathrm{SGP,\,jet},F}^{qg\to \mathrm{jet}(q)} & = & -\frac{6 v^4 w^4}{(1-v)^3}+\frac{4 v^3 (2 v-3) w^3}{(1-v)^3}-\frac{v^2 \left(3 v^2-8 v+9\right) w^2}{(1-v)^3}+2\frac{1+v^2}{(1-v)^3}\,,\nn\\
A_{4,\mathrm{SGP,\,jet},F}^{qg\to \mathrm{jet}(q)} & = & -\frac{10 v^4 w^4}{(1-v)^3}+4\frac{v^3 (3 v-4) w^3}{(1-v)^3}-\frac{v^2 \left(5 v^2-8v+19\right) w^2}{(1-v)^3}\nn\\
&& -\frac{4 v \left(v^2+2v+3\right) w}{(1-v)^3}-4\frac{(1+v)^3}{(1-v)^3}\,.\label{eq:AsjetFq}
\eea 
We note that the explicit form of the coefficient $A_{0,\mathrm{SGP,\,jet},F}^{qg\to \mathrm{jet}(q)}$ depends on the jet algorithm adopted \cite{Mukherjee:2012uz}, and the
one given here applies to the anti-$k_T$ algorithm.\\

The specific form of the contribution from  former gluon fragmentation reads,
\bea
\hat{\sigma}_{\mathrm{SGP,\,jet},F}^{qg\to \mathrm{jet}(g)}(w,R,\mu) & = & A_{0,\mathrm{SGP,\,jet},F}^{qg\to \mathrm{jet}(g)}(R,\mu)\,\delta(1-w)+ A_{1,\mathrm{SGP,\,jet},F}^{qg\to \mathrm{jet}(g)}\,\frac{1}{(1-w)_+}\nn\\
&&+A_{2,\mathrm{SGP,\,jet},F}^{qg\to \mathrm{jet}(g)}\,
\ln\left(R^2\,v^2(1-w)^2\,\tfrac{t\,u}{s\,\mu^2}\right)
+A_{3,\mathrm{SGP,\,jet},F}^{qg\to \mathrm{jet}(g)},\label{eq:ASGPgjet}
\eea
with
\bea
A_{0,\mathrm{SGP,\,jet},F}^{qg\to \mathrm{jet}(g)}(R,\mu) & = & -\frac{v \left(1+v^2\right)}{(1-v)^3}\left(\ln \left(R^2 \tfrac{t\, u}{s\,\mu^2}\right)+2 \ln (v)+1\right)\,,\nn\\
A_{1,\mathrm{SGP,\,jet},F}^{qg\to \mathrm{jet}(g)} & = & -\frac{2 v \left(1+v^2\right)}{(1-v)^3}\,,\nn\\
A_{2,\mathrm{SGP,\,jet},F}^{qg\to \mathrm{jet}(g)} & = & \frac{6 v^4 w^4}{(1-v)^3}-\frac{8 v^4 w^3}{(1-v)^3}+\frac{\left(3 v^2-2 v+3\right) v^2 w^2}{(1-v)^3}+\frac{4}{1-v+v\, w}-\frac{2 (1-v)}{(1-v+v\, w)^2}-\frac{2}{1-v}\,,\nn
\eea
\bea
A_{3,\mathrm{SGP,\,jet},F}^{qg\to \mathrm{jet}(g)} & = & \frac{10 v^4 w^4}{(1-v)^3}-\frac{4 v^3 (3 v-2) w^3}{(1-v)^3}+\frac{v^2 \left(5 v^2-10 v+9\right) w^2}{(1-v)^3}-\frac{2 v \left(v^2-4 v+1\right) w}{(1-v)^3}\nn\\
 && -\frac{2
   \left(v^3-6 v^2+5 v-2\right)}{(1-v)^3} -\frac{4(1-v)}{1-v+v\, w}\,.\label{eq:AsjetFg}
\eea 
The coefficient for the derivative term is simpler, contains only distributions, and reads,
\bea
\hspace{-0.7cm}\hat{\sigma}_{\mathrm{SGP,\,jet},F^\prime}^{qg\to \mathrm{jet}(q)}(w,R,\mu)&=& C_F\left(A_{0,\mathrm{SGP,\,jet},F^\prime}^{qg\to \mathrm{jet}(q)}(R,\mu)\delta(1-w)+A_{1,\mathrm{SGP,\,jet},F^\prime}^{qg\to \mathrm{jet}(q)}\left(\frac{\ln(1-w)}{1-w}\right)_+\right.\nn\\
&&\hspace{2cm}\left.+A_{2,\mathrm{SGP,\,jet},F^\prime}^{qg\to \mathrm{jet}(q)}\frac{1}{(1-w)_+}\right)\,,\label{eq:AdSGPqjet}
\eea
with
\bea
A_{0,\mathrm{SGP,\,jet},F^\prime}^{qg\to \mathrm{jet}(q)}(R,\mu) & = & -\frac{1+v^2}{(1-v)^3}\left[2\ln ^2(v)+(\tfrac{3}{2}+2\,\ln(v))\ln \left(R^2\, \tfrac{t\, u}{s\,\mu^2}\right)-\tfrac{13}{2}+\tfrac{2}{3}\pi^2\right]\,,\nn\\
A_{1,\mathrm{SGP,\,jet},F^\prime}^{qg\to \mathrm{jet}(q)} & = & -\frac{4 \left(1+v^2\right)}{(1-v)^3}\,,\nn\\
A_{2,\mathrm{SGP,\,jet},F^\prime}^{qg\to \mathrm{jet}(q)} & = &-\frac{2 \left(1+v^2\right)}{(1-v)^3} \ln \left(R^2\,\tfrac{t\, u}{s\,\mu^2}\right)-\frac{4 \left(1+v^2\right)}{(1-v)^3} \ln (v)\,.\label{eq:AsjetdFq}
\eea 
Again, $A_{0,\mathrm{SGP,\,jet},F^\prime}^{qg\to \mathrm{jet}(q)}$ is specific to the anti-$k_T$ algorithm.

\section{Model Input for the Quark-Gluon-Quark Correlation Functions \texorpdfstring{$F$}{FFT} and \texorpdfstring{$G$}{GFT}\label{appsec:Model}}

In this appendix, we provide details on the model input for the quark-gluon-quark correlation functions $F(x,x^\prime)$ and $G(x,x^\prime)$ that we use for our numerical studies discussed in Section \ref{sec:Numerics}.
We start with a discussion of the soft-gluon pole matrix element $F^q(x,x)$. It can be related to the first TMD moment of the Sivers function using \eqref{eq:SiversSGP}, which in turn has been extracted in the literature from global data fits, in particular from SIDIS. For more information on the status of the fits, we refer the reader to the recent comprehensive review of TMDs \cite{Boussarie:2023izj}. For our work, we rely on one of the earliest extractions of the first moment of the Sivers function, given in~\cite{Anselmino:2008sga}, the reason being its simple numerical implementation. We stress once more that the purpose of our numerical study is to explore the impact of NLO effects on the single-spin observable $A_{RL}$ in \eqref{eq:cmrightleft} rather than to provide fully phenomenologically relevant predictions. For this purpose, the early parameterization of \cite{Anselmino:2008sga} suffices.

In Ref.~\cite{Anselmino:2008sga}, the first TMD moment of the Sivers function is parameterized at a scale $\mu_0=1.55\,\mathrm{GeV}$ as follows:
\be
\pi \,F^{q}(x,x,\mu_0) =f_{1T}^{\perp (1),q}(x,\mu_0)= - \tfrac{1}{2}N^q(x)\,f_1^q(x,\mu_0)\,\sqrt{2\mathrm{e}}\,\frac{M_1^3\,\langle k_T^2\rangle}{M\,(M_1^2+\langle k_T^2 \rangle)^2}\,,\label{eq:Anselminofit}
\ee
with $M$ the nucleon mass, the flavor-independent mass parameters $M_1=0.583\,\mathrm{GeV}$ and $\langle k_T^2 \rangle = 0.25\,\mathrm{GeV}^2$, $\mathrm{e}=2.7182\ldots$ the Euler constant, and $f_1^q(x,\mu_0)$ the MSTW2008 quark PDF for flavor $q$ \cite{Martin:2009iq} evaluated at the scale $\mu_0$. The flavor-dependent factor $N^q(x)$ has the form
\[N^q(x)=N^q\,x^{\alpha_q}(1-x)^{\beta_q}\,\frac{(\alpha_q+\beta_q)^{\alpha_q+\beta_q}}{\alpha_q^{\alpha_q}\,\beta_q^{\beta_q}}\,,\]
with flavor-dependent parameters $N^q$, $\alpha_q$, $\beta_q$ whose values can be found in Ref.~\cite{Anselmino:2008sga}. The flavors included in \eqref{eq:Anselminofit} are $q=u,d,s$. The SGP matrix element for negative $x$ can be related via charge conjugation (see the discussion below \eqref{eq:DefPhiF}) to the first TMD moment of the antiquark Sivers function ($\bar{q}=\bar{u},\bar{d},\bar{s}$),
\be
\pi \,F^{q}(-x,-x,\mu_0)=f_{1T}^{\perp (1),\bar{q}}(x,\mu_0)\,.\label{eq:antiquark}
\ee
As can be seen from Eq. \eqref{eq:LO}, the SGP input \eqref{eq:Anselminofit} is already sufficient to produce predictions for the right-left asymmetry $A_{RL}$ at LO. 
However, at NLO accuracy we need input for $F$ and $G$ on their full support. Both functions are essentially unknown on the ``off-diagonal'' support $x\neq x^\prime$ and, to the best of our knowledge, have never been extracted from data. In order to study the NLO effects, we therefore have to resort to models for these correlation functions.

As a first step, we rewrite a specific point $(x,x^\prime)$ within the support of the functions $F^q$ and $G^q$ in terms of ``polar coordinates'', i.e. $x(r,\varphi)=r\,\cos(\varphi+\tfrac{\pi}{4})$, $x^\prime(r,\varphi)=r\,\sin(\varphi+\tfrac{\pi}{4})$, with
\bea
r & = & \sqrt{x^2+(x^\prime)^2}\,,\nn\\
\varphi & = & \left\{\begin{aligned}
-\tfrac{\pi}{4}&+\arctan(x^\prime/x),\quad x\ge0\,,x^\prime\ge x\,,\\
\tfrac{3\pi}{4}&+\arctan(x^\prime/x),\quad x<0\,,\\
\tfrac{7\pi}{4}&+\arctan(x^\prime/x),\quad x\ge0 \,,x^\prime < x\,.\end{aligned}\right.\label{eq:PolarCoord}
\eea
Note that we count the polar angle $\varphi$ from the ``diagonal'' axis of support ($x^\prime=x$) instead of from the $x$-axis. We then consider the $qgq$ correlation functions as functions of $r$ and $\varphi$, with $r\in \,]0,\sqrt{2}]$, $\varphi\in [0,2\pi]$. The next observation is that both $F(r,\varphi)$ and $G(r,\varphi)$ are $2\pi$-periodic in $\varphi$, that is, $F(r,\varphi)=F(r,\varphi+2\pi)$ and $G(r,\varphi)=G(r,\varphi+2\pi)$. This feature allows us to express both functions as Fourier series,
\bea
F^q(r,\varphi) & = & \sum_{n=0}^\infty\,\left[A^q_n(r)\,\cos(n\varphi)+D^q_n(r)\,\sin(n\varphi)\right]\,,\nn\\
G^q(r,\varphi) & = & \sum_{n=0}^\infty\,\left[C^q_n(r)\,\cos(n\varphi)+B^q_n(r)\,\sin(n\varphi)\right]\,.\label{eq:FourierSeries1}
\eea
The important symmetry constraints \eqref{eq:Symmetry} of the correlation functions under $x\leftrightarrow x^\prime$ can be conveniently implemented into the Fourier Series \eqref{eq:FourierSeries1}. In particular, it is easy to see that the symmetry of $F$, i.e. $F(x,x^\prime)=+F(x^\prime,x)\Leftrightarrow F(r,\varphi)=+F(r,2\pi-\varphi)$, enforces that all Fourier coefficients $D_n(r)$ vanish. Similarly, antisymmetry of $G$, that is, $G(x,x^\prime)=-G(x^\prime,x)\Leftrightarrow G(r,\varphi)=-G(r,2\pi-\varphi)$, means that all Fourier coefficients $C_n(r)$ also vanish.

Furthermore, we can easily implement the constraint \eqref{eq:SiversSGP} as follows,
\bea
F^q(r,\varphi=0) & = & \tfrac{1}{\pi}\,f_{1T}^{\perp (1),q}(\tfrac{r}{\sqrt{2}})= A^q_0(r)+A^q_1(r)+A^q_2(r)+A^q_3(r)+...\,,\nn\\
F^q(r,\varphi=\pi) & = & \tfrac{1}{\pi}\,f_{1T}^{\perp (1),\bar{q}}(\tfrac{r}{\sqrt{2}})= A^q_0(r)-A^q_1(r)+A^q_2(r)-A^q_3(r)+...\,.\nn
\eea
We can solve these constraints for the Fourier coefficients for $A_0$ and $A_1$, and obtain,
\bea
A_0^q(r) & = & \tfrac{1}{2\pi}\left(f_{1T}^{\perp (1),q}(\tfrac{r}{\sqrt{2}})+f_{1T}^{\perp (1),\bar{q}}(\tfrac{r}{\sqrt{2}})\right)-A_2^q(r)-A_4^q(r)-A_6^q(r)-\ldots\,,\nn\\
A_1^q(r) & = & \tfrac{1}{2\pi}\left(f_{1T}^{\perp (1),q}(\tfrac{r}{\sqrt{2}})-f_{1T}^{\perp (1),\bar{q}}(\tfrac{r}{\sqrt{2}})\right)-A_3^q(r)-A_5^q(r)-A_7^q(r)-\ldots\,.\label{eq:FourierCoeffConstraint}
\eea
Implementing \eqref{eq:FourierCoeffConstraint} into \eqref{eq:FourierSeries1} yields, ($q\pm\bar{q}$ denotes the sum/difference of quark and antiquark distributions as in \eqref{eq:FourierCoeffConstraint})
\bea
F^q(r,\varphi) & = & \tfrac{1}{2\pi}f_{1T}^{\perp (1),q+\bar{q}}(\tfrac{r}{\sqrt{2}})+\tfrac{1}{2\pi}f_{1T}^{\perp (1),q-\bar{q}}(\tfrac{r}{\sqrt{2}})\,\cos(\varphi)\nn\\
&&\hspace{-2cm} +\sum_{n=1}^{\infty}\left[A_{2n}^q(r)\,(\cos(2n\varphi)-1)\right]+\sum_{n=1}^{\infty}\left[A_{2n+1}^q(r)\,(\cos((2n+1)\varphi)-\cos(\varphi))\right]\,,\nn\\
G^q(r,\varphi) & = & \sum_{n=1}^\infty\,\left[B^q_n(r)\,\sin(n\varphi)\right]\,.\label{eq:FourierSeries2}
\eea
Up to this point, the two Fourier expansions in \eqref{eq:FourierSeries2} are exact and model-independent. In particular, the Fourier coefficients $A_n(r)$, $B_n(r)$ depend on $r=\sqrt{x^2+(x^\prime)^2}$. In other words, for every Fourier component $n$ there are two (unknown) functions $A_n(r)$, $B_n(r)$ that should ideally be fitted to experimental data. This is of course an impossible task to do for all Fourier components. However, the series \eqref{eq:FourierSeries2} are quite useful for building a parameterization that can be used as input for the numerical study of the observable $A_{RL}$ in \eqref{eq:cmrightleft} at NLO. For this, we need to apply some simplifying assumptions.

First, we note that often Fourier series converges reasonably fast on an interval $[0,2\pi]$. Whether it does, depends of course on the series, but let us assume that we achieve a reasonable approximation to the ``true'' correlation functions already with the first six Fourier components for each flavor. To be specific, we assume that all Fourier coefficients for $n\ge8$ in case of $F$ and for $n\ge7$ in case of $G$ vanish, i.e., $A^q_{n\ge 8}(r)=0$, $B^q_{n\ge 7}(r)=0$. If this assumption turns out to be wrong for some reason, for instance, because the explanation of experimental data may require a higher precision, one is free to add more Fourier components and truncate the series \eqref{eq:FourierSeries2} at higher $n$.

Secondly, we may make further assumptions about the functional form of the Fourier coefficients $A^q_n(r)$, $B^q_n(r)$. In a first step, we may introduce modified coefficients $a_n^q(r)$, $b_n^q(r)$ according to
\bea
A^q_{n=2,4,6,...}(r) & \equiv & \tfrac{1}{2\pi}f_{1T}^{\perp (1),q+\bar{q}}(\tfrac{r}{\sqrt{2}}) \,a^q_{n=2,4,6,...}(r)\,,\nn\\
A^q_{n=1,3,5,...}(r) & \equiv & \tfrac{1}{2\pi}f_{1T}^{\perp (1),q-\bar{q}}(\tfrac{r}{\sqrt{2}}) \,a^q_{n=1,3,5,...}(r)\,,\nn\\
B^q_{n}(r) & \equiv & -\tfrac{1}{\pi}f_{1T}^{\perp (1),q+\bar{q}}(\tfrac{r}{\sqrt{2}}) \,b^q_{n}(r)\,.\label{eq:FouriermodCoeff}
\eea
The idea of this modification is the assumption that the ``size'' or ``scale'' of the quark-gluon-quark correlation functions $F$ and $G$ is roughly set by the SGP diagonal determined by the extraction \eqref{eq:Anselminofit}. Under this assumption, the modified coefficients $a_n(r)$, $b_n(r)$ may vary on the order of magnitude of 1, but not, say, 1000.

Third, we may even go one step further and approximate the modified coefficients in \eqref{eq:FouriermodCoeff} to be effectively constants. We replace the functions $a_n^q(r)$, $b_n^q(r)$ by their mean values $a_n^q(r)\to \langle a_n^q(r)\rangle=a_n^q$, $b_n^q(r)\to \langle b_n^q(r)\rangle=b_n^q$. 

Applying these assumptions, we end up with the following model ansÃ¤tze for the quark-gluon-quark correlation functions:
\bea
F^q(r,\varphi)\Big|_{\mathrm{model}} & = & \tfrac{1}{2\pi}f_{1T}^{\perp (1),q+\bar{q}}(\tfrac{r}{\sqrt{2}})\,\left[1+\sum_{n=1}^{3}\left[a_{2n}^q\,(\cos(2n\varphi)-1)\right]\right]\nn\\
&&\hspace{-2cm}+\tfrac{1}{2\pi}f_{1T}^{\perp (1),q-\bar{q}}(\tfrac{r}{\sqrt{2}})\,\left[\cos(\varphi)+\sum_{n=1}^{3}\left[a_{2n+1}^q\,(\cos((2n+1)\varphi)-\cos(\varphi))\right]\right]\,,\nn\\
G^q(r,\varphi)\Big|_{\mathrm{model}} & = & -\tfrac{1}{\pi}f_{1T}^{\perp (1),q+\bar{q}}(\tfrac{r}{\sqrt{2}})\,\sum_{n=1}^6\,\left[b^q_n\,\sin(n\varphi)\right]\,.\label{eq:FourierSeriesModel}
\eea
Effectively, the models \eqref{eq:FourierSeriesModel} allow us to describe each of the functions $F$, $G$ by six parameters, separately for each flavor. 
We may collect these parameters as entries in a vector in the following way (note the specific ordering of the even and odd Fourier coefficients for $F$):
\bea
\bm{a}^q & = & \left(a_2^q,a_4^q,a_6^q;a_3^q,a_5^q,a_7^q\right)\,,\nn\\
\bm{b}^q & = & \left(b_1^q,b_2^q,b_3^q,b_4^q,b_5^q,b_6^q\right)\,.\label{eq:FourierModelCoeff}
\eea
Note that no matter the values the vectors $\bm{a}^q$, $\bm{b}^q$ take, the soft-gluon pole is always provided by the extraction \eqref{eq:Anselminofit}. It is the only constraint available from data for the functions $F$ and $G$. 

In order to smoothen the transition of the functions $F$ and $G$ at the boundaries $|x|=1$, $|x^\prime|=1$, $|x-x^\prime|=1$  of their support it is helpful to introduce an envelope function, for example of the following form:
\bea
e(x,x^\prime) &=& \left(\frac{2}{1+\mathrm{e}^{50\,(x^2-1)^3}}-1\right)\,\left(\frac{2}{1+\mathrm{e}^{50\,((x^\prime)^2-1)^3}}-1\right)\times\nn\\
&& \left(\frac{2}{1+\mathrm{e}^{50\,((x-x^\prime)^2-1)^3}}-1\right)\theta(1-|x|)\theta(1-|x^\prime|)\theta(1-|x-x^\prime|)\,.\label{eq:Envelope}
\eea
The function $e(x,x^\prime)$ in \eqref{eq:Envelope} is approximately unity, except in the vicinity of the boundary $|x|=1$, $|x^\prime|=1$, $|x-x^\prime|=1$. Multiplying the function $e(x,x^\prime)$ in \eqref{eq:Envelope} with the model expressions \eqref{eq:FourierModelCoeff} does not significantly alter the model but ensures smoothness of the functions $F$ and $G$ even at the boundary. In particular, it ensures $F(x,1)=F(1,x^\prime)=G(x,1)=G(1,x^\prime)=0$ and $(\partial_2F)(x,1)=(\partial_1F)(1,x^\prime)=(\partial_2G)(x,1)=(\partial_1G)(1,x^\prime)=0$.
Therefore, in the following, we perform the replacement
\[F^q(r,\varphi)\Big|_{\mathrm{model}}\to F^q(r,\varphi)\Big|_{\mathrm{model}}e(x,x^\prime)\quad ;\quad G^q(r,\varphi)\Big|_{\mathrm{model}}\to G^q(r,\varphi)\Big|_{\mathrm{model}}e(x,x^\prime).\]

\paragraph{Constraints from Lattice QCD}
There exists another source of (somewhat indirect) information on $F$ from lattice QCD that one may apply here as well. It turns out that one can express the second moment of the genuine twist-3 part of the DIS structure function $\bar{g}_2$, the so-called $d_2$ moment, in terms of the fully integrated quark-gluon-quark function $F$. This feature has been discussed, for example, in~\cite{Shuryak:1981pi,Jaffe:1989xx}. Interestingly, one can interpret the moment $d_2$ as a probe of the color Lorentz force, mediated by the strong force in the nucleon \cite{Burkardt:2008ps,Aslan:2019jis}.
The connection between $d_2$ and $F$ is as follows,
\be
d_2^q = -\int_{-1}^1\dd x\int_{x-1}^1\dd x^\prime\,F^q(x,x^\prime)\,.\label{eq:d2Def}
\ee
In Refs.~\cite{Gockeler:2005vw,Burger:2021knd,Bickerton:2020hjo,Crawford:2024wzx} the $d_2$ moments for up and down quarks were computed on the lattice. 
The values found in~\cite{Bickerton:2020hjo} were reported as
\bea
d_2^u = -0.00365(25)\; ; & d_2^d = 0\,.\label{eq:d2Lattice}
\eea
Despite the limitations related to unphysical pion masses or renormalization schemes, these values give us another hint at the size of the function $F$. 
Due to the linearity of the model \eqref{eq:FourierSeriesModel} in the Fourier coefficients $a^q$, it is easy to implement the lattice constraint \eqref{eq:d2Lattice}. We simply insert Eq.~\eqref{eq:FourierSeriesModel} into \eqref{eq:d2Def} and obtain
\be
A_0^q+A_2^q\,a_2^q+A_4^q\,a_4^q+A_6^q\,a_6^q+A_3^q\,a_3^q+A_5^q\,a_5^q+A_7^q\,a_7^q=d_2^q\,,\label{eq:d2Constr}
\ee
where
\bea
A_0^q & \equiv & -\int_{-1}^1\dd x\int_{-1}^1\dd x^\prime\,F^q(r,\varphi)\Big|_{\mathrm{model},\,\bm{a}^q=\bm{0}}\,,\nn\\
A_i^q & \equiv &-A_0^q-\int_{-1}^1\dd x\int_{-1}^1\dd x^\prime\,F^q(r,\varphi)\Big|_{\mathrm{model},\,a_i^q=1,\,a^q_{j\neq i}=0}\,.\label{eq:d2ConstAs}
\eea
By solving the constraints \eqref{eq:d2Constr} for the Fourier coefficient $a_2^q$ (for example) we can express it as a function of the remaining coefficients and thus remove one degree of freedom,
\be
a_2^q=(d_2^q-(A_0^q+A_4^q\,a_4^q+A_6^q\,a_6^q+A_3^q\,a_3^q+A_5^q\,a_5^q+A_7^q\,a_7^q))/A_2^q\,.\label{eq:d2a2}
\ee
We emphasize that no constraint similar to \eqref{eq:d2Constr} is known for the correlation function $G$.

In the following we will consider three scenarios for the correlation functions based on  three specific choices of Fourier coefficients \eqref{eq:FourierModelCoeff}. Each of the following scenarios is consistent with the constraint \eqref{eq:d2Def} for the $d_2^u,d_2^d$. 
We note that all our scenarios are formulated
in terms of the first TMD moments of the Sivers functions $f_{1T}^{\perp (1),q+\bar{q}}$ and hence, via Eq.~\eqref{eq:Anselminofit}, 
in terms of the unpolarized twist-2 PDFs $f_1^q$. The corresponding relations can strictly hold only at one scale $\mu_0$, 
since the evolutions of the $f_{1T}^{\perp (1),q+\bar{q}}$ and $f_1^q$ differ. Nevertheless, as discussed in Sec.~\ref{sec:Numerics},
in our numerical studies we assume $f_{1T}^{\perp (1),q+\bar{q}}$ and $f_1^q$ to be related in the same way at all scales,
simply as a rough means of mimicking the scale evolution of the twist-3 quark correlation functions.

\begin{figure}[tbp]
\centering
\includegraphics[width=0.49\textwidth]{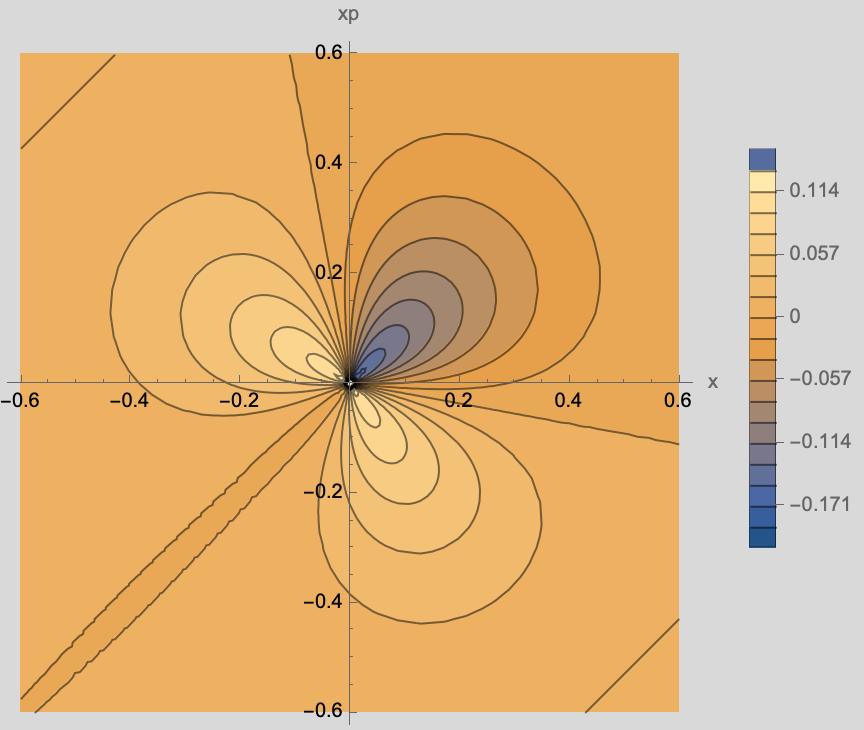}
\includegraphics[width=.49\textwidth]{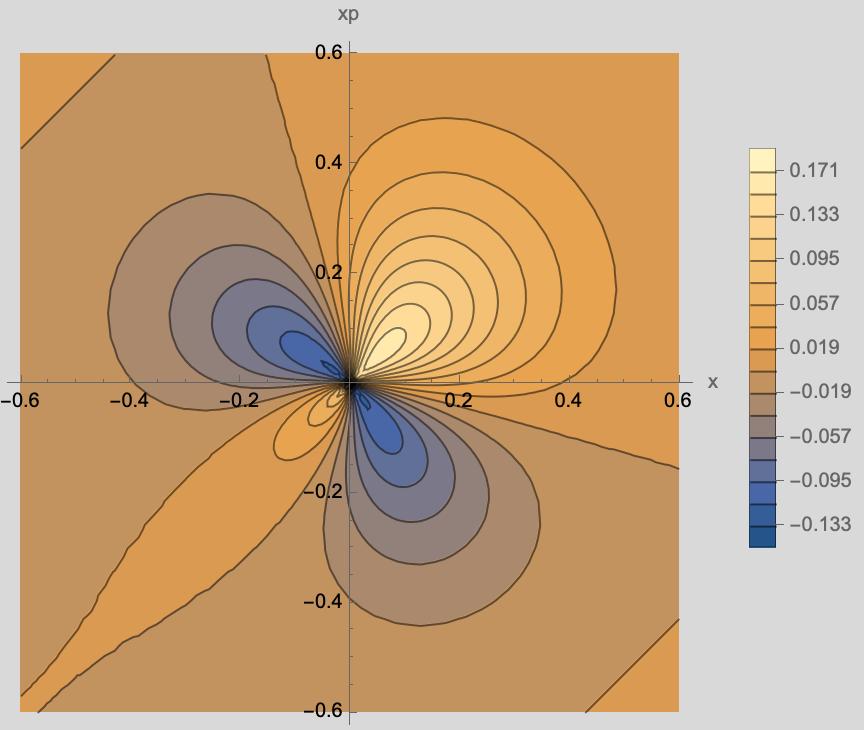}
\caption{Correlation functions $F^u(x,x^\prime)$ \textbf{(left)} and $F^d(x,x^\prime)$ \textbf{(right)} for Scenario 0.\label{fig:Scen0F}}
\end{figure}

\begin{figure}[tbp]
\centering
\includegraphics[width=0.49\textwidth]{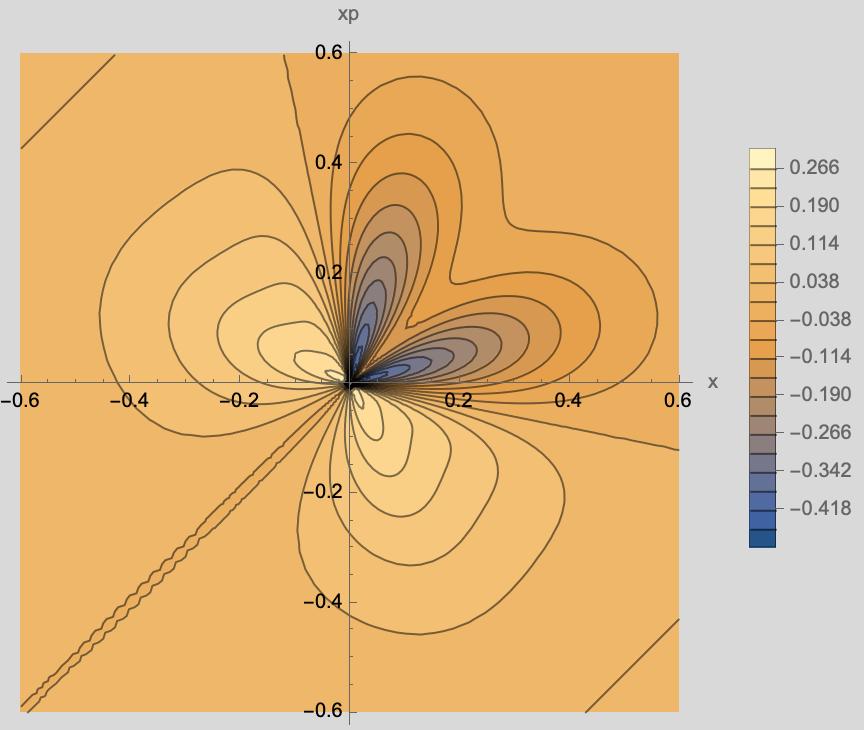}
\includegraphics[width=.49\textwidth]{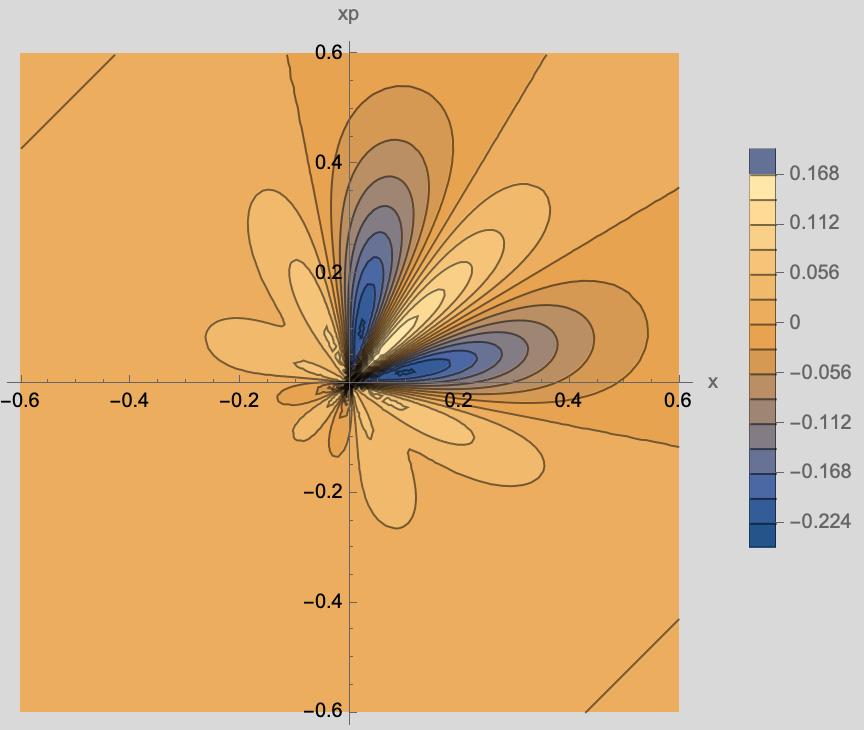}
\caption{Same as Fig.~\ref{fig:Scen0F}, but for Scenario 1.\label{fig:Scen1F}}
\end{figure}

\subsection{Scenario 0\label{appsub:Scen0}}

For the first scenario we choose the Fourier coefficients \eqref{eq:FourierModelCoeff} as follows:
\bea
\bm{a}^u & = & \left(1.1578,0,0;0,0,0\right)\,,\nn\\
\bm{a}^d & = & \left(1.0173,0,0;0,0,0\right)\,,\nn\\
\bm{a}^s & = & \left(0,0,0;0,0,0\right)\,,\nn\\
\bm{b}^u & = & \left(0,0,0,0,0,0\right)\,,\nn\\
\bm{b}^d & = & \left(0,0,0,0,0,0\right)\,,\nn\\
\bm{b}^s & = & \left(0,0,0,0,0,0\right)\,.\label{eq:Scenario0}
\eea
This choice sets the correlation function $G$ basically to zero, and provides a correlation function $F$ that is as ``levelled'' as possible by utilizing only the first three Fourier components $\cos(0\varphi)$, $\cos(1\varphi)$, $\cos(2\varphi)$ in the model ansatz \eqref{eq:FourierSeriesModel}. Note that an additional choice $a_2^q=0$ would violate the $d_2$-constraint \eqref{eq:d2Constr}. We show the resulting $F^q$ as contour plots in Fig.~\ref{fig:Scen0F}. We expect that NLO contributions to the asymmetry $A_{RL}$ within this scenario are minimal, since $G=0$ and the plots in Fig.~\ref{fig:Scen0F} show little ``structure'' of $F^q$. 

\subsection{Scenario 1\label{appsub:Scen1}}

\begin{figure}[tbp]
\centering
\includegraphics[width=0.49\textwidth]{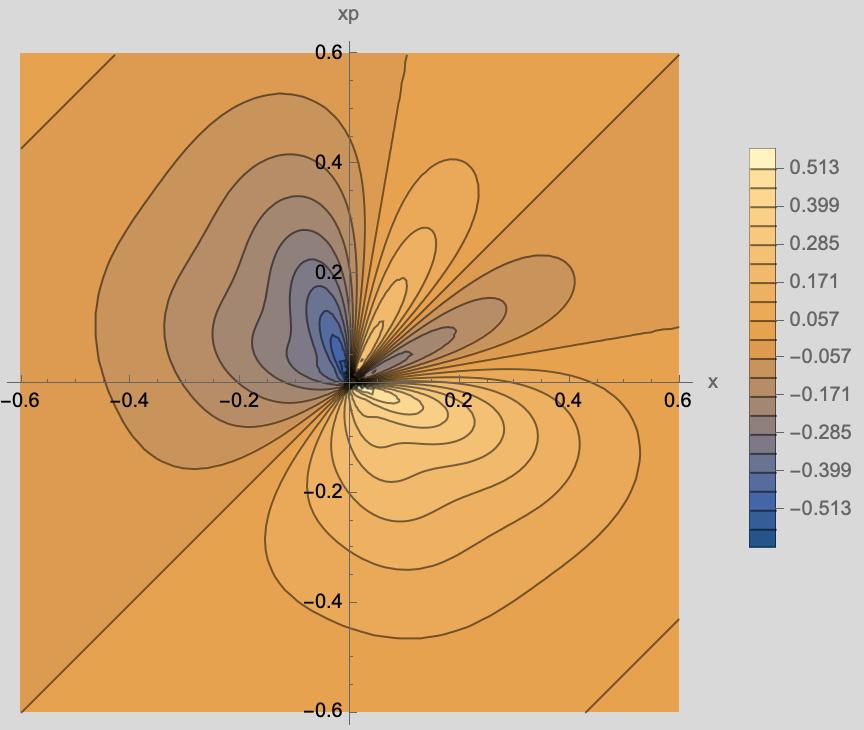}
\includegraphics[width=.49\textwidth]{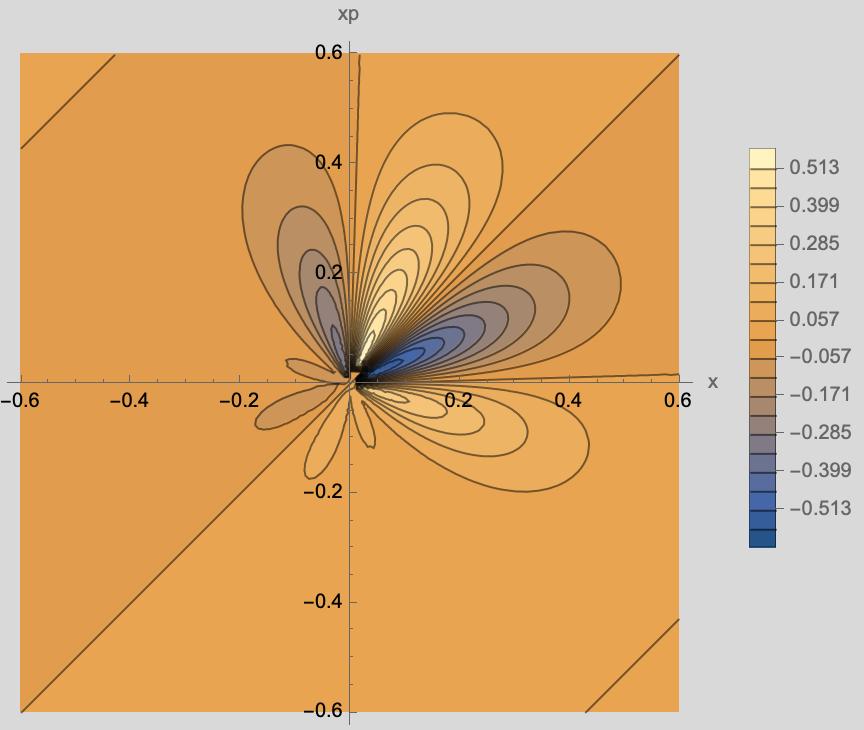}
\caption{Correlation functions $G^u(x,x^\prime)$ \textbf{(left)} and $G^d(x,x^\prime)$ \textbf{(right)} for Scenario 1.\label{fig:Scen1G}}
\end{figure}

In this scenario we populate most of the Fourier coefficients \eqref{eq:FourierModelCoeff} for the $u$- and $d$-quark correlation functions with relatively moderate values that vary between $-1$ and $1$. Doing so, we also generate a nonvanishing correlation function $G$. As before we choose the values of the Fourier coefficients such that the $d_2$-constraint from Lattice QCD \eqref{eq:d2ConstAs} is satisfied. We consider this scenario somewhat more ``realistic'' as far as the sizes of the correlation functions are concerned. 
To be specific, we choose the Fourier coefficients in Scenario 1 as
\bea
\bm{a}^u & = & \left(2.5308,-\tfrac{2}{3},-\tfrac{2}{3};-\tfrac{1}{3},-1,-\tfrac{1}{3}\right)\,,\nn\\
\bm{a}^d & = & \left(-0.3429,\tfrac{2}{3},\tfrac{2}{3};\tfrac{1}{3},1,\tfrac{1}{3}\right)\,,\nn\\
\bm{a}^s & = & \left(0,0,0;0,0,0\right)\,,\nn\\
\bm{b}^u & = & \left(-2.5308,\tfrac{1}{3},\tfrac{2}{3},1,\tfrac{2}{3},\tfrac{1}{3}\right)\,,\nn\\
\bm{b}^d & = & \left(0.3429,-\tfrac{1}{3},-\tfrac{2}{3},-1,-\tfrac{2}{3},-\tfrac{1}{3}\right)\,,\nn\\
\bm{b}^s & = & \left(0,0,0,0,0,0\right)\,.\label{eq:Scenario1}
\eea
The contour plots for the resulting $u$- and $d$-quark correlation functions $F^q(x,x^\prime)$ are shown in Fig.~\ref{fig:Scen1F}.
Figure~\ref{fig:Scen1G} shows the corresponding correlation functions $G^q$ for up and down quarks. Note that the symmetry and antisymmetry properties of $F$ and $G$ under exchange $x\leftrightarrow x^\prime$ are well visible in Figs.~\ref{fig:Scen1F} and \ref{fig:Scen1G}.

\subsection{Scenario 2\label{appsub:Scen2}}

\begin{figure}[tbp]
\centering
\includegraphics[width=0.49\textwidth]{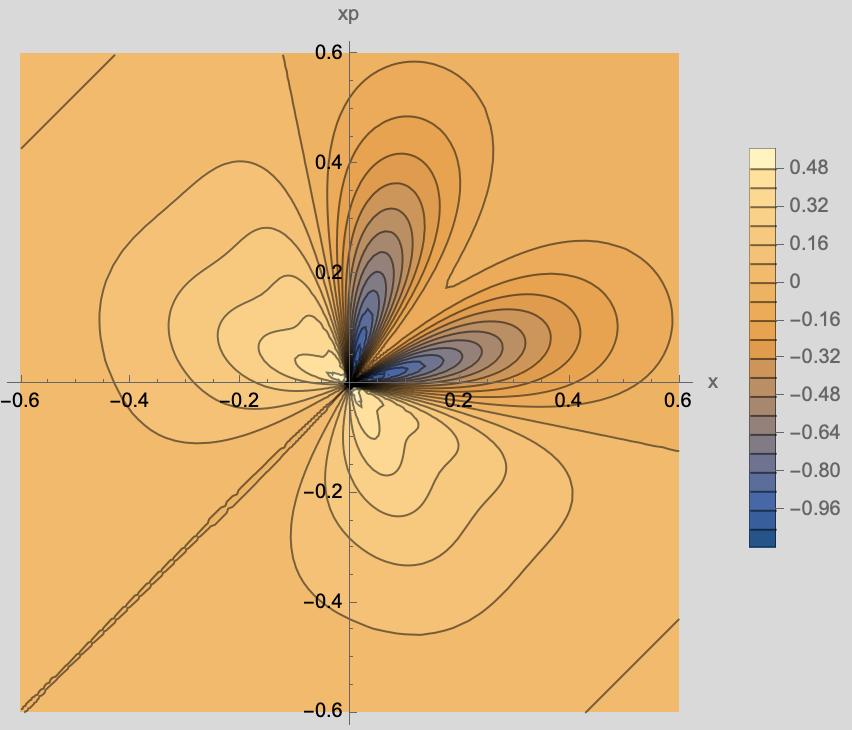}
\includegraphics[width=.49\textwidth]{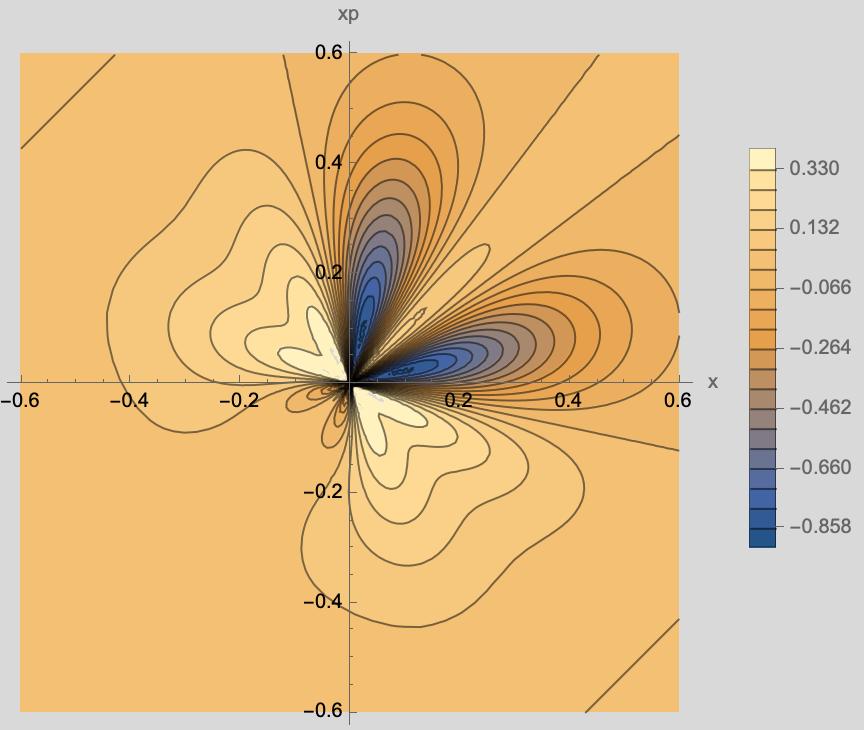}
\caption{Same as Fig.~\ref{fig:Scen0F}, but for Scenario 2.\label{fig:Scen2F}}
\end{figure}

\begin{figure}[tbp]
\centering
\includegraphics[width=0.49\textwidth]{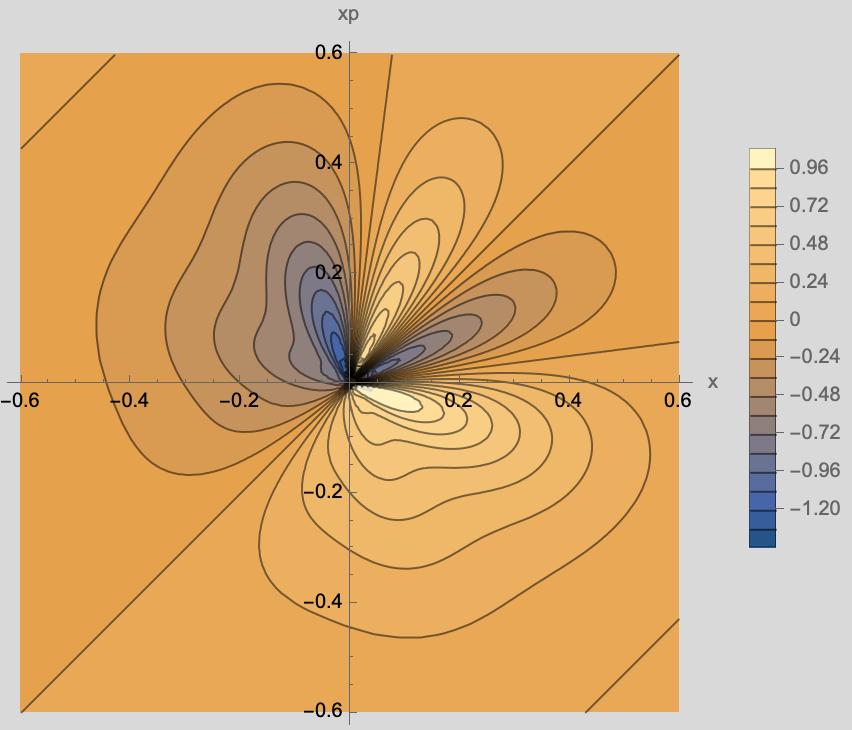}
\includegraphics[width=.49\textwidth]{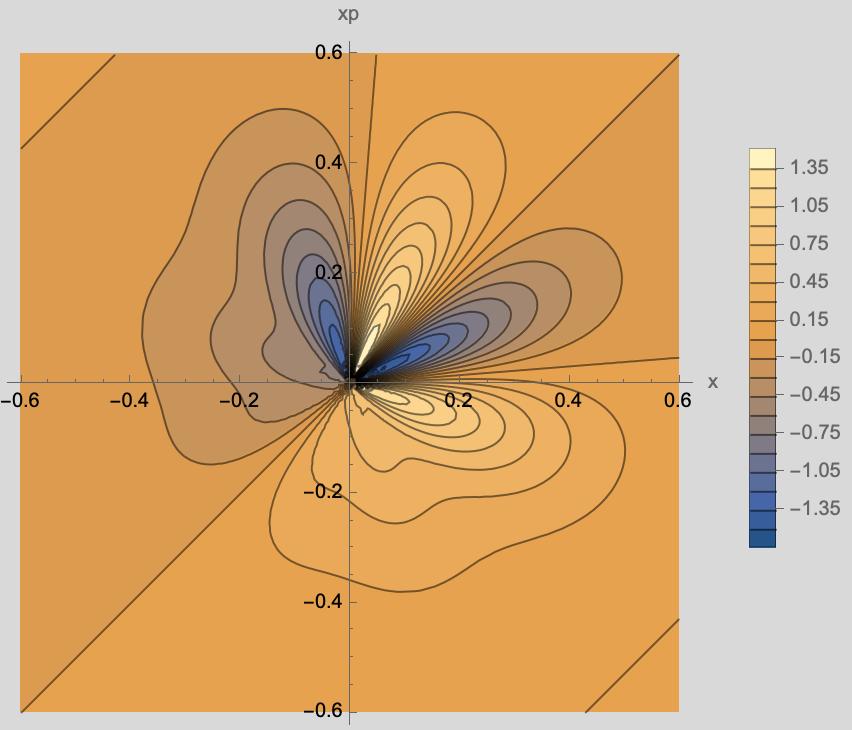}
\caption{Same as Fig.~\ref{fig:Scen1G}, but for Scenario 2.\label{fig:Scen2G}}
\end{figure}

Finally, Scenario 2 is similar to Scenario 1, but with Fourier coefficients inflated by a factor of three in order to magnify the effects of the NLO corrections on 
$A_{RL}$. Specifically, we choose
\bea
\bm{a}^u & = & \left(5.2767,-2,-2;-1,-3,-1\right)\,,\nn\\
\bm{a}^d & = & \left(-3.0634,2,2;1,3,1\right)\,,\nn\\
\bm{a}^s & = & \left(0,0,0;0,0,0\right)\,,\nn\\
\bm{b}^u & = & \left(-5.2767,1,2,3,2,1\right)\,,\nn\\
\bm{b}^d & = & \left(3.0634,-1,-2,-3,-2,-1\right)\,,\nn\\
\bm{b}^s & = & \left(0,0,0,0,0,0\right)\,.\label{eq:Scenario2}
\eea
As for the other two scenarios, the constraint \eqref{eq:d2Def} on $d_2$ by lattice QCD is satisfied also here. The resulting contour plots for the $u$- and $d$-quark correlation functions $F^q$ are shown in Fig.~\ref{fig:Scen2F}, while Figure~\ref{fig:Scen2G} presents the corresponding contour plots for $G$.

\bibliography{Referenzen}

\end{document}